\documentclass[titlepage,oneside,a4paper,11pt]{book}

\usepackage{axodraw}
\input{pix.sty}
\input{psfig.sty}
\usepackage{color}
\usepackage[nottoc]{tocbibind}
\usepackage[margin=10pt,font=small,labelfont=bf]{caption}
\usepackage{microtype,amsmath,amssymb,amsfonts,amscd,setspace,palatino,mathpazo,graphicx,fancyhdr}
\usepackage[multiple,stable]{footmisc}
\usepackage[amsmath,amsthm,thmmarks]{ntheorem}
\allowdisplaybreaks[4]
\usepackage[noconfig]{refstyle}
\usepackage[square,comma,numbers,compress]{natbib}
\usepackage[hyperfootnotes=false,linktocpage=true,linkbordercolor={.1 .1 1},%
         urlbordercolor={.1 .1 1},citebordercolor={.1 .1 1},menubordercolor={.1 .1 1}]{hyperref}
\hypersetup{
	pdftitle={A gauge-invariant reorganization of thermal gauge theory},
	pdfauthor={Nan Su},
	pdfsubject={Theoretical Physics},
	pdfkeywords={Reummation, thermal field theory, QCD, QED, renormalization, hard-thermal-loop, SPT, HTLpt},
	bookmarksnumbered}
\usepackage{geometry}
\let\eqref=\relax
\newref{eq}{name={eq.~},Name={Eq.~},names={eqs.~},Names={Eqs.~},rngtxt={-},refcmd=(\ref{#1})}
\newref{f}{name={footnote~},Name={Footnote~},names={footnotes~},Names={Footnotes~}}
\newref{app}{name={appendix~},Name={Appendix~},names={appendixes~},Names={Appendixes~}}
\newref{tab}{name={table~},Name={Table~},names={tables~},Names={Tables~}}
\newref{ch}{name={chapter~},Name={Chapter~},names={chapters~},Names={Chapters~}}
\newref{sec}{name={section~},Name={Section~},names={sections~},Names={Sections~}}
\newref{fig}{name={figure~},Name={Figure~},names={figures~},Names={Figures~}}
\numberwithin{equation}{section}
\def\square{\vcenter{\vbox{\hrule height.4pt
          \hbox{\vrule width.4pt height8pt
          \kern8pt\vrule width.4pt}\hrule height.4pt}}}

\def\ranglec{\rangle_{\!\!c}}
\def\ranglex{\rangle_{\!\!x}}
\def\ranglecx{\rangle_{\!\!c,x}}

\newcommand{\beq}{\begin{equation}}
\newcommand{\eeq}{\end{equation}}
\newcommand{\bqa}{\begin{eqnarray}}
\newcommand{\eqa}{\end{eqnarray}}

\protect

\def\sumint{\hbox{$\sum$}\!\!\!\!\!\!\,{\int}}

\begin{document}

\setlength{\baselineskip}{16.5pt plus 1pt minus 0.1pt} % suppress under-full vbox warnings

\pagestyle{empty}

\begin{titlepage}

\begin{center}
\huge \sc{A Gauge-Invariant Reorganization of Thermal Gauge Theory}

\end{center}

\vspace{2cm}
\begin{center}
\Large Dissertation \\
zur Erlangung des Doktorgrades \\ 
der Naturwissenschaften \\

\vspace{2cm}
vorgelegt beim Fachbereich Physik \\
der Johann Wolfgang Goethe-Universit{\"a}t \\
in Frankfurt am Main \\

\vspace{2cm}
von \\
Nan Su\\
aus Beijing, China

\vspace{2cm}
Frankfurt am Main, 2010 \\ 
(D 30)
\end{center}

\end{titlepage}

\newpage

\begin{titlepage}

\vspace*{12cm}
\large{\flushleft vom Fachbereich Physik der Johann Wolfgang Goethe-Universit{\"a}t in Frankfurt am Main als Dissertation angenommen. \\

\vspace{2cm}
Dekan: Prof. Dr. Dirk-Hermann Rischke

\vspace{1cm}
Gutachter: Prof. Dr. Horst St{\"o}cker, Prof. Dr. Michael Strickland

\vspace{1cm}
Datum der Disputation: August 16, 2010
}

\end{titlepage}

\cleardoublepage
\thispagestyle{empty}
\begin{titlepage}

\vspace*{5cm}
\begin{center}
\Large{\it For my parents}
\end{center}

\end{titlepage}

\cleardoublepage
\thispagestyle{empty}
\vspace*{\stretch{1}}
\begin{quotation}
``When you are solving a problem, don't worry. Now, after you have solved the problem, then that's the time to worry.''
\begin{flushright}
--- \textsc{Richard Feynman}
\end{flushright}
\end{quotation}
\vspace*{\stretch{3}}

\pagestyle{fancy}

\cleardoublepage
\pagenumbering{roman}
\setcounter{page}{1}

%%%%%%%%%%%%%%%%%%%%%%%%%%%%%%%%%%%%%%%%%%%%%%%%%%%%%%%%%%%%%
%
%	Include File:			DON'T COMPILE !!!
%
%%%%%%%%%%%%%%%%%%%%%%%%%%%%%%%%%%%%%%%%%%%%%%%%%%%%%%%%%%%%%

\chapter*{Abstract}

This dissertation is devoted to the study of thermodynamics for quantum gauge theories. The poor convergence of quantum field theory at finite temperature has been the main obstacle in the practical applications of thermal QCD for decades. In this dissertation I apply hard-thermal-loop perturbation theory, which is a gauge-invariant reorganization of the conventional perturbative expansion for quantum gauge theories to the thermodynamics of QED and Yang-Mills theory to three-loop order. For the Abelian case, I present a calculation of the free energy of a hot gas of electrons and photons by expanding in a power series in $m_D/T$, $m_f/T$ and $e^2$, where $m_D$ and $m_f$ are the photon and electron thermal masses, respectively, and $e$ is the coupling constant. I demonstrate that the hard-thermal-loop perturbation reorganization improves the convergence of the successive approximations to the QED free energy at large coupling, $e \sim 2$. For the non-Abelian case, I present a calculation of the free energy of a hot gas of gluons by expanding in a power series in $m_D/T$ and $g^2$, where $m_D$ is the gluon thermal mass and $g$ is the coupling constant. I show that at three-loop order hard-thermal-loop perturbation theory is compatible with lattice results for the pressure, energy density, and entropy down to temperatures $T \sim 2-3\;T_c$. The results suggest that HTLpt provides a systematic framework that can be used to calculate static and dynamic quantities for temperatures relevant at LHC.

\cleardoublepage

\fancyhead[RE,LO]{Acknowledgements}
%%%%%%%%%%%%%%%%%%%%%%%%%%%%%%%%%%%%%%%%%%%%%%%%%%%%%%%%%%%%%
%
%	Include File:			DON'T COMPILE !!!
%
%%%%%%%%%%%%%%%%%%%%%%%%%%%%%%%%%%%%%%%%%%%%%%%%%%%%%%%%%%%%%

\chapter*{Acknowledgements}

First of all, I would like to express my deep gratitude to my supervisor Horst St\"ocker. It is his innumerous encouragement and support made my doctoral study joyful and creative. I still clearly remember when first met him in China, he motivated me to talk as much as I can to as many people as possible, do not afraid of making mistakes, because no mistakes no progress. His wisdom keeps shining inspirations on me!

I am greatly grateful to my external advisor Michael Strickland who introduced me to this fascinating and exciting topic which opened my eyes and sharpened my teeth. This dissertation would never have been completed without his insightful guidance and countless patience on my naiveness and ignorance. Gratitude also goes to my collaborator Jens Andersen who always has tried to give me a hand whenever needed. Without his commitment and hospitality, this project could never have been as smooth as it is. The collaboration with Mike and Jens is such an enjoyment!

I would like to acknowledge Dani\"el Boer without whose encouragement I might have never been able to continue a doctoral study. I am very thankful to Qun Wang who supported me during my hard time which finally led to a successful collaboration. I am indebted to Qun's effort which made it possible for me to continue physics in Frankfurt and hospitality whenever I visited USTC. I thank my undergraduate supervisor Hai-Yang Yang for his continuous moral support over the past few years.

I appreciate the fruitful and enlightening interactions with Marcus Bleicher, Eric Braaten, Tomas Brauner, Adrian Dumitru, Jean-Sebastien Gagnon, Carsten Greiner, Miklos Gyulassy, Mei Huang, Jiang-Yong Jia, Yu Jia, Mikko Laine, Axel Maas, Jorge Noronha, Robert Pisarski, Jian-Wei Qiu, Dirk Rischke, Chihiro Sasaki, York Schr\"oder, Lorenz von Smekal, Harmen Warringa, Zhe Xu.

The time at FIAS and ITP could not be more lively and lovely thanks to my dear colleagues: Maximilian Attems, Wei-Tian Deng, Veronica Dexheimer, Qing-Guo Feng, Michael Hauer, Lian-Yi He, Xu-Guang Huang, Benjamin Koch, Michaela Koller, Fritz Kretzschmar, Qing-Feng Li, Hossein Malekzadeh, Mauricio Martinez, Sophie Nahrwold, Jaki Noronha-Hostler, Basil Sa'd, Bj\"orn Schenke, Mehmet Suzen, Laura Tolos, Giorgio Torrieri, Tian Zhang, Yu-Zhong Zhang, Jun-Mei Zhu.

I would like to thank Gabriela Meyer for all her kindness and help in the institute. Thanks also to Alexander Achenbach, Walburga Bergmann, Claudia Gressler, Michael Lehmann, Eike Sch\"adel for administrative and computing supports. The financial supports from FIGSS and HGS-HIRe are gratefully acknowledged.

Special thanks to Lars Leganger for help with typesettings.

Last but not least, I am indebted to the countless supports from my parents and all members in the family over years.

\cleardoublepage

\pagestyle{plain}

\tableofcontents

\cleardoublepage

%%%%%%%%%%%%%%%%%%%%%%%%%%%%%%%%%%%%%%%%%%%%%%%%%%%%%%%%%%%%%%%%%%%%%%
\pagestyle{fancy}
\fancyhead{}
\fancyfoot{}
\lhead{\leftmark}
\cfoot{\thepage}

\pagenumbering{arabic}
\setcounter{page}{1}

%%%%%%%%%%%%%%%%%%%%%%%%%%%%%%%%%%%%%%%%%%%%%%%%%%%%%%%%%%%%%
%
%	Include File:			DON'T COMPILE !!!
%
%%%%%%%%%%%%%%%%%%%%%%%%%%%%%%%%%%%%%%%%%%%%%%%%%%%%%%%%%%%%%

\chapter{Introduction}\label{chapter:introduction}

The beginning of experiments at the Relativistic Heavy-Ion Collider (RHIC) at Brookhaven National Laboratory (BNL) in 1999 marked the beginning of a new era in ultrarelativistic heavy-ion collisions. One of the primary goals of the RHIC program is to discover and study the quark-gluon plasma (QGP) whose existence is predicted by quantum chromodynamics (QCD). In addition, looking forward, ultrarelativistic heavy-ion collision experiments are part of the Large Hadron Collider (LHC) program at European Organization for Nuclear Research (CERN). The LHC experiments, for which full beam runs are scheduled in 2011, will provide data on heavy ion collisions at center of mass energies of 5.5 TeV per nucleon pair collision and will open a new chapter in the study of partonic matter under extreme conditions. 

For the RHIC and LHC experiments to have the greatest possible impact on science, it is essential to make as close a connection to the fundamental theory of QCD as possible. There is an urgent need for theoretical analysis that is based rigorously on QCD but which can also make contact with more phenomenological approaches, particularly in the area of equilibrium and non-equilibrium dynamics of QCD at \emph{intermediate coupling}, $g\sim2$, or equivalently $\alpha_s=g^2/(4\pi)\sim0.3$. 

We have to be extremely careful when dealing with this intermediately coupled region. Naively, $g\sim2$ seems to suggest the breakdown of perturbation theory in this region. This is also in line with the observations from the early RHIC data that the state of matter created there behaved more like a strongly coupled fluid than a weakly coupled plasma~\cite{rhicexperiment}. As a result, the term ``quark-gluon plasma'' might need to be modified to ``quark-gluon liquid'', and a description in terms of hydrodynamics or AdS/CFT correspondence might be more appropriate. However on the other hand, $g\sim2$ is not huge especially when considering that $\alpha_s=g^2/(4\pi)\sim0.3$ is still a small number. So people have not yet totally lost faith in perturbation theory and as a payback observables like jet quenching~\cite{pert} and elliptic flow~\cite{Xu:2007jv} have been able to be described using a perturbative formalism. Therefore it seems that a complete understanding of QGP would require knowledge from both strong-coupling and weak-coupling formalisms, and in this dissertation I focus on the latter approach.

Thermodynamics describes the bulk properties of matter in or near equilibrium which are theoretically clean and well defined. The calculation of thermodynamic functions for finite temperature field theories has a long history. In the early 1990s the free energy was calculated to order $g^4$ for massless scalar $\phi^4$ theory \cite{Frenkel:1992az,AZ-95}, quantum electrodynamics (QED) \cite{Parwani:1994xi,AZ-95} and QCD \cite{AZ-95}, respectively. The corresponding calculations to order $g^5$ were obtained soon afterwards \cite{Parwani:1994zz, Braaten:1995cm,Parwani:1994je,Andersen:1995ej,Zhai:1995ac,BN-96}. Recent results have extended the calculation of the QCD free energy by determining the coefficient of the $g\log g$ contribution \cite{Kajantie:2002wa}. For massless scalar theories the perturbative free energy is now known to order $g^6$ \cite{Gynther:2007bw} and $g^8 \log g$ \cite{Andersen:2009ct}.

Unfortunately, for all the above-mentioned theories the resulting weak-coupling approximations, truncated order-by-order in the coupling constant, are poorly convergent unless the coupling constant is tiny. Therefore a straightforward perturbative expansion in powers of $\alpha_s$ for QCD does not seem to be of any quantitative use even at temperatures many orders of magnitude higher than those achievable in heavy-ion collisions. 

The poor convergence of finite-temperature perturbative expansions of thermodynamic functions stems from the fact that at high temperature the classical solution is not described by massless gluonic states. Instead one must include plasma effects such as the screening of electric fields and Landau damping via a self-consistent hard-thermal-loop (HTL) resummation~\cite{Braaten:1989mz}. The inclusion of plasma effects can be achieved by reorganizing perturbation theory.

There are several ways of systematically reorganizing the finite-temperature perturbative expansion~\cite{Blaizot:2003tw,Kraemmer:2003gd,Andersen:2004fp}. In this dissertation I will focus on the hard-thermal-loop perturbation theory (HTLpt) method~\cite{htl1,htl2,Andersen:2009tw,Andersen:2009tc,Andersen:2010ct}. The HTLpt method is inspired by variational perturbation theory~\cite{opt,lde,vpt}. HTLpt is a gauge-invariant extension of screened perturbation theory (SPT)~\cite{K-P-P-97,CK-98,Andersen:2000yj,Andersen:2001ez,Andersen:2008bz}, which is a perturbative reorganization for finite-temperature massless scalar field theory. In the SPT approach, one introduces a single variational parameter which has a simple interpretation as a thermal mass. In SPT a mass term is added to and subtracted from the scalar Lagrangian, with the added piece kept as part of the free Lagrangian and the subtracted piece associated with the interactions. The mass parameter is then required to satisfy a variational equation which is obtained by a principle of minimal sensitivity. This naturally led to the idea that one could apply a similar technique to gauge theories by adding and subtracting a mass in the Lagrangian. However, in gauge theories, one cannot simply add and subtract a local mass term since this would violate gauge invariance. Instead, one adds and subtracts an HTL improvement term which modifies the propagators and vertices self-consistently so that the reorganization is manifestly gauge invariant~\cite{Braaten:1991gm}.

This dissertation focuses on the study of thermodynamics for gauge theories. In the rest of this chapter, a brief introduction to statistical physics and thermal QCD is provided. In Chapter~\ref{chapter:resum}, we show the emergence of infrared divergences in thermal field theory and how the weak-coupling expansion treats them systematically. HTLpt is introduced in Chapter~\ref{chapter:htl}, where we discuss its formalism as well as techniques that make HTLpt calculations tractable. Chapters~\ref{chapter:qed} and~\ref{chapter:ym} are devoted to the study of thermodynamics to three-loop order using HTLpt for QED and Yang-Mills theory, respectively. We summarize in Chapter~\ref{chapter:sum} together with a brief outlook for the real-time application of HTLpt.

\section{Statistical physics and quantum partition function}

For a relativistic system which can freely exchange energy and particles with its surroundings, the most important function in thermodynamics is the grand canonical partition function
\bqa
Z \;=\; \sum_{\rm states} e^{-{\cal E}_i/T}
    \;=\; \sum_{\rm states} \langle {\cal E}_i | e^{-H/T} | {\cal E}_i \rangle
    \;=\; {\rm Tr}\,e^{-{\cal H}/T}\;.
\label{Z}
\eqa
Here ${\cal E}_i$ is the energy of the state $| {\cal E}_i \rangle$ and ${\cal H}$ is the Hamiltonian of the system. The temperature of the system is denoted by $T$, and since this dissertation only concerns with high temperature physics, the chemical potential of the particles in the system is set to zero for simplicity. All of the thermodynamic properties can be determined from~(\ref{Z}). For example, the pressure, entropy and energy are given by
\bqa
{\cal P} \!\!&=&\!\! {\partial (T \log Z) \over \partial V} \;, \\
{\cal S} \!\!&=&\!\! {\partial (T \log Z) \over \partial T} \;, \\
{\cal E} \!\!&=&\!\! - {\cal P}V + T{\cal S}\;,
\eqa 
where $V$ is the volume of the system. Typically, the width $L$ of a system is much larger than the inverse temperature, (i.e. $L\gg2\pi/T$), such that one can use the infinite volume limit to describe the thermodynamics of a finite volume to good approximation. The advantage of the infinite volume limit is that field theoretic calculations simplify. In all calculations performed in this thesis, this infinite volume limit is taken. Then it turns out that $\log Z$ becomes proportional to $V$, such that the pressure becomes
\bqa
{\cal P} \;=\; {T \log Z \over V} \;.
\eqa

The extension to field theory is straightforward. If ${\cal H}$ is the Hamiltonian of a quantum field theory in $d$-dimensional space and hence $(d+1)$-dimensional spacetime, then the partition function~(\ref{Z}) is
\bqa
Z \;=\; {\rm Tr}\,e^{-{\cal H}/T} \;=\; \int {\cal D}\varphi\,e^{-\int_0^{1/T}\!\!d\tau\int d^dx\,{\cal L}(\varphi)}\;,
\label{Z-QFT}
\eqa
with ${\cal L}$ the Lagrangian density of the theory and \emph{periodic} boundary conditions 
\bqa
\varphi(0,{\bf x}) \;=\; \varphi(1/T,{\bf x})\;.
\eqa
for bosonic fields $\varphi$. For fermionic fields, it turns out that to implement Pauli statistics one must impose \emph{anti-periodic} boundary conditions
\bqa
\varphi(0,{\bf x}) \;=\; - \varphi(1/T,{\bf x})\;.
\eqa

\section{QCD at finite temperature}

Quantum chromodynamics is a gauge theory for the strong interaction describing the interactions between quarks and gluons. The QCD Lagrangian density in Minkowski space reads
\bqa
{\cal L}_{\rm QCD}\;=\;
-{1\over2}{\rm Tr}\left[G_{\mu\nu}G^{\mu\nu}\right]
+\sum_i\bar{\psi}_i \left[i\gamma^\mu D_\mu - m_i\right] \psi_i
+{\cal L}_{\rm gf}
+{\cal L}_{\rm ghost} \;.
\label{L}
\eqa
The gluon field strength is $G_{\mu\nu}= \partial_{\mu}A_{\nu}-\partial_{\nu}A_{\mu} -ig[A_{\mu},A_{\nu}]$. The gluon field is $A_\mu = A_\mu^a t^a$, with generators $t^a$ of the fundamental representation of SU(3) normalized so that ${\rm Tr}\,t^a t^b=\delta^{ab}/2$. In the quark sector there is an explicit sum over the $N_f$ quark flavors with masses $m_i$ and $D_\mu = \partial_\mu - i g A_\mu$ is the covariant derivative in the fundamental representation. The Lagrangian~(\ref{L}) is mathematically simple and beautiful, however in order to carry out a physical calculation with it, a gauge fixing is needed to remove unphysical degrees of freedom. The ghost term ${\cal L}_{\rm ghost}$ depends on the choice of the gauge-fixing term ${\cal L}_{\rm gf}$. One popular choice for the gauge-fixing term that depends on an arbitrary gauge parameter $\xi$ is the general covariant gauge:
\bqa
{\cal L}_{\rm gf}\!\!&=&\!\!
-{1\over\xi}{\rm Tr}\left[\left(\partial^{\mu}A_{\mu}\right)^2\right] \;.
\eqa
The corresponding ghost term in the general covariant gauge reads
\bqa
{\cal L}_{\rm ghost} \;=\; - \bar\eta^a \partial^2 \eta^a + g f^{abc} \bar\eta^a \partial^\mu(A_\mu^b \eta^c)\;,	
\eqa
where $\eta$ and $\bar\eta$ are anti-commuting ghosts and anti-ghosts respectively and $f^{abc}$ is structure constant of SU(3).

The finite temperature QCD partition function is obtained by a Wick rotation of the theory from Minkowski space to Euclidean space. It is achieved by the substitution $t=i\tau$ with $t$ being the Minkowski time and $\tau$ being the Euclidean one. The resulting Euclidean partition function is
\bqa
Z \;=\; \int {\cal D}A_\mu {\cal D}\bar{\psi} {\cal D}\psi {\cal D}\bar{\eta} {\cal D}\eta \exp\left[-\int_0^{1/T}\!\!d\tau\int d^3x\,{\cal L}_{\rm QCD}^E\right]\;,
\eqa
with ${\cal L}_{\rm QCD}^E$ the Wick-rotated Lagrangian denity. Feynman rules are exactly the same as in zero-temperature field theory except that the imaginary time $\tau$ is now compact with extent $1/T$. To go from $\tau$ to frequency space, we should perform a Fourier series decomposition rather than a Fourier transform. The only difference with zero-temperature Feynman rules will then be that loop frequency integrals are replaced by loop frequency sums:
\begin {equation}
  \int \frac{d^4p}{(2\pi)^4} \; \rightarrow \; T \sum_\omega \int \frac{d^3p}{(2\pi)^3}
\end {equation}
with the sum over discrete imaginary-time frequencies known as Matsubara frequencies
\bqa
  \omega_n \!\!& = &\!\! 2 n \pi T \hspace{2cm}\mbox{bosons} \;, \\
  \omega_n \!\!& = &\!\! (2n+1) \pi T \hspace{1cm}\mbox{fermions} \;.
\eqa
to implement the periodic or anti-periodic boundary conditions. A detailed explanation of the imaginary-time formalism is given in Appendix~\ref{app:ITF}.

\section{Beta function and asymptotic freedom}

The beta function $\beta(g)$ of a quantum field theory encodes the dependence of a coupling parameter $g$ on the energy scale $\mu$ of a given physical process. It is defined by the relation:
\bqa
\beta(g) \;=\; \mu{\partial g \over \partial \mu} \;.
\label{beta}
\eqa
This dependence on the energy scale is known as the running of the coupling parameter, and theory of this kind of scale-dependence in quantum field theory is described by the renormalization group which refers to a mathematical apparatus that allows one to investigate the changes of a physical system as one views it at different distance scales.

To lowest order in the coupling constant a beta function is either positive indicating the growth of charge at short distance or negative indicating the decrease of charge at short distance. Until 1973, only examples of the former were known~\footnote{'t Hooft reported a similar discovery at the Marseille conference on renormalization of Yang-Mills fields and applications to particle physics in 1972 without publishing it~\cite{'t Hooft}.}. The discovery that only non-Abelian gauge theories allow for a negative beta function is usually credited to Gross and Wilczek~\cite{GW73}, and to Politzer~\cite{Politzer73}. The solution to~({\ref{beta}}) for QCD reads
\bqa
\alpha_s(\mu) \;=\; {g(\mu)^2 \over 4\pi} \;=\; {2\pi \over \left(11 - {2 \over 3}N_f\right) \log\left(\mu / \Lambda_{\rm QCD}\right)}\;,
\eqa
which clearly shows \emph{asymptotic freedom}~\cite{GW73,Politzer73}, i.e. $\alpha \rightarrow 0$ as $\mu \rightarrow \infty$. The parameter $\Lambda_{\rm QCD}$ is a scale above which the theory works ``chosen'' by the  world in which we live. It is well known that QCD exhibits confinement at large distances or low energies which terminates the validity of perturbation theory due to the infrared growth of the coupling. However it is precisely the asymptotic freedom that ensures the possibility of a perturbative treatment for the ultraviolet sector of the theory which sets the stage to study the high temperature phase of non-Abelian theory in this dissertation. 
%%%%%%%%%%%%%%%%%%%%%%%%%%%%%%%%%%%%%%%%%%%%%%%%%%%%%%%%%%%%%
%
%	Include File:			DON'T COMPILE !!!
%
%%%%%%%%%%%%%%%%%%%%%%%%%%%%%%%%%%%%%%%%%%%%%%%%%%%%%%%%%%%%%

\chapter{The Need for Resummation}
\label{chapter:resum}

In this chapter, and in the rest of the dissertation, we consider thermal field theories at high temperatures, which means temperatures much higher than all zero-temperature masses or any mass scales generated at zero temperature. 

It has been known for many years that naive perturbation theory, or the loop expansion breaks down at high temperature due to infrared divergences. Diagrams which are nominally of higher order in the coupling constant contribute to leading order in $g$. A consistent perturbative expansion requires the resummation of an infinite subset of diagrams from all orders of perturbation theory. We discuss these issues next.

\section{Scalar field theory}

We start our discussion by considering the simplest interacting thermal field theory, namely that of a single massless scalar field with a $\phi^4$ interaction. The Euclidean Lagrangian is
\bqa
{\mathcal L}\;=\;{1\over2}(\partial_{\mu}\phi)^2+{g^2\over24}\phi^4\;.
\label{sl}
\eqa
Perturbative calculations at zero temperature proceed by dividing the Lagrangian into a free part and an interacting part according to
\bqa
{\mathcal L}_{\rm free}\!\!&=&\!\!{1\over2}(\partial_{\mu}\phi)^2\;,\\
{\mathcal L}_{\rm int}\!\!&=&\!\!{g^2\over24}\phi^4\;.
\eqa
Radiative corrections are then calculated in a loop expansion which is equivalent to a power series in $g^2$. We shall see that the perturbative expansion breaks down at finite temperature and the weak-coupling expansion becomes an expansion in $g$ rather than $g^2$.

\begin{figure}[htb]
\begin{center}
%$\mbox{1-loop}:\hspace{5mm} \oneloopse \hspace{1cm}\mbox{2-loop}:\hspace{5mm} \twoloopsea \;\;\; \twoloopseb$
\includegraphics{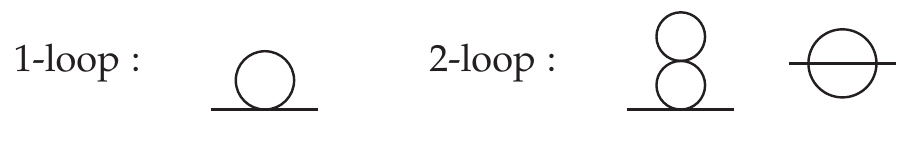}
\end{center}
\caption[a]{One- and two-loop scalar self-energy graphs.}
\label{thmass}
\end{figure}

We will first calculate the self-energy by evaluating the relevant diagrams. The Feynman diagrams that contribute to the self-energy up to two loops are shown in Fig.~\ref{thmass}.

The one-loop diagram is independent of the external momentum and the resulting integral expression is
\bqa\nonumber
\Pi^{(1)}\!\!&=&\!\!{1\over2}g^2\sumint_P{1\over P^2} \; , \\ \nonumber
\!\!&=&\!\!{g^2\over24}T^2 \; , \\
\!\!&\equiv&\!\!m^2 \;,
\label{fm}
\eqa
where the superscript indicates the number of loops. The notation $P=(P_0,{\bf p})$ represents the Euclidean four-momentum. The Euclidean energy $P_0$ has discrete values: $P_0=2n\pi T$ for bosons and $P_0=(2n+1)\pi T$ for fermions, where $n$ is an integer. Eq.~(\ref{fm}) represents the leading order thermal mass of our scalar field. The sum-integral $\Sigma\!\!\!\!{\int}_{P}$, which is defined in Eq.~(\ref{sumint-def}), represents a summation over Matsubara frequencies and integration of spatial momenta in $d=3-2\epsilon$ dimensions~\footnote{For an introduction to thermal field theory and the imaginary time formalism see Refs.~\cite{kap} and \cite{bellac}.}. The above sum-integral has ultraviolet power divergences that are set to zero in dimensional regularization. We are then left with the finite result~(\ref{fm}), which shows that thermal fluctuations generate a mass for the scalar field of order $gT$. This thermal mass is analogous to the Debye mass which is well-known from the nonrelativistic QED plasma.

\begin{figure}[htb]
\begin{center}
%\begin{tabular}{ccccccccccc}
%$\oneloopseb = $ &
%$\oneloopse$ & + &
%$\twoloopsea$ & + &
%$\threeloopsea$ & + \hspace{5mm} $\cdots$
%\\
%    &
%{\scriptsize (a)} &&
%{\scriptsize(b)} &&
%{\scriptsize(c)} &&
%\end{tabular}
\includegraphics{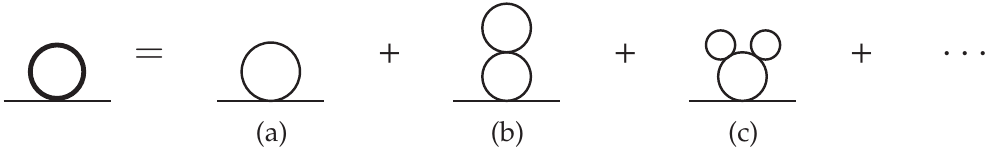}
\caption[a]{Bubble diagrams contributing to the scalar self-energy.}
\label{2lself}
\end{center}
\end{figure}

We next focus on the two-loop diagrams and first consider the double-bubble in Fig.~\ref{2lself} (b). 

\noindent
This diagram is also independent of the external momentum and gives the following sum-integral
\bqa
\Pi^{(2b)}\;=\;-{1\over4}g^4\sumint_{PQ}{1\over P^2}{1\over Q^4} \;.
\eqa
This integral is infrared divergent. The problem stems from the middle loop with two propagators. In order to isolate the source of the divergence, we look at the contribution from the zeroth Matsubara mode to the $Q$ integration
\bqa
-{1\over4}g^4\sumint_P{1\over P^2}T\int_{\bf q}{1\over q^4} \;,
\label{g3}
\eqa 
with $\int_{\bf q}$ defined in~Eq.~(\ref{3d-int-def}). The integral $\int_{\bf q} 1/ q^4$ behaves like $1/q$, as a result Eq.~(\ref{g3}) is linearly infrared divergent as $q \rightarrow 0$. This infrared divergence indicates that naive perturbation theory breaks down at finite temperature. However, in practice this infrared divergence is screened by a thermally generated mass and we must somehow take this into account. The thermal mass can be incorporated by using an effective propagator:
\bqa
\Delta(P)\;=\;{1\over P^2+m^2}\;,
\label{impprop}
\eqa
with $m\sim gT \ll T$.

\begin{figure}[htb]
\begin{center}
% $\threeloopseb$
\includegraphics{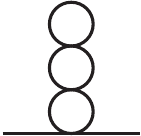}
\end{center}
\caption[a]{ Subleading three-loop self-energy diagram.}
\label{fig:3l}
\end{figure}

If the momenta of the propagator is of order $T$ or {\it hard}, clearly the thermal mass is a perturbation and can be omitted. However, if the momenta of the propagator is of order $gT$ or {\it soft}, the thermal mass is as large as the bare inverse propagator and cannot be omitted. The mass term in the propagator~(\ref{impprop}) provides an infrared cutoff of order $gT$. The contribution from~(\ref{g3}) would then be
\bqa
-{1\over4}g^4\sumint_P{1\over P^2}T\int_{\bf q}{1\over(q^2+m^2)^2}
\;=\;-{1\over4}g^4\left({T^2\over 12}\right)\left({T\over8\pi m}\right)
+{\mathcal O}\left(g^4mT\right)\;.
\eqa
Since $m\sim gT$, this shows that the double-bubble contributes at order $g^3T^2$ to the self-energy and not at order $g^4T^2$ as one might have expected. Similarly, one can show that the diagrams with any number of bubbles like Fig.~\ref{2lself}c are all of order $g^3$. Clearly, naive perturbation theory breaks down since the order-$g^3$ correction to the thermal mass receives contributions from all loop orders. On the other hand, the three-loop diagram shown in Fig.~\ref{fig:3l}, is of order $g^4T^2$ and thus subleading. Therefore, we only need to resum a subset of all possible Feynman graphs in order to obtain a consistent expansion in $g$.

\vspace{6mm}

If we use the effective propagator to recalculate the one-loop self-energy, we obtain
\bqa\nonumber
\Pi^{(1)}(P)\!\!&=&\!\!{1\over2}g^2\sumint_P{1\over P^2+m^2} \\ \nonumber
\!\!&=&\!\!{1\over2}g^2\left[T\int_{\bf p}{1\over p^2+m^2}+\sumint_P^{\prime}{1\over P^2}
+{\mathcal O}\left(m^2\right)\right] \\
\!\!&=&\!\!{g^2\over24}T^2\left[1-{g\sqrt{6}\over4\pi}+{\mathcal O}\left(g^2\right)\right]\;.
\eqa
where here, and in the following, the prime on the sum-integral indicates that we have excluded the $n=0$ mode from the sum over the Matsubara frequencies. The order $g^3$ corresponds to the summation of the bubble diagrams in Fig.~\ref{2lself}, which can be verified by expanding the effective propagator~(\ref{impprop}) around $m=0$. Thus by taking the thermal mass into account, one is resumming an infinite set of diagrams from all orders of perturbation theory.

The self-energy~(\ref{fm}) is the first example of a {\it hard thermal loop} (HTL). Hard thermal loops are loop corrections which are $g^2T^2/P^2$ times the corresponding tree-level amplitude, where $P$ is a momentum that characterizes the external lines. From this definition, we see that, whenever $P$ is hard, the loop correction is suppressed by $g^2$ and is thus a perturbative correction. However, for soft $P$, the hard thermal loop is ${\mathcal O}(1)$ and is therefore as important as the tree-level contribution to the amplitude. These loop corrections are called ``hard'' because the relevant integrals are dominated by momenta of order $T$.  Also note that the hard thermal loop in the two-point function is finite since it is exclusively due to thermal fluctuations. Quantum fluctuations do not enter. Both properties are shared by all hard thermal loops.

What about higher-order $n$-point functions in scalar thermal field theory? One can show that within scalar theory the one-loop correction to the four-point function for high temperature behaves as~\cite{Pisarski:1990ds}
\bqa
\Gamma^{(4)}\;\propto\;g^4\log\left(T/p\right)\;,
\eqa
where $p$ is the external momentum. Thus the loop correction to the four-point function increases logarithmically with temperature. It is therefore always down by $g^2\log(1/g)$, and it suffices to use a bare vertex. More generally, it can be shown that the only hard thermal loop in scalar field theory is the tadpole diagram in Fig.~\ref{thmass} and resummation is taken care of by including the thermal mass in the propagator. In gauge theories, the situation is much more complicated as we shall see in the next section.

\section{Gauge theories}
\label{gres}

In the previous section, we demonstrated the need for resummation in a hot scalar theory. For scalar theories, resummation simply amounts to including the thermal mass in the propagator and since the running coupling depends logarithmically on the temperature, corrections to the bare vertex are always down by powers of $g^2\log{1/g}$. In gauge theories, the situation is more complicated. The equivalent HTL self-energies are no longer local, but depend in a nontrivial way on the external momentum. In addition, it is also necessary to use effective vertices that also depend on the external momentum. It turns out that all hard thermal loops are gauge-fixing independent~\cite{Braaten:1989mz,Frenkel:1989br,Braaten:1989kk,Braaten:1990az,Taylor:1990ia,Kobes:1990dc}. This was shown explicitly in covariant gauges, Coulomb gauges, and axial gauges. They also satisfy tree-level like Ward identities. Furthermore, there exists a gauge invariant effective Lagrangian, found independently by Braaten and Pisarski \cite{Braaten:1991gm} and by Taylor and Wong \cite{Taylor:1990ia}, that generates all of the hard thermal loop $n$-point functions. From a renormalization group point of view this is an effective Lagrangian for the soft scale $gT$ that is obtained by integrating out the hard scale $T$. We return to the HTL Lagrangian in Chapter~\ref{chapter:htl}.

\subsection{Polarization tensor}

\begin{figure}[t]
\begin{center}
\vspace{6mm}
% $\oneloopQEDse$
\includegraphics{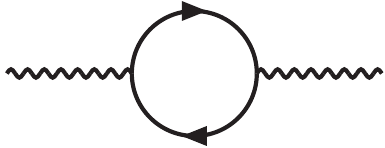}
\end{center}
\caption[a]{
One-loop photon self-energy diagram.
}
\label{qed}
\end{figure}

We next discuss in some detail the hard thermal loop for the vacuum polarization tensor $\Pi^{\mu\nu}$. For simplicity, we focus on QED here. The Feynman diagram for the one-loop photon self-energy is shown in Fig.~\ref{qed} and results in the following sum-integral
\bqa
\Pi^{\mu\nu}(P)\;=\;e^2\,\sumint_{\{K\}}{\rm Tr}\left[{K\!\!\!\!/\gamma^{\mu}(K\!\!\!\!/-P\!\!\!\!/)\gamma^{\nu}\over K^2(K-P)^2}\right]\;,
\eqa
where ${\rm Tr}$ denotes the trace over Dirac indices. After taking the trace, the self-energy becomes
\bqa\nonumber
\Pi^{\mu\nu}(P)\!\!&=&\!\!8e^2\sumint_{\{K\}}{K^{\mu}K^{\nu}\over K^2(K-P)^2}-4\delta^{\mu\nu}e^2\sumint_{\{K\}}{1\over K^2}
\\ &&
+\;2\delta^{\mu\nu}P^2e^2\sumint_{\{K\}}{1\over K^2(K-P)^2}-4e^2\sumint_{\{K\}}{K^{\mu}P^{\nu}+K^{\nu}P^{\mu}\over K^2(K-P)^2} \;,
\eqa
where we have assumed, for now, that $d=3$. Since we are interested in the high-temperature limit, we may assume that $K \gg P$ because the leading contribution in $T$ to the loop integral is given by the region $K \sim T$. With this assumption, the self-energy simplifies to
\bqa
\Pi^{\mu\nu}(P)\;=\;8e^2\sumint_{\{K\}}{K^{\mu}K^{\nu}\over K^2(K-P)^2}-4\delta^{\mu\nu}e^2\sumint_{\{K\}}{1\over K^2}\;.
\label{htl-self}
\eqa
%%%%%%%%%%%%%%%%%%%%%%%%%%%%%%%%%%%%%%%%
%%%%%%%%%%%%%%%%%%%%%%%%%%%%%%%%%%%%%%%%
We first consider the spatial components of $\Pi^{\mu\nu}(P)$. The sum over Matsubara frequencies can be evaluated using
\bqa
&& T\sum_{\{K_0\}}{1 \over K^2(P-K)^2} \nonumber \\ 
\!\!&=&\!\! {1 \over 4k|{\bf p}-{\bf k}|} \Bigg\{\Big(1-n_F(k)-n_F(|{\bf p}-{\bf k}|)\Big)\left[{-1 \over iP_0-k-|{\bf p}-{\bf k}|}+{1 \over iP_0+k+|{\bf p}-{\bf k}|}\right] \nonumber \\ 
&& -\;\Big(-n_F(k)+n_F(|{\bf p}-{\bf k}|)\Big)\left[{-1 \over iP_0-k+|{\bf p}-{\bf k}|}+{1 \over iP_0+k-|{\bf p}-{\bf k}|}
\right]\Bigg\} \, ,
\eqa 
which is derived from a contour integral in the complex energy plane. The second term in Eq.~(\ref{htl-self}) is rather simple. We obtain 
\bqa\nonumber
\Pi^{ij}(P)\!\!&=&\!\!-2e^2\delta^{ij}\int_{\bf k}{1\over k}\left(1-2n_F(k)\right)+2e^2\int_{\bf k}{k^ik^j\over k|{\bf k}-{\bf p}|} \\ 
&&\nonumber
\times\;\Bigg\{\Big(1-n_F(k)-n_F(|{\bf k}-{\bf p}|)\Big)\left[{-1 \over iP_0-k-|{\bf k}-{\bf p}|}+{1 \over iP_0+k+|{\bf k}-{\bf p}|}\right] \nonumber \\ 
&& -\;\Big(-n_F(k)+n_F(|{\bf k}-{\bf p}|)\Big)\left[{-1 \over iP_0-k+|{\bf k}-{\bf p}|}+{1 \over iP_0+k-|{\bf k}-{\bf p}|}
\right]\Bigg\} \, , \nonumber \\
\label{omself}
\eqa 
where $n_F(x)=1/(\exp(\beta x)+1)$ is the Fermi-Dirac distribution function. The zero-temperature part of Eq.~(\ref{omself}) is logarithmically divergent in the ultraviolet. This term depends on the external momentum and is cancelled by standard zero-temperature wavefunction renormalization. We next consider the terms that depend on temperature. In the case that the loop momentum is soft, the Fermi-Dirac distribution functions can be approximated by a constant. The contribution from the integral over the magnitude of $k$ is then of order $g^3$ and subleading. When the loop momentum is hard, one can expand the terms in the integrand in powers of the external momentum. We can then make the following approximations 
\bqa
n_F(|{\bf k}-{\bf p}|)\!\!&\approx&\!\!n_F(k)-{dn_F(k)\over dk}{\bf p}\!\cdot\!\hat{\bf k}\;,\\
|{\bf k}-{\bf p}|\!\!&\approx&\!\!k-{\bf p}\!\cdot\!\hat{\bf k}
\;,
\eqa
where $\hat{\bf k}={\bf k}/k$ is a unit vector. Thus the angular integration decouples from the 
integral over the magnitude $k$. This implies
\bqa
\Pi^{ij}(P)\!\!&=&\!\!-{2e^2 \over \pi^2}\int_0^{\infty}dk\;k^2\,{dn_F(k) \over dk}\int{d\Omega\over4\pi}{-iP_0\over{-iP_0+\bf p}\!\cdot\!\hat{\bf k}} \hat{k}^i\hat{k}^j \;,
\nonumber \\
\!\!&=&\!\! {e^2 T^2 \over 3}\int{d\Omega\over4\pi}{-iP_0\over{-iP_0+\bf p}\!\cdot\!\hat{\bf k}} \hat{k}^i\hat{k}^j \;.
\eqa

The other components of the self-energy tensor $\Pi^{\mu\nu}(P)$ are derived in the same manner or obtained using the transversality of polarization tensor:
\bqa
P^{\mu}\Pi^{\mu\nu}(P)\;=\;0\;.
\eqa
One finds~\cite{bellac}
\bqa
\Pi^{00}(P)\!\!&=&\!\! {e^2 T^2 \over 3} \left(\int{d\Omega\over4\pi}{iP_0\over-iP_0+{\bf p}\!\cdot\!\hat{\bf k}} +1 \right)\;, 
\\ 
\Pi^{0j}(P)\!\!&=&\!\! {e^2 T^2 \over 3} \int{d\Omega\over4\pi}{-P_0 \over -iP_0+{\bf p}\!\cdot\!\hat{\bf k}}\hat{k}^j\;.
\eqa
In $d$ dimensions, we can compactly write the self-energy tensor as
\bqa
\label{a1}
\Pi^{\mu\nu}(P)\;=\;m_D^2\left[{\mathcal T}^{\mu\nu}(P,-P)-N^{\mu}N^{\nu}\right]\;,
\label{scomp}
\eqa
where $N$ specifies the thermal rest frame is canonically given by $N = (-i,{\bf 0})$. We have defined 
\bqa
m_D^2\;=\;-4(d-1)e^2\sumint_{\{K\}}{1\over K^2}\;=\;{e^2 T^2 \over 3}\;,
\eqa
and the tensor ${\mathcal T}^{\mu\nu}(P,Q)$, which is defined only for momenta that satisfy $P+Q=0$, is
\bqa
{\mathcal T}^{\mu\nu}(P,-P)\;=\;\left \langle Y^{\mu}Y^{\nu}{P\!\cdot\!N \over P\!\cdot\!Y}\right\rangle_{\bf\hat{y}} \;.
\label{T2-def}
\eqa
The angular brackets indicate averaging over the spatial directions of the light-like vector $Y=(-i,\hat{\bf y})$. The tensor ${\mathcal T}^{\mu\nu}$ is symmetric in $\mu$ and $\nu$. Because of transverality and the rotational symmetry around the $\hat{\bf p}$-axis, one can express the self-energy in terms of two independent functions, $\Pi_T(P)$ and $\Pi_L(P)$:
\bqa\nonumber
\Pi^{\mu\nu}(P)\!\!&=&\!\!\Pi_{L}(P){P^2\delta^{\mu\nu}-P^{\mu}P^{\nu}\over p^2} \\ 
&&+\;\left[\Pi_T(P)+{P^2\over p^2}\Pi_L(P)\right]\delta^{\mu i}\left(\delta^{ij}-\hat{p}^i\hat{p}^j\right)\delta^{j\nu} \;,
\label{defself}
\eqa
where the functions $\Pi_T(P)$ and $\Pi_L(P)$ are
\bqa
\Pi_T(P)\!\!&=&\!\!{1\over2}(\delta^{ij}-\hat{p}^i\hat{p}^j)\Pi^{ij}(P)
\label{pit}
\;,\\
\Pi_L(P)\!\!&=&\!\!-\Pi^{00}(P)\;.
\label{pil}
\eqa
In three dimensions, the self-energies $\Pi_T(P)$ and $\Pi_L(P)$
reduce to
\bqa
\Pi_T(P)\!\!&=&\!\!-{m_D^2\over2}{P_0^2\over p^2}\left[1+{P^2\over2 iP_0 p}\log{iP_0+p\over iP_0-p}\right]\;,
\label{redt} \\
\Pi_L(P)\!\!&=&\!\!m_D^2\left[1- {iP_0\over2p}\log{iP_0+p\over iP_0-p}\right]\;.
\label{redl}
\eqa
The hard thermal loop in the photon propagator was first calculated by Silin more than forty years ago \cite{silin}. The hard thermal loop in the gluon self-energy was first calculated by Klimov and Weldon \cite{klim,Weldon:1982aq}. It has the same form as in QED, but where the Debye mass $m_D$ is replaced by
\bqa
m_D^2\;=\;g^2 \left[(d-1)^2 C_A\sumint_{K}{1\over K^2}-2(d-1) N_f \sumint_{\{K\}}{1\over K^2}\right]\;,
\label{qcdmd}
\eqa
where $C_A=N_c$ is the number of colors and $N_f$ is the number of flavors. When $d=3$ the QCD gluon Debye mass becomes
\bqa
m_D^2\;=\;{1\over3} \left( C_A + {1\over2}N_f \right) g^2 T^2 \; .
\label{qcdmd3}
\eqa

\subsection{Fermionic self-energy}

The electron self-energy is given by
\bqa
\label{selfq}
\Sigma(P)\;=\;m_f^2{\mathcal T}\!\!\!\!/(P)
\;,
\eqa
where
\bqa
\label{deftf}
{\mathcal T}^{\mu}(P) \;=\; -\left\langle{Y^{\mu}\over P\cdot Y}\right\rangle_{\hat{\bf y}}\;,
\eqa
and $m_f$ is the thermal electron mass
\bqa
m_f^2\;=\;-3e^2\sumint_{\{K\}}{1\over K^2}\;.
\eqa
In QCD, the quark mass is given by
\bqa
m_q^2\;=\;-3C_Fg^2\sumint_{\{K\}}{1\over K^2}\;.
\eqa

\subsection{Higher $n$-point functions}

In gauge theories, there are also hard thermal loops involving vertices. For instance, the one-loop correction to the three-point function in QED, can compactly be written as
\bqa
\Gamma^{\mu}(P,Q,R)\;=\;\gamma^{\mu}-m_f^2\tilde{{\mathcal T}}^{\mu}(P,Q,R)\;,
\eqa
where the tensor in the HTL correction term is only defined for $P-Q+R=0$:
\bqa
\tilde{{\mathcal T}}^{\mu}(P,Q,R)\;=\;\left\langle Y^{\mu}\left({Y\!\!\!\!/\over (Q\!\cdot\!Y)(R\!\cdot\!Y)}\right)\right\rangle_{\hat{\bf y}}\;.
\label{T3-def}
\eqa
The quark-gluon vertex satisfies the Ward identity
\bqa
P^{\mu}\Gamma^{\mu}(P,Q,R)\;=\;S^{-1}(Q)-S^{-1}(R)\;,
\label{qward1}
\eqa
where $S(q)$ is the resummed effective fermion propagator. 

In QED there are, in fact, infinitely many amplitudes with hard thermal loops. To be precise, there are hard thermal loops in all $n$-point functions with two fermion lines and $n-2$ photon lines. In non-Abelian gauge theories such as QCD, there are in addition hard thermal loops in amplitudes with $n$ gluon lines \cite{Braaten:1989mz}.

%%%%%%%%%%%%%%%%%%%%%%%%%%%%%%%%%%%%%%%%%%%%%%%%%%%%%%%%%%%%%
%
%	Include File:			DON'T COMPILE !!!
%
%%%%%%%%%%%%%%%%%%%%%%%%%%%%%%%%%%%%%%%%%%%%%%%%%%%%%%%%%%%%%

\section{Weak-coupling expansion}
\label{weak}

The Braaten-Pisarski resummation program has been used to calculate the thermodynamic functions as a weak-coupling expansion in $g$. They have now been calculated explicitly through order $g^8 \log g$ for massless $\phi^4$ theory~\cite{Frenkel:1992az,AZ-95,Parwani:1994zz,Braaten:1995cm,Gynther:2007bw,Andersen:2009ct}, through order $e^5$ for QED~\cite{AZ-95,Parwani:1994xi,Parwani:1994je,Andersen:1995ej,Zhai:1995ac}, and through order $g^6 \log g$ for QCD~\cite{AZ-95,Zhai:1995ac,BN-96,Kajantie:2002wa}. In this section, we review these calculations in some detail.

\subsection{Scalar field theory}

The simplest way of dealing with the infrared divergences in scalar field theory is to reorganize perturbation theory in such a way that it incorporates the effects of the thermally generated mass $m$ into the free part of the Lagrangian. One possibility is to divide the Lagrangian~(\ref{sl}) into free and interacting parts according to
\bqa
{\mathcal L}_{\rm free}\!\!&=&\!\!{1\over2}(\partial_{\mu}\phi)^2+{1\over2}m^2\phi^2\,, \\
{\mathcal L}_{\rm int}\!\!&=&\!\!{g^2\over24}\phi^4-{1\over2}m^2\phi^2\,.
\label{sint}
\eqa
Both terms in Eq.~(\ref{sint}) are treated as interaction terms of the same order, namely $g^2$.  However, the resummation implied by the above is rather cumbersome when it comes to calculating Green's function with zero external energy. Static Green's functions can always be calculated directly in imaginary time without having to analytically continue them back to real time. This implies that we can use Euclidean propagators with discrete energies when analyzing infrared divergences which greatly simplifies the treatment. In particular, since only propagators with zero Matsubara frequency have no infrared cutoff of order $T$, only for these modes is the thermal mass of order $gT$ relevant as an IR cutoff. Thus, another possibility is to add and subtract a mass term only for the zero-frequency mode. This approach has the advantage that we do not need to expand the sum-integrals in powers of $m^2/T^2$ in order to obtain the contribution from a given term in powers of $g^2$. We will follow this path in the remainder of this section and write
\bqa
{\mathcal L}_{\rm free}\!\!&=&\!\!{1\over2}(\partial_{\mu}\phi)^2+{1\over2}m^2\phi^2\delta_{P_0,0}\;,
\label{L-free-phi}
\\
{\mathcal L}_{\rm int}\!\!&=&\!\!{g^2\over24}\phi^4-{1\over2}m^2\phi^2\delta_{P_0,0}\;.
\label{L-int-phi}
\eqa
The free propagator then takes the form
\bqa
\Delta(P)\;=\;{1-\delta_{P_0,0}\over P^2}+{\delta_{P_0,0}\over p^2+m^2}\;.
\eqa
This way of resumming is referred to as {\it static resummation}~\cite{static-resum}. It is important to point out that this simplified resummation scheme can only be used to calculate static quantities such as the pressure or screening masses. Calculation of dynamical quantities requires the full Braaten-Pisarski resummation program. The problem is that the calculation of correlation functions with zero external frequencies cannot unambiguously be analytically continued to real time~\cite{Kraemmer:1994az}.

\begin{figure}[t]
\centering
%
%\begin{tabular}{ccccccc}
%$\oneloop$ &
%$\figureeight$ &
%$\oneloopX$ &
%$\triplebubble$ &
%$\basketball$ &
%$\figureeightX$ &
%$\oneloopXX$
%\vspace{5mm}
%\\
%${\mathcal F}_{\rm 0}$ &
%${\mathcal F}_{\rm 1a}$ &
%${\mathcal F}_{\rm 1b}$ &
%${\mathcal F}_{\rm 2a}$ &
%${\mathcal F}_{\rm 2b}$ &
%${\mathcal F}_{\rm 2c}$ &
%${\mathcal F}_{\rm 2d}$ 
%\end{tabular}
\includegraphics{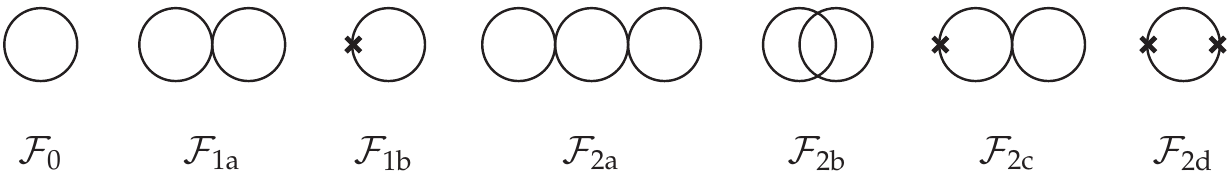}
\caption[a]{Diagrams which contribute up to three-loop order in scalar perturbation theory. A boldfaced $\times$ indicates an insertion of $m^2$.}  
\label{scalarpertgraphs}

\end{figure}

\subsubsection{Perturbative expansion}

Following the decomposition in~(\ref{L-free-phi}) and (\ref{L-int-phi}), the partition function for $\phi^4$ theory reads
\bqa
Z \;=\; \int {\cal D}\phi \, e^{ - ( S_{\rm free} + S_{\rm int} ) } \;,
\label{Z-phi}
\eqa
with the free action $S_{\rm free}$ and the action due to interactions $S_{\rm int}$ defined by
\bqa
S_{\rm free} \!\!&=&\!\! \int_0^{1/T}\!\!d\tau\int d^3x\,{\cal L}_{\rm free} \;, \\
S_{\rm int} \!\!&=&\!\! \int_0^{1/T}\!\!d\tau\int d^3x\,{\cal L}_{\rm int} \;.
\eqa
Expanding in powers of $S_{\rm int}$, (\ref{Z-phi}) becomes
\bqa
Z \;=\; \int {\cal D}\phi \, e^{ - S_{\rm free} } \sum_{n=0}^\infty {(-S_{\rm int})^n \over n!} \;.
\eqa
Taking logarithm for both sides, we get
\bqa
\log Z \!\!&=&\!\! \log \left[ \int {\cal D}\phi \, e^{ - S_{\rm free} } \right] + \log \left[ 1+ \sum_{n=0}^\infty {\int {\cal D}\phi \, e^{ - S_{\rm free} } (-S_{\rm int})^n \over n! \int {\cal D}\phi \, e^{ - S_{\rm free} }} \right] 
\nonumber \\
&=&\!\! \log Z_0 + \log Z_{\rm I}
\;.
\eqa
In this way, the partition function has been divided into two parts. The free part $\log Z_0$ describes the classical limit of the theory, i.e. a gas of non-interacting scalar particles, which can be evaluated analytically due to the fact that it is a quadratic or Gaussian function in the field $\phi$. The interacting part $\log Z_{\rm I}$ contains quantum corrections to the classical theory, which is accessed perturbatively as a power series in $S_{\rm int}$. Using the relation $\log\det A = {\rm Tr}\log A$ it is easy to show that
%\bqa
%\log Z_0 \;=\; \vspace{1cm}\includegraphics{figures/bubble} \;=\; -{1 \over 2} {\rm Tr} \log[-\partial^2] \;=\; -{V \over 2T} \sumint_P \log P^2 \;,
%\eqa
\bqa
\includegraphics{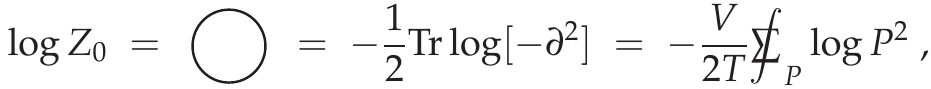}
\eqa
where the single closed loop denotes the corresponding Feynman diagram contributing to $\log Z_0$. As an example of the perturbative expansion of static resummation, we next consider the calculation of the free energy ${\cal F}$, which is the negative pressure ${\cal F} \!=\! - {\cal P}$, through order $g^5$ in scalar field theory. This involves the evaluation of vacuum graphs up to three-loop order shown in Fig.~\ref{scalarpertgraphs}.

\subsubsection{One loop}

The one-loop contribution to the free energy is
\bqa
{\mathcal F}_0\;=\;{1\over2}T\int_{\bf p}\log\left(p^2+m^2\right)+{1\over2}\sumint_P^{\prime}\log P^2\;.
\label{pf1}
\eqa
Using the integrals and sum-integrals contained in the appendices the result for this diagram in the limit $\epsilon\rightarrow0$ is
\bqa
{\mathcal F}_0\;=\;-{\pi^2\over90}T^4-{Tm^3\over12\pi}\;.
\eqa

\subsubsection{Two loops }

The two-loop contribution to the free-energy is given by 
\bqa
{\mathcal F}_1\;=\;{\mathcal F}_{\rm 1a} + {\mathcal F}_{\rm 1b}\;,
\label{pf2}
\eqa
with
\bqa
{\mathcal F}_{\rm 1a}\!\!&=&\!\!{1\over8}g^2\left(T\int_{\bf p}{1\over p^2+m^2}+\sumint_P^{\prime}{1\over P^2}\right)^2\;, \\ 
{\mathcal F}_{\rm 1b}\!\!&=&\!\!-{1\over2}m^2T\int_{\bf p}{1\over p^2+m^2}\;.
\eqa
The result for these diagrams in the limit $\epsilon\rightarrow0$ is
\bqa\nonumber
{\mathcal F}_{\rm 1a}\!\!&=&\!\!{\pi^2T^4\over90}\left\{{5\over4}\left(g \over 4\pi\right)^2\left[1+\epsilon\left(4+4{\zeta^{\prime}(-1)\over\zeta(-1)}\right)\right]\left({\mu\over4\pi T}\right)^{4\epsilon}
\right. \\ &&
\left.
-\;{5\sqrt{6}\over2}\left(g \over 4\pi\right)^3\left[1+\epsilon\left(4+2{\zeta^{\prime}(-1)\over\zeta(-1)}\right)\right]\left({\mu\over4\pi T}\right)^{2\epsilon}\left({\mu\over2m}\right)^{2\epsilon}+{15\over2}\alpha^2\right\}, \nonumber \\
&& \\
{\mathcal F}_{\rm 1b}\!\!&=&\!\!{\pi^2T^4\over90}{5\sqrt{6}\over2}\left(g \over 4\pi\right)^3\;,
\eqa
where we have kept all terms that contribute through order $\epsilon$, because they are needed for the counterterm diagrams in the three-loop free energy.

\subsubsection{Three loops }

The three-loop contribution is given by
\bqa
{\mathcal F}_{\rm 2}\;=\;{\mathcal F}_{\rm 2a}+{\mathcal F}_{\rm 2b}+{\mathcal F}_{\rm 2c}+{\mathcal F}_{\rm 2d}+{{\mathcal F}_{\rm 1a} \over g^2} \Delta_1 g^2\;,
\label{pf3}
\eqa
where the expressions for the diagrams are 
\bqa
{\mathcal F}_{\rm 2a}\!\!&=&\!\!-{1\over16}g^4\left(T\int_{\bf p}{1\over p^2+m^2}+\sumint_P^{\prime}{1\over P^2}\right)^2\left(T\int_{\bf p}{1\over(p^2+m^2)^2}+\sumint_P^{\prime}{1\over P^4}\right)\!, \nonumber \\
&& \\ \nonumber
{\mathcal F}_{\rm 2b}\!\!&=&\!\!-{1\over48}g^4\sumint_{PQR}^{\prime}{1\over P^2}{1\over Q^2}{1\over R^2}{1\over (P+Q+R)^2} \\ &&
-\;{1\over48}g^4T^3\int_{\bf pqr}{1\over p^2+m^2}{1\over q^2+m^2}{1\over r^2+m^2}{1\over ({\bf p}+{\bf q}+{\bf r})^2+m^2}
\label{32b}
\;,\\ \nonumber && \\
{\mathcal F}_{\rm 2c}\!\!&=&\!\!{1\over4}g^2m^2\left(T\int_{\bf p}{1\over(p^2+m^2)}+\sumint_P^{\prime}{1\over P^2}\right)\left(T\int_{\bf p}{1\over(p^2+m^2)^2}\right)\;,\\ \nonumber && \\
{\mathcal F}_{\rm 2d}\!\!&=&\!\!-{1\over4}m^4T\int_{\bf p}{1\over(p^2+m^2)^2}\;.
\eqa
Note that the basketball diagram ${\cal F}_{\rm 2b}$ is the only diagram through three loops that cannot be decomposed to solely a set of one loop integrals, which therefore needs special treatment~\cite{Andersen:2000zn}. The result for these diagrams in the limit $\epsilon\rightarrow0$ is
\bqa\nonumber
{\mathcal F}_{\rm 2a\!\!}&=&\!\!{\pi^2T^4\over90}\left\{-{5\sqrt{6}\over8}\left(g \over 4\pi\right)^3-{5\over8}\left(g \over 4\pi\right)^4\left[{1\over\epsilon}+2\gamma-8+4{\zeta^{\prime}(-1)\over\zeta(-1)}\right]\left({\mu\over4\pi T}\right)^{6\epsilon}\right. \\ &&
\left.
+\;{5\sqrt{6}\over4}\left(g \over 4\pi\right)^5\left[{1\over\epsilon}+2\gamma+1+2{\zeta^{\prime}(-1)\over\zeta(-1)}\right]\left({\mu\over4\pi T}\right)^{4\epsilon}\left({\mu\over2m}\right)^{2\epsilon}\right\}
\label{3f1}
\;,\\ \nonumber && \\ \nonumber
{\mathcal F}_{\rm 2b}\!\!&=&\!\!{\pi^2T^4\over90}\left\{-{5\over4}\left(g \over 4\pi\right)^4\left[{1\over\epsilon}+8{\zeta^{\prime}(-1)\over\zeta(-1)}-2{\zeta^{\prime}(-3)\over\zeta(-3)}+{91\over15}\right]\left({\mu\over4\pi T}\right)^{6\epsilon}
\right. \\ &&
\left.
+\;{5\sqrt{6}\over2}\left(g \over 4\pi\right)^5\left[{1\over\epsilon}+8-4\log2\right]\left({\mu\over2m}\right)^{6\epsilon}\right\}\;, \\ \nonumber && \\
{\mathcal F}_{\rm 2c}\!\!&=&\!\!{\pi^2T^4\over90}\left[{5\sqrt{6}\over4}\left(g \over 4\pi\right)^3-{15\over2}\left(g \over 4\pi\right)^4\right] \;,\\ \nonumber && \\
{\mathcal F}_{\rm 2d}\!\!&=&\!\!-{\pi^2T^4\over90}{5\sqrt{6}\over8}\left(g \over 4\pi\right)^3 
\label{3f4}
\;.
\\ \nonumber
\eqa

\subsubsection{Free energy through $g^5$}

Combining the one-, two-, and three-loop contributions given by Eqs.~(\ref{pf1}),~(\ref{pf2}), and~(\ref{pf3}), respectively, gives the free energy through order $g^5$
\bqa\nonumber
{{\cal F} \over {\cal F}_0}\!\!&=&\!\!1-{5\over4}\left(g \over 4\pi\right)^2+{5\sqrt{6}\over3}\left(g \over 4\pi\right)^3+{15\over4}\left[\log{\mu\over4\pi T}-{59\over45}+{1\over3}\gamma+{4\over3}{\zeta^{\prime}(-1)\over\zeta(-1)}
\right. \\ \nonumber && \left.
-\;{2\over3}{\zeta^{\prime}(-3)\over\zeta(-3)}\right]\left(g \over 4\pi\right)^4-{15\sqrt{6}\over2}\left[\log{\mu\over4\pi T}-{4\over3}\log{\left(g \over 4\pi\right)}+{5\over6}
\right. \\ && \left.
-\;{2\over3}\log{2\over3}+{1\over3}\gamma-{2\over3}{\zeta^{\prime}(-1)\over\zeta(-1)}\right]\left(g \over 4\pi\right)^5 \;,
\label{fscaw}
\eqa
where ${\cal F}_0=-\pi^2T^4/90$ is the free energy of an ideal gas of non-interacting scalar bosons. The pressure through order $g^5$ was first calculated using resummation by Parwani and Singh~\cite{Parwani:1994zz} and later by Braaten and Nieto using effective field theory~\cite{Braaten:1995cm}.

With $\alpha=(g/4\pi)^2$, the renormalization group equation for the coupling $g^2$ reads
\bqa
\mu{d\alpha\over d\mu}\;=\;{3\alpha^2}\;.
\label{rsca}
\eqa
Using Eq.~(\ref{rsca}), one can verify that the free energy (\ref{fscaw}) is RG-invariant up to corrections of order $g^6\log g$.

%%%%%%%%%%%%%%%%%%%%%%%%%%%%%%%%%%%%%%%%%%%%%%%%%%%%%%%%%%%%%%%%
\vspace{6mm}
\begin{figure}[htb]
\begin{center}
%\centerline{\pdfig{file=figures/scalarpertpressure,width=10cm}}
\includegraphics[width=10cm]{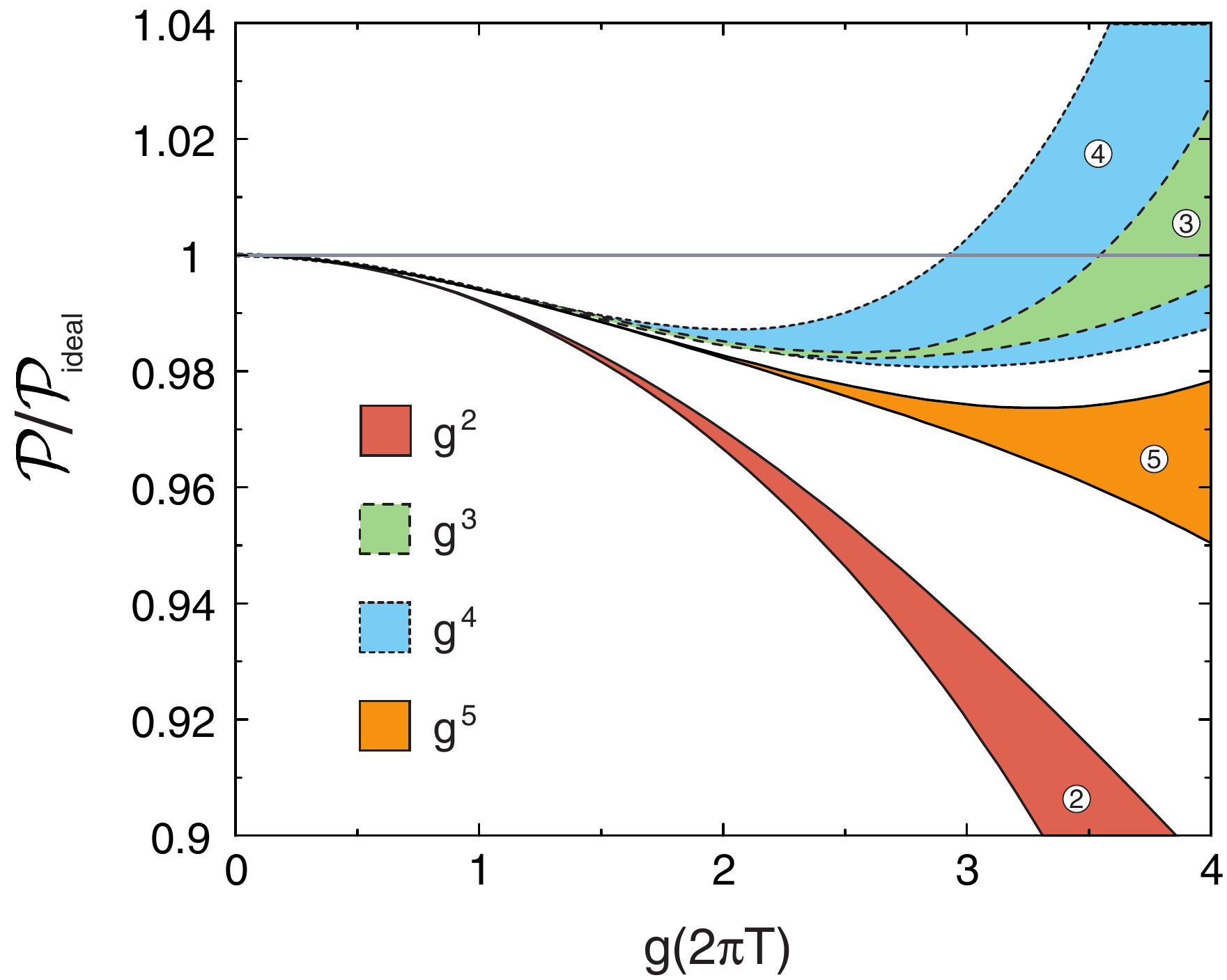}
\end{center}
\vspace*{8pt}
\caption{Weak-coupling expansion for the free energy of massless $\phi^4$ theory normalized to that of an ideal gas as a function of $g(2\pi T)$ to orders $g^2$, $g^3$, $g^4$, and $g^5$ are shown as bands that correspond to varying the renormalization scale $\mu$ by a factor of 2 around $2\pi T$. This figure is adapted from Ref.~\cite{Andersen:2004fp}.}
\label{fpert}
\end{figure}
%%%%%%%%%%%%%%%%%%%%%%%%%%%%%%%%%%%%%%%%%%%%%%%%%%%%%%%%%%%%%%%%
In Fig.~\ref{fpert}, we show the successive perturbative approximations to the free energy as a function of $g(2\pi T)$. The bands are obtained by varying the renormalization scale $\mu$ by a factor of 2 around the central value $\mu=2\pi T$. The lack of convergence of the weak-coupling expansion is evident from this figure. The band obtained by varying $\mu$ by a factor of 2 is not necessarily a good measure of the error, but it is certainly a lower bound on the theoretical error. Another indicator of the theoretical error is the deviation between successive approximations.  We can infer from Fig.~\ref{fpert} that the error grows rapidly when $g(2 \pi T)$ exceeds 1.5.

\subsection{Gauge theories}

In this section, we discuss the application of weak-coupling expansion to gauge theories. The Euclidean Lagrangian for an SU($N_c$) gauge theory with $N_f$ fermions in the fundamental representation is
\bqa
{\mathcal L}\;=\;{1\over4}G_{\mu\nu}^aG_{\mu\nu}^a
+\bar{\psi}\gamma_{\mu}D_{\mu}\psi
\;,
\eqa
where $G_{\mu\nu}^a=\partial_{\mu}A_{\nu}^a-\partial_{\nu}A_{\mu}^a+gf^{abc}A_{\mu}^bA_{\nu}^c$ is the field strength, $g$ is the gauge coupling and $f^{abc}$ are the structure constants. The covariant derivative is $D_{\mu}=\partial_{\mu}-igA_{\mu}^aT^a$, where $T^a$ are the generators in the fundamental representation.

The constants $d_A$ and $C_A$ are the dimension and quadratic Casimir invariant of the adjoint representation, with
\begin{equation}
\delta^{aa}=d_A\,,\;\;\;\;\;\;\;\;\;\;\;\;\;\;\;
f^{abc}f^{dbc}=C_A\delta^{ad}\,.
\end{equation}
$d_F$ is the dimension of the total fermion representation
, and $S_F$ and $S_{2F}$
are defined in terms of the generators $T^a$ for the total fermion
representation as
\begin{equation}
S_F ={1 \over d_A}{\rm tr}(T^2)\,,\;\;\;\;\;\;\;\;\;\;\;\;\;\;
S_{2F} = {1 \over d_A}{\rm tr} [(T^2)^2]\,,
\end{equation}
where $T^2 = T^a T^a$.
For SU($N_c$) with $N_f$ fermions in the fundamental representation,
the standard normalization of the coupling gives
\begin{equation}
\label{SUNnotation}
d_A = N_c^2-1 \,,
\;\;\;\;\;
C_A = N_c \,,
\;\;\;\;\;
d_F = N_cN_f \,,
\;\;\;\;\;
S_F = {1\over2} N_f \,,
\;\;\;\;\;
S_{2F} = {N_c^2-1 \over 4N_c} N_f \,.
\end{equation}
For U(1) theory, relabel $g$ as $e$ and let the charges of the $N_f$
fermions be $q_i e$.  Then
\begin{equation}
d_A = 1 \,,\;\;\;\;\;
C_A = 0 \,,\;\;\;\;\;
d_F = N_f \,,\;\;\;\;\;
S_F = \sum_i q_i^2 \,,\;\;\;\;\;
S_{2F} = \sum_i q_i^4 \,.
\label{QEDnotation}
\end{equation}

If we are only interested in static quantities, we can apply the same simplified resummation scheme also to gauge theories~\cite{AZ-95,Zhai:1995ac}. Thus we are interested in the static limit of the polarization tensor $\Pi_{\mu\nu}$. In that limit $\Pi_T$ vanishes and $\Pi_L=m_D^2$. In analogy with the scalar field theory, we rewrite the Lagrangian by adding and subtracting a mass term ${1\over2}m^2_DA_0^aA_0^a\delta_{P_0,0}$. One of the mass terms is then absorbed into the propagator for the timelike component of the gauge field $A_0$, while the other is treated as a perturbation. 

The free energy through $g^5$ requires the evaluation of diagrams up to three loops. The strategy is the same in the scalar case, where one distinguishes between hard and soft loop momenta. The result reads
\begin{eqnarray}
{\cal F} \!\!&=&\!\! -d_A{\pi^2T^4 \over 45} \Biggr \{1 + {7 \over 4}{d_F \over d_A} - 5 \left(C_A+{5\over2}S_F\right)\left ({g \over 4 \pi} \right )^2+240 \left({C_A+S_F \over 3}\right)^{3 \over 2}\left ({g \over 4 \pi} \right )^3
\nonumber\\
&& \qquad
+\;240 C_A (C_A+S_F)\left ({g \over 4 \pi} \right )^4\log\left({g\over 2\pi}\sqrt{C_A+S_F \over 3}\right)
\nonumber\\
&& %\qquad
-\;5 \Biggr [
C_A^2
\left(
{22 \over 3} \log{{\mu \over 4\pi T}}
+{38 \over 3} {\zeta'(-3) \over \zeta(-3)}
-{148 \over 3} {\zeta'(-1) \over \zeta(-1)}
-4 \gamma
+{64 \over 5}
\right)
\nonumber\\
&& \qquad
+\;C_A S_F
\left(
{47 \over 3} \log{{\mu \over 4\pi T}}
+{1 \over 3} {\zeta'(-3) \over \zeta(-3)}
-{74 \over 3} {\zeta'(-1) \over \zeta(-1)}
-8 \gamma
+{1759 \over 60}
+{37\over5}\log{2}
\right)
\nonumber\\
&& \qquad
+\;S_F^2
\left(
-{20 \over 3} \log{{\mu \over 4\pi T}}
+{8 \over 3} {\zeta'(-3) \over \zeta(-3)}
-{16 \over 3} {\zeta'(-1) \over \zeta(-1)}
-4 \gamma
-{1 \over 3}
+{88 \over 5}\log{2}
\right)
\nonumber\\
&& \qquad
+\;S_{2F}
\left(
-{105 \over 4}
+24\log{2}
 \right)
\Biggr ]\left (g \over 4 \pi\right )^4
\nonumber\\
&&
+\;5 \sqrt{C_A + S_F \over 3} \Biggr [C_A^2
\left ( 176 \log{{\mu \over 4\pi T}}
+176 \gamma
-24 \pi^2
-494
+264 \log{2}
\right )
\nonumber\\
&& \qquad \qquad \qquad \qquad
+\;C_A S_F \left (112 \log{{\mu \over 4\pi T}}
+112 \gamma
+72
-128 \log{2}
\right )
\nonumber\\
&& \qquad \qquad \qquad \qquad
+\;S_F^2 \left (-64 \log{{\mu \over 4\pi T}}
-64 \gamma
+32
-128 \log{2}
\right )
\nonumber\\
&& \qquad \qquad \qquad \qquad
-\:144 S_{2F}
\Biggr ]\left (g \over 4 \pi \right )^5
\Biggr \} \,.
\label{f-qcd-w}
\end{eqnarray}

The one-loop beta function for an SU($N_c$) gauge theory with $N_f$ fermions reads
\bqa
\mu{d\alpha_s\over d\mu}\;=\;-\left({11 \over 3}N_c - {2 \over 3}N_f\right){\alpha_s^2 \over 2\pi}\;,
\label{rg-qcd}
\eqa
written in terms of $\alpha_s=g^2/4\pi$. Using Eq.~(\ref{rg-qcd}), one can verify that the free energy (\ref{f-qcd-w}) is RG-invariant up to corrections of order $g^6\log g$.

%%%%%%%%%%%%%%%%%%%%%%%%%%%%%%%%%%%%%%%%%%%%%%%%%%%%%%%%%%%%%%%%
\begin{figure}[htb]
\begin{center}
%\centerline{\psfig{file=figures/FpertQ.eps,width=10cm}} 
\includegraphics[width=10cm]{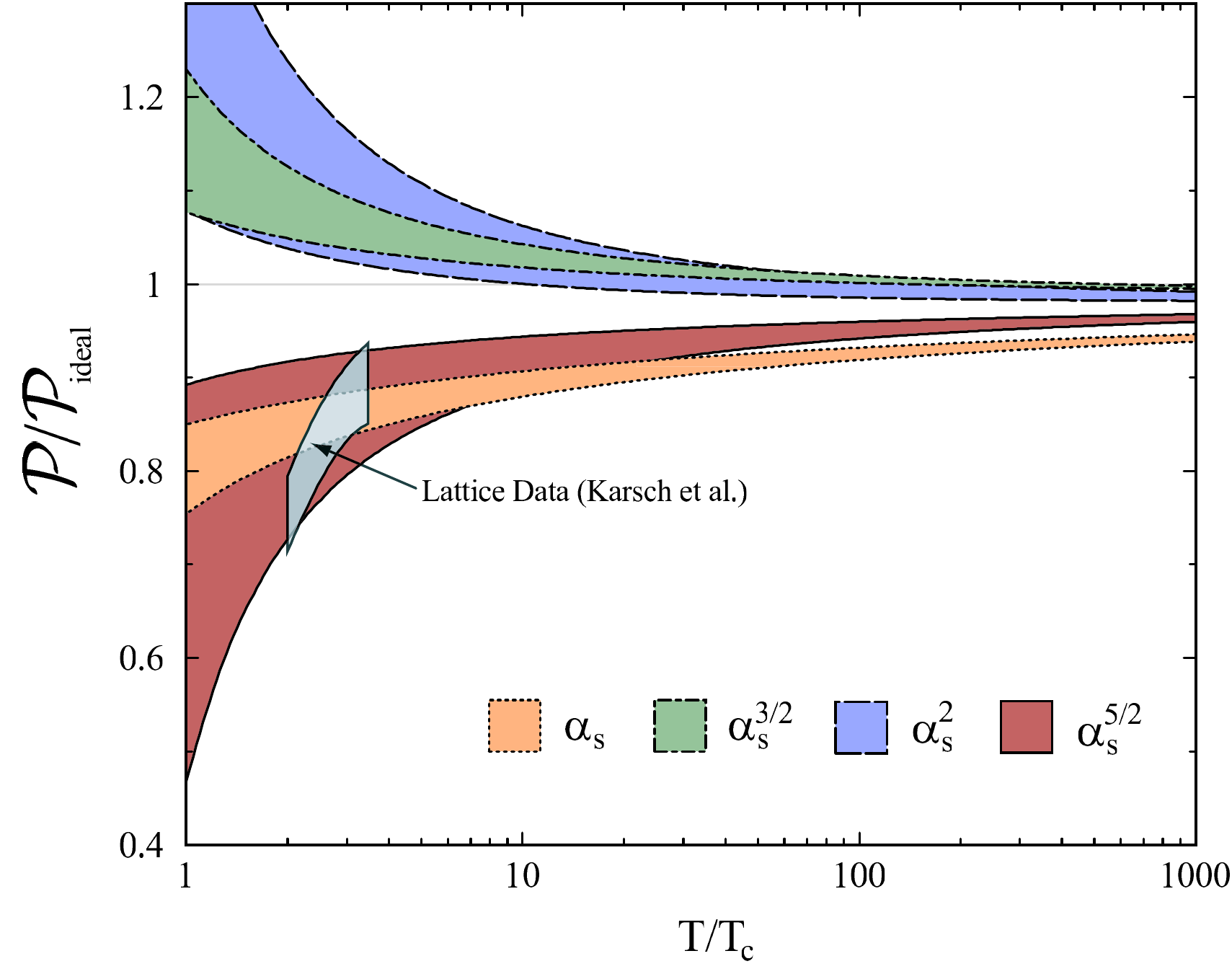}
\end{center}
\vspace*{8pt}
\caption{Weak-coupling expansion for the free energy of SU(3) gauge theory with $N_f=2$ normalized to that of an ideal gas as a function of $T/T_c$ to orders $g^2$, $g^3$, $g^4$, and $g^5$ are shown as bands that correspond to varying the renormalization scale $\mu$ by a factor of 2 around $2\pi T$~\cite{htl2}. Also shown is the lattice estimate by Karsch et al.~\cite{Karsch:2000ps} for the free energy. The band indicates the estimated systematic error of their result which is reported as $(15\pm5)\%$. This figure is adapted from Ref.~\cite{htl2}.}
\label{fpertqcd}
\end{figure}
%%%%%%%%%%%%%%%%%%%%%%%%%%%%%%%%%%%%%%%%%%%%%%%%%%%%%%%%%%%%%%%%

The free energy for QCD through order $g^4$ was first derived by Arnold and Zhai~\cite{AZ-95}. Later it was extended to order $g^5$ by Zhai and Kastening~\cite{Zhai:1995ac} using the above resummation techniques, and by Braaten and Nieto using effective field theory~\cite{BN-96}. The order-$g^5$ contribution is the last contribution that can be calculated using perturbation theory. At order $g^6$, although the electric $g^6 \log g$ contribution is still perturbatively accessible~\cite{Kajantie:2002wa}, however perturbation theory breaks down due to infrared divergences in the magnetic sector~\cite{Linde:1980ts,Gross:1980br}.

In Fig.~\ref{fpertqcd}, the free energy of QCD ($N_c=3$) with $N_f=2$ is shown as a function of the temperature $T/T_c$, where $T_c$ is the critical temperature for the deconfinement transition. In the plot we have scaled the free energy by the free energy of an ideal gas of quarks and gluons which for arbitrary $N_c$ and $N_f$ is
\beq
{\cal F}_{\rm ideal}\;=\;-{\pi^2 \over 45} T^4 \left( N_c^2-1 + {7\over 4} N_c N_f \right) \, .
\label{fideal}
\eeq
The weak-coupling expansions through orders $g^2$, $g^3$, $g^4$, and $g^5$ are shown as bands that correspond to varying the renormalization scale, $\mu$, by a factor of 2 around the central value $\mu=2\pi T$. As successive terms in the weak-coupling expansion are added, the predictions change wildly and the sensitivity to the renormalization scale grows. It is clear that a reorganization of the perturbation series is essential if perturbative calculations are to be of any quantitative use at temperatures accessible in heavy-ion collisions.

%%%%%%%%%%%%%%%%%%%%%%%%%%%%%%%%%%%%%%%%%%%%%%%%%%%%%%%%%%%%%
%
%	Include File:			DON'T COMPILE !!!
%
%%%%%%%%%%%%%%%%%%%%%%%%%%%%%%%%%%%%%%%%%%%%%%%%%%%%%%%%%%%%%

\chapter{Hard-Thermal-Loop Perturbation Theory}
\label{chapter:htl}

In this chapter, we introduce the hard-thermal-loop perturbation theory, which is a gauge-invariant resummation scheme for thermal gauge theories. We discuss its formalism and properties, as well as some technicalities which make the evaluation of loop diagrams tractable analytically. This chapter forms the basis for the discussions in the rest of the dissertation.

\section{Introduction}

One possible conclusion from the bad convergence of the weak-coupling expansion of QCD free energy in Section~\ref{weak} is that the quark-gluon plasma is completely nonperturbative, and that it can only be studied by nonperturbative methods like lattice gauge theory. This would be a very unfortunate conclusion from the perspective of the search for the quark-gluon plasma. Real-time processes can serve as the signatures for a quark-gluon plasma at intermediate coupling. While lattice gauge theory can be used to calculate thermodynamic properties \cite{Boyd:1996bx,Karsch:2000ps}, its application to dynamical quantities currently suffers from large systematic errors \cite{Meyer:2007ic,Laine:2009dd}.

There is another possible interpretation of the failure of the conventional perturbation series. It could simply be a signal that we are using the \emph{wrong degrees of freedom}. Naive perturbation theory is an expansion around an ideal gas of massless quarks and gluons. This generates infrared divergences that must be rendered finite either by resumming infinite classes of diagrams or by nonperturbative methods. While such a procedure gives a well-defined weak-coupling expansion, in practice the coefficients seem to be too large for the expansion to be of any use. It is possible that another choice for the degrees of freedom would generate diagrams with better infrared behavior and successive approximations with better convergence properties.

The high-temperature limit of QCD provides a clue as to what those degrees of freedom might be. In this limit, quarks and gluons are \emph{quasiparticles} with temperature-dependent masses \cite{Kalashnikov:1979cy,Klimov:1982bv,Weldon:1982aq}. Furthermore, quarks and gluons acquire additional propagating degrees of freedom: in addition to the two usual transverse polarization modes of the gluon, there is a collective mode with longitudinal polarization called the plasmon; in addition to the two usual spin states of a quark, there is a collective mode with two spin states called the plasmino. The quasiparticle mass of the gluon is also intimately tied to the screening properties of the plasma. Chromoelectric fields are screened by the Debye mechanism beyond a screening length of $\sim  1/m_D$ where $m_D$ is the gluon quasiparticle mass. Oscillating chromomagnetic fields are also screened, with a screening length that scales like $(m_D^2\omega)^{-1/3}$, where $\omega$ is the frequency. At very low frequencies ($\omega$ of order $\alpha_s^2T$), nonperturbative effects take over, so that static chromomagnetic fields have a screening length of order $1/(\alpha_sT)$.

Quasiparticle masses, collective modes, and screening are all tied together by gauge invariance. The problem is therefore how to incorporate plasma effects into the perturbation expansion for QCD while preserving gauge invariance. This problem was solved at leading order in $g$ by Braaten and Pisarski \cite{Braaten:1989mz}. They developed a method called \emph{hard-thermal-loop} (HTL) resummation for summing all Feynman diagrams that are leading order in $g$ for amplitudes involving soft external momenta of order $gT$. This method can be used to systematically calculate higher order corrections as an expansion in powers of $g$. 

As one step further for resumming graphs, Andersen, Braaten and Strickland introduced \emph{hard-thermal-loop perturbation theory} (HTLpt) \cite{htl1}, which is essentially a reorganization of the conventional perturbation expansion for QCD that selectively resums higher order effects related to quasiparticles and screening.

\section{Formalism}
\label{formalism}

HTLpt is formulated in Minkowski space, therefore it applies to both thermodynamics and real-time dynamics straightforwardly. We use pure-glue QCD next as an example to show the formalism of the theory. Note that all the discussions here apply equally to the case with quarks. The Minkowskian Lagrangian density that generates the perturbative expansion for pure-glue QCD can be expressed in the form
\bqa
{\mathcal L}_{\rm QCD}\;=\;-{1\over2}{\rm Tr}\left(G_{\mu\nu}G^{\mu\nu}\right)+{\mathcal L}_{\rm gf}+{\mathcal L}_{\rm ghost}+\Delta{\mathcal L}_{\rm QCD},
\label{L-QCD}
\eqa
where $G_{\mu\nu}=\partial_{\mu}A_{\nu}-\partial_{\nu}A_{\mu}-ig[A_{\mu},A_{\nu}]$ is the gluon field strength and $A_{\mu}$ is the gluon field expressed as a matrix in the SU($N_c$) algebra. The ghost term ${\mathcal L}_{\rm ghost}$ depends on the choice of the gauge-fixing term ${\mathcal L}_{\rm gf}$.

The perturbative expansion in powers of $g$ generates ultraviolet divergences. The renormalizability of perturbative QCD guarantees that all divergences in physical quantities can be removed by renormalization of the coupling constant $\alpha_s= g^2/4 \pi$. There is no need for wavefunction renormalization, because physical quantities are independent of the normalization of the field. There is also no need for renormalization of the gauge parameter, because physical quantities are independent of the gauge parameter.

Hard-thermal-loop perturbation theory is a reorganization of the perturbation series for thermal gauge theories with the Lagrangian density written as
\bqa
{\mathcal L}\;=\;\left({\mathcal L}_{\rm QCD} + {\mathcal L}_{\rm HTL} \right) \Big|_{g \to \sqrt{\delta} g} + \Delta{\mathcal L}_{\rm HTL} \; .
\label{L-HTLQCD}
\eqa
The HTL-improvement term appearing above is $(1-\delta)$ times the isotropic HTL effective action which generates all HTL $n$-point functions~\cite{Braaten:1991gm}
\bqa
\label{L-HTL-IMP}
{\mathcal L}_{\rm HTL}\;=\;-{1\over2}(1-\delta)m_D^2 {\rm Tr}\left(G_{\mu\alpha}\left\langle {y^{\alpha}y^{\beta}\over(y\cdot D)^2}\right\rangle_{\!\!\hat{\bf y}}G^{\mu}_{\;\;\beta}\right) \; ,
\eqa
where $D_{\mu}$ is the covariant derivative in the adjoint representation, $y^{\mu}=(1,\hat{{\bf y}})$ is a light-like four-vector, and $\langle\ldots\rangle_{\hat{\bf y}}$ represents the average over the directions of $\hat{{\bf y}}$. The term~(\ref{L-HTL-IMP}) has the form of the effective Lagrangian that would be induced by a rotationally-invariant ensemble of charged sources with infinitely high momentum. Note that the covariant derivatives in the denominators make the HTL-improvement terms gauge invariant by modifying all $n$-point functions self-consistently. The parameter $m_D$ can be identified with the Debye screening mass.

HTLpt is defined by treating $\delta$ as a formal expansion parameter~\cite{delta}~\footnote{For applications of this so-called linear delta expansion other than HTLpt, please see Ref.~\cite{delta-application} for a broad but far from complete list.}. Physical observables are calculated in HTLpt by expanding them in powers of $\delta$, truncating at some specified order, and then setting $\delta=1$. This defines a reorganization of the perturbation series in which the effects of the $m_D^2$ term in~(\ref{L-HTL-IMP}) are included to all orders but then systematically subtracted out at higher orders in perturbation theory by the $\delta m_D^2$ term in~(\ref{L-HTL-IMP}). If we set $\delta=1$, the Lagrangian (\ref{L-HTLQCD}) reduces to the QCD Lagrangian (\ref{L-QCD}); while the free Lagrangian, which reads ${\cal L}_{\rm QCD} +{\cal L}_{\rm HTL}$, is obtained by setting $\delta=0$ and describes gluon quasiparticles with screening masses $m_D$. 

We stress here that the $\delta$ expansion is equivalent to loop expansion, i.e. leading order (LO) $\delta$ expansion is one loop, next-to-leading order (NLO) is two loops, next-to-next-to-leading order (NNLO) is three loops, and so on. If the expansion in $\delta$ could be calculated to all orders, the final result would not depend on $m_D$ when we set $\delta=1$. However, any truncation of the expansion in $\delta$ produces results that depend on $m_D$. Some prescription is required to determine $m_D$ as a function of $T$ and $\alpha$. In the next two chapters we will discuss different mass prescriptions.

The HTL perturbation expansion generates ultraviolet divergences. In QCD perturbation theory, renormalizability constrains the ultraviolet divergences to have a form that can be cancelled by the counterterm Lagrangian $\Delta{\cal L}_{\rm QCD}$. We will demonstrate that renormalized perturbation theory can be implemented by including a counterterm Lagrangian $\Delta{\cal L}_{\rm HTL}$ among the interaction terms in (\ref{L-HTLQCD}). There is no proof that the HTL perturbation expansion is renormalizable, so the general structure of the ultraviolet divergences is not known; however, it was shown in previous papers \cite{htl2} that it was possible to renormalize the NLO HTLpt prediction for the free energy of QCD using only a vacuum counterterm, a Debye mass counterterm, and a fermion mass counterterm. In this dissertation we will show that renormalization is also possible at NNLO.

The free Lagrangian in general covariant gauge is obtained by setting $\delta=0$ in~(\ref{L-HTLQCD}):
\bqa
\nonumber
{\mathcal L}_{\rm free}\!\!&=&\!\!-{\rm Tr}\left(\partial_{\mu}A_{\nu}\partial^{\mu}A^{\nu}-\partial_{\mu}A_{\nu}\partial^{\nu}A^{\mu}\right)-{1\over\xi}{\rm Tr}\left[\left(\partial^{\mu}A_{\mu}\right)^2\right] \\
&&-\;{1\over2}m_D^2\mbox{Tr}\left[(\partial_{\mu}A_{\alpha}-\partial_{\alpha}A_{\mu})\left\langle{y^{\alpha} y^{\beta}\over(y\cdot\partial)^2}\right\rangle_{\!\!\hat{\bf y}}(\partial^{\mu}A_{\beta}-\partial_{\beta}A^{\mu})\right].
\label{L-free}
\eqa
The resulting propagator is the HTL gluon propagator and the remaining terms in~(\ref{L-HTLQCD}) are treated as perturbations. The propagator can be decomposed into transverse and longitudinal pieces which in Minkowski space are given by
\bqa
\Delta_T(p)\!\!&=&\!\!{1 \over p^2 - \Pi_T(p)} \;,
\label{Delta_T}
\\
\Delta_L(p)\!\!&=&\!\!{1 \over -n_p^2 p^2+\Pi_L(p)} \;,
\label{Delta_L}
\eqa
where $n_p^{\mu} = n^{\mu} - p^{\mu}(n\!\cdot\!p / p^2)$ with $n = (1,{\bf 0})$ being the vector that specifies the thermal rest frame, and $\Pi_T$ and $\Pi_L$ are the transverse and longitudinal self-energies, respectively, and read
\bqa
\Pi_T(p)\!\!&=&\!\! {m_D^2 \over (d-1)n_p^2} \left[ {\cal T}^{00}(p,-p) - 1 + n_p^2 \right] \;,
\label{Pi_T}
\\
\Pi_L(p)\!\!&=&\!\! m_D^2 \left[ 1 - {\cal T}^{00}(p,-p) \right]\;,
\label{Pi_L}
\eqa
with ${\cal T}^{00}(p,-p)$ defined in~(\ref{T00-int}). Note that there are also HTL vertex corrections which are given by similar but somewhat more complicated expressions which can be found in Refs.~\cite{Braaten:1989mz,htl2,Mrowczynski:2004kv}.

As mentioned above, HTLpt is a systematic framework for performing calculations in thermal gauge theories which is gauge invariant by construction. It systematically includes several physical effects of the plasma such as the propagation of massive particles, screening of interactions, and Landau damping. We briefly comment on these issues next.
\begin{figure}
\begin{center}
\includegraphics[width=10cm]{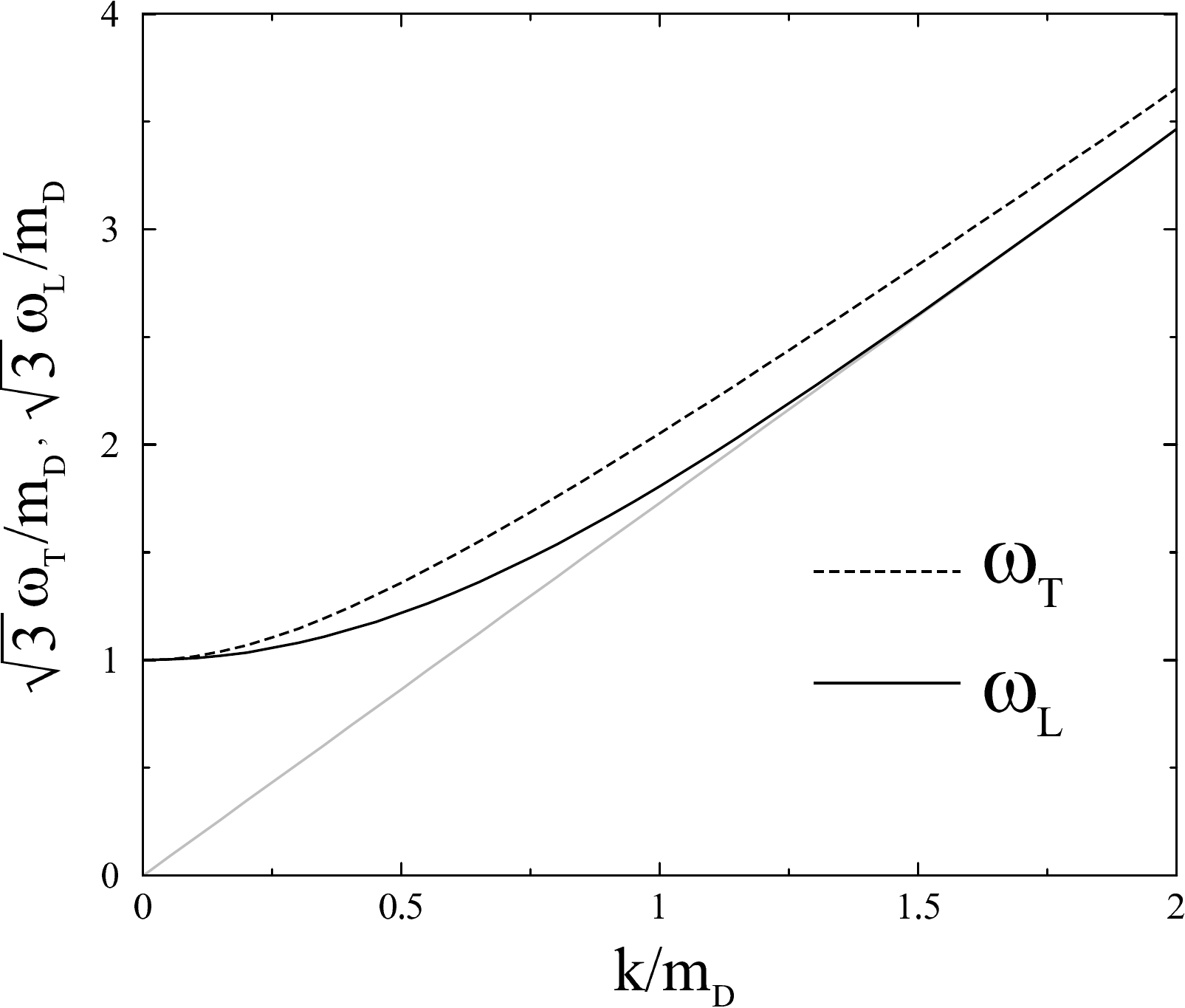}
%\centerline{\psfig{file=figures/dispersion,width=10cm}}
\end{center}
\vspace*{8pt}
\caption{Longitudinal and transverse dispersion relations. This figure is adapted from Ref.~\cite{Andersen:2004fp}.}
\label{gdisp}
\end{figure}

\subsection{Massive quasiparticles}

The HTL self-energies are included in the zeroth order propagators which results in the resummed HTL propagator for gluons. The dispersion relations for the transverse and longitudinal gluonic degrees of freedom are determined by locating the zeros of the inverse propagator which gives the following two equations
\bqa
\omega_T^2-k^2-\Pi_T(\omega_T,k)\!\!&=&\!\!0 \;, \\
k^2+\Pi_L(\omega_L,k)\!\!&=&\!\!0 \;.
\eqa
The dispersion relations for transverse and longitudinal gluons are shown in Fig.~\ref{gdisp}.  As can been seen from this Figure both modes approach a constant in the limit of small momentum and approach the light-cone in the limit of large momentum.

\subsection{Screening}

HTLpt also includes screening of interactions which can be seen by examining the static limit of the HTL propagators. For instance, the inclusion of the longitudinal self-energy changes the Coulomb potential of two static charges in the plasma to a Yukawa potential:
\bqa
\lim_{\omega\rightarrow0}\Delta_L(\omega,k)\;=\;{1 \over k^2+m_D^2} \;,
\eqa
This result shows that chromoelectric fields are screened on a scale $r \sim m_D^{-1}$.
Likewise, the screening of long wavelength chromomagnetic fields is determined by the transverse propagator for small frequencies
\bqa
\Delta_T(\omega,k)\;\sim\;{1\over k^2 + i {\pi\over 4} m_D^2 \omega/k}\;.
\label{ms}
\eqa
This shows that there is no screening of static magnetic fields meaning that the magnetic mass problem of non-Abelian gauge theories at high temperature is not solved by HTL resummation.  However, HTL resummation does give {\it dynamical} screening at a scale $r\sim\left(m^2_D\omega\right)^{-{1 \over 3}}$. Note that the divergences associated with the absence of static magnetic screening do not pose a problem until four-loop order; however, at four loops the lack of static magnetic screening gives rise to infrared divergences that cause perturbation theory to break down.

\subsection{Landau damping}

The transverse and longitudinal HTL self-energies also contain the physics of Landau damping.  Landau damping represents a transfer of energy from the soft modes to the hard modes for spacelike momentum. This can be seen from the analytic structure of the self-energies given by Eqs.~(\ref{redt}) and (\ref{redl}).  Because of the logarithms appearing in these functions there is an imaginary contribution to the self-energies for $-k<\omega<k$ which gives the rate of energy transfer from the soft to hard modes. Note that ignoring this contribution leads to gauge variant and   unrenormalizable results.

\section{Technicalities}

Calculations in HTLpt are much more difficult than in ordinary perturbative QCD, because the Feynman rules are more complicated. In spite of the complexity of the Feynman rules, calculations do appear to be tractable with the help of the following techniques. Due to the fact that only thermodynamic quantities are considered in this dissertation, from now on we switch the discussions to Euclidean space for convenience.

\subsection{Mass expansion}

The calculation of the free energy in HTLpt involves the evaluation of vacuum diagrams. In Refs.~\cite{htl1,htl2}, the free energy was reduced to scalar sum-integrals. The one-loop free energy were evaluated exactly by replacing the sums by contour integrals, extracting the poles in $\epsilon$, and then reducing the momentum integrals to integrals that were at most two-dimensional and could therefore be easily evaluated numerically. Evaluating two-loop free energy exactly would involve the evaluation of five-dimensional numerical integrals which turned out to be intractable. Therefore attacking the third loop in this way is hopeless. 

The fact that $m_D \sim gT$ suggests that $m_D/T$ can be treated as an expansion parameter of order $g$ in terms of which the sum-integrals can be further expanded~\cite{Andersen:2001ez}. It was shown that the first few terms in the $m_D/T$ expansion of the sum-integrals gave a surprisingly accurate approximation to the exact result~\cite{htl1,Andersen:2001ez}. We will adopt this mass expansion trick in the calculation of three-loop HTL free energy in the next two chapters. We will carry out the $m_D/T$ expansion to high enough order to include all terms through order $g^5$ if $m_D /T$ is taken to be of order $g$. The two-loop approximation will be perturbatively accurate to order $g^3$ and the three-loop approximation accurate to order $g^5$. We demonstrate next how the mass expansion works by using the simplest example of one-loop photon diagram (1a) in Fig.~\ref{fig:dia1}.

The expression of the one-loop photon diagram (1a) in Fig.~\ref{fig:dia1} after taking into account the ghost contribution is,
\bqa
{\cal F}_{\rm 1a}\;=\;-{1\over2}\sumint_{P}\!\left\{(d-1)\log\left[-\Delta_T(P)\right]+\log\Delta_L(P)\right\}.
\eqa
After plugging in~(\ref{Delta_T}) and (\ref{Delta_L}) for $\Delta_{T/L}$ with $\Pi_{T/L}$ defined in~(\ref{Pi_T}) and (\ref{Pi_L}) and expanding to second order in $m_D^2$, the hard contribution from (1a) reads,
\bqa\nonumber
{\cal F}_{\rm 1a}^{(h)}\!\!&=&\!\!{d-1\over2}\sumint_P\log\left(P^2\right)+{1\over2}\sumint_P\log\left(p^2\right)+{m_D^2\over2}\sumint_P{1\over P^2}-{m_D^4\over4(d-1)}
\\ &&
\times\sumint_P\left[{1\over P^4}+{d\over p^4}-{2\over p^2P^2}-{2d\over p^4}{\cal T}_P+{2\over p^2P^2}{\cal T}_P+{d\over p^4}\left({\cal T}_P\right)^2\right]\;.
\eqa
Note that the integrands $\log\!\left(p^2\right)$ and $1/p^4$ have no scale, so the corresponding sum-integrals vanish in dimensional regularization. Finally, the hard contribution from (1a) becomes
\bqa\nonumber
{\cal F}_{\rm 1a}^{(h)}\!\!&=&\!\!{d-1\over2}\sumint_P\log\left(P^2\right)+{m_D^2\over2}\sumint_P{1\over P^2}
\\ &&
-\;{m_D^4\over4(d-1)}\sumint_P\left[{1\over P^4}-{2\over p^2P^2}-{2d\over p^4}{\cal T}_P+{2\over p^2P^2}{\cal T}_P+{d\over p^4}\left({\cal T}_P\right)^2\right]\;,
\eqa
with sum-integrals listed in App.~\ref{app:sumint}. All the other diagrams are to be evaluated in the same spirit.

\subsection{Simplified $\delta$ expansion}

We have introduced the $1-\delta$ description in~(\ref{L-HTL-IMP}). The purpose of doing so is to distinguish interactions from the free part in the Lagrangian by associating every interaction term with a label $\delta$. The subtracted $m_D^2$ term in~(\ref{L-HTLQCD}) generates self-energy and vertex insertions that systematically eliminate the effects of the added $m_D^2$ term from lower orders. The number of the diagrams with self-energy and vertex insertions grows exponentially as we go to higher and higher orders in the $\delta$ expansion. Therefore evaluating each diagram individually would become hopeless at higher loop order. Since all the self-energy and vertex insertions originate from the $(1-\delta)m_D^2$ term in~(\ref{L-HTL-IMP}), the diagrams with self-energy and vertex insertions can be obtained by substituting $m_D^2\rightarrow(1-\delta)m_D^2$ in the original diagrams and expanding to appropriate order in $\delta$. The $\delta$ expansion for hard contributions is trivial since the $m_D$ dependence in hard modes only enters as multiplicative factors which are of even order in $m_D$. The $\delta$ expansion for soft contributions are much more involved due to the fact that $m_D$ also appears in denominators for the soft contributions. We use again the one-loop photon diagram (1a) in Fig.~\ref{fig:dia1} next to show how to carry out the $\delta$ expansion.

The soft contribution of the one-loop photon diagram (1a) in Fig.~\ref{fig:dia1} reads,
\bqa
{\cal F}^{(s)}_{\rm 1a}\!\!&=&\!\!{1\over2}T\int_{\bf p}\log\left(p^2+m_D^2\right)\;.
\eqa
After substituting $m_D^2\rightarrow(1-\delta)m_D^2$ and expanding to order $\delta^2$ to include all terms through $g^5$, we obtain
\bqa\nonumber
&&\!\!{1\over2}T\int_{\bf p}\log\left[p^2+(1-\delta)m_D^2\right]\\
&=&\!\! {1\over2}T\int_{\bf p}\log\left(p^2+m_D^2\right) - {\delta\over2}m_D^2T\int_{\bf p}{1\over p^2+m_D^2} - {\delta^2\over4}m_D^4T\int_{\bf p}{1\over(p^2+m_D^2)^2} \nonumber \\
&=&\!\!{\cal F}^{(s)}_{\rm 1a}+\delta{\cal F}^{(s)}_{\rm 2c}+\delta^2{\cal F}^{(s)}_{\rm 3h}\;,
\eqa
from which the one-loop photon diagram with one and two self-energy insertions, i.e. (2c) in Fig.~\ref{fig:dia1} and (3h) in Fig.~\ref{fig:dia2}, are generated  systematically. In the following chapters we will show that with the help of $\delta$ expansion, the evaluation of the diagrams with self-energy and vertex insertions becomes incredibly simple and straightforward.

%%%%%%%%%%%%%%%%%%%%%%%%%%%%%%%%%%%%%%%%%%%%%%%%%%%%%%%%%%%%%
%
%	Include File:			DON'T COMPILE !!!
%
%%%%%%%%%%%%%%%%%%%%%%%%%%%%%%%%%%%%%%%%%%%%%%%%%%%%%%%%%%%%%

\chapter{QED Thermodynamics to Three Loops}\label{chapter:qed}

The thermodynamics of QED is studied in this chapter using the hard-thermal-loop perturbation theory reorganization of finite-temperature gauge theory. We calculate the free energy through three loops by a dual expansion in $m_D/T$, $m_f/T$ and $e^2$, where $m_D$ and $m_f$ are thermal masses and $e$ is the coupling constant. The results demonstrate that the hard-thermal-loop perturbation reorganization improves the convergence of the successive approximations to the QED free energy at large coupling, $e \sim 2$. The reorganization is gauge invariant by construction, and due to cancellation among various contributions during renormalization, we obtain a completely analytic result for the resummed thermodynamic potential at three loops. This chapter is based on: {\it Three-loop HTL free energy for QED}, J.~O.~Andersen, M.~Strickland and N.~Su, Phys.\ Rev.\  D {\bf 80}, 085015 (2009). 

\section{Introduction}

\begin{figure}[t]
\begin{center}
\vspace{6mm}
\includegraphics[width=10cm]{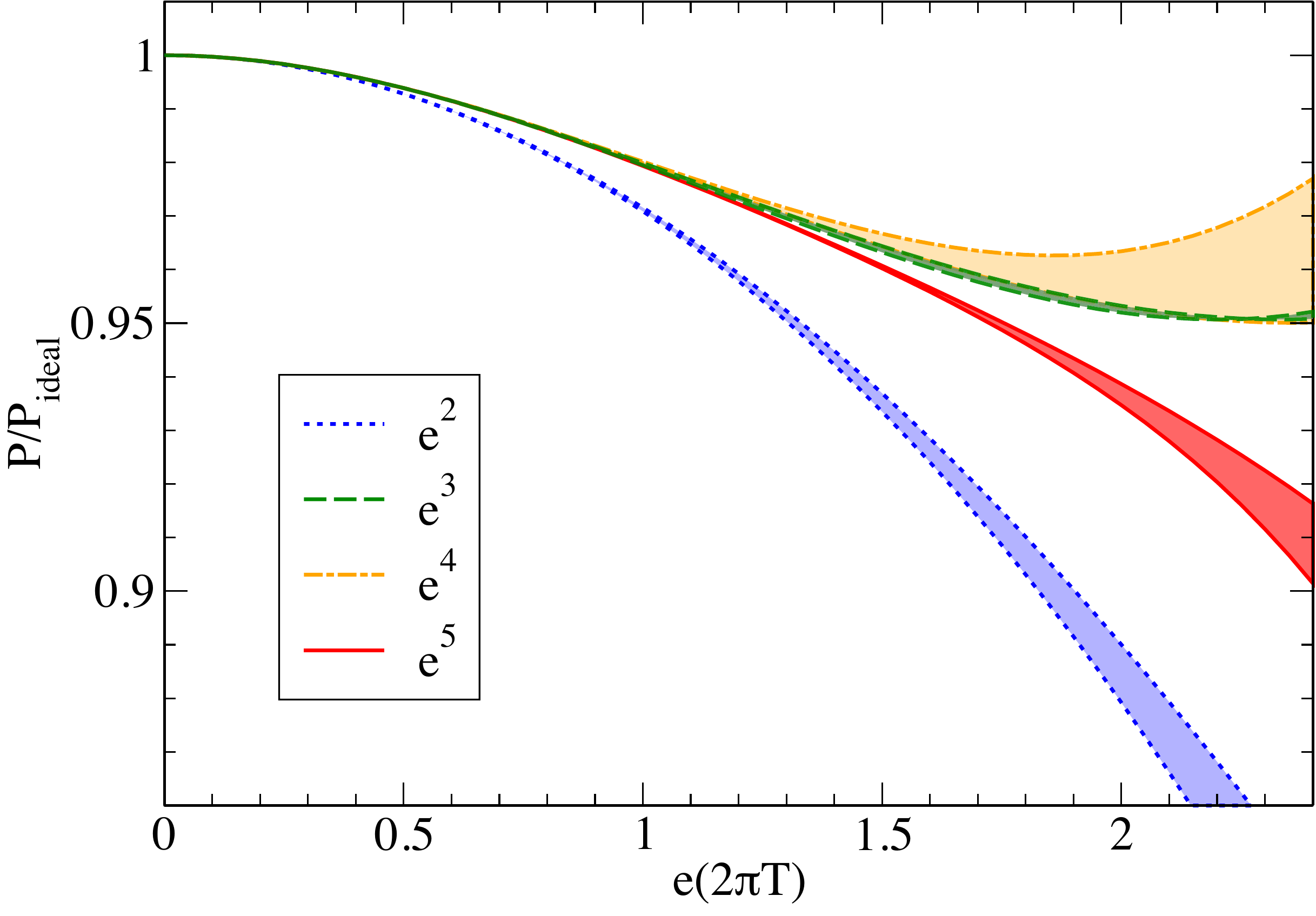}
\end{center}
\vspace{-3mm}
\caption{Successive perturbative approximations to the QED pressure (negative of the free energy). Each band corresponds to a truncated weak-coupling expansion accurate to order $e^2$, $e^3$, $e^4$, and $e^5$, respectively. Shaded bands correspond to variation of the renormalization scale $\mu$ between $\pi T$ and $4 \pi T$.}
\label{fig:pertpressure}
\end{figure}

The weak-coupling expansion of the QED free energy is known to order $e^5$~\cite{AZ-95,Parwani:1994xi,Parwani:1994je,Andersen:1995ej,Zhai:1995ac}. In Fig.~\ref{fig:pertpressure} we show the successive perturbative approximations to the QED free energy. As can be seen from this figure, at couplings larger than $e \sim 1$ the QED weak-coupling approximations also exhibit poor convergence which is as bad as its counterpart in QCD. 

In spite of the complexity of the Feynman rules, calculations with HTLpt do appear to be tractable. Andersen, Braaten, Petitgirard, and Strickland have demonstrated this by calculating the next-to-leading order (NLO) free energy for QCD~\cite{htl2}. Although their results showed striking improvement of convergence comparing to the naive weak-coupling expansion, there were still problems remained at two-loop order. First, both the leading order (LO) and NLO free energies have a wrong curvature below 2$T_c$. Instead of going down the results rise up towards $T_c$. This is due to the fact that the truncation order in the dual expansion was $g^5$ and the NLO approximation is only perturbatively accurate to order $g^3$. The missing $g^4$ and $g^5$ terms will enter at three loops. Second, in the NLO renormalization only vacuum and mass counterterms were needed, therefore the self-consistent running coupling could not be derived systematically from the calculation and it had to be added by hand in the results. The coupling constant renormalization also enters at three-loop order. Therefore it is clear that in order to complete the calculation, we need to attack the third loop. However, comparing to Abelian case, a direct three-loop non-Abelian calculation might cause unnecessary complications which should not be the main concern, we therefore decided to use QED as a test case to develop the necessary techniques for attacking QCD.

%%%%%%%%%%%%%%%%%%%%%%%%%%%%%%%%%%%%%%%%%%%%%%%%%%%%%%%%%%%%%%%%
%%%%%%%%%%%%%%%%%%%%%%%%%%%%%%%%%%%%%%%%%%%%%%%%%%%%%%%%%%%%%%%%

\section{HTL perturbation theory}

\label{HTLpt}

The Lagrangian density for massless QED in Minkowski space is
\bqa
{\cal L}_{\rm QED}\;=\;-{1\over4}F_{\mu\nu}F^{\mu\nu}+i \bar\psi \gamma^\mu D_\mu \psi+{\cal L}_{\rm gf}+{\cal L}_{\rm gh}+\Delta{\cal L}_{\rm QED}\;.
\label{L-QED}
\eqa
Here the field strength is $F^{\mu\nu}=\partial^{\mu}A^{\nu}-\partial^{\nu}A^{\mu}$ and the covariant derivative is $D^{\mu}=\partial^{\mu}+ieA^{\mu}$. The ghost term ${\cal L}_{\rm gh}$ depends on the gauge-fixing term ${\cal L}_{\rm gf}$. In this chapter we choose the class of covariant gauges where the gauge-fixing term is
\bqa
{\cal L}_{\rm gf}\;=\;-{1\over2\xi}\left(\partial_{\mu}A^{\mu}\right)^2\;,
\eqa
with $\xi$ being the gauge-fixing parameter. In this class of gauges, the ghost term decouples from the other fields.

The HTLpt Lagrangian density for QED is written as
\bqa
{\cal L}\;=\;\left({\cal L}_{\rm QED} + {\cal L}_{\rm HTL} \right) \Big|_{e \to \sqrt{\delta} e} + \Delta{\cal L}_{\rm HTL}\;,
\label{L-HTLQED}
\eqa
where the HTL-improvement term reads
\bqa
{\cal L}_{\rm HTL}\;=\;-{1\over4}(1-\delta)m_D^2 F_{\mu\alpha}\left\langle {y^{\alpha}y^{\beta}\over(y\cdot\partial)^2}\right\rangle_{\!\!\hat{\bf y}}F^{\mu}_{\;\;\beta}+(1-\delta)\,i m_f^2 \bar{\psi}\gamma^\mu\left\langle {y_{\mu}\over y\cdot D}\right\rangle_{\!\!\hat{\bf y}}\psi
	\;,
\label{L-HTL}
\eqa
with the parameter $m_D$ identified with the Debye screening mass and the parameter $m_f$ identified as the induced finite-temperature electron mass.

Although the renormalizability of the HTL perturbation expansion has not yet been proven, the renormalization was achieved at NLO for the free energy of QCD using only a vacuum energy counterterm, a Debye mass counterterm, and a quark mass counterterm~\cite{htl2}. In this chapter we will show that this is also possible at next-to-next-to-leading order (NNLO) with the introduction of a new coupling constant counterterm which coincides with its perturbative value at zero temperature giving rise to the standard one-loop running. The necessary counterterms are
\bqa
\delta\Delta\alpha\!\!&=&\!\!N_f{\alpha^2\over3\pi\epsilon}\delta^2\;,
\label{delalpha} \\ 
\Delta m_D^2\!\!&=&\!\!N_f\left({\alpha\over3\pi\epsilon}+{\cal O}(\delta^2\alpha^2)
\right)(1-\delta)m_D^2\;,
\label{delmd} \\ 
\Delta m_f^2\!\!&=&\!\!\left(-{3\alpha\over4\pi\epsilon}+{\cal O}(\delta^2\alpha^2)
\right)(1-\delta)m_f^2\;,
\label{delmf}\\
\Delta{\cal E}_0\!\!&=&\!\!\left({1\over128\pi^2\epsilon}+{\cal O}(\delta\alpha)
\right)(1-\delta)^2m_D^4\;.
\label{del1e0}
\eqa

As discussed in Section~\ref{formalism} a prescription is required to determine $m_D$ and $m_f$ as a function of $T$ and $\alpha$ when truncating at a finite order in $\delta$. As one possibility we will treat both as variational parameters that should be determined by minimizing the free energy. If we denote the free energy truncated at some order in $\delta$ by $\Omega(T,\alpha,m_D,m_f,\mu,\delta)$, our prescription is
\bqa
{\partial \ \ \over \partial m_D}\Omega(T,\alpha,m_D,m_f,\mu,\delta=1)\!\!&=&\!\!0 \;, 
\label{gapmd}\\
{\partial \ \ \over \partial m_f}\Omega(T,\alpha,m_D,m_f,\mu,\delta=1)\!\!&=&\!\!0 \;.
\label{gapmf}
\eqa
Since $\Omega(T,\alpha,m_D,m_f,\mu,\delta=1)$ is a function of the variational parameters $m_D$ and $m_f$, we will refer to it as the {\it thermodynamic potential}. We will refer to the variational equations (\ref{gapmd}) and (\ref{gapmf}) as the {\it gap equations}. The free energy ${\cal F}$ is obtained by evaluating the thermodynamic potential at the solution to the gap equations (\ref{gapmd}) and (\ref{gapmf}). Other thermodynamic functions can then be obtained by taking appropriate derivatives of ${\cal F}$ with respect to $T$.

\section{Diagrams for the thermodynamic potential}

\begin{figure*}[t]
\begin{center}
\includegraphics[width=7.5cm]{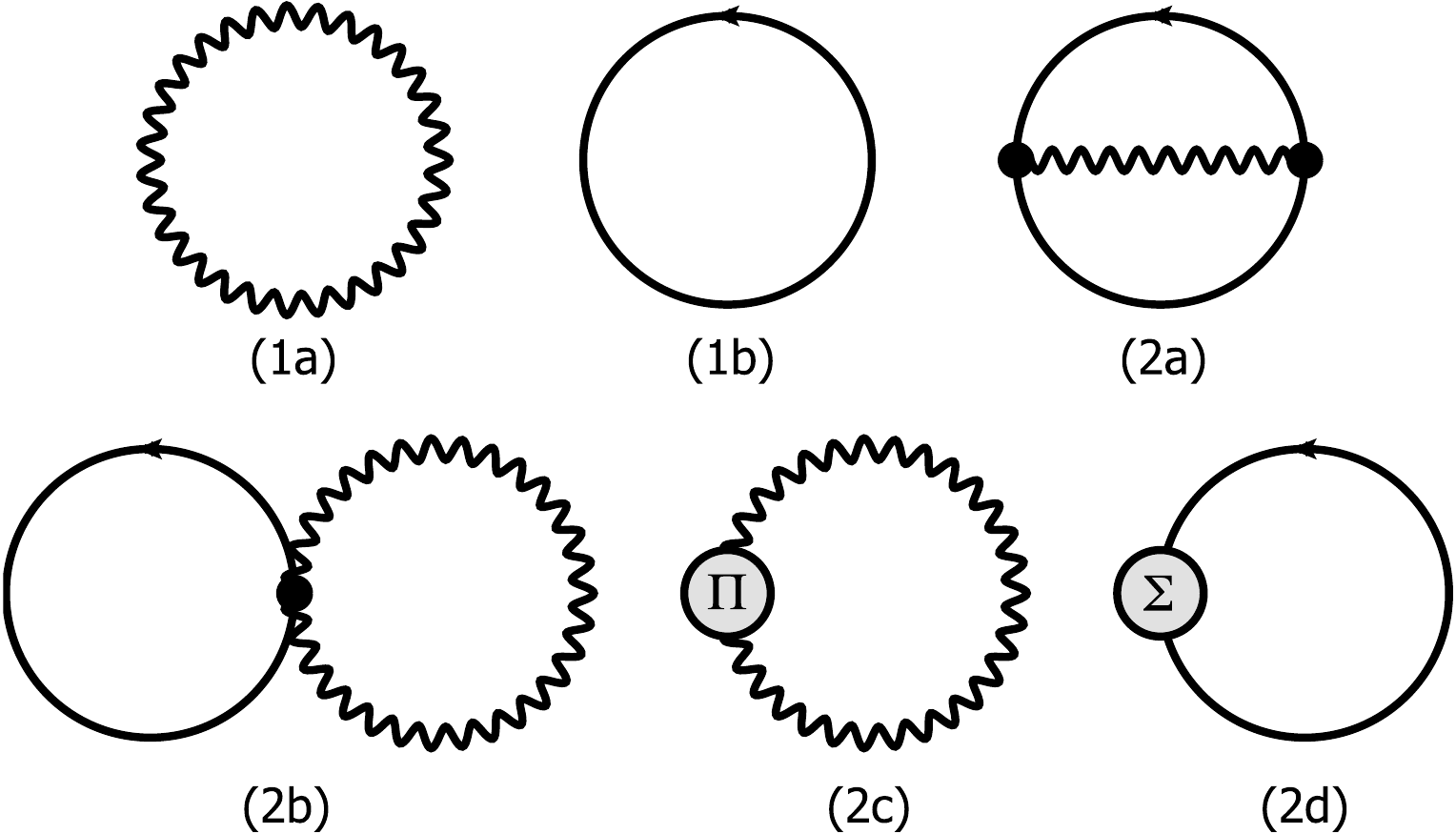}
\end{center}
\caption{Diagrams contributing through NLO in HTLpt. The undulating lines are photon propagators and the solid lines are fermion propagators. A circle with a $\Pi$ indicates a photon self-energy insertion and a circle with a $\Sigma$ indicates a fermion self-energy insertion. All propagators and vertices shown are HTL-resummed propagators and vertices.}
\label{fig:dia1}
\end{figure*}

\begin{figure*}[t]
\begin{center}
\includegraphics[width=14.2cm]{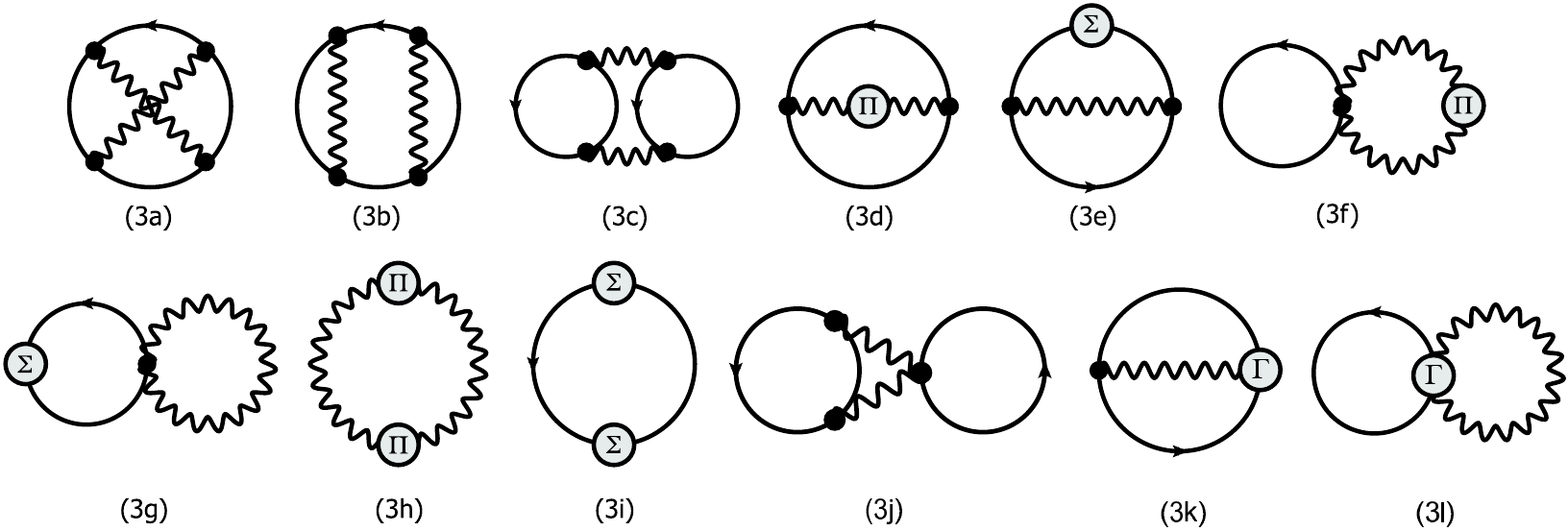}
\end{center}
\caption{Diagrams contributing to NNLO in HTLpt through order $e^5$. The undulating lines are photon propagators and the solid lines are fermion propagators. A circle with a $\Pi$ indicates a photon self-energy insertion and a circle with a $\Sigma$ indicates a fermion self-energy insertion. The propagators are HTL-resummed propagators and the black dots indicate HTL-resummed vertices. The lettered vertices indicate that only the HTL correction is included.}
\label{fig:dia2}
\end{figure*}

\begin{figure*}[t]
\begin{center}
\includegraphics[width=11cm]{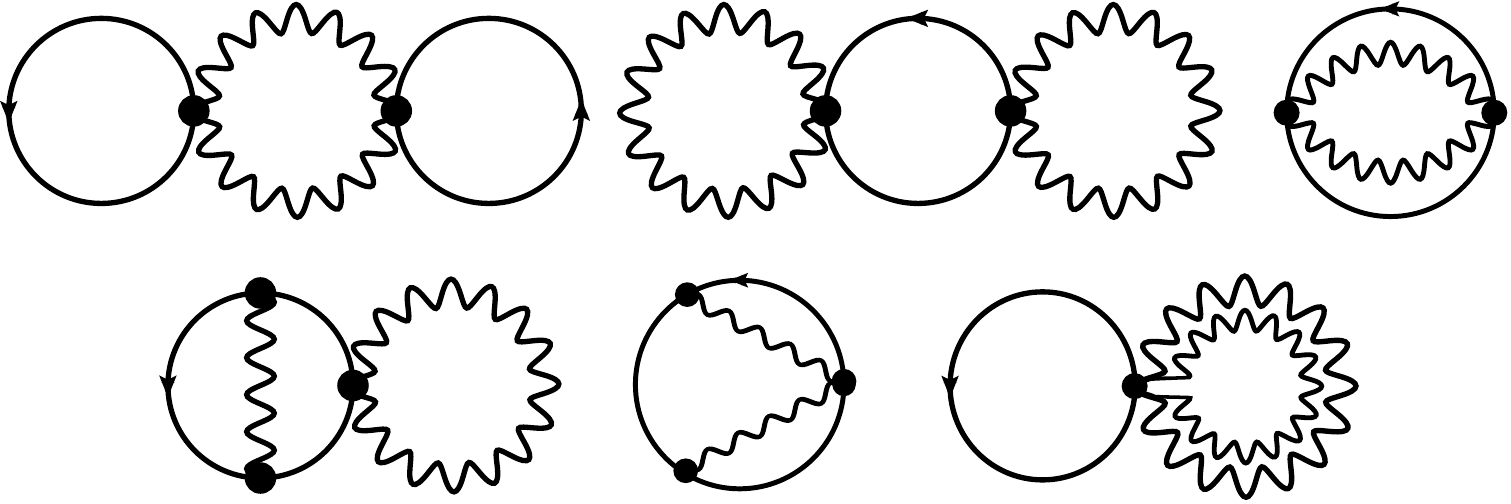}
\end{center}
\caption{Diagrams contributing to NNLO in HTLpt beyond order $e^5$.  The diagrams in the first line above first contribute at order $e^8$ and the second line at order $e^6$. The undulating lines are photon propagators and the solid lines are fermion propagators. All propagators and vertices shown are HTL-resummed propagators and vertices.}
\label{fig:dia3}
\end{figure*}

In this section, we list the expressions for the diagrams that contribute to the thermodynamic potential through order $\delta^2$, aka NNLO, in HTL perturbation theory. The diagrams are shown in Figs.~\ref{fig:dia1}, \ref{fig:dia2}, and \ref{fig:dia3}. Because of our dual truncation in $m_D$, $m_f$, and $e$ the diagrams listed in Fig.~\ref{fig:dia3} do not contribute to our final expression so we will not explicitly list their integral representations. The expressions here will be given in Euclidean space; however, in Appendix~\ref{app:rules} we present the HTLpt Feynman rules in Minkowski space.

%%%%%%%%%%%%%%%%%%%%%%%%%%%%%%%%%%%%%%%%%%%%%%%%%%%%%%%%%%%%%%%%
%%%%%%%%%%%%%%%%%%%%%%%%%%%%%%%%%%%%%%%%%%%%%%%%%%%%%%%%%%%%%%%%

%%%%%%%%%%%%%%%%%%%%%%%%%%%%%%%%%%%%%%%%%%%%%%%%%%%%%%%%%%%%%%%%
%%%%%%%%%%%%%%%%%%%%%%%%%%%%%%%%%%%%%%%%%%%%%%%%%%%%%%%%%%%%%%%%

%%%%%%%%%%%%%%%%%%%%%%%%%%%%%%%%%%%%%%%%%%%%%%%%%%%%%%%%%%%%%%%%
%%%%%%%%%%%%%%%%%%%%%%%%%%%%%%%%%%%%%%%%%%%%%%%%%%%%%%%%%%%%%%%%

The thermodynamic potential at leading order in HTL perturbation theory for QED with $N_f$ massless electrons is
\bqa
\Omega_{\rm LO}\;=\;{\cal F}_{\rm 1a} + N_f{\cal F}_{\rm 1b}+\Delta_0{\cal E}_0\;.
\eqa
Here, ${\cal F}_{\rm 1a}$ is the contribution from the photons
\bqa
{\cal F}_{\rm 1a}\;=\;-{1\over2}\sumint_{P}\!\left\{(d-1)\log\left[-\Delta_T(P)\right]+\log\Delta_L(P)\right\}.
\eqa
The transverse and longitudinal HTL propagators $\Delta_T(P)$ and $\Delta_L(P)$ are given in (\ref{Delta-T}) and (\ref{Delta-L}). The electron contribution is
\bqa
\label{lq}
{\cal F}_{\rm 1b}\;=\;-\sumint_{\{P\}}\log\det\left[P\!\!\!\!/-\Sigma(P)\right]\;.
\eqa
The leading-order vacuum energy counterterm $\Delta_0{\cal E}_0$ is given by
\bqa
\Delta_0{\cal E}_0\;=\;{1\over128\pi^2\epsilon} m_D^4 \;.
\label{lovac}
\eqa

The thermodynamic potential at NLO in HTL perturbation theory can be written as
\bqa\nonumber
\Omega_{\rm NLO}\!\!&=&\!\!\Omega_{\rm LO} + N_f \left({\cal F}_{\rm 2a} + {\cal F}_{\rm 2b} + {\cal F}_{\rm 2d}\right) + {\cal F}_{\rm 2c} + \Delta_1{\cal E}_0 \\
&& + \; \Delta_1 m_D^2{\partial\over\partial m_D^2}\Omega_{\rm LO} + \Delta_1 m_f^2{\partial\over\partial m_f^2}\Omega_{\rm LO}\;,
\label{OmegaNLO}
\eqa
where $\Delta_1{\cal E}_0$, $\Delta_1m_D^2$, and $\Delta_1m_f^2$ are the terms of order $\delta$ in the vacuum energy and mass counterterms:
\bqa
\label{dvac1}
\Delta_1{\cal E}_0\!\!&=&\!\!-{1\over64\pi^2\epsilon}m_D^4\;, \\
\label{dmd1}
\Delta_1m_D^2\!\!&=&\!\!N_f{\alpha\over3\pi\epsilon}m_D^2\;, \\ 
\Delta_1m_f^2\!\!&=&\!\!-{3\alpha\over4\pi\epsilon}m_f^2\;.
\label{dmf1}
\eqa
The contributions from the two-loop diagrams with electron-photon three- and four-point vertices are
\bqa
{\cal F}_{\rm 2a}\!\!&=&\!\!{1\over2}e^2\sumint_{P\{Q\}}\mbox{Tr}\left[\Gamma^{\mu}(P,Q,R)S(Q)\Gamma^{\nu}(P,Q,R)S(R)\right]\Delta^{\mu\nu}(P)\;,
\label{3qg}
\\ 
{\cal F}_{\rm 2b}\!\!&=&\!\!{1\over2}e^2\sumint_{P\{Q\}}\mbox{Tr}\left[\Gamma^{\mu\nu}(P,-P,Q,Q)S(Q)\right]\Delta^{\mu\nu}(P)\;,
\label{4qg}
\eqa
where $R=Q-P$.
The contribution from the HTL photon counterterm diagram with a single photon self-energy insertion is
\bqa 
{\cal F}_{\rm 2c}\;=\;{1\over2}\sumint_{P}\Pi^{\mu\nu}(P)\Delta^{\mu\nu}(P)\;.
\eqa
The contribution from the HTL electron counterterm diagram with a single electron self-energy insertion is
\bqa
\label{Sigma}
{\cal F}_{\rm 2d}\;=\;-\sumint_{\{P\}}\mbox{Tr}\left[\Sigma(P)S(P)\right]\;.
\eqa
The role of the counterterm diagrams (2c) and (2d) is to avoid overcounting of diagrams when using effective propagators in (1a) and (1b). Similarly, the role of counterterm diagram (3k) is to avoid overcounting when using effective vertices in (2a).

The thermodynamic potential at NNLO in HTL perturbation theory can be written as
\bqa\nonumber
\Omega_{\rm NNLO}\!\!&=&\!\!\Omega_{\rm NLO}+N_f^2({\cal F}_{\rm 3c}+{\cal F}_{\rm 3j})
+N_f({\cal F}_{\rm 3a}+{\cal F}_{\rm 3b}
+{\cal F}_{\rm 3d}
+{\cal F}_{\rm 3e}
+{\cal F}_{\rm 3f}+{\cal F}_{\rm 3g}
+{\cal F}_{\rm 3i}
\\&&\nonumber
+\;{\cal F}_{\rm 3k}+{\cal F}_{\rm 3l})
+{\cal F}_{\rm 3h}
+\Delta_2{\cal E}_0
+\Delta_2 m_D^2{\partial\over\partial m_D^2}
\Omega_{\rm LO}+\Delta_2 m_f^2{\partial\over\partial m_f^2}\Omega_{\rm LO}
\\ && \nonumber
+\;\Delta_1 m_D^2{\partial\over\partial m_D^2}
\Omega_{\rm NLO}
+\Delta_1 m_f^2{\partial\over\partial m_f^2}\Omega_{\rm NLO}
+{1\over2}\left({\partial^2\over(\partial m_D^2)^2}\Omega_{\rm LO}\right)\left(\Delta_1m_D^2\right)^2
\\ &&
+\;{1\over2}\left({\partial^2\over(\partial m_f^2)^2}\Omega_{\rm LO}\right)\left(\Delta_1m_f^2\right)^2
+{F_{\rm 2a+2b}\over\alpha}\Delta_1\alpha
\;.
\label{OmegaNNLO}
\eqa
where $\Delta_2{\cal E}_0$, $\Delta_2m_D^2$, $\Delta_2m_f^2$, and $\Delta_1\alpha$ are terms of order $\delta^2$ in the vacuum energy, mass and coupling constant counterterms:
\bqa
\label{dvac2} 
\Delta_2{\cal E}_0\!\!&=&\!\!{1\over128\pi^2\epsilon}m_D^4\;, \\
\label{dmd2} 
\Delta_2m_D^2\!\!&=&\!\!-N_f{\alpha\over3\pi\epsilon}m_D^2\;, \\ 
\Delta_2m_f^2\!\!&=&\!\!{3\alpha\over4\pi\epsilon}m_f^2\;.
\label{dmf2} 
\eqa

The contributions from the three-loop diagrams are given by
\bqa
\nonumber
{\cal F}_{\rm 3a}\!\!&=&\!\!{1\over4}e^4
\sumint_{P\{QR\}}{\rm Tr}\left[\Gamma^{\mu}(-P,Q-P,Q)S(Q)
\Gamma^{\alpha}(Q-R,Q,R)S(R)\Gamma^{\nu}(P,R,R-P)
\right.\\ && \left.
\!\!\times S(R-P)\Gamma^{\beta}(-Q+R,R-P,Q-P)S(Q-P)\right]
\Delta^{\mu\nu}(P)\Delta^{\alpha\beta}(Q-R)\;,
\\
\label{ph1}
{\cal F}_{\rm 3b}\!\!&=&\!\!
{1\over2}e^4\sumint_{P\{QR\}}{\rm Tr}
\left[\Gamma^{\mu}(P,P+Q,Q)S(Q)\Gamma^{\beta}(-R+Q,Q,R)S(R)
\Gamma^{\alpha}(R-Q,R,Q)
\right.\nonumber\\ && \left.
\!\!\times S(Q)\Gamma^{\nu}(-P,Q,P+Q)S(P+Q)\right]
\Delta^{\mu\nu}(P)\Delta^{\alpha\beta}(R-Q)\;,
\\
\label{ph2}
{\cal F}_{\rm 3c}\!\!&=&\!\!-{1\over4}
e^4\sumint_{P\{QR\}}
{\rm Tr}\left[
\Gamma^{\mu}(P,P+Q,Q)S(Q)\Gamma^{\beta}(-P,Q,P+Q)S(P+Q)
\right]
 \nonumber\\ &&
\!\!\times
{\rm Tr}\left[
\Gamma^{\nu}(-P,R,P+R)S(P+R)
\Gamma^{\alpha}(P,P+R,R)S(R)
\right]\Delta^{\mu\nu}(P)
\Delta^{\alpha\beta}(P)
\;,
\label{ring} \\ \nonumber
{\cal F}_{\rm 3j}\!\!&=&\!\!-{1\over2}
e^4\sumint_{P\{QR\}}
{\rm Tr}\left[
\Gamma^{\alpha\beta}(P,-P,R,R)
S(R)\right]\Delta^{\alpha\mu}(P)\Delta^{\beta\nu}(P)
\\ 
&&\!\!\times
{\rm Tr}\left[
\Gamma^{\mu}(P,P+Q,Q)S(Q)\Gamma^{\nu}(-P,Q,P+Q)S(P+Q)\right]\,.
\eqa

The contributions from the two-loop diagrams with electron-photon three- and four-point vertices with an insertion of a photon self-energy 
\bqa\nonumber
{\cal F}_{\rm 3d}\!\!&=&\!\!-{1\over2}e^2\sumint_{P\{Q\}}\mbox{Tr}\left[\Gamma^{\mu}(P,Q,R)S(Q)\Gamma^{\nu}(P,Q,R)S(R)\right]
\Delta^{\mu\alpha}(P)\Pi^{\alpha\beta}(P)\Delta^{\beta\nu}(P)\;,
\\ &&
\label{3qgpi}
\\ 
{\cal F}_{\rm 3f}\!\!&=&\!\!-{1\over2}e^2\sumint_{P\{Q\}}\mbox{Tr}\left[\Gamma^{\mu\nu}(P,-P,Q,Q)S(Q)\right]
\Delta^{\mu\alpha}(P)\Pi^{\alpha\beta}(P)\Delta^{\beta\nu}(P)\;,
\label{4qgpi}
\eqa
where $R=Q-P$.

The contributions from the two-loop diagrams with the electron-photon three and four-point vertices with an insertion of an electron self-energy are
\bqa \nonumber
{\cal F}_{\rm 3e}\!\!&=&\!\!-e^2\sumint_{P\{Q\}}\Delta^{\alpha\beta}(P){\rm Tr}\left[\Gamma^{\alpha}(P,Q,R)S(Q)
\Sigma(Q)S(Q)\Gamma^{\beta}(P,Q,R)S(R)\right]\;,
\\ &&
\label{3qgs}
\\
{\cal F}_{\rm 3g}\!\!&=&\!\!-{1\over2}e^2\sumint_{P\{Q\}}\Delta^{\mu\nu}(P)
{\rm Tr}\left[\Gamma^{\mu\nu}(P,-P,Q,Q)S(Q)\Sigma(Q)S(Q)\right]\;,
\label{4qgs}
\eqa
where $R=Q-P$.

The contribution from the HTL photon counterterm diagram with two photon self-energy insertions is
\bqa
{\cal F}_{\rm 3h}\;=\;-{1\over4}\sumint_{P}\Pi^{\mu\nu}(P)\Delta^{\nu\alpha}(P)\Pi^{\alpha\beta}(P)\Delta^{\beta\mu}(P)\;.
\eqa

The contribution from HTL electron counterterm with two electron self-energy insertions is
\bqa
{\cal F}_{\rm 3i}\;=\;{1\over2}\sumint_{\{P\}}{\rm Tr}\left[S(P)\Sigma(P)S(P)\Sigma(P)\right]\;.
\label{sigma2}
\eqa

The remaining three-loop diagrams involving HTL-corrected vertex terms are given by
\bqa
{\cal F}_{\rm 3k}\!\!&=&\!\!e^2m_f^2\sumint_{P\{Q\}}\mbox{Tr}\!\left[\tilde{\cal T}^{\mu}(P,Q,R)S(Q)\Gamma^{\nu}(P,Q,R)S(R)\right]
\Delta^{\mu\nu}(P)\;,
\label{3k}
\\
{\cal F}_{\rm 3l}\!\!&=&\!\!-{1\over2}e^2\sumint_{P\{Q\}}\mbox{Tr}\left[\Gamma^{\mu\nu}(P,-P,Q,Q)S(Q)\right]
\Delta^{\mu\nu}(P)\;,
\label{3l}
\eqa
where $\tilde{\cal T}^{\mu}$ is the HTL correction term given in Eq.~(\ref{T3-def}). Note also that diagram (3l) is the same as (2b) since there is no tree-level electron-photon four-vertex.

In the remainder of this chapter, we work in Landau gauge ($\xi=0$), but we emphasize that the HTL perturbation theory method of reorganization is gauge-fixing independent to all orders in $\delta$ (loop expansion) by construction.\\

\section{Expansion in the mass parameters}

In this section we carry out the mass expansion for all the diagrams listed in the last section to high enough order to include all terms through order $e^5$ if $m_D$ and $m_f$ are taken to be of order $e$. The NLO approximation will be perturbatively accurate to order $e^3$ and the NNLO approximation accurate to order $e^5$.

The free energy can be divided into contributions from hard and soft momenta. In the one-loop diagrams, the contributions are either hard $(h)$ or soft $(s)$, while at two-loop level, there are hard-hard $(hh)$ and hard-soft $(hs)$ contributions. There are no soft-soft $(ss)$ contributions since one of the loop momenta is fermionic and always hard. At three loops there are hard-hard-hard $(hhh)$, hard-hard-soft $(hhs)$, and hard-soft-soft $(hss)$ contributions. There are no soft-soft-soft $(sss)$ contributions, again due to the hard fermionic lines.

In the process of the calculation we will see that there are many cancellations between the lower-order HTL-improved diagrams and the higher-order HTL-improved counterterm diagrams. This is by construction and is part of the systematic way in which HTLpt converges to the known perturbative expansion. For example, one can see that diagrams (2c) and (3h) subtract out the modification of the hard gluon propagator due to the HTL-improvement of the propagator in diagram (1a). Likewise, one expects cancellations to occur between diagrams (1b), (2d) and (3i); (2a), (3d), (3e) and (3k); and (2b), (3f), (3g), and (3l). Below we will explicitly demonstrate how these cancellations occur.

\subsection{One-loop sum-integrals}

\subsubsection{Hard contribution}

For hard momenta, the self-energies are suppressed by $m_D/T$ and $m_f/T$ relative to the inverse free propagators, so we can expand in powers of $\Pi_T(P)$, $\Pi_L(P)$, and $\Sigma(P)$.

For the one-loop graph (1a), we need to expand to second order in $m^2_D$:

\bqa\nonumber
{\cal F}_{\rm 1a}^{(h)}\!\!&=&\!\!{1\over2}(d-1)\sumint_P\log\left(P^2\right)+{1\over2}m_D^2\sumint_P{1\over P^2}
\\ && \nonumber
-\;{1\over4(d-1)}m_D^4\sumint_P\left[{1\over P^4}-2{1\over p^2P^2}-2d{1\over p^4}{\cal T}_P+2{1\over p^2P^2}{\cal T}_P+d{1\over p^4}\left({\cal T}_P\right)^2\right]
\\ \!\!&=&\!\! \nonumber
- {\pi^2 \over 45} T^4 + {1 \over 24} \left[1+ \left( 2 + 2{\zeta'(-1) \over \zeta(-1)} \right) \epsilon \right]\left( {\mu \over 4 \pi T} \right)^{2\epsilon} m_D^2 T^2
\\&& 
- \;{1 \over 128 \pi^2}\left( {1 \over \epsilon} - 7 + 2 \gamma + {2 \pi^2\over 3} \right)\left( {\mu \over 4 \pi T} \right)^{2\epsilon} m_D^4 \,.
\label{Flo-h}
\eqa
The one-loop graph with a photon self-energy insertion (2c) has an explicit factor of $m_D^2$ and so we need only to expand the sum-integral to first order in $m_D^2$:
\bqa\nonumber
{\cal F}_{\rm 2c}^{(h)}\!\!&=&\!\!-{1\over2}m_D^2\sumint_P{1\over P^2}
\\ && \nonumber
+\;{1\over2(d-1)}m_D^4\sumint_P\left[{1\over P^4}-2{1\over p^2P^2}-2d{1\over p^4}{\cal T}_P+2{1\over p^2P^2}{\cal T}_P+d{1\over p^4}\left({\cal T}_P\right)^2\right]\\
\!\!&=&\!\!
-{1 \over 24} \left[1 + \left( 2 + 2{\zeta'(-1) \over \zeta(-1)} \right) \epsilon \right]
\left( {\mu \over 4 \pi T} \right)^{2\epsilon} m_D^2 T^2
\nonumber\\&& 
+\;{1 \over 64 \pi^2}
\left( {1 \over \epsilon} - 7 + 2 \gamma + {2 \pi^2\over 3} \right)\left( {\mu \over 4 \pi T} \right)^{2\epsilon} m_D^4 \,.
\label{ct1}
\eqa
The one-loop graph with two photon self-energy insertions (3h) must be expanded to zeroth order in $m_D^2$:
\bqa\nonumber
{\cal F}_{\rm 3h}^{(h)}\!\!&=&\!\!
-{1\over4(d-1)}m_D^4\sumint_P\left[{1\over P^4}-2{1\over p^2P^2}-2d{1\over p^4}{\cal T}_P+2{1\over p^2P^2}{\cal T}_P+d{1\over p^4}\left({\cal T}_P\right)^2\right]\\
\!\!&=&\!\!
-{1 \over 128\pi^2}\left( {1 \over \epsilon} - 7 + 2 \gamma + {2 \pi^2\over 3} \right)\left( {\mu \over 4 \pi T} \right)^{2\epsilon} m_D^4 \,.
\label{ct2}
\eqa
The sum of Eqs.~(\ref{Flo-h})-(\ref{ct2}) is very simple:
\bqa\nonumber
{\cal F}_{\rm 1a+2c+3h}^{(h)}\!\!&=&\!\!
{1\over2}(d-1)\sumint_P\log\left(P^2\right)\\
\!\!&=&\!\!-{\pi^2\over45}T^4
\;.
\eqa
This is the free energy of an ideal gas of photons.

The one-loop graph (1b) needs to be expanded to second order in $m^2_f$:
\bqa\nonumber
{\cal F}_{\rm 1b}^{(h)}\!\!&=&\!\!-2\sumint_{\{P\}}\log P^2-4m_f^2\sumint_{\{P\}}{1\over P^2}
\\ && \nonumber 
+\;2m_f^4\sumint_{\{P\}}\left[{2\over P^4}-{1\over p^2P^2}+{2\over p^2P^2}{\cal T}_P-{1\over p^2P_0^2}\left({\cal T}_P\right)^2
\right]
\\ \nonumber
\!\!&=&\!\!-{7\pi^2\over180}T^4
+{1\over6}
\left[1+\left(2-2\log2+2{\zeta^{\prime}(-1)\over\zeta(-1)}\right)\epsilon
\right]
\left({\mu\over4\pi T}\right)^{2\epsilon}
m_f^2T^2
\\ &&
+\;{1\over12\pi^2}\left(\pi^2-6\right) 
m_f^4\,.
\label{fc0}
\eqa
The one-loop fermion loop with a fermion self-energy insertion (2d) must be expanded to first order in $m_f^2$:
\bqa\nonumber
{\cal F}_{\rm 2d}^{(h)}\!\!&=&\!\!4m_f^2\sumint_{\{P\}}{1\over P^2}-4m_f^4\sumint_{\{P\}}\left[{2\over P^4}-{1\over p^2P^2}+{2\over p^2P^2}{\cal T}_P-{1\over p^2P_0^2}\left({\cal T}_P\right)^2\right]
\\ \nonumber \!\!&=&\!\!
-{1\over6}\left[1+\left(2-2\log2+2{\zeta^{\prime}(-1)\over\zeta(-1)}\right)\epsilon\right]\left({\mu\over4\pi T}\right)^{2\epsilon}m_f^2T^2
\\ &&
-\;{1\over6\pi^2}\left(\pi^2-6\right) m_f^4\,.
\label{fc1}
\eqa
The one-loop fermion loop with two self-energy insertions (3i) must be expanded to zeroth order in $m_f^2$:
\bqa\nonumber
{\cal F}_{\rm 3i}^{(h)}\!\!&=&\!\!2m_f^4\sumint_{\{P\}}\left[{2\over P^4}-{1\over p^2P^2}+{2\over p^2P^2}{\cal T}_P-{1\over p^2P_0^2}\left({\cal T}_P\right)^2\right]
\\ \!\!&=&\!\!
{1\over12\pi^2}\left(\pi^2-6\right)m_f^4\,.
\label{fc2}
\eqa
The sum of Eqs.~(\ref{fc0})-(\ref{fc2}) is particularly simple:
\bqa\nonumber
{\cal F}_{\rm 1b+2d+3i}^{(h)}\!\!&=&\!\!-2\sumint_{\{P\}}\log P^2 \\
\!\!&=&\!\!-{7\pi^2\over180}T^4
\;.
\eqa
This is the free energy of an ideal gas of a single massless fermion.

\subsubsection{Soft contribution}

The soft contributions in the diagrams (1a), (2c), and (3h) arise from the $P_0=0$ term in the sum-integral. At soft momentum $P=(0,{\bf p})$, the HTL self-energy functions reduce to $\Pi_T(P) = 0$ and $\Pi_L(P) = m_D^2$. The transverse term vanishes in dimensional regularization because there is no momentum scale in the integral over ${\bf p}$. Thus the soft contributions come from the longitudinal term only and read
\bqa\nonumber
{\cal F}^{(s)}_{\rm 1a}
\!\!&=&\!\!{1\over2}T\int_{\bf p}\log\left(p^2+m_D^2\right)\\
\!\!&=&\!\! - {m_D^3T\over12\pi}
\left( {\mu \over 2 m} \right)^{2 \epsilon}\left[
1+{8\over3}\epsilon
\right]\;, \nonumber\\
\label{count11}
\\ \nonumber
{\cal F}^{(s)}_{\rm 2c}\!\!&=&\!\!
-{1\over2}m_D^2T\int_{\bf p}{1\over p^2+m_D^2}
\\
\!\!&=&\!\!
{m^3_DT\over 8\pi} \left( {\mu \over 2 m_D} \right)^{2 \epsilon}
\left[1 + 2 \epsilon \right]
\label{count12}
\;, \nonumber \\
\\ \nonumber
{\cal F}^{(s)}_{\rm 3h}\!\!&=&\!\! - {1\over4}m_D^4T\int_{\bf p}{1\over(p^2+m_D^2)^2}
\\
\!\!&=&\!\! - {m^3_DT\over32\pi}
\;.
\eqa
Note that we have kept the terms through order $\epsilon$ in Eqs.~(\ref{count11}) and~(\ref{count12}) as they are required in the calculation of the contributions due to counterterms. There is no soft contribution from the leading-order fermion term~(\ref{lq}) or from the HTL counterterms~(\ref{Sigma}) and (\ref{sigma2}).

\subsection{Two-loop sum-integrals}

For hard momenta, the self-energies are suppressed by $m_D/T$ and $m_f/T$ relative to the inverse free propagators, so we can expand in powers of $\Pi_T$, $\Pi_L$, and $\Sigma$.

\subsubsection{$(hh)$ contribution}

The $(hh)$ contribution from~(\ref{3qg}) and~(\ref{4qg}) was calculated in Ref.~\cite{htl2} and reads
\bqa\nonumber
{\cal F}_{\rm 2a+2b}^{(hh)}\!\!&=&\!\!(d-1)e^2\left[\sumint_{\{PQ\}}{1\over P^2Q^2}
-\sumint_{P\{Q\}}{2\over P^2Q^2}\right] 
+2m_D^2e^2\sumint_{P\{Q\}}\left[{1\over p^2P^2Q^2}
{\cal T}_P
\right.\\ \nonumber &&\left.
+\;{1\over (P^2)^2Q^2}
- {d-2\over d-1}{1\over p^2P^2Q^2}
\right]
+m_D^2e^2\sumint_{\{PQ\}}
\left[ {d+1\over d-1}{1\over P^2Q^2r^2}
\right.\\ \nonumber &&\left.
-\;{4d\over d-1}{q^2\over P^2Q^2r^4}-{2d\over d-1}
{P\!\cdot\!Q\over P^2Q^2r^4}\right]{\cal T}_R  
+m_D^2e^2\sumint_{\{PQ\}}\left[ {3-d\over d-1}{1\over P^2Q^2R^2}
\right.\\ \nonumber &&\left.
+\;{2d\over d-1}{P\!\cdot\! Q\over P^2Q^2r^4}
-{d+2\over d-1}
{1\over P^2Q^2r^2} 
+{4d\over d-1}{q^2\over P^2Q^2r^4}
-{4\over d-1}{q^2\over P^2Q^2r^2R^2} 
\right] 
\\ \nonumber
&&
+\;2m_f^2e^2(d-1)\sumint_{\{PQ\}}\left[ {1\over P^2Q_0^2Q^2}
+{p^2-r^2\over P^2q^2Q_0^2R^2}
\right] {\cal T}_Q
\\ \nonumber &&
+\;2m_f^2e^2(d-1)\sumint_{P\{Q\}} \left[{2\over P^2(Q^2)^2}
-{1\over P^2Q_0^2Q^2}{\cal T}_Q\right]
\\ \nonumber
&&
+\;2m_f^2e^2(d-1)\sumint_{\{PQ\}}\left[ {d+3\over d-1}{1\over P^2Q^2R^2}
- {2\over P^2(Q^2)^2} 
+{r^2-p^2\over q^2P^2Q^2R^2}\right] \\ \nonumber
\!\!& = &\!\! {5\pi^2\over72}{\alpha\over\pi}T^4\left[ 1 + \left(3 - {12\over5}\log2 
+4{\zeta'(-1)\over\zeta(-1)}\right)\epsilon\right]
\left({\mu\over4\pi T}\right)^{4\epsilon} 
\\ \nonumber
&& 
-\;{1\over72}\left[{1\over\epsilon} \;+\; 1.30107 
\right]
{\alpha\over\pi}
\left({\mu\over4\pi T}\right)^{4\epsilon}m_D^2T^2
\\ 
&&
+\;{1\over8}\left[{1\over\epsilon} \;+\; 8.97544 
\right]
{\alpha\over\pi}
\left({\mu\over4\pi T}\right)^{4\epsilon}m_f^2T^2
\;.
\label{33}
\eqa
Consider next the $(hh)$ contribution from~(\ref{3qgpi}) and~(\ref{4qgpi}). The easiest way to calculate this term, is to expand the two-loop diagrams (2a) and (2b) to first order in $m_D^2$. This yields
\bqa\nonumber
{\cal F}_{\rm 3d+3f}^{(hh)}\!\!&=&\!\!
- 2m_D^2e^2\sumint_{P\{Q\}}
\left[{1\over p^2P^2Q^2}{\cal T}_P+{1\over (P^2)^2Q^2}
- {d-2\over d-1}{1\over p^2P^2Q^2}
\right]
\\ \nonumber &&
-\;m_D^2e^2\sumint_{\{PQ\}}
\left[ {d+1\over d-1}{1\over P^2Q^2r^2}
-{4d\over d-1}{q^2\over P^2Q^2r^4}-{2d\over d-1}
{P\!\cdot\!Q\over P^2Q^2r^4}\right]{\cal T}_R  
\\ \nonumber
&&
-\;m_D^2e^2\sumint_{\{PQ\}}\left[ {3-d\over d-1}{1\over P^2Q^2R^2}+
{2d\over d-1}{P\!\cdot\! Q\over P^2Q^2r^4}
-{d+2\over d-1}
{1\over P^2Q^2r^2} 
\right.\\ \nonumber &&\left.
+\;{4d\over d-1}{q^2\over P^2Q^2r^4}
-{4\over d-1}{q^2\over P^2Q^2r^2R^2} 
\right] \\ 
\!\!&=&\!\! {1\over72}\left[{1\over\epsilon} \;+\; 1.30107
\right]
{\alpha\over\pi}
\left({\mu\over4\pi T}\right)^{4\epsilon}m_D^2T^2 \;.
\eqa
We also need the $(hh)$ contributions from the diagrams (3e), (3g), (3k), and (3l) The first two diagrams are given by~(\ref{3qgs}), ~(\ref{4qgs}), while the last remaining ones are given by ~(\ref{3k}) and~(\ref{3l}). The easiest way to calculate these contributions is to expand the two-loop diagrams (2a) and (2b) to first order in $m_f^2$. This yields 
\bqa\nonumber
{\cal F}_{\rm 3e+3g+3k+3l}^{(hh)}\!\!&=&\!\!
- 2m_f^2e^2(d-1)\sumint_{\{PQ\}}\left[ {1\over P^2Q_0^2Q^2}
+{p^2-r^2\over P^2q^2Q_0^2R^2}
\right] {\cal T}_Q 
\\ \nonumber &&
-\;2m_f^2e^2(d-1)\sumint_{P\{Q\}} \left[{2\over P^2(Q^2)^2}
-{1\over P^2Q_0^2Q^2}{\cal T}_Q\right] 
\\ \nonumber 
&&
-\;2m_f^2e^2(d-1)\sumint_{\{PQ\}}\left[ {d+3\over d-1}{1\over P^2Q^2R^2}
- {2\over P^2(Q^2)^2} 
+{r^2-p^2\over q^2P^2Q^2R^2}\right] \\
\!\!&=&\!\!  - {1\over8}\left[{1\over\epsilon} \;+\; 8.97544
\right]
{\alpha\over\pi}
\left({\mu\over4\pi T}\right)^{4\epsilon}m_f^2T^2 \;.
\label{44}
\eqa
The sum of the terms in~(\ref{33})--(\ref{44}) is very simple
\bqa\nonumber
{\cal F}_{\rm 2a+2b+3d+3e+3f+3g+3k+3l}^{(hh)}\!\!&=&\!\!
(d-1)e^2\left[\sumint_{\{PQ\}}{1\over P^2Q^2}
-\sumint_{P\{Q\}}{2\over P^2Q^2}\right]
\\ \!\!&=&\!\! {5 \pi^2\over72} {\alpha\over\pi} T^4\,. 
\eqa
This is the two-loop contribution from the perturbative expansion of the free energy in QED.

\subsubsection{$(hs)$ contribution}

In the $(hs)$ region, the momentum $P$ is soft. The momenta $Q$ and $R$ are always hard. The function that multiplies the soft propagator $\Delta_T(0,{\bf p})$, $\Delta_L(0,{\bf p})$, or $\Delta_X(0,{\bf p})$ can be expanded in powers of the soft momentum ${\bf p}$. The soft propagators $\Delta_T(0,{\bf p})$, $\Delta_L(0,{\bf p})$, and $\Delta_X(0,{\bf p})$ are defined in Eqs.~(\ref{Delta-T:M}), (\ref{Delta-L:M}) and (\ref{Delta-X}), respectively. In the case of $\Delta_T(0,{\bf p})$, the resulting integrals over ${\bf p}$ have no scale and they vanish in dimensional regularization. The integration measure $\int_{\bf p}$ scales like $m_D^3$, the soft propagators $\Delta_L(0,{\bf p})$ and $\Delta_X(0,{\bf p})$ scale like $1/m_D^2$, and every power of $p$ in the numerator scales like $m_D$.

The terms that contribute through order $e^2 m_D^3 T$ and $e^2m_f^2m_DT$ from ~(\ref{3qg}) and~(\ref{4qg}) were calculated in Ref.~\cite{htl2} and read
\bqa\nonumber
{\cal F}_{\rm 2a+2b}^{(hs)}\!\!&=&\!\!2e^2T\int_{\bf p}{1\over p^2+m^2_D}
\sumint_{\{Q\}}\left[
{1\over Q^2}-{2q^2\over Q^4}\right]
\\ \nonumber &&
+\;2m_D^2e^2T\int_{\bf p}{1\over p^2+m_D^2}
\sumint_{\{Q\}}
\left[{1\over Q^4}
-{2\over d}(3+d){q^2\over Q^6}+{8\over d}{q^4\over Q^8}
\right]
\\ 
&&
-\;4m_f^2e^2T\int_{\bf p}{1\over p^2+m_D^2}
\sumint_{\{Q\}}\left[{3\over Q^4}
-{4q^2\over Q^6} -{4\over Q^4} {\cal T}_Q
-{2\over Q^2}\bigg\langle {1\over(Q\!\cdot\!Y)^2} \bigg\rangle_{\!\!\bf \hat y}
\right] \nonumber \\
\!\!&=&\!\!-{1\over6}\alpha m_DT^3\left[1+\left(3-2\log2
+2{\zeta^{\prime}(-1)\over\zeta(-1)}\right)
\epsilon\right] \nonumber
\left({\mu\over4\pi T}\right)^{2\epsilon}
\left({\mu\over2m_D}\right)^{2\epsilon}
\\&& \nonumber
+\;{\alpha\over24\pi^2}\left[{1\over\epsilon}
+\left(1+2\gamma+4\log2\right)
\right]
\left({\mu\over4\pi T}\right)^{2\epsilon}
\left({\mu\over2m_D}\right)^{2\epsilon}
m_D^3T
\\ &&
-\;{\alpha\over2\pi^2}m_f^2m_DT
\;.
\label{first2}
\eqa
The $(hs)$ contribution from ~(\ref{3qgpi}) and~(\ref{4qgpi}) can again be calculated from the diagrams (2a) and (2b) by Taylor expanding their contribution to first order in $m_D^2$. This yields
\bqa\nonumber
{\cal F}_{\rm 3d+3f}^{(hs)}\!\!&=&\!\!
2m_D^2e^2T\int_{\bf p}{1\over(p^2+m_D^2)^2}
\sumint_{\{Q\}}\left[{1\over Q^2}-{2q^2\over Q^4}\right]
\\ \nonumber &&
-\;2m_D^2e^2T\int_{\bf p}{p^2\over(p^2+m^2_D)^2}\sumint_{\{Q\}}
\left[{1\over Q^4}-{2\over d}(3+d){q^2\over Q^6}+{8\over d}{q^4\over Q^8}
\right]
\\ && \nonumber
-\;4m_D^2m_f^2e^2T\int_{\bf p}{1\over(p^2+m_D^2)^2}\sumint_{\{Q\}}
\left[{3\over Q^4}
-{4q^2\over Q^6}
-{4\over Q^4}{\cal T}_{Q}-{2\over Q^2}
\bigg\langle
 {1\over(Q\!\cdot\!Y)^2} \bigg\rangle_{\!\!\bf \hat y}
\right] \\ \nonumber
\!\!&=&\!\! {1\over12}\alpha m_DT^3
-{\alpha\over16\pi^2}\left[{1\over\epsilon}
+\left({1\over3}+2\gamma+4\log2\right)
\right]
\left({\mu\over4\pi T}\right)^{2\epsilon}\left({\mu\over2m_D}\right)^{2\epsilon}
m_D^3T
\\ &&
+\;{\alpha\over4\pi^2}m_f^2m_DT
\;.
\label{hs3d3f}
\eqa
We also need the $(hs)$ contributions from the diagrams (3e), (3g), (3k), and (3l). Again we calculate their contributions by expanding the two-loop diagrams (2a) and (2b) to first order in $m_f^2$. This yields
\bqa\nonumber
{\cal F}_{\rm 3e+3g+3k+3l}^{(hs)}\!\!&=&\!\!4m_f^2e^2T\int_{\bf p}{1\over p^2+m_D^2}
\sumint_{\{Q\}}\left[{3\over Q^4}
-{4q^2\over Q^6} -{4\over Q^4} {\cal T}_Q
-{2\over Q^2}\bigg\langle {1\over(Q\!\cdot\!Y)^2} \bigg\rangle_{\!\!\bf \hat y}
\right] \\
\!\!&=&\!\!{\alpha\over2\pi^2}m_f^2m_DT
\;.
\label{last2}
\eqa

\subsubsection{$(ss)$ contribution}
There are no contributions from the $(ss)$ sector since fermionic momenta are always hard.

\subsection{Three-loop sum-integrals}

\subsubsection{$(hhh)$ contribution}

If all three loop momenta are hard, we can expand the propagators in powers of $\Pi_{\mu\nu}(P)$ and $\Sigma(P)$. Through order $e^5$, we can use bare propagators and vertices. The diagrams (3a), (3b), and (3c) were calculated in Refs.~\cite{Parwani:1994xi,AZ-95} and their contribution is

\bqa\nonumber
{\cal F}^{(hhh)}_{\rm 3a+3b+3c}\!\!&=&\!\!{1\over2}(d-1)(d-5)e^4\sumint_{\{PQR\}}{1\over P^2Q^2R^2(P+Q+R)^2}
\\ && \nonumber
 -\;(d-1)(d-3)e^4\sumint_{PQ\{R\}}{1\over P^2Q^2R^2(P+Q+R)^2}
\\ \nonumber &&
+\;(d-1)^2e^4\sumint_{\{P\}}{1\over P^4}\left[\sumint_{Q}{1\over Q^2}-\sumint_{\{Q\}}{1\over Q^2}\right]^2 
\\ && \nonumber
+\;(d-1)^2e^4\sumint_{PQ\{R\}}{1\over P^2Q^2R^2(P+Q+R)^2}
\\ && \nonumber
-\;2(d-1)^2e^4\sumint_{\{P\}QR}{Q\!\cdot\!R\over P^2Q^2R^2(P+Q)^2(P+R)^2}
\\ && \nonumber 
-\;4e^4(d-3)\sumint_{P\{QR\}}{1\over P^4Q^2R^2}
\\ && \nonumber
-\;(d-3)e^4\sumint_{\{PQR\}}{1\over P^2Q^2R^2(P+Q+R)^2}
\\ &&
-\;16e^4\sumint_{P\{QR\}}{(Q\!\cdot\!R)^2\over P^4Q^2R^2(P+Q)^2(P+R)^2}\;.
\eqa
Using the expression for the sum-integrals in Appendix~\ref{app:sumint}, we obtain
\bqa\nonumber
{\cal F}_{\rm N_f(3a+3b)+N_f^23c}^{(hhh)}\!\!&=&\!\!
-N_f^2{5\pi^2\over216}\left({\alpha\over\pi}\right)^2T^4
\left({\mu\over4\pi T}\right)^{6\epsilon}
\left[
{1\over\epsilon}+{31\over10}+{6\over5}\gamma
\right. \\ && \nonumber \left.
-\;{192\over25}\log2
+{28\over5}{\zeta^{\prime}(-1)\over\zeta(-1)}
-{4\over5}{\zeta^{\prime}(-3)\over\zeta(-3)}\right] 
\\ &&
+\;N_f{\pi^2\over192}\left({\alpha\over\pi}\right)^2T^4\left[35-32\log2
\right]\;.
\label{qedf}
\eqa

\subsubsection{$(hhs)$ contribution}

The diagrams (3a) and (3b) are both infrared finite in the limit $m_D\rightarrow0$. This implies that the $e^5$ contribution is given by using a dressed longitudinal propagator and bare vertices. The ring diagram (3c) is infrared divergent in that limit. The contribution through $e^5$ is obtained by expanding in powers of self-energies and vertices. Finally, the diagram (3j) also gives a contribution of order $e^5$. Since the electron-photon four-vertex is already of order $e^2m_f^2$, we can use a dressed longitudinal propagator and bare fermion propagators as well as bare electron-photon three-vertices. Note that both (3c) and (3j) are proportional to $N_f^2$ and so it is more convenient to calculate their sum. One finds
\bqa\nonumber
{\cal F}_{\rm 3a}^{(hhs)}\!\!&=&\!\!
2(d-1)(d-3)e^4T\int_{\bf p}{1\over p^2+m_D^2}
\sumint_{\{Q\}}{1\over Q^4}\left[
\sumint_{R}{1\over R^2}-\sumint_{\{R\}}{1\over R^2}
\right] 
\\ &&
+\;8(d-1)e^4T\int_{\bf p}{1\over p^2+m_D^2}
\sumint_{Q\{R\}}{q_0r_0\over Q^2R^4(Q+R)^2}\;, 
\\ 
{\cal F}_{\rm 3b}^{(hhs)}\!\!&=&\!\!
-8(d-1)e^4T\int_{\bf p}{1\over p^2+m_D^2}
\sumint_{Q\{R\}}{q_0r_0\over Q^2R^4(Q+R)^2}\;, 
\\ \nonumber
{\cal F}_{\rm 3c+3j}^{(hhs)}\!\!&=&\!\!
-4e^4T\int_{\bf p}{1 \over (p^2+m_D^2)^2}
\left[\sumint_{\{Q\}}
{1\over Q^2}-{2q^2\over Q^4}
\right]^2
\\ && \nonumber
+\;8e^4T\int_{\bf p}{p^2 \over (p^2+m_D^2)^2}
\sumint_{\{Q\}}\left[{1\over Q^2}-{2q^2\over Q^4}\right]
\\ && \nonumber
\times\sumint_{\{R\}}\left[
{1\over R^4}-{2\over d}(3+d)
{r^2\over R^6}+{8\over d}{r^4\over R^8}
\right]
\\ && \nonumber +\;16
m_f^2e^4T\int_{\bf p}{1\over(p^2+m_D^2)^2}
\sumint_{\{Q\}}\left[{1\over Q^2}-{2q^2\over Q^4}\right]
\\ &&
\times\sumint_{\{R\}}\left[
{3\over R^4}-{4r^2\over R^6}
-{4\over R^4}{\cal T}_R-{2\over R^2}
\bigg\langle{1\over(R\!\cdot\!Y)^2} \bigg\rangle_{\!\!\bf \hat y}
\right]\;.
\label{hhs3c}
\eqa
Using the expressions for the integrals and sum-integrals listed in Appendixes~\ref{app:sumint} and \ref{app:int}, we obtain
\bqa\nonumber
{\cal F}_{\rm N_f(3a+3b)+N_f^2(3c+3j)}^{(hhs)}\!\!&=&\!\!
-N_f^2{\pi\alpha^2T^5\over18m_D}+N_f^2{\alpha^2m_DT^3\over12\pi}
\left[{1\over\epsilon}+{4\over3}+2\gamma
+2\log2
\right. \\ && \nonumber \left.
+\;2{\zeta'(-1)\over\zeta(-1)}
\right]\left({\mu\over4\pi T}\right)^{4\epsilon}
\left({\mu\over2m_D}\right)^{2\epsilon}
+N_f{\alpha^2m_DT^3\over4\pi}
\\&&
-\;N_f^2{\alpha^2\over3\pi m_D}m_f^2T^3
\;.
\eqa

\subsubsection{$(hss)$ contribution}

The $(hss)$ modes first start to contribute at order $e^6$, and therefore at our truncation order the $(hss)$ contributions vanish.  

\subsubsection{$(sss)$ contribution}
There are no contributions from the $(sss)$ sector since fermionic momenta are always hard.

\section{Thermodynamic potentials}

In this section we present the final renormalized thermodynamic potential explicitly through order $\delta^2$, aka NNLO. The final NNLO expression is completely analytic; however, there are some numerically determined constants which remain in the final expressions at NLO.

\subsection{Leading order}

The complete expression for the leading order thermodynamic potential is given by the sum of Eqs.~(\ref{Flo-h}),~(\ref{fc0}), and~(\ref{count11}) plus the leading vacuum energy counterterm~(\ref{lovac}):
\bqa\nonumber
\Omega_{\rm LO} \!\!&=&\!\! 
- {\pi^2 T^4\over45}
\left\{ 1 + {7\over4} N_f- {15 \over 2} \hat m_D^2 - 30 
N_f\hat m_f^2
+ 30 \hat m_D^3
\right.\\ && \left. 
+\; {45 \over 4}
\left( \log {\hat \mu \over 2}
        - {7\over2} + \gamma + {\pi^2\over 3} \right)
        \hat m_D^4
	- 60 N_f\left(\pi^2-6\right)\hat m_f^4
	\right\} \;.
\label{Omega-LO}
\eqa
where $\hat m_D$, $\hat m_f$, and $\hat \mu$ are dimensionless variables:
\bqa
\hat m_D \!\!&=&\!\! {m_D \over 2 \pi T}  \;,
\\
\hat m_f \!\!&=&\!\! {m_f \over 2 \pi T}  \;,
\\
\hat \mu \!\!&=&\!\! {\mu \over 2 \pi T}  \;. 
\eqa

\subsection{Next-to-leading order}

The renormalization contributions at first order in $\delta$ are
\bqa
\Delta_1\Omega\;=\;\Delta_1{\cal E}_0
+\Delta_1m_D^2{\partial\over\partial m_D^2}\Omega_{\rm LO}+
\Delta_1m_f^2{\partial\over\partial m_f^2}\Omega_{\rm LO}\;.
\eqa
Using the results listed in Eqs.~(\ref{dvac1}), (\ref{dmd1}), and (\ref{dmf1}), the complete contribution from the counterterm at first order in $\delta$ is
\bqa
\Delta_1\Omega \!\!&=&\!\! 
-{\pi^2 T^4\over45} \Bigg\{ {45\over4\epsilon} \hat m_D^4 
	+N_f {\alpha \over \pi} \Bigg[
-{5\over2}
	\left({1\over\epsilon}+2\log{\hat\mu\over 2} + 
2 {\zeta'(-1)\over\zeta(-1)}+2 \right) \hat m_D^2
\nonumber \\ &&
+\;15 \left({1\over\epsilon}+2\log{\hat\mu\over 2} - 
2 \log \hat m_D +2 \right) \hat m_D^3
\nonumber \\ &&
	+\;{45 \over 2} 
		\left({1\over\epsilon}+2+2\log{\hat\mu\over 2} - 
2\log2 + 2{\zeta'(-1)\over\zeta(-1)}\right) 
			\hat m_f^2 \Bigg] \Bigg\} \;. 
\label{OmegaVMct1}
\eqa
Adding the NLO counterterms (\ref{OmegaVMct1}) to the contributions from the various NLO diagrams, we obtain the renormalized NLO thermodynamic potential
\bqa
\Omega_{\rm NLO}\!\!&=&\!\!
- {\pi^2 T^4\over45} \Bigg\{ 
	1 + {7\over4}N_f - 15 \hat m_D^3 
	- {45\over4}\left(\log\hat{\mu\over2}-{7\over2}+\gamma+{\pi^2\over3}
\right)\hat m_D^4
\nonumber \\ &&
	+\;60N_f\left(\pi^2-6\right)\hat m_f^4
	+ N_f{\alpha\over\pi} \Bigg[ -{25\over8}
	+ 15 \hat m_D
\nonumber \\ &&
	+\;5\left(\log{\hat\mu \over 2}-2.33452\right)\hat m_D^2
	-45\left(\log{\hat\mu \over 2}+2.19581\right)\hat m_f^2
\nonumber \\ &&
	-\;30\left(\log{\hat\mu \over 2}-{1\over2}
+\gamma+2\log2\right)\!\!\hat m_D^3
	+ 180\hat m_D \hat m_f^2 \Bigg]
\Bigg\} \;.
\label{Omega-NLO}
\eqa

\subsection{Next-to-next-to-leading order}

The renormalization contributions at second order in $\delta$ are
\bqa\nonumber
\Delta_2\Omega\!\!&=&\!\!\Delta_2{\cal E}_0
+\Delta_2m_D^2{\partial\over\partial m_D^2}\Omega_{\rm LO}+
\Delta_2m_f^2{\partial\over\partial m_f^2}\Omega_{\rm LO}
+\Delta_1m_D^2{\partial\over\partial m_D^2}\Omega_{\rm NLO}
\\ &&\nonumber
+\;\Delta_1m_f^2{\partial\over\partial m_f^2}\Omega_{\rm NLO}
+{1\over2}\left({\partial^2\over(\partial m_D^2)^2}
\Omega_{\rm LO}\right)\left(\Delta_1m_D^2\right)^2
\\ &&
+\;{1\over2}\left({\partial^2\over(\partial m_f^2)^2}
\Omega_{\rm LO}\right)\left(\Delta_1m_f^2\right)^2
+{F_{2a+2b}\over\alpha}\Delta_1\alpha
\;.
\eqa
Using the results listed in Eqs.~(\ref{dvac2}), (\ref{dmd2}), and (\ref{dmf2}), the complete contribution from the counterterms at second order in $\delta$ is
\bqa
\Delta_2\Omega\!\!&=&\!\! 
-{\pi^2 T^4\over45} \Bigg\{ -{45\over8\epsilon} \hat m_D^4 
	+ N_f {\alpha \over \pi} \Bigg[
{5\over2}\left({1\over\epsilon}+2\log{\hat\mu\over 2} 
+ 2 {\zeta'(-1)\over\zeta(-1)}+2 \right) \hat m_D^2
\nonumber \\ &&
-\; {45\over2}\left({1\over\epsilon}+2\log{\hat\mu\over 2} - 
2 \log \hat m_D +{4\over3} \right) \hat m_D^3
\nonumber \\ &&
-\;{45 \over 2} \left({1\over\epsilon}+2+2\log{\hat\mu\over 2} - 2\log2 + 
2{\zeta'(-1)\over\zeta(-1)}\right) 
			\hat m_f^2 \Bigg] \nonumber \\ &&
	+\;N_f^2 \left({\alpha \over \pi}\right)^2 \Bigg[ - {25\over24}
\left({1\over\epsilon} + 4\log{\hat\mu\over 2} + 3 - {12\over5}\log2 + 
4{\zeta'(-1)\over\zeta(-1)}\right) \nonumber \\ &&
	+\;{15\over2}\left({1\over\epsilon}+4\log{\hat\mu\over 2}
	- 2 \log \hat m_D + {7\over3} -2\log2 + 2{\zeta'(-1)\over\zeta(-1)}\right) 
\hat m_D	
	\Bigg]
			\Bigg\} \;.
\label{OmegaVMct2}
\eqa
Adding the NNLO counterterms (\ref{OmegaVMct2}) to the contributions from the various NNLO diagrams, we obtain the renormalized NNLO thermodynamic potential. We note that at NNLO all numerically determined coefficients of order $\epsilon^0$ drop out and we are left with a final result which is completely analytic. The resulting NNLO thermodynamic potential is
\begin{eqnarray}
\Omega_{\rm NNLO}\!\!&=&\!\!
- {\pi^2 T^4\over45} \Bigg\{ 
	1 + {7\over4}N_f - {15\over4} \hat m_D^3 
\nonumber \\ &&
	+\;N_f {\alpha\over\pi} \Bigg[ -{25\over8}
	+ {15\over2} \hat m_D
	+15 \left(\log{\hat\mu \over 2}-{1\over2}
+\gamma+2\log2\right)\!\!\hat m_D^3
	- 90\hat m_D \hat m_f^2 \Bigg]
\nonumber \\ &&
+\;N_f \left({\alpha\over\pi}\right)^2 
\Bigg[{15\over64}(35-32\log2)-{45\over2} \hat m_D\Bigg] 
\nonumber \\ &&
+\;N_f^2 \left({\alpha\over\pi}\right)^2 \Bigg[{25\over12}
\left(\log{\hat\mu \over 2}+{1\over20}+{3\over5}\gamma-{66\over25}\log2
+{4\over5}{\zeta^{\prime}(-1)\over\zeta(-1)}
-{2\over5}{\zeta^{\prime}(-3)\over\zeta(-3)}
\right)
\nonumber \\ &&
+\;{5\over4}{1\over\hat m_D} - 15\left(\log{\hat\mu \over 2}-{1\over2}
+\gamma+2\log2\right)\!\!\hat m_D
+{30}{\hat{m}_f^2\over\hat{m}_D}
\Bigg]
\Bigg\} \;.
\label{Omega-NNLO}
\end{eqnarray}

We note that the coupling constant counterterm listed in Eq.~(\ref{delalpha}) coincides with the known one-loop running of the QED coupling constant
\beq
\mu \frac{d e^2}{d \mu} \;=\; \frac{N_fe^4}{6 \pi^2} \, .
\label{runningcoupling}
\eeq
Below we will present results as a function of $e$ evaluated at the renormalization scale $2 \pi T$.  Note that when the free energy is evaluated at a scale different than $\mu = 2 \pi T$ we use Eq.~(\ref{runningcoupling}) to determine the value of the coupling at $\mu = 2 \pi T$.

We have already seen that there are several cancellations that take place algebraically, irrespective of the values of $m_D$ and $m_f$. For example the $(hh)$ contribution from the two-loop diagrams (2a) and (2b) cancel against the $(hh)$ contribution from the diagrams (3d), (3e), (3f), and (3g). As long as only hard momenta are involved, these cancellations will always take place once the relevant sum-integrals are expanded in powers of $m_D/T$ and $m_f/T$. This is no longer the case when soft momenta are involved. However, further cancellations do take place if one chooses the weak-coupling values for the mass parameters. For example, if one uses the weak-coupling value for the Debye mass,
\bqa\nonumber
m_D^2\!\!&=&\!\!4 N_f e^2\sumint_{\{Q\}}\left[{1\over Q^2}-{2q^2\over Q^4}\right]\\
\!\!&=&\!\!{4\pi\over3}N_f\alpha T^2\;,
\eqa
the terms proportional to $m_f^2$ in $\Omega_{\rm NNLO}$ cancel algebraically and HTLpt reduces to the weak-coupling result for the free energy through $e^5$. This reduction is by construction in HTLpt which also provides a consistency check that in the weak-coupling limit, HTLpt coincidences with weak-coupling expansion.

\section{Free energy}

The mass parameters $m_D$ and $m_f$ in hard-thermal-loop perturbation theory are in principle completely arbitrary. To complete a calculation, it is necessary to specify $m_D$ and $m_f$ as functions of $e$ and $T$. In this section we will consider two possible mass prescriptions in order to see how much the results vary given the two different assumptions. First we will consider the variational solution resulting from the thermodynamic potential, Eqs.~(\ref{gapmd}) and (\ref{gapmf}), and second we will consider using the $e^5$ perturbative expansion of the Debye mass \cite{Blaizot:1995kg,Andersen:1995ej} and the $e^3$ perturbative expansion of the fermion mass~\cite{Carrington:2008dw}.

\subsection{Variational Debye mass}

The NLO and NNLO variational Debye mass is determined by solving Eqs.~(\ref{gapmd}) and (\ref{gapmf}) using the NLO and NNLO expressions for the thermodynamic potential, respectively. The free energy is then obtained by evaluating the NLO and NNLO thermodynamic potentials, (\ref{Omega-NLO}) and (\ref{Omega-NNLO}), at the solution to the gap equations (\ref{gapmd}) and (\ref{gapmf}). Note that at NNLO the gap equation for the fermion mass is trivial and gives $m_f=0$. The NNLO gap equation for $m_D$ reads
\bqa
&& \hspace{-0.7cm}
{5 \over 4} N_f^2 \left( {\alpha \over \pi} \right)^2 
\;=\; \left[ - {45 \over 4} 
+ 45 N_f {\alpha\over\pi} 
    \left( \log{\hat\mu \over 2} - {1\over2} + \gamma + 2\log2 \right) 
\right] \hat m_D^4
\nonumber \\
&& \hspace{1.2cm}
+\; \left[ {15 \over 2} N_f {\alpha \over \pi} - {45 \over 2} N_f \left( {\alpha \over \pi} \right)^2 
- 15 N_f^2 \left( {\alpha \over \pi} \right)^2 
    \left( \log{\hat\mu \over 2} - {1\over2} + \gamma + 2\log2 \right)
\right] \hat m_D^2 \;.
\nonumber \\
\label{gap-qed}
\eqa

\begin{figure}[h]
\vspace{3mm}
\begin{center}
\includegraphics[width=10cm]{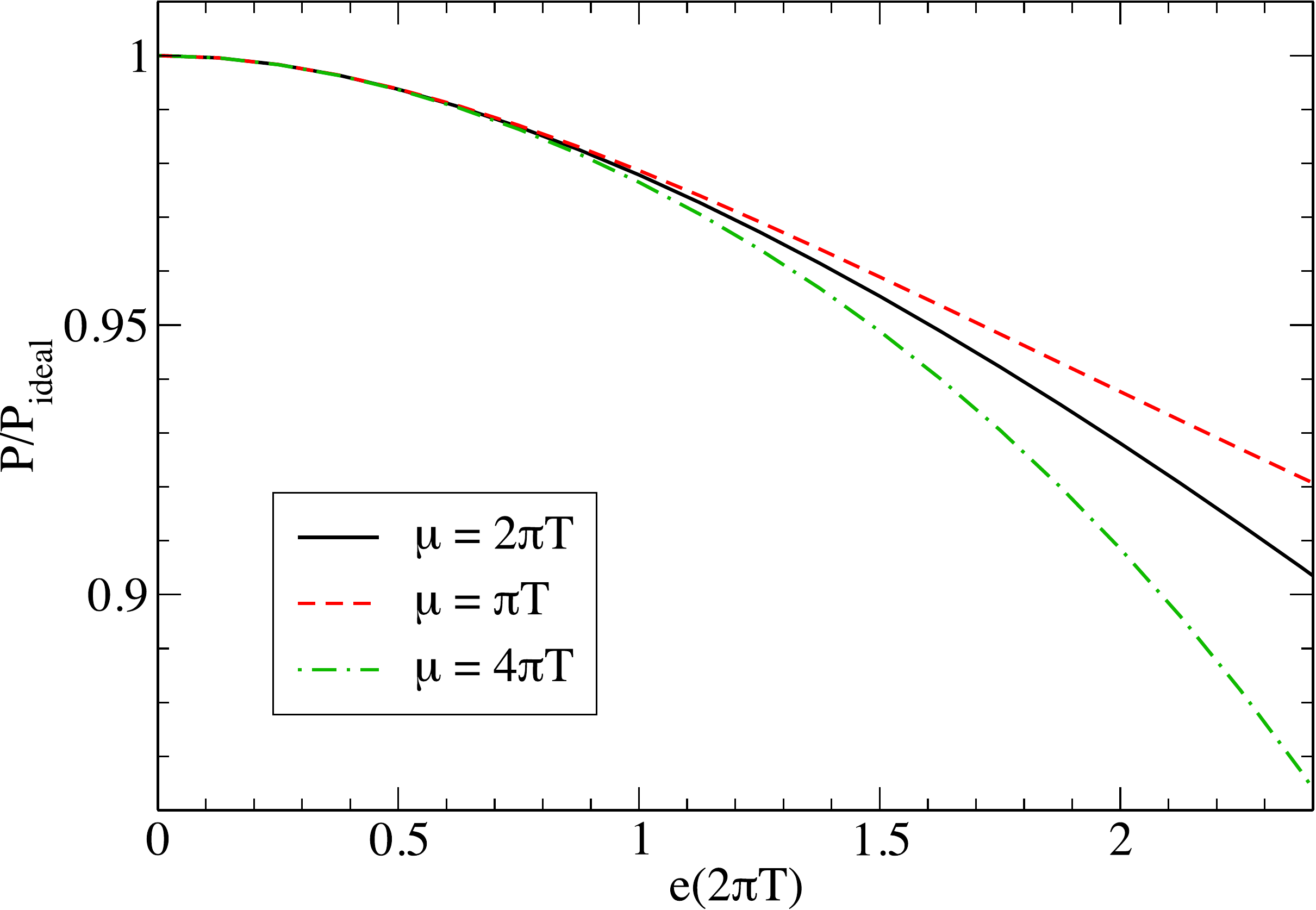}
\end{center}
\vspace{-3mm}
\caption{NLO HTLpt predictions for the free energy of QED with $N_f=1$ and the variational Debye mass. Different curves correspond to varying the renormalization scale $\mu$ by a factor of 2 around $\mu=2\pi T$. }
\label{fig:NLO}
\end{figure}

\begin{figure}[h]
\begin{center}
\vspace{3mm}
\includegraphics[width=10cm]{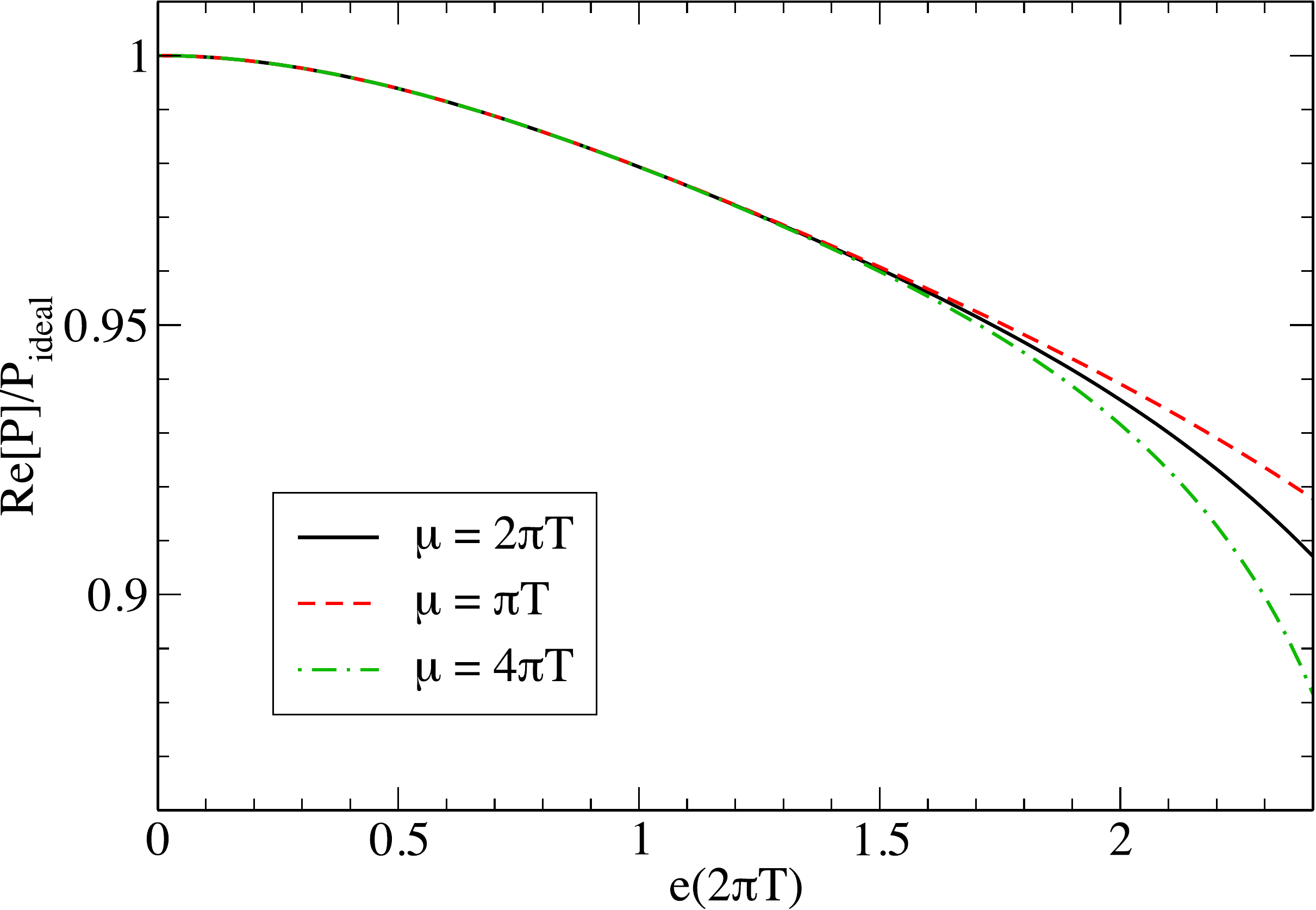}\\
\vspace{6mm}
\includegraphics[width=10cm]{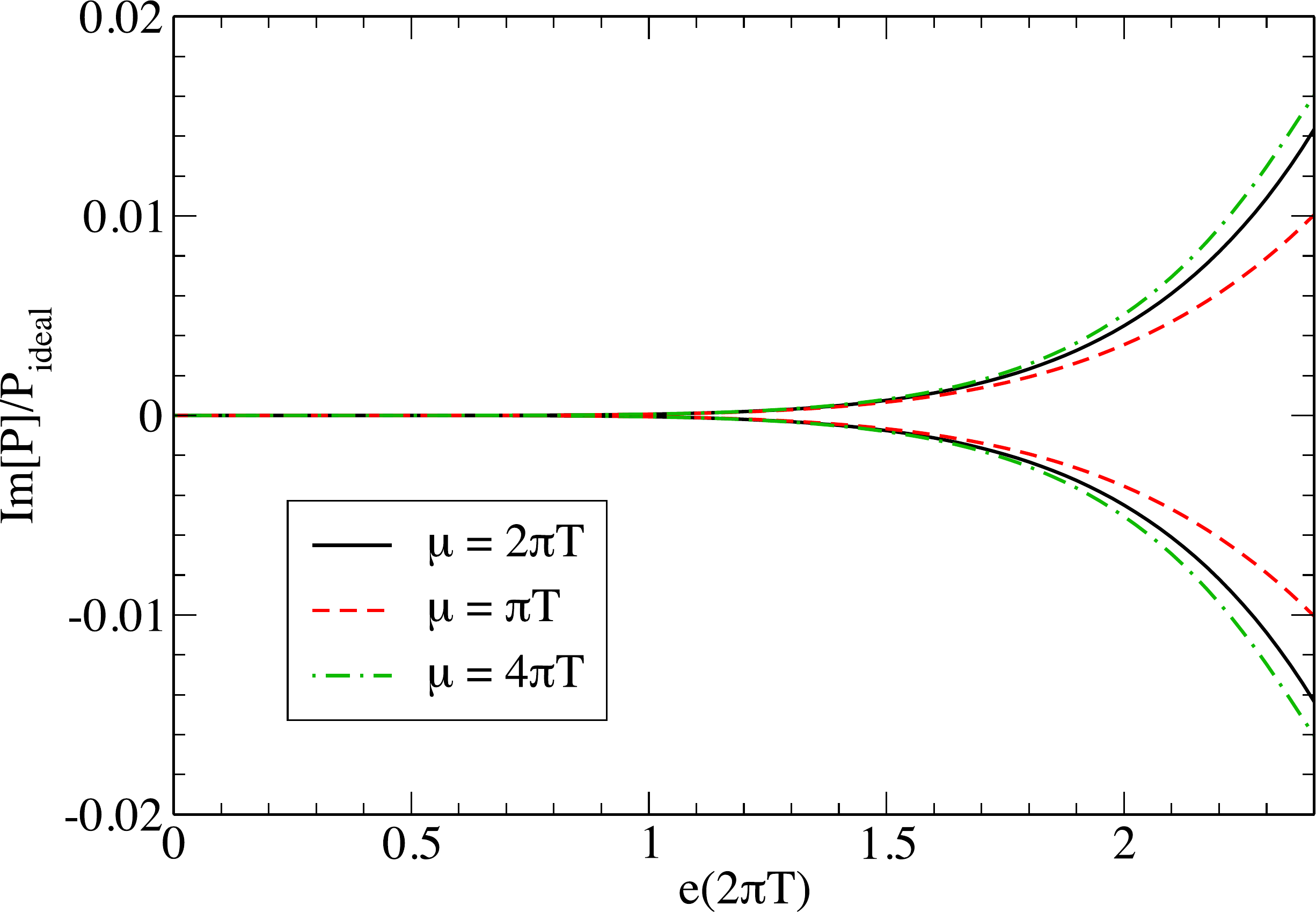}
\end{center}
\vspace{-3mm}
\caption{Real (top panel) and imaginary (bottom panel) parts of NNLO HTLpt predictions for the free energy of QED with $N_f=1$ and the variational Debye mass. Different curves correspond to varying the renormalization scale $\mu$ by a factor of 2 around $\mu=2\pi T$. }
\label{fig:NNLO}
\end{figure}

\begin{figure}[h]
\begin{center}
\vspace{3mm}
\includegraphics[width=10cm]{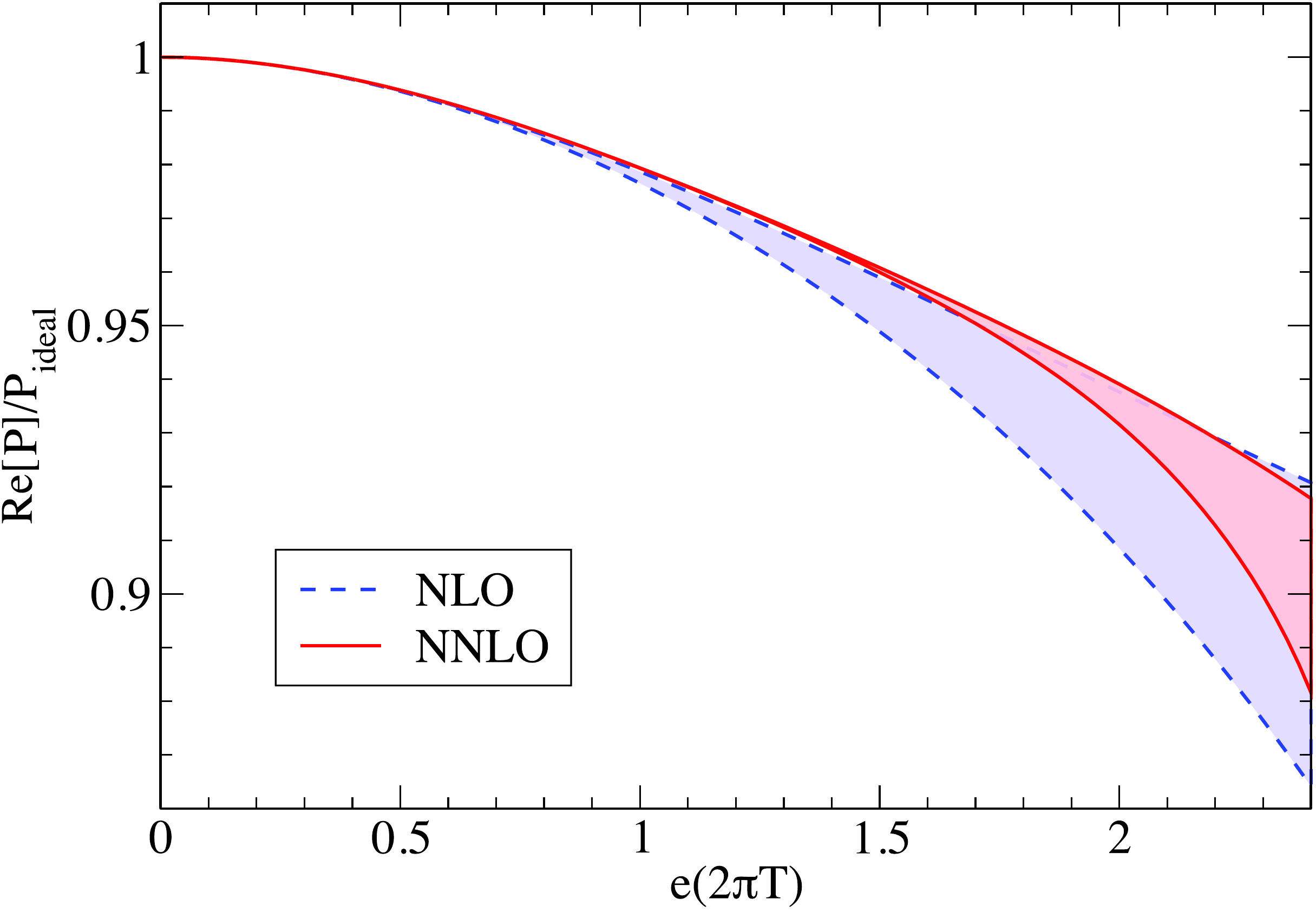}
\end{center}
\vspace{-3mm}
\caption{A comparison of the renormalization scale variations between NLO and NNLO HTLpt predictions for the free energy of QED with $N_f=1$ and the variational Debye mass. The bands correspond to varying the renormalization scale $\mu$ by a factor of 2 around $\mu=2\pi T$.}
\label{fig:NLONNLO}
\end{figure}

In Figs.~\ref{fig:NLO}, \ref{fig:NNLO} and \ref{fig:NLONNLO} we plot the NLO and NNLO HTLpt predictions for the free energy of QED with $N_f=1$. As can be seen in Fig.~\ref{fig:NLONNLO} the renormalization scale variation of the results decreases as one goes from NLO to NNLO. This is in contrast to weak-coupling expansions for which the scale variation can increase as the truncation order is increased. 
%%%%%%%%%%%%%%%%%%%%%%%%%%%%%%%%%%%%%%%%%%%%%%%%%%%%%%%%%%%%%%%%
%%%%%%%%%%%%%%%%%%%%%%%%%%%%%%%%%%%%%%%%%%%%%%%%%%%%%%%%%%%%%%%%
%%%%%%%%%%%%%%%%%%%%%%%%%%%%%%%%%%%%%%%%%%%%%%%%%%%%%%%%%%%%%%%%
%%%%%%%%%%%%%%%%%%%%%%%%%%%%%%%%%%%%%%%%%%%%%%%%%%%%%%%%%%%%%%%%
%%%%%%%%%%%%%%%%%%%%%%%%%%%%%%%%%%%%%%%%%%%%%%%%%%%%%%%%%%%%%%%%
%%%%%%%%%%%%%%%%%%%%%%%%%%%%%%%%%%%%%%%%%%%%%%%%%%%%%%%%%%%%%%%%

One troublesome issue with the variational Debye mass is that at NNLO the solutions to~(\ref{gap-qed}) have a small imaginary part. The NNLO gap equation~(\ref{gap-qed}) is quadratic in $m_D^2$, so it has two complex conjugate solutions for $m_D^2$. Although the solutions are real for large couplings, they become complex when the coupling is smaller than a critical value, e.g. the critical value for $N_f = 1$ and $\mu = 2 \pi T$ is $e(2\pi T) = 3.38946$. For small coupling and finite $N_f$, the variational Debye masses can be expanded as follow:
\beq
m_D^2 \;=\; {N_f \over 3} e^2 T^2 \,\pm\, i {N_f \over \pi \sqrt{6}} e^3 T^2  \,+\, {\cal O}(e^4) \;,
\eeq
which reproduces the weak-coupling result~(\ref{mass}) at leading order, however the $e^3$ term becomes imaginary which starts to deviate from the weak-coupling result. It should be mentioned that as $N_f \rightarrow \infty$, the critical value below which $m_D$ becomes complex goes to zero. We plot the imaginary part of the free energy which results from these imaginary contributions to the variational Debye mass in Fig.~\ref{fig:NNLO} (bottom panel). The imaginary contributions to the variational Debye mass come with both a positive and negative sign corresponding to the two possible solutions to the quadratic variational gap equation. The positive sign would indicate an unstable solution while the negative sign would indicate a damped solution. These imaginary parts are most likely an artifact of the dual truncation at order $e^5$; however, without extending the truncation to higher order, it is difficult to say. They do not occur at NLO in HTLpt in either QED or QCD. We note that a similar effect has also been observed in NNLO scalar $\phi^4$ theory \cite{Andersen:2000yj} and Yang-Mills theory~\cite{Andersen:2009tc,Andersen:2010ct}. Because of this complication, in the next subsection we will discuss a different mass prescription in order to assess the impact of these small imaginary parts.

\subsection{Perturbative Debye and fermion masses}

\begin{figure}[h]
\begin{center}
\vspace{3mm}
\includegraphics[width=10cm]{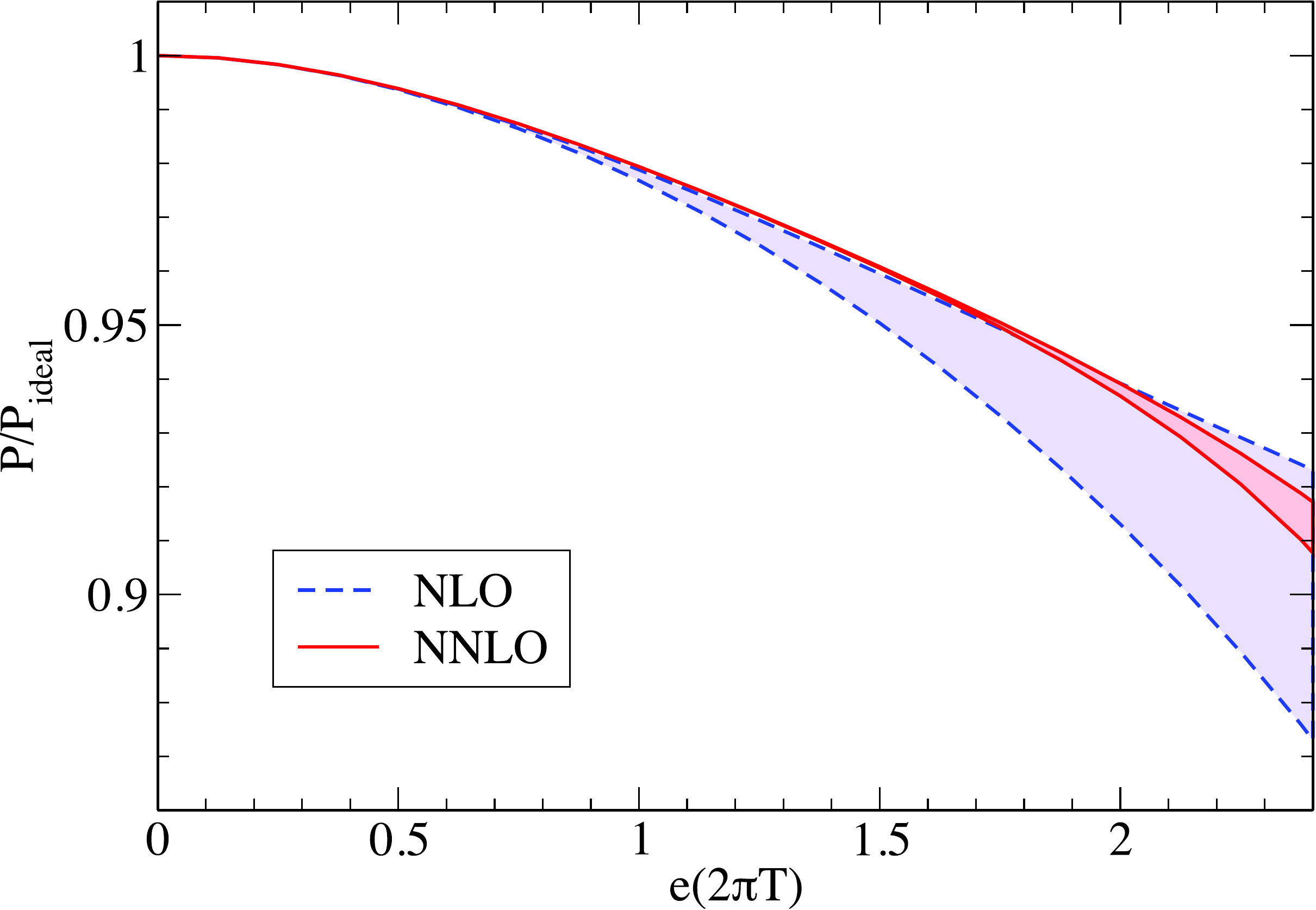}
\end{center}
\vspace{-3mm}
\caption{A comparison of the renormalization scale variations between NLO and NNLO HTLpt predictions for the free energy of QED with $N_f=1$ using the perturbative thermal masses given in Eqs.~(\ref{mass}) and (\ref{fmass}). The bands correspond to varying the renormalization scale $\mu$ by a factor of 2 around $\mu=2\pi T$.}
\label{fig:NLONNLOpmd}
\end{figure}

The perturbative Debye and fermion masses for QED have been calculated  through order $e^5$~\cite{Blaizot:1995kg,Andersen:1995ej} and $e^3$~\cite{Carrington:2008dw}, respectively:
\begin{eqnarray}
\label{mass}
m_D^{2}\!\!&=&\!\!{1\over3}N_fe^2T^{2}\left [1
-\frac{e^{2}}{24\pi^{2}}\left(4\gamma + 7  
+4\log \frac{\hat\mu}{2}+8\log2\right) 
+ {e^3\sqrt{3}\over4\pi^3} \right] \; ,\\
m_f^2\!\!&=&\!\!{1\over8}N_fe^2T^2\left[
1-{2.854\over4\pi}e\right]\;.
\label{fmass}
\end{eqnarray}
Plugging~(\ref{mass}) and~(\ref{fmass}) into the NLO and NNLO thermodynamic potentials, (\ref{Omega-NLO}) and (\ref{Omega-NNLO}), we obtain the results shown in Fig.~\ref{fig:NLONNLOpmd}. The renormalization scale variation is quite small in the NNLO result. 

%%%%%%%%%%%%%%%%%%%%%%%%%%%%%%%%%%%%%%%%%%%%%%%%%%%%%%%%%%%%%%%%
%%%%%%%%%%%%%%%%%%%%%%%%%%%%%%%%%%%%%%%%%%%%%%%%%%%%%%%%%%%%%%%%

\subsection{Comparison with the $\Phi$-derivable approach}

\begin{figure}[t]
\begin{center}
\vspace{3mm}
\includegraphics[width=10cm]{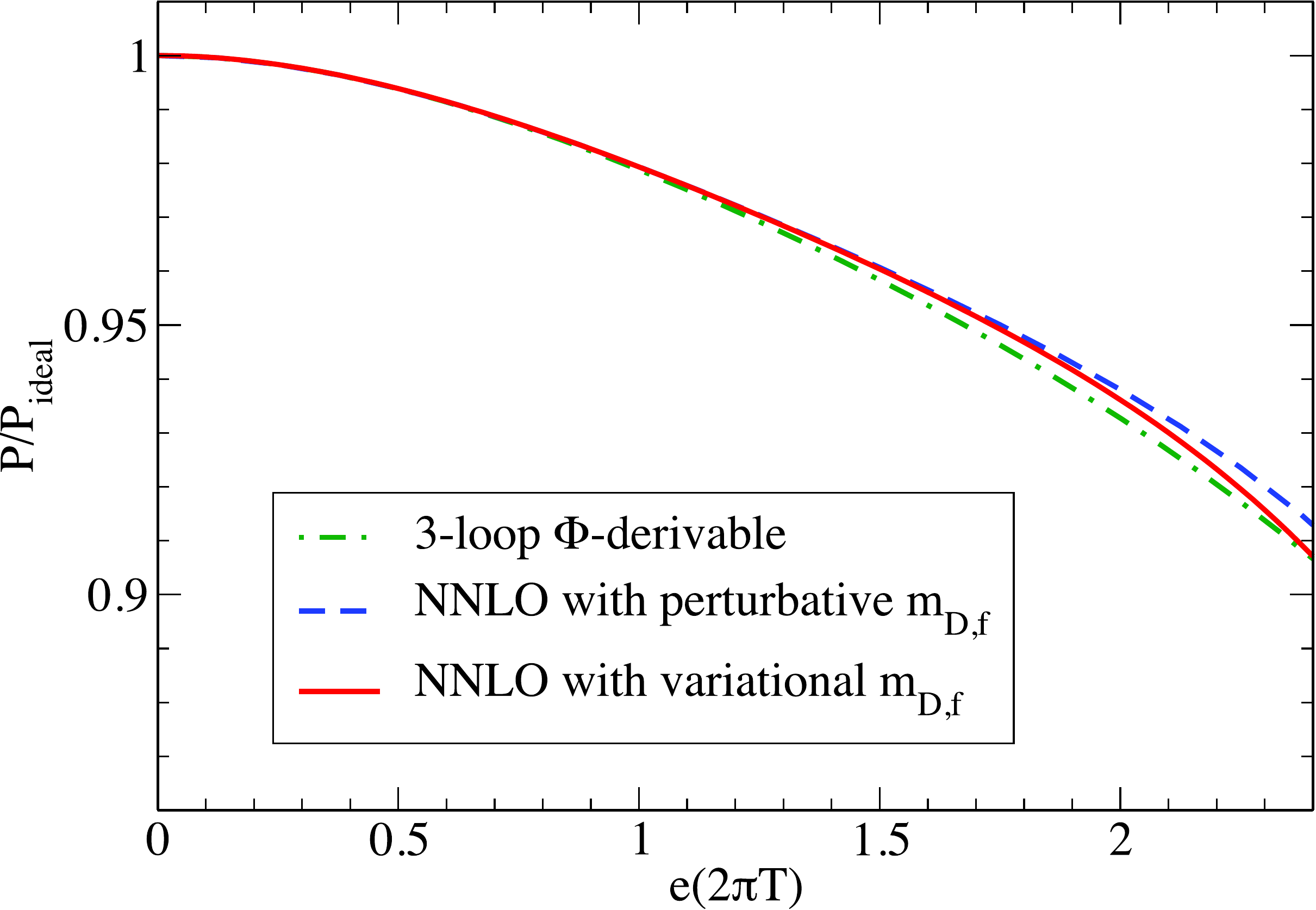}
\end{center}
\vspace{-3mm}
\caption{A comparison of the predictions for the free energy of QED with $N_f=1$ between three-loop $\Phi$-derivable approximation \cite{Andersen:2004re} and NNLO HTLpt at $\mu=2\pi T$.}
\label{fig:PhivsNNLO}
\end{figure}

Having obtained the NNLO HTLpt result for the free energy we can now compare the results obtained using this reorganization with results obtained within the $\Phi$-derivable approach. In Fig.~\ref{fig:PhivsNNLO} we show a comparison of our NNLO HTLpt results with a three-loop calculation obtained previously using a truncated three-loop $\Phi$-derivable approximation \cite{Andersen:2004re}. For the NNLO HTLpt prediction we show the results obtained using both the variational and perturbative mass prescriptions. As can be seen from this figure, there is very good agreement between the NNLO $\Phi$-derivable and HTLpt approaches out to large coupling. The difference between these two predictions at $e=2.4$ is merely $0.6\%$. In all cases we have chosen the renormalization scale to be $\mu = 2\pi T$.

%%%%%%%%%%%%%%%%%%%%%%%%%%%%%%%%%%%%%%%%%%%%%%%%%%%%%%%%%%%%%%%%
%%%%%%%%%%%%%%%%%%%%%%%%%%%%%%%%%%%%%%%%%%%%%%%%%%%%%%%%%%%%%%%%
%%%%%%%%%%%%%%%%%%%%%%%%%%%%%%%%%%%%%%%%%%%%%%%%%%%%%%%%%%%%%%%%
\begin{figure}[h]
\begin{center}
\vspace{3mm}
\includegraphics[width=10cm]{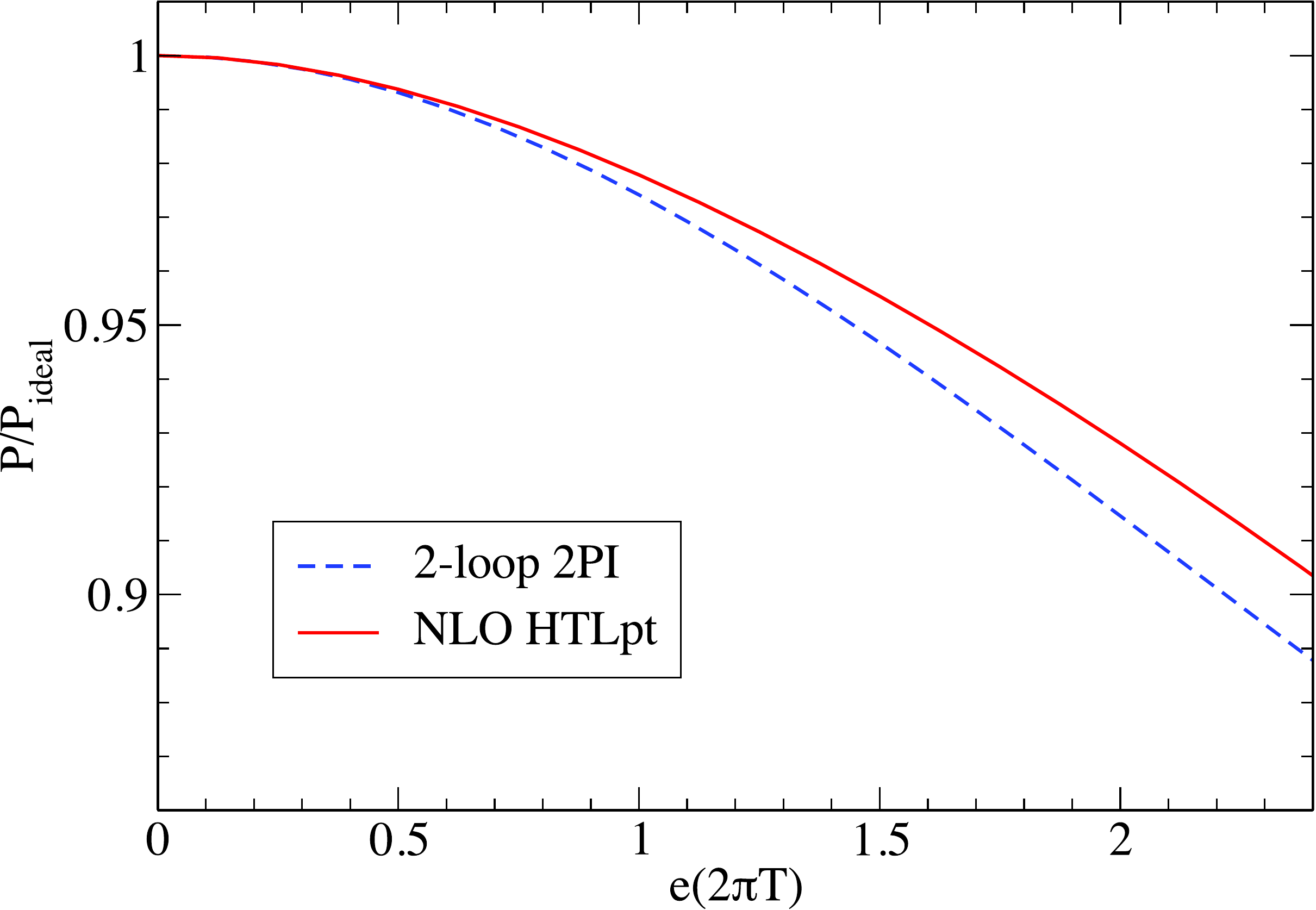}
\end{center}
\vspace{-3mm}
\caption{A comparison of the predictions for the free energy of QED with $N_f=1$ between the two-loop 2PI approximation in Landau gauge \cite{Borsanyi:2007bf} and NLO HTLpt at $\mu=2\pi T$.}
\label{fig:2PIvsNLO}
\end{figure}
%%%%%%%%%%%%%%%%%%%%%%%%%%%%%%%%%%%%%%%%%%%%%%%%%%%%%%%%%%%%%%%%

As a further consistency check, in Fig.~\ref{fig:2PIvsNLO} we show a comparison between the untruncated two-loop numerical $\Phi$-derivable approach calculation of Ref.~\cite{Borsanyi:2007bf} and our NLO HTLpt result using the variational mass. In both cases we have chosen the renormalization scale to be $\mu = 2\pi T$. From this figure we see that there is a reasonable agreement between the NLO numerical $\Phi$-derivable and NLO HTLpt results; however, the agreement is not as good as the corresponding NNLO results shown in Fig.~\ref{fig:PhivsNNLO}.  

We note that the results of \cite{Borsanyi:2007bf} were computed in the Landau gauge ($\xi=0$).  As detailed in their paper, their result is gauge dependent. Such gauge dependence is unavoidable in the 2PI $\Phi$-derivable approach since it only uses dressed propagators. In Ref.~\cite{Andersen:2004re} it was explicitly shown that the two-loop $\Phi$-derivable Debye mass is gauge independent only up to order $e^2$, resulting in gauge variation of the free energy at order $e^4$. This is in agreement with general theorems stating that the gauge variance appears at one order higher than the truncation \cite{Arrizabalaga:2002hn}.

\subsection{QCD free energy at large $N_f$}

\begin{figure}[htb]
\begin{center}
\vspace{3mm}
\includegraphics[width=10cm]{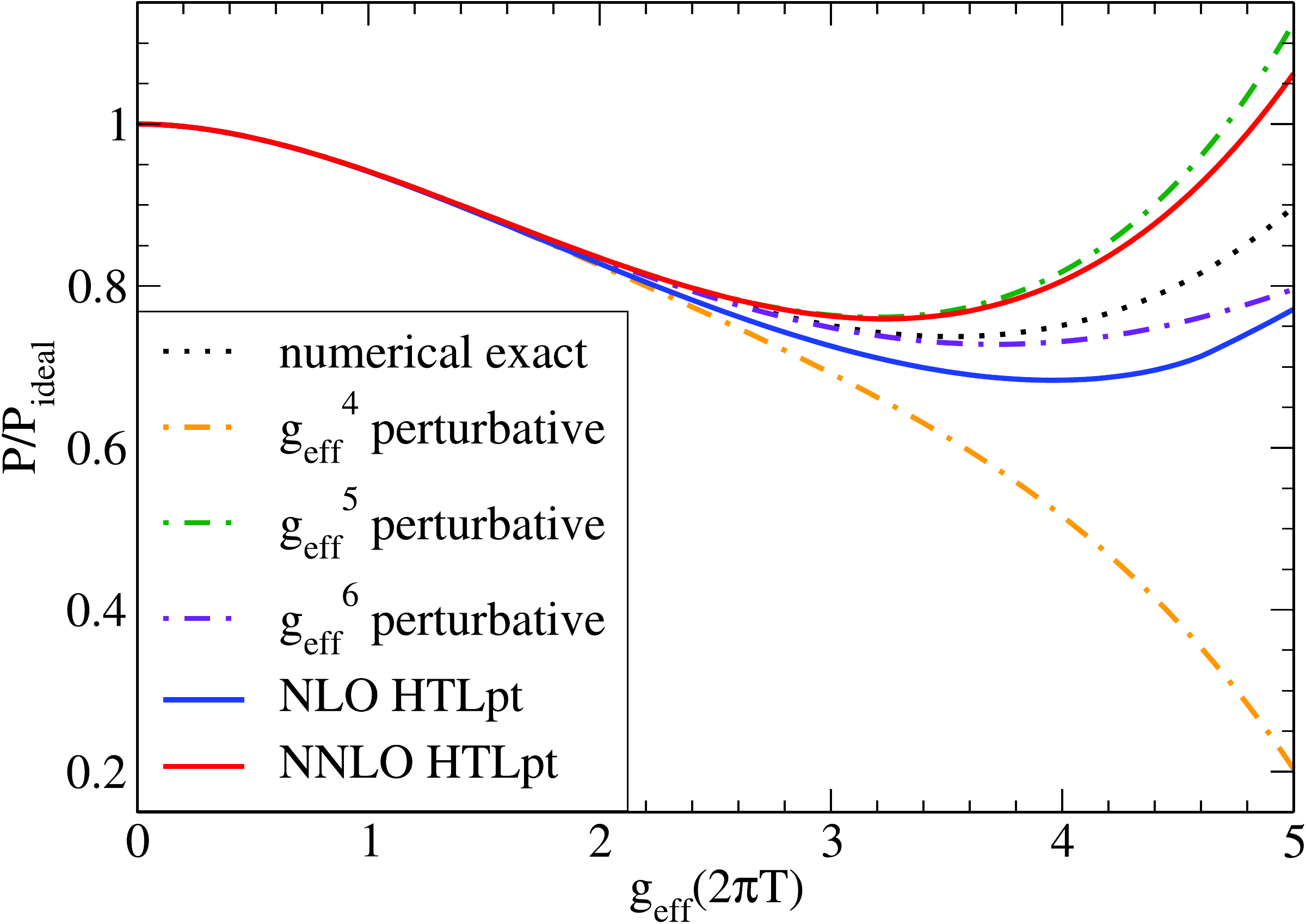}
\end{center}
\vspace{-3mm}
\caption{A comparison of the predictions for the large $N_f$ free energy of QED between the numerical result from~\cite{Ipp:2003zr}, NLO and NNLO HTLpt, and perturbative $g_{\rm eff}^4$ through $g_{\rm eff}^6$~\cite{Gynther:2009qf} results at $\mu = e^{-\gamma} \pi T$.}
\label{fig:N_f}
\end{figure}

The large $N_f$ limit is achieved by taking $N_f$ to be large while holding $e^2N_f$ fixed. The large $N_f$ coupling for QED is defined by $g_{\rm eff} \equiv e \sqrt{N_f}$. By power counting, it is easily to see that in perturbation theory only ring diagrams survive in the large $N_f$ limit, which indicates the equivalence of QED and QCD at large $N_f$. In the large $N_f$ limit, it is possible to solve for the ${\cal O}(N_f^0)$ contribution to the free energy exactly~\cite{Moore:2002md,Ipp:2003zr}. In Fig.~\ref{fig:N_f} we plot the NLO and NNLO HTLpt predictions for the free energy at large $N_f$ along with the numerical result of Ref.~\cite{Ipp:2003zr}, as well as the perturbative $g_{\rm eff}^4$, $g_{\rm eff}^5$ and newly obtained $g_{\rm eff}^6$~\cite{Gynther:2009qf} predictions at $\mu = e^{-\gamma} \pi T$. The NLO HTLpt result seem to diverge from the exact result around $g_{\rm eff} = 2$, while the NNLO one from $g_{\rm eff} = 2.8$, however both of their large coupling behaviors qualitatively fit that of the numerical result.  
%%%%%%%%%%%%%%%%%%%%%%%%%%%%%%%%%%%%%%%%%%%%%%%%%%%%%%%%%%%%%%%%
%%%%%%%%%%%%%%%%%%%%%%%%%%%%%%%%%%%%%%%%%%%%%%%%%%%%%%%%%%%%%%%%

\section{Conclusions}

In this chapter we calculated the three-loop HTLpt thermodynamic potential in QED. Having obtained this we applied two mass prescriptions, variational and perturbative, to fix the {\em a priori} undetermined parameters $m_D$ and $m_f$ that appear in the HTL-improved Lagrangian.  We found that the resulting expressions for the free energy were the same to an accuracy of 0.6\% at $e=2.4$ giving us confidence in the prediction. We also compared the HTLpt three-loop result with a three-loop $\Phi$-derivable approach \cite{Andersen:2004re} and found agreement at the subpercentage level at large coupling. Besides, the large $N_f$ HTLpt three-loop result is in reasonable agreement with the exact numerical one~\cite{Ipp:2003zr}.

In addition, we showed that the HTLpt NLO and NNLO approximations have improved convergence at large coupling compared to the naively truncated weak-coupling expansion and that the renormalization scale variation at NNLO using both the variational and perturbative mass prescriptions was quite small. Therefore, the NNLO HTLpt method result seems to be quite reliable. This is important since, unlike the $\Phi$-derivable approach, the HTLpt reorganization is gauge invariant by construction and is formulated directly in Minkowski space allowing it to, in principle, also be applied to the calculation of dynamical quantities.  

The renormalization of the three-loop thermodynamic potential required only known vacuum energy, mass, and coupling constant counterterms, and the resulting running coupling was found to coincide with the canonical QED one-loop running. This provides further evidence that the HTLpt framework is renormalizable despite the new divergences which are introduced during HTL improvement.

Finally, we note that at three loops we could obtain an entirely analytic expression for the renormalized NNLO thermodynamic potential. There were a number of cancellations that took place during renormalization which resulted in an expression that was independent of any numerically determined subleading coefficients in the sum-integrals. With the confidence in the techniques, we are ready to step into non-Abelian theories.

%%%%%%%%%%%%%%%%%%%%%%%%%%%%%%%%%%%%%%%%%%%%%%%%%%%%%%%%%%%%%
%
%	Include File:			DON'T COMPILE !!!
%
%%%%%%%%%%%%%%%%%%%%%%%%%%%%%%%%%%%%%%%%%%%%%%%%%%%%%%%%%%%%%

\chapter{Yang-Mills Thermodynamics to Three Loops}\label{chapter:ym}

In this chapter, we study the thermodynamics of Yang-Mills theory using the hard-thermal-loop perturbation theory in the same spirit of Chapter~\ref{chapter:qed}. We show that at three-loop order hard-thermal-loop perturbation theory is compatible with lattice results for the pressure, energy density, and entropy down to temperatures $T\sim2-3\,T_c$. This chapter is based on: {\it Gluon Thermodynamics at Intermediate Coupling}, J.~O.~Andersen, M.~Strickland and N.~Su, Phys.\ Rev.\ Lett.\  {\bf 104}, 122003 (2010), and {\it Three-loop HTL gluon thermodynamics at intermediate coupling}, J.~O.~Andersen, M.~Strickland and N.~Su, JHEP {\bf 1008}, 113 (2010).

\section{Introduction}

The goal of ultrarelativistic heavy-ion collision experiments is to generate energy densities and temperatures high enough to create a quark-gluon plasma. One of the chief theoretical questions which has emerged in this area is whether it is more appropriate to describe the state of matter generated during these collisions using weakly-coupled quantum field theory or a strong-coupling approach based on the AdS/CFT correspondence. Early data from the Relativistic Heavy Ion Collider (RHIC) at Brookhaven National Labs indicated that the state of matter created there behaved more like a fluid than a plasma and that this ``quark-gluon fluid'' is strongly coupled \cite{rhicexperiment}.

In the intervening years, however, work on the perturbative side has shown that observables like jet quenching \cite{pert} and elliptic flow \cite{Xu:2007jv} can also be described using a perturbative formalism. Since in phenomenological applications predictions are complicated by the modeling required to describe, for example, initial-state effects, the space-time evolution of the plasma, and hadronization of the plasma, there are significant theoretical uncertainties remaining.  Therefore, one is hard put to conclude whether the plasma is strongly or weakly coupled based solely on RHIC data. To have a cleaner testing ground one can compare theoretical calculations with results from lattice quantum chromodynamics (QCD).

Looking forward to the upcoming heavy-ion experiments scheduled to take place at the Large Hadron Collider (LHC) at the European Organization for Nuclear Research (CERN) it is important to know if, at the higher temperatures generated, one expects a strongly-coupled (liquid) or weakly-coupled (plasma) description to be more appropriate. At RHIC, initial temperatures on the order of one to two times the QCD critical temperature, $T_c \sim 190$ MeV, were obtained.  At LHC, initial temperatures on the order of $4-5\,T_c$ are expected. The key question is, will the generated matter behave more like a plasma of quasiparticles at these higher temperatures.

\begin{figure}[t]
\begin{center}
\includegraphics[width=10cm]{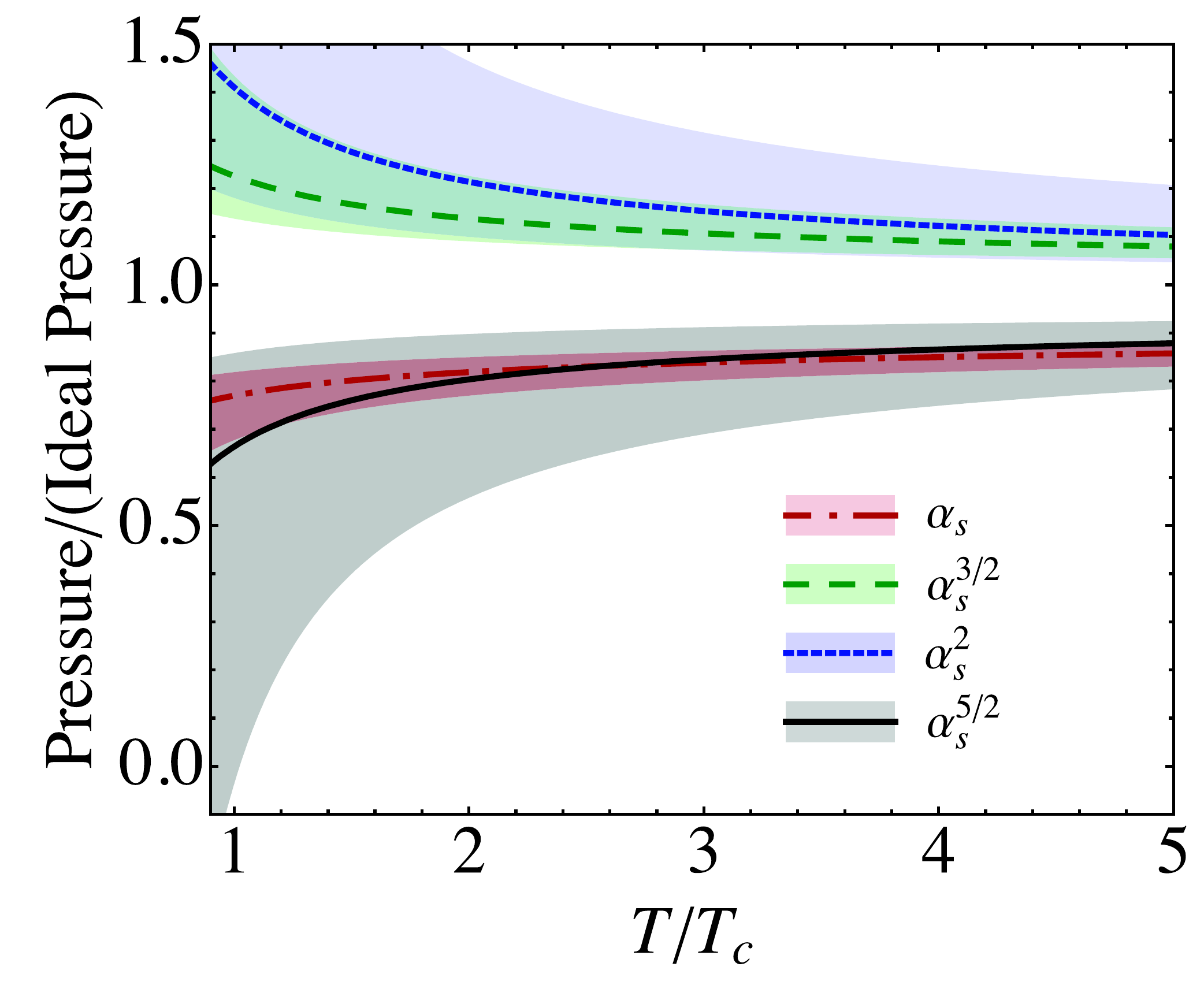}
\end{center}
\vspace{-3mm}
\caption{Weak-coupling expansion for the scaled pressure of pure-glue QCD. Shaded bands show the result of varying the renormalization scale $\mu$ by a factor of 2 around $\mu = 2 \pi T$.}
\label{pertpressure-YM}
\end{figure}

As is well known, the weak-coupling expansion for the free energy of SU(3) pure-glue QCD fails to   converge at phenomenologically relevant temperatures that are being created in the colliders~\cite{AZ-95,Zhai:1995ac,BN-96,Kajantie:2002wa}. In Fig.~\ref{pertpressure-YM} we show the weak-coupling expansion results through order $\alpha_s^{5/2}$ from which one sees severe oscillations as higher orders are included in the expansion. Equipped with the techniques as well as the confidence from the QED calculation in Chapter~\ref{chapter:qed}, we are ready to generalize HTLpt to Yang-Mills theory.

%%%%%%%%%%%%%%%%%%%%%%%%%%%%%%%%%%%%%%%%%%%%%%%%%%%%
%%%%%%%%%%%%%%%%%%%%%%%%%%%%%%%%%%%%%%%%%%%%%%%%%%%%

\section{HTL perturbation theory}

\label{HTLpt}

The Lagrangian density for SU($N_c$) Yang-Mills theory in Minkowski space is
\bqa
{\cal L}_{\rm YM}\;=\;-{1\over2}{\rm Tr}\left[G_{\mu\nu}G^{\mu\nu}\right] 
+{\cal L}_{\rm gf}
+{\cal L}_{\rm gh}
+\Delta{\cal L}_{\rm YM}\;.
\label{L-YM}
\eqa
Here the field strength is 
$G^{\mu\nu}=\partial^{\mu}A^{\nu}-\partial^{\nu}A^{\mu}-ig[A^{\mu},A^{\nu}]$, with $A^\mu$ an element of the SU($N_c$) gauge group. The ghost term ${\cal L}_{\rm gh}$ depends on the gauge-fixing term ${\cal L}_{\rm gf}$. In this chapter we choose the class of covariant gauges where the gauge-fixing term is
\bqa
{\cal L}_{\rm gf}\;=\-{1\over\xi}{\rm tr}\left[\left(\partial_{\mu}A^{\mu}\right)^2\right]\;,
\eqa
with $\xi$ being the gauge-fixing parameter.

The HTLpt Lagrangian density for Yang-Mills theory is written as
\bqa
{\cal L}= \left({\cal L}_{\rm YM}
+ {\cal L}_{\rm HTL} \right) \Big|_{g \to \sqrt{\delta} g}
+ \Delta{\cal L}_{\rm HTL}\;.
\label{L-HTLYM}
\eqa
The HTL-improvement term is
\bqa
{\cal L}_{\rm HTL}=-{1\over2}(1-\delta)m_D^2 {\rm Tr}
\left(G_{\mu\alpha}\left\langle {y^{\alpha}y^{\beta}\over(y\cdot D)^2}
	\right\rangle_{\!\!\hat{\bf y}}G^{\mu}_{\;\;\beta}\right)\;,
\label{L-HTL-YM}
\eqa
where the covariant derivative reads $D^{\mu}=\partial^{\mu}-igA^{\mu}$.

Similar to the QED case, as we will show HTLpt for Yang-Mills theory is also renormalizable at NNLO with a coupling constant counterterm, a Debye mass counterterm and a vacuum energy counterterm which read
\bqa
\delta\Delta\alpha_s\!\!&=&\!\!-{11N_c\over12\pi\epsilon}\alpha_s^2\delta^2+{\cal O}(\delta^3\alpha^3_s)\;,
\label{delalpha-YM}
\\ 
\Delta m_D^2\!\!&=&\!\!\left(-{11N_c\over12\pi\epsilon}\alpha_s\delta+{\cal O}(\delta^2\alpha_s^2)
\right)(1-\delta)m_D^2\;,
\label{delmd-YM} \\ 
\Delta{\cal E}_0&=&\left({N_c^2-1\over128\pi^2\epsilon}+{\cal O}(\delta\alpha_s)\right)(1-\delta)^2m_D^4\;.
\label{del1e0-YM}
\eqa

In the following we will first obtain the thermodynamic potential $\Omega$ which is a function of $T$, $\alpha_s$ and $m_D$. In order to get the free energy ${\cal F}$, some prescription is required to determine $m_D$ as a function of $T$ and $\alpha_s$. We will discuss several prescriptions in Section~\ref{functions}.

\section{Diagrams for the thermodynamic potential}

\begin{figure*}[t]
\begin{center}
\includegraphics[width=9cm]{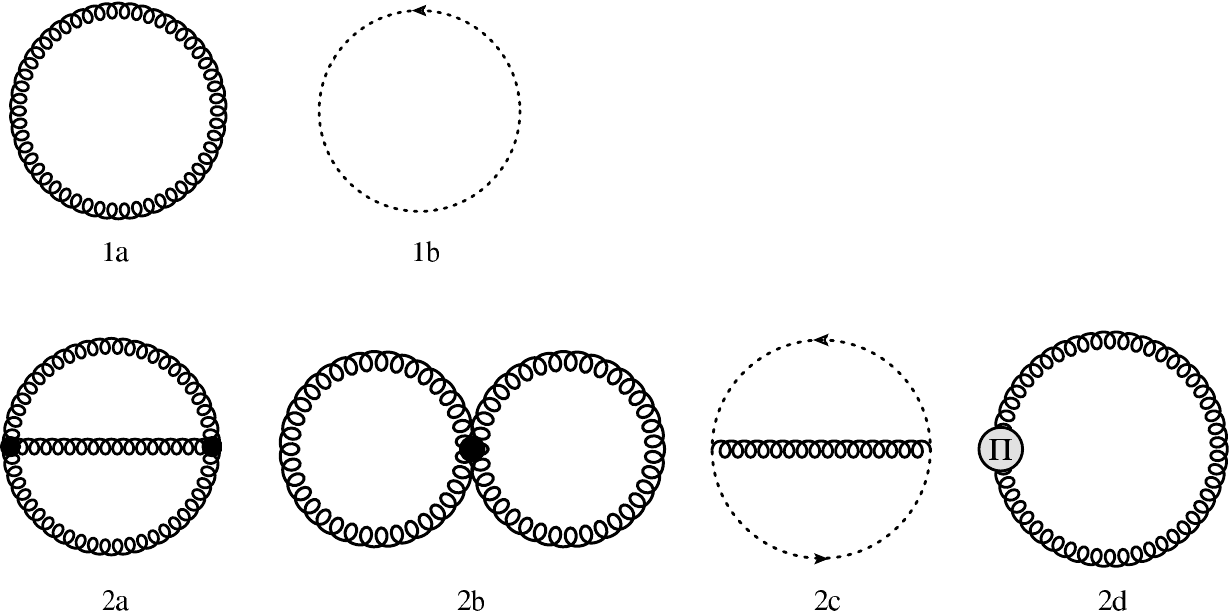}
\end{center}
\caption{Diagrams contributing through NLO in HTLpt. The spiral lines are gluon propagators and the dotted lines are ghost propagators. A circle with a $\Pi$ indicates a gluon self-energy insertion. All propagators and vertices shown are HTL-resummed propagators and vertices.}
\label{fig:dia1-YM}
\end{figure*}

\begin{figure*}[t]
\begin{center}
\includegraphics[width=14.4cm]{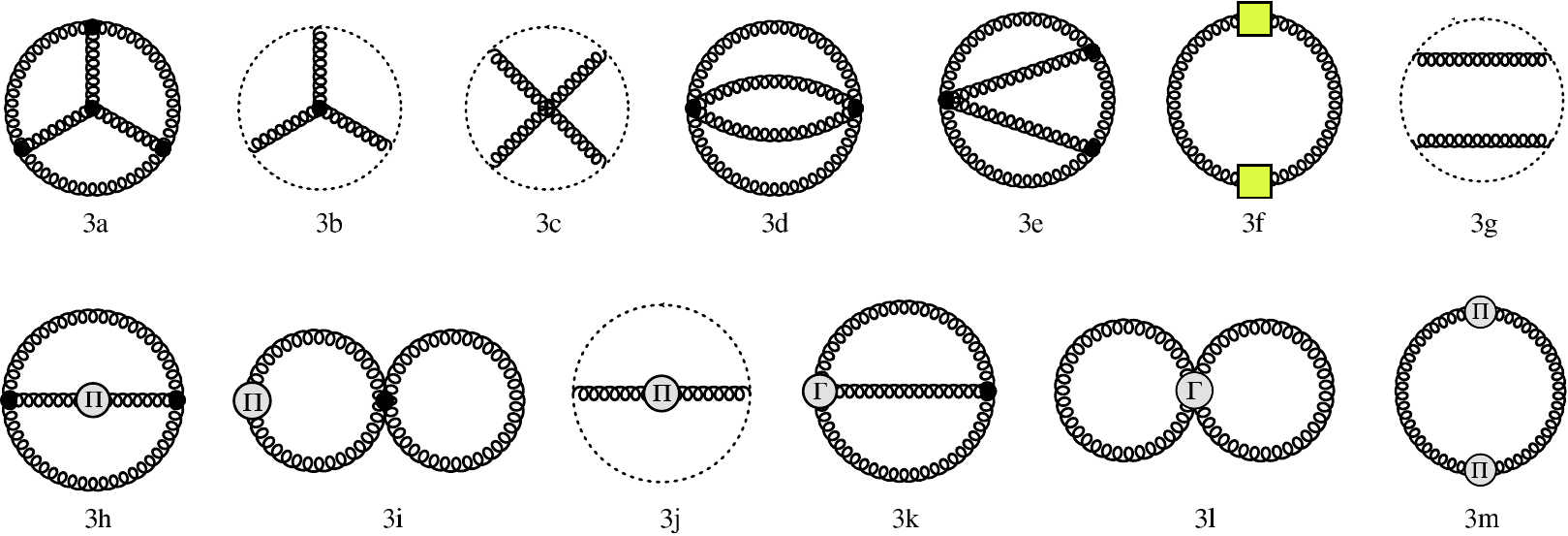}
\end{center}
\caption{Diagrams contributing to NNLO in HTLpt through order $g^5$. The spiral lines are gluon propagators and the dotted lines are ghost propagators. A circle with a $\Pi$ indicates a gluon self-energy insertion. The propagators are HTL-resummed propagators and the black dots indicate HTL-resummed vertices. The lettered vertices indicate that only the HTL correction is included. The yellow box in (3f) denotes the insertion of the one-loop self-energy defined in Fig.~\ref{keydiagrams}.}
\label{fig:dia2-YM}
\end{figure*}

In this section, we list the expressions for the diagrams that contribute
to the thermodynamic potential through order $\delta^2$, aka NNLO, in HTL
perturbation theory. The diagrams are shown in Figs.~\ref{fig:dia1-YM},
and \ref{fig:dia2-YM}.  A key to the diagrams is given in Fig.~\ref{keydiagrams}.
The expressions here will be given in Euclidean space; however, in Appendix
\ref{app:rules} we present the HTLpt Feynman rules in Minkowski space.

%%%%%%%%%%%%%%%%%%%%%%%%%%%%%%%%%%%%%%%%%%%%%%%%%%%%%%%%%%%%%%%%
%%%%%%%%%%%%%%%%%%%%%%%%%%%%%%%%%%%%%%%%%%%%%%%%%%%%%%%%%%%%%%%%
%%%%%%%%%%%%%%%%%%%%%%%%%%%%%%%%%%%%%%%%%%%%%%%%%%%%%%%%%%%%%%%%
%%%%%%%%%%%%%%%%%%%%%%%%%%%%%%%%%%%%%%%%%%%%%%%%%%%%%%%%%%%%%%%%
\begin{figure*}[t]
\begin{center}
\includegraphics[width=10cm]{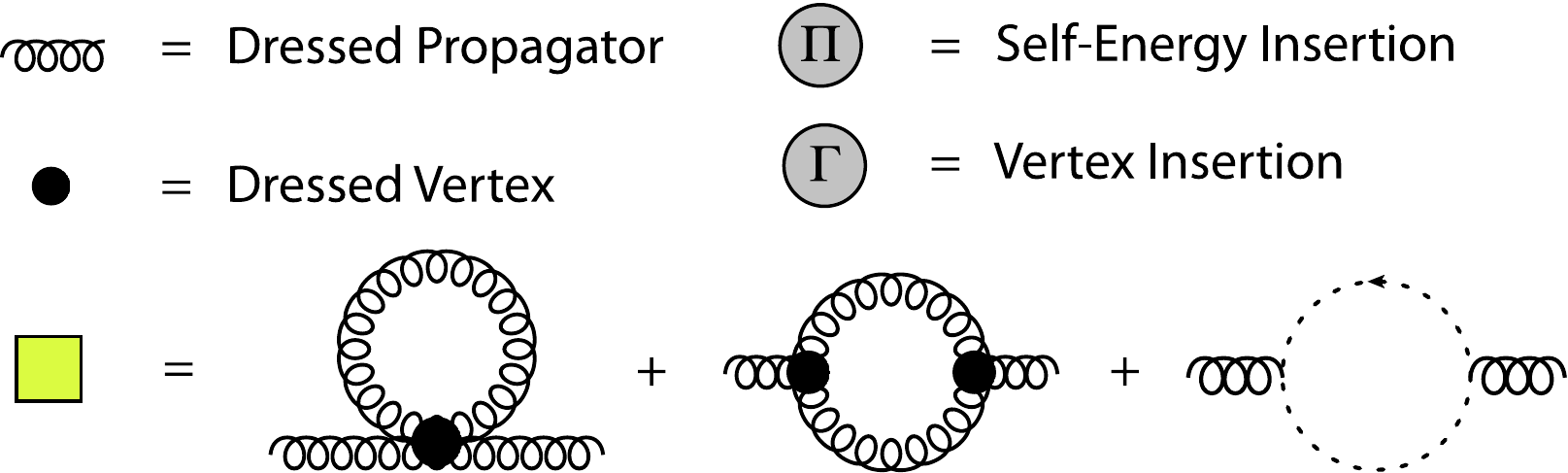}
\end{center}
\caption{Key to the diagrams in Figs.~\ref{fig:dia1-YM} and~\ref{fig:dia2-YM}.}
\label{keydiagrams}
\end{figure*}
%%%%%%%%%%%%%%%%%%%%%%%%%%%%%%%%%%%%%%%%%%%%%%%%%%%%%%%%%%%%%%%%

The thermodynamic potential at leading order in HTL perturbation theory for
QCD 
\bqa
\Omega_{\rm LO}\;=\;(N_c^2-1){\cal F}_{\rm 1a+1b}+\Delta_0{\cal E}_0\;.
\eqa
Here, ${\cal F}_{\rm 1a+1b}$ is the contribution from the gluons
and ghost shown on the first line of Fig.~\ref{fig:dia1-YM}
\bqa
{\cal F}_{\rm 1a+1b}\!\!&=&\!\!-{1\over2}\sumint_{P}\!\left\{(d-1)\log\left[-\Delta_T(P)\right]+\log\Delta_L(P)
\right\}.
\eqa
The transverse and longitudinal HTL propagators $\Delta_T(P)$ and $\Delta_L(P)$ are given in (\ref{Delta-T}) and (\ref{Delta-L}). The leading-order vacuum energy counterterm $\Delta_0{\cal E}_0$ was determined in Ref.~\cite{htl1}:
\bqa
\Delta_0{\cal E}_0\;=\;{N_c^2-1\over128\pi^2\epsilon} m_D^4 \;.
\label{lovac-YM}
\eqa

The thermodynamic potential at NLO in HTL perturbation theory can be written as
\bqa\nonumber
\Omega_{\rm NLO\!\!}&=&\!\!\Omega_{\rm LO}+(N_c^2-1)[{\cal F}_{\rm 2a}+{\cal F}_{\rm 2b}+{\cal F}_{\rm 2c}+{\cal F}_{\rm 2d}] \\ &&
+\;\Delta_1{\cal E}_0+\Delta_1 m_D^2{\partial\over\partial m_D^2}\Omega_{\rm LO} \;,
\label{OmegaNLO-YM}
\eqa
where $\Delta_1{\cal E}_0$ and $\Delta_1m_D^2$ are the terms of order $\delta$ in the vacuum energy density and mass counterterms:
\bqa
\label{del111-YM}
\Delta_1{\cal E}_0\!\!&=&\!\!-{N_c^2-1\over64\pi^2\epsilon}m_D^4 \;, \\
\Delta_1m_D^2\!\!&=&\!\!-{11N_c\over12\pi\epsilon}\alpha_s m_D^2 \;.
\label{del333-YM}
\eqa
The contributions from the two-loop diagrams with the three-gluon and four-gluon vertices are
\bqa
{\cal F}_{\rm 2a}
\!\!&=&\!\!{N_c\over12}g^2\sumint_{PQ}\Gamma^{\mu\lambda\rho}(P,Q,R)\Gamma^{\nu\sigma\tau}(P,Q,R)\Delta^{\mu\nu}(P) 
\Delta^{\lambda\sigma}(Q)\Delta^{\rho\tau}(R) \;,\\
{\cal F}_{\rm 2b} 
\!\!&=&\!\!{N_c\over8}g^2\sumint_{PQ}\Gamma^{\mu\nu,\lambda\sigma}(P,-P,Q,-Q)\Delta^{\mu\nu}(P) 
\Delta^{\lambda\sigma}(Q) \;,
\eqa
where $R=-Q-P$. The contribution from the ghost diagram is
\bqa
{\cal F}_{\rm 2c}\;=\;{N_c\over2}g^2\sumint_{PQ}{1\over Q^2}{1\over R^2}Q^{\mu}R^{\nu}\Delta^{\mu\nu}(P) \;.
\eqa
The contribution from the HTL gluon counterterm diagram with a single gluon self-energy insertion is
\bqa 
{\cal F}_{\rm 2d}\;=\;{1\over2}\sumint_{P}\Pi^{\mu\nu}(P)\Delta^{\mu\nu}(P)\;.
\eqa

The thermodynamic potential at NNLO in HTL perturbation theory can be written as
\bqa\nonumber
\Omega_{\rm NNLO}\!\!&=&\!\!\Omega_{\rm NLO}
+(N_c^2-1)\left[
{\cal F}_{\rm 3a}+{\cal F}_{\rm 3b}+{\cal F}_{\rm 3c}
+{\cal F}_{\rm 3d}+{\cal F}_{\rm 3e}
+{\cal F}_{\rm 3f}+{\cal F}_{\rm 3g}+{\cal F}_{\rm 3h}+{\cal F}_{\rm 3i}\right. \\ 
&& \left. \nonumber
+\;{\cal F}_{\rm 3j}+{\cal F}_{\rm 3k}+{\cal F}_{\rm 3l}
+{\cal F}_{\rm 3m}
\right]
+\Delta_2{\cal E}_0
+\Delta_2 m_D^2{\partial\over\partial m_D^2}
\Omega_{\rm LO}
+\Delta_1 m_D^2{\partial\over\partial m_D^2}
\Omega_{\rm NLO} \\ 
&& +\;{1\over2}\left({\partial^2\over(\partial m_D^2)^2}
\Omega_{\rm LO}\right)\left(\Delta_1m_D^2\right)^2
+(N_c^2-1){F_{\rm 2a+2b+2c}\over\alpha_s}\Delta_1\alpha_s \;,
\label{OmegaNNLO-YM}
\eqa
where $\Delta_1\alpha_s$, $\Delta_2m_D^2$, and $\Delta_2{\cal E}_0$ are terms of order $\delta^2$ in the coupling constant, mass, and vacuum energy counterterms:
\bqa
\Delta_1\alpha_s\!\!&=&\!\!-{11N_c\over12\pi\epsilon}\alpha_s^2\;,
\label{delalpha2} \\ 
\Delta_2m_D^2\!\!&=&\!\!{11N_c\over12\pi\epsilon}\alpha_sm_D^2\;,
\label{delmd2} \\ 
\Delta_2{\cal E}_0\!\!&=&\!\!{N_c^2-1\over128\pi^2\epsilon}m_D^4\;.
\label{del1e2}
\eqa

The contributions from the three-loop diagrams are given by
\bqa\nonumber
{\cal F}_{\rm 3a}\!\!&=&\!\!
{N_c^2\over24}g^4\sumint_{PQR}\Gamma^{\alpha\beta\gamma}(P,Q,-P-Q)\Delta^{\alpha\theta}(P)\Delta^{\beta\mu}(Q)\Delta^{\gamma\sigma}(P+Q)
\\ && \nonumber \times
\Gamma^{\mu\nu\delta}(-Q,-R,Q+R)\Delta^{\pi\nu}(R)\Delta^{\delta\lambda}(Q+R) 
\\ && \times 
\Gamma^{\sigma\lambda\rho}(P+Q,-Q-R,R-P)\Delta^{\rho\phi}(R-P)\Gamma^{\theta\phi\pi}(-P,P-R,R) \;, \\ \nonumber
{\cal F}_{\rm 3b}\!\!&=&\!\!
{N_c^2\over3}g^4\sumint_{PQR}{R^{\alpha}(P+Q+R)^{\beta}(P+R)^{\gamma}\over R^2(P+R)^2(P+Q+R)^2}\Gamma^{\mu\lambda\nu}(-P,-Q,P+Q)
\\ && \times
\Delta^{\alpha\mu}(P)\Delta^{\beta\nu}(P+Q)\Delta^{\gamma\lambda}(Q) \;, \\
{\cal F}_{\rm 3c}\!\!&=&\!\!
-{N_c^2\over4}g^4\sumint_{PQR}{(Q+R)^{\alpha}(R-P)^{\beta}(Q+R-P)^{\mu}R^{\nu}\over R^2(Q+R)^2(Q+R-P)^2(R-P)^2}\Delta^{\alpha\beta}(P)\Delta^{\mu\nu}(Q) \;, \\
{\cal F}_{\rm 3d}\!\!&=&\!\! \nonumber
{N_c^2\over48}\sumint_{PQR}\Gamma^{\alpha\beta,\mu\nu}(P,Q,R,S)\Gamma^{\gamma\delta,\sigma\lambda}(P,Q,R,S)\Delta^{\alpha\gamma}(P)\Delta^{\beta\delta}(Q)\Delta^{\mu\sigma}(R)\Delta^{\nu\lambda}(S) \;,
\\ && \\ \nonumber
{\cal F}_{\rm 3e}\!\!&=&\!\!
-{N_c^2\over4}\sumint_{PQR}\Gamma^{\alpha\mu,\gamma\sigma}(P,Q,R,S)\Delta^{\alpha\beta}(P)\Delta^{\mu\nu}(Q)\Delta^{\gamma\delta}(R)\Delta^{\sigma\phi}(S)\Delta^{\theta\lambda}(P+Q)
\\ &&\times
\Gamma^{\beta\nu\theta}(-P,-Q,P+Q)\Gamma^{\lambda\delta\phi}(-P-Q,-R,-S) \;,
\\
{\cal F}_{\rm 3f}\!\!&=&\!\!
\sumint_{P}\bar{\Pi}^{\mu\nu}(P)\Delta^{\nu\alpha}(P)
\bar{\Pi}^{\alpha\beta}(P)\Delta^{\beta\mu}(P)\;,
\\
{\cal F}_{\rm 3g}\!\!&=&\!\!
{N_c^2\over2}g^4\sumint_{PQR}{P^{\alpha}(P+Q)^{\mu}P^{\nu}(P+R)^{\beta}\over P^4(P+Q)^2(P+R)^2}\Delta^{\mu\nu}(Q)\Delta^{\alpha\beta}(R) \;,
\eqa
where $S=-(P+Q+R)$ and $\bar{\Pi}^{\mu\nu}(P)$ is the one-loop gluon self-energy with HTL-resummed propagators and vertices defined by the yellow box in Fig.~\ref{keydiagrams}:
\bqa\nonumber
\bar{\Pi}^{\mu\nu}(P)
\!\!&=&\!\!{1\over2}N_cg^2\sumint_Q\Gamma^{\mu\nu,\alpha\beta}(P,-P,Q,-Q)\Delta^{\alpha\beta}(Q)+{1\over2}N_cg^2\sumint_Q\Gamma^{\mu\alpha\beta}(P,Q,-R)
\\&&
\times\Delta^{\alpha\gamma}(Q)\Gamma^{\nu\gamma\delta}(P,Q,-R)\Delta^{\beta
\delta}(R)
+N_cg^2\sumint_Q{Q^{\mu}R^{\nu}\over Q^2R^2}\;,
\eqa
where $R=P+Q$. The contributions from the two-loop diagrams with a single self-energy insertion are
\bqa\nonumber
{\cal F}_{\rm 3h}\!\!&=&\!\!
-{N_c\over4}g^2\sumint_{PQ}\Gamma^{\alpha\mu\nu}(P,Q,-Q-P)\Gamma^{\beta\gamma\delta}(P,Q,-Q-P) 
\\ && 
\hspace{1.8cm} \times
\Delta^{\alpha\sigma}(P)\Pi^{\sigma\lambda}(P)\Delta^{\lambda\beta}(P)\Delta^{\mu\gamma}(Q)\Delta^{\nu\delta}(-Q-P)\;,\\
{\cal F}_{\rm 3i}\!\!&=&\!\!
-{N_c\over4}g^2\sumint_{PQ}\Gamma^{\alpha\beta,\mu\nu}(P,-P,Q,-Q)\Delta^{\alpha\gamma}(P)\Pi^{\gamma\delta}(P)\Delta^{\delta\beta}(P)\Delta^{\mu\nu}(Q)\;,\\
{\cal F}_{\rm 3j}\!\!&=&\!\!
-{N_c\over2}g^2\sumint_{PQ}{P^{\alpha}(P+Q)^{\beta}\over P^2(P+Q)^2}\Delta^{\alpha\mu}(Q)\Pi^{\mu\nu}(Q)\Delta^{\nu\beta}(Q)\;.
\eqa
The two-loop diagrams with a subtracted vertex is
\bqa\nonumber
{\cal F}_{\rm 3k}
\!\!&=&\!\!{N_c\over6}g^2m_D^2\sumint_{PQ}{\cal T}^{\mu\lambda\rho}(P,Q,-Q-P)\Gamma^{\nu\sigma\tau}(P,Q,-Q-P)
\\ && \hspace{2.1cm} \times
\Delta^{\mu\nu}(P)\Delta^{\lambda\sigma}(Q)\Delta^{\rho\tau}(-Q-P)\;,\\
{\cal F}_{\rm 3l}
\!\!&=&\!\!{N_c\over8}g^2m_D^2\sumint_{PQ}{\cal T}^{\mu\nu,\lambda\sigma}(P,-P,Q,-Q)\Delta^{\mu\nu}(P)\Delta^{\lambda\sigma}(Q)\;.
\eqa
The contribution from the HTL gluon counterterm diagram with two gluon self-energy insertions is
\bqa
{\cal F}_{\rm 3m}\!\!&=&\!\!
-{1\over4}\sumint_{P}\Pi^{\mu\nu}(P)\Delta^{\nu\alpha}(P)\Pi^{\alpha\beta}(P)\Delta^{\beta\mu}(P)\;.
\eqa

\section{Expansion in the mass parameter}

In this section we carry out the mass expansion for all the diagrams listed in the last section to high enough order to include all terms through order $g^5$ if $m_D$ is taken to be of order $g$. The NLO approximation will be perturbatively accurate to order $g^3$ and the NNLO approximation accurate to order $g^5$.

The free energy can be divided into contributions from hard and soft momenta. In the one-loop diagrams, the contributions are either hard $(h)$ or soft $(s)$, while at the two-loop level, there are hard-hard $(hh)$, hard-soft $(hs)$, and soft-soft $(ss)$ contributions. At three loops there are hard-hard-hard $(hhh)$, hard-hard-soft $(hhs)$, hard-soft-soft $(hss)$, and soft-soft-soft $(sss)$ contributions.

\subsection{Leading order}

\subsubsection{Hard contribution}

For hard momenta, the self-energies are suppressed by $m_D/T$ relative to the propagators, so we can expand in powers of $\Pi_T(P)$ and $\Pi_L(P)$.

For the one-loop graphs (1a) and (1b), we need to expand to second order in $m^2_D$:
\bqa\nonumber
{\cal F}_{\rm 1a+1b}^{(h)}\!\!&=&\!\!
{1\over2}(d-1)\sumint_P\log\left(P^2\right)+{1\over2}m_D^2\sumint_P{1\over P^2}
\\ && \nonumber
-\;{1\over4(d-1)}m_D^4\sumint_P\left[
{1\over P^4}-2{1\over p^2P^2}-2d{1\over p^4}{\cal T}_P
+2{1\over p^2P^2}{\cal T}_P+d{1\over p^4}\left({\cal T}_P\right)^2
\right]
\\ \nonumber
\!\!&=&\!\!
- {\pi^2 \over 45} T^4
+ {1 \over 24} \left[ 1
        + \left( 2 + 2{\zeta'(-1) \over \zeta(-1)} \right) \epsilon \right]
\left( {\mu \over 4 \pi T} \right)^{2\epsilon} m_D^2 T^2
\\&&
-\;{1 \over 128 \pi^2}
\left( {1 \over \epsilon} - 7 + 2 \gamma + {2 \pi^2\over 3} 
\right)
\left( {\mu \over 4 \pi T} \right)^{2\epsilon} m_D^4 \;.
\label{Flo-h-YM}
\eqa

\subsubsection{Soft contribution}

The soft contribution in the diagrams (1a) and (1b) arises from the $P_0=0$ term in the sum-integral. At soft momentum $P=(0,{\bf p})$, the HTL self-energy functions reduce to $\Pi_T(P) = 0$ and $\Pi_L(P) = m_D^2$. The transverse term vanishes in dimensional regularization because there is no momentum scale in the integral over ${\bf p}$. Thus the soft contributions come from the longitudinal term only and read
\bqa\nonumber
{\cal F}^{(s)}_{\rm 1a+1b}
\!\!&=&\!\!{1\over2}T\int_{\bf p}\log\left(p^2+m_D^2\right)\\
\!\!&=&\!\!- {m_D^3T\over12\pi}\left( {\mu \over 2 m} \right)^{2 \epsilon}\left[1+{8\over3}\epsilon\right]\;.
\label{count11-YM}
\eqa

We have kept the order $\epsilon$ terms in the $m_D^2$ and $m_D^3$ terms, respectively in Eqs.~(\ref{Flo-h-YM}) and~(\ref{count11-YM}) since they contribute in the counterterms at next-to-leading order.

\subsection{Next-to-leading order}

\subsubsection{Hard contribution}
The one-loop graph with a gluon self-energy insertion (2d) has an explicit factor of $m_D^2$ and so we need only to expand the sum-integal to first order in $m_D^2$:
\bqa\nonumber
{\cal F}_{\rm 2d}^{(h)}\!\!&=&\!\!-{1\over2}
m_D^2\sumint_P{1\over P^2}
\\ && \nonumber
+\;{1\over2(d-1)}m_D^4\sumint_P\left[
{1\over P^4}-2{1\over p^2P^2}-2d{1\over p^4}{\cal T}_P
+2{1\over p^2P^2}{\cal T}_P
+d{1\over p^4}\left({\cal T}_P\right)^2
\right]\\
\!\!&=&\!\!
-{1 \over 24} \left[ 1
        + \left( 2 + 2{\zeta'(-1) \over \zeta(-1)} \right) \epsilon \right]
\left( {\mu \over 4 \pi T} \right)^{2\epsilon} m_D^2 T^2
\nonumber\\&&
+\;{1 \over 64 \pi^2}
\left( {1 \over \epsilon} - 7 + 2 \gamma_E + {2 \pi^2\over 3} 
\right)
\left( {\mu \over 4 \pi T} \right)^{2\epsilon} m_D^4 \;.
\label{ct1-YM}
\eqa

\subsubsection{Soft contribution}
The soft contribution from (2d) arises from the $P_0=0$ term in the sum-integral. Only the longitudinal part $\Pi_L(P)$ of the self-energy contributes and reads
\bqa\nonumber
{\cal F}^{(s)}_{\rm 2d}\!\!&=&\!\!
-{1\over2}m_D^2T\int_{\bf p}{1\over p^2+m_D^2} \\
\!\!&=&\!\!
{m^3_DT\over 8\pi} \left( {\mu \over 2 m_D} \right)^{2 \epsilon}\left[1 + 2 \epsilon \right]\;.
\label{count12-YM}
\eqa

\subsubsection{$(hh)$ contribution}

For hard momenta, the self-energies are suppressed by $m_D/T$ relative to the propagators, so we can expand in powers of $\Pi_T$ and $\Pi_L$. The two-loop contribution was calculated in Ref.~\cite{htl2} and reads
\bqa
\nonumber
{\cal F}^{(hh)}_{\rm 2a+2b+2c}\!\!&=&\!\!
{N_c\over4}g^2(d-1)^2\sumint_{PQ}\left[
{1\over P^2}{1\over Q^2}\right]+
{N_c\over4}g^2m_D^2\sumint_{PQ}\left[
-2(d-1){1\over P^2}{1\over Q^4}
\right.\\ \nonumber
&&\left.
+\;2(d-2){1\over P^2}{1\over q^2Q^2}+(d+2){1\over Q^2R^2r^2}
-2d{P\cdot Q\over P^2Q^2r^4}-4d{q^2\over P^2Q^2r^4}
\right. \\&& \nonumber
+\;4{q^2\over P^2Q^2r^2R^2}-2(d-1){1\over P^2}{1\over q^2Q^2}{\cal T}_Q
-(d+1){1\over P^2Q^2r^2}{\cal T}_R
\left.
\right.\\ && \left.
+\;4d{q^2\over P^2Q^2r^4}{\cal T}_R
+2d{P\cdot Q\over P^2Q^2r^4}{\cal T}_R
\right]\;.
\label{hh2loop-YM}
\eqa
Using the expressions for the sum-integrals listed in Appendix~\ref{app:sumint}, we obtain
\bqa\nonumber
{\cal F}^{(hh)}_{\rm 2a+2b+2c}\!\!&=&\!\!
{\pi^2\over12}{N_c\alpha_s\over3\pi}
\left[1+\left(2+4{\zeta^{\prime}(-1)\over\zeta(-1)}\right)\epsilon
\right]\left({\mu\over4\pi T}\right)^{4\epsilon}T^4
\\ &&
-\;{7\over96}\left[{1\over\epsilon}+4.621
\right]{N_c\alpha_s\over3\pi}\left({\mu\over4\pi T}\right)^{4\epsilon}m^2_DT^2\;.
\label{hhcount-YM}
\eqa

\subsubsection{$(hs)$ contribution}

In the $(hs)$ region, the momentum $P$ is soft. The momenta $Q$ and $R$ are always hard. The function that multiplies the soft propagator $\Delta_T(0,{\bf p})$, $\Delta_L(0,{\bf p})$ or $\Delta_X(0,{\bf p})$ can be expanded in powers of the soft momentum ${\bf p}$. In the case of $\Delta_T(0,{\bf p})$, the resulting integrals over ${\bf p}$ have no scale and they vanish in dimensional regularization. The integration measure $\int_{\bf p}$ scales like $m_D^3$, the soft propagators $\Delta_L(0,{\bf p})$ and $\Delta_X(0,{\bf p})$ scale like $1/m_D^2$, and every power of $p$ in the numerator scales like $m_D$. The two-loop contribution was calculated in Ref.~\cite{htl2} and reads
\bqa\nonumber
{\cal F}_{\rm 2a+2b+2c}^{(hs)}\!\!&=&\!\!{N_c\over2}g^2T\int_{\bf p}{1\over p^2+m_D^2}
\sumint_Q\left[-(d-1){1\over Q^2}+2(d-1){q^2\over Q^4}\right]
\\ && \nonumber
+\;N_cg^2m_D^2T\int_{\bf p}{1\over p^2+m_D^2}
\sumint_Q\left[-(d-4){1\over Q^4}
\right. \\ &&\left.
+\;{(d-1)(d+2)\over d}{q^2\over Q^6}
-{4(d-1)\over d}{q^4\over Q^8}
\right]\;.
\eqa

In order to facilitate the calculations, it proves useful to isolate the terms that are speficic to HTL perturbation theory. After integrating by parts and using the results from Zhai and Kastening~\cite{Zhai:1995ac}, we can write
\bqa\nonumber
{\cal F}_{\rm 2a+2b+2c}^{(hs)}\!\!&=&\!\!
{N_c\over2}g^2T(d-1)^2\int_{\bf p}{1\over p^2+m_D^2}\sumint_Q{1\over Q^2}
\\ && \nonumber
+\;{N_c\over12}[d^2-5d+16]g^2Tm_D^2\int_{\bf p}{1\over p^2+m_D^2}\sumint_Q{1\over Q^4}
\\ &&
-\;{N_c\over2}(d-5)g^2Tm_D^2\int_{\bf p}{1\over p^2+m_D^2}\sumint_Q{1\over Q^4}\;.
\eqa
Using the expressions for the integrals and sum-integrals in Appendices~\ref{app:sumint} and \ref{app:int}, we obtain
\bqa\nonumber
{\cal F}^{(hs)}_{\rm 2a+2b+2c}\!\!&=&\!\!
-{\pi\over2}{N_c\alpha_s\over3\pi}m_DT^3\left[
1+\left(2+2{\zeta^{\prime}(-1)\over\zeta(-1)}\right)\epsilon
\right]\left({\mu\over4\pi T}\right)^{2\epsilon}
\left({\mu\over2m_D}\right)^{2\epsilon}
\\ &&
-\;{11\over32\pi}\left({1\over\epsilon}+{27\over11}+2\gamma\right)
{N_c\alpha_s\over3\pi}
\left({\mu\over4\pi T}\right)^{2\epsilon}
\left({\mu\over2m_D}\right)^{2\epsilon}m_D^3T\;.
\label{hscount-YM}
\eqa

\subsubsection{$(ss)$ contribution}

The $(ss)$ contribution was obtained by Braaten and Nieto by a two-loop calculation in electrostatic QCD (EQCD) in three dimensions~\cite{BN-96}. Alternatively, one can isolate the $(ss)$ contributions from the two-loop diagrams which were calculated by Arnold and Zhai in Ref.~\cite{AZ-95}. In Ref.~\cite{htl2}, this contribution was calculated and agrees with earlier results. One finds
\bqa\nonumber
{\cal F}^{(ss)}_{\rm 2a+2b+2c}\!\!&=&\!\!
{1\over4}N_cg^2T^2\int_{\bf pq}{p^2+4m_D^2\over p^2(q^2+m_D^2)(r^2+m_D^2)}
\\ \!\!&=&\!\!{3\over16}\left[{1\over\epsilon}+3\right]{N_c\alpha_s\over3\pi}
\left({\mu\over2m_D}\right)^{4\epsilon}m_D^2T^2
\;.
\eqa
We have kept the order $\epsilon$ in Eqs.~(\ref{ct1-YM}),~(\ref{count12-YM}),~(\ref{hhcount-YM}), and~(\ref{hscount-YM}) since they contribute in the counterterms at NNLO.

\subsection{Next-to-next-to-leading order}

\subsubsection{Hard contribution}

The one-loop graph with two gluon self-energy insertions (3m) is proportional to $m_D^4$ and so must be expanded to zeroth order in $m_D^2$
\bqa\nonumber
{\cal F}_{\rm 3m}^{(h)}\!\!&=&\!\!
-{1\over4(d-1)}m_D^4\sumint_P\left[{1\over P^4}-2{1\over p^2P^2}-2d{1\over p^4}{\cal T}_P
+2{1\over p^2P^2}{\cal T}_P+d{1\over p^4}\left({\cal T}_P\right)^2\right]\\
\!\!&=&\!\!
-{1 \over 128\pi^2}\left( {1 \over \epsilon} - 7 + 2 \gamma + {2 \pi^2\over 3} \right)
\left( {\mu \over 4 \pi T} \right)^{2\epsilon} m_D^4 \,.
\label{ct2-YM}
\eqa

\subsubsection{Soft contribution}

The soft contribution from (3m) arises from the $P_0=0$ term in the sum-integral. Only the longitudinal part $\Pi_L(P)$ of the self-energy contributes and reads
\bqa\nonumber
{\cal F}^{(s)}_{\rm 3m}\!\!&=&\!\! - {1\over4}m_D^4T\int_{\bf p}{1\over(p^2+m_D^2)^2} \\
\!\!&=&\!\! - {m^3_DT\over32\pi}\;.
\eqa

\subsubsection{$(hh)$ contribution}

We also need the $(hh)$ contribution from the diagrams (3h)-(3l). We calculate their contributions by expanding the two-loop diagrams (2a)-(2c) to first order in $m_D^2$. This yields
\bqa\nonumber
{\cal F}^{(hh)}_{\rm 3h-3l}\!\!&=&\!\!
-{N_c\over4}g^2m_D^2\sumint_{PQ}\left[
-2(d-1){1\over P^2}{1\over Q^4}+2(d-2){1\over P^2}{1\over q^2Q^2}+(d+2){1\over Q^2R^2r^2}
\right.\\ \nonumber&&\left.
+\;4{q^2\over P^2Q^2r^2R^2}
-2(d-1){1\over P^2}{1\over q^2Q^2}{\cal T}_Q
-2d{P\cdot Q\over P^2Q^2r^4}-4d{q^2\over P^2Q^2r^4}
\right. \\ \nonumber&&\left.
-\;(d+1){1\over P^2Q^2r^2}{\cal T}_R
+4d{q^2\over P^2Q^2r^4}{\cal T}_R
+2d{P\cdot Q\over P^2Q^2r^4}{\cal T}_R
\right]\\
\!\!&=&\!\!
{7\over96}\left[
{1\over\epsilon}+4.621
\right]{N_c\alpha_s\over3\pi}\left({\mu\over4\pi T}\right)^{4\epsilon}
m^2_DT^2\;.
\label{hh2loopself-YM}
\eqa

\subsubsection{$(hs)$ contribution}

We also need the $(hs)$ contribution from the diagrams (3h)-(3l). Again we calculate their contributions by expanding the two-loop diagrams (2a)-(2c) to first order in $m_D^2$. This yields
\bqa\nonumber
{\cal F}_{\rm 3h-3l}^{(hs)}
\!\!&=&\!\!
{N_c\over2}g^2(d-1)^2m_D^2T\int_{\bf p}{1\over(p^2+m_D^2)^2}\sumint_Q{1\over Q^2}
\\ && \nonumber 
-\;{N_c\over12}g^2m_D^2T\left[d^2-5d+16\right]\int_{\bf p}{p^2\over(p^2+m_D^2)^2}
\sumint_Q{1\over Q^4}
\\ &&
+\;{N_c\over2}g^2(d-5)m_D^2T\int_{\bf p}{p^2\over(p^2+m_D^2)^2}
\sumint_Q{1\over Q^4}\;.
\eqa
Using the expressions in Appendices~\ref{app:sumint} and \ref{app:int}, we obtain
\bqa\nonumber
{\cal F}_{\rm 3h-3l}^{(hs)}\!\!&=&\!\!
{\pi\over4}{N_c\alpha_s\over3\pi}m_DT^3
+{33\over64\pi}\left({1\over\epsilon}+{59\over33}+2\gamma\right)
{N_c\alpha_s\over3\pi}
\left({\mu\over4\pi T}\right)^{2\epsilon}
\left({\mu\over2m_D}\right)^{2\epsilon}m_D^3T\;. \\
\eqa

\subsubsection{$(ss)$ contribution}

The $(ss)$ contribution from the two-loop diagrams with a single self-energy insertion can be easily obtained by expanding the two-loop result in powers of $m_D^2$. This yields
\bqa\nonumber
{\cal F}^{(ss)}_{\rm 3h-3l}\!\!&=&\!\!-{1\over4}N_cg^2m_D^2T^2
\int_{\bf pq}\left[{4\over p^2(q^2+m^2_D)(r^2+m_D^2)}
-{2(p^2+4m_D^2)\over p^2(q^2+m^2_D)^2(r^2+m_D^2)}\right] \\
\!\!&=&\!\! -{3\over16}
\left[{1\over\epsilon}+1\right]{N_c\alpha_s\over3\pi}
\left({\mu\over2m_D}\right)^{4\epsilon}m_D^2T^2\;.
\eqa
We have verified this by explicitly calculating the relevant diagrams.

\subsubsection{$(hhh)$ contribution}

If all the three loop momenta are hard, we can obtain the $m_D/T$ expansion simply by expanding in powers of $m_D^2$. To obtain the expansion through order $g^5$, we can use bare propagators and vertices. The contributions from the three-loop diagrams were first calculated by Arnold and Zhai in Ref.~\cite{AZ-95}, and later by Braaten and Nieto~\cite{BN-96}.
One finds
\bqa\nonumber
{\cal F}^{(hhh)}_{\rm 3a-3g}\!\!&=&\!\!
{N_c^2\over4}g^4(d-1)^2
\sumint_{PQR}\left[
-(d-5){1\over P^2Q^2R^4}
-{1\over2}{1\over P^2Q^2R^2(P+Q+R)^2}
\right. \\ && \nonumber
\left. -{(P-Q)^4\over P^2Q^2R^4(Q-R)^2(R-P)^2}
\right]\\
\!\!&=&\!\!-{25\pi^2\over48}\left({N_c\alpha_s\over3\pi}\right)^2
\left[{1\over\epsilon}
+{238\over125}
+{12\over25}\gamma
+{176\over25}{\zeta^{\prime}(-1)\over\zeta(-1)}
-{38\over25}{\zeta^{\prime}(-3)\over\zeta(-3)}
\right]\left({\mu\over4\pi T}\right)^{6\epsilon}T^4\;. \nonumber \\
\eqa

\subsubsection{$(hhs)$ contributions}

All the diagrams except (3f) are infrared finite in the limit $m_D\rightarrow0$. This implies that the $g^5$ contribution is given by using a dressed longitudinal propagator and bare vertices. The ring diagram (3f) is infrared divergent in that limit. The contribution through $g^5$ is obtained by expanding in powers of self-energies and vertices and one obtains
\bqa\nonumber
{\cal F}^{(hhs)}_{\rm 3a-3g}\!\!&=&\!\!
-{N_c^2\over4}g^4T(d-1)^4\int_{\bf p}{1\over(p^2+m^2)^2}\sumint_{QR}{1\over Q^2R^2}
\\ && \nonumber
+\;{N_c^2\over12}g^4(d-1)^2
\left[d^2-11d+46\right]\int_{\bf p}{p^2\over(p^2+m^2)^2}\sumint_{QR}
{1\over Q^2R^4}
\\ \nonumber \!\!&=&\!\!
-{\pi^3\over2}\left({N_c\alpha_s\over3\pi}\right)^2
{T^5\over m_D}
\\ &&
-\;{33\pi\over16}\left({N_c\alpha_s\over3\pi}\right)^2
\left[
{1\over\epsilon}+{59\over33}+2\gamma+2{\zeta^{\prime}(-1)\over\zeta(-1)}
\right]m_DT^3
\left({\mu\over2m_D}\right)^{2\epsilon}\left({\mu\over4\pi T}\right)^{4\epsilon}
\;. \nonumber \\
\eqa

\subsubsection{$(hss)$ contribution}

For all the diagrams that are infrared safe, the $(hss)$ contribution is of order $g^4m^2$, i.e. $g^6$ and can be ignored. The infrared divergent diagrams contribute as follows
\bqa\nonumber
{\cal F}^{(hss)}_{\rm 3a-3g}\!\!&=&\!\!{1\over4}g^4T^2N_c^2(d-1)^2T^2\sumint_{R}{1\over R^2}
\\ && \nonumber
\times\int_{\bf pq}
\left[{4\over p^2(q^2+m^2_D)(r^2+m_D^2)}
-{2(p^2+4m_D^2)\over p^2(q^2+m^2_D)^2(r^2+m_D^2)}\right] 
\\ \!\!&=&\!\!{3\pi^2\over4}
\left[{1\over\epsilon}+1+2{\zeta^{\prime}(-1)\over\zeta(-1)}\right]
\left({N_c\alpha_s\over3\pi}\right)^2
\left({\mu\over2m_D}\right)^{4\epsilon}\left({\mu\over4\pi T}\right)^{2\epsilon}
T^4\;. 
\eqa

\subsubsection{$(sss)$ contribution}
 
The $(sss)$ contribution is given by a three-loop calculation of the free energy of EQCD in three dimensions. This calculation was performed in Ref.~\cite{BN-96}. Alternatively, one can isolate the $(sss)$ contributions from the diagrams listed in Ref.~\cite{AZ-95}. The result is 
\bqa\nonumber
{\cal F}^{(sss)}_{\rm 3a-3g}\!\!&=&\!\!
N_c^2g^4T^3\int_{\bf pqr}\left\{
-{1\over4}{1\over(p^2+m^2_D)(q^2+m^2_D)(r^2+m^2_D)^2} 
\right. \\&&\left. \nonumber
+\;{2\over(p^2+m^2_D)(q^2+m^2_D)(r^2+m^2_D)({\bf q}-{\bf r})^2}
\right. \\&&\left. \nonumber
-\;{2m_D^2\over(p^2+m^2_D)(q^2+m^2_D)(r^2+m^2_D)^2({\bf q}-{\bf r})^2}
\right. \\&&\left. \nonumber
-\;{m_D^2\over(p^2+m^2_D)(q^2+m^2_D)(r^2+m^2_D)({\bf q}-{\bf r})^4}
\right. \\&&\left. \nonumber
-\;{1\over4}{({\bf p}-{\bf q})^2\over(p^2+m^2_D)(q^2+m^2_D)(r^2+m^2_D)({\bf q}-{\bf r})^2({\bf r}-{\bf p})^2}
\right. \\&&\left. \nonumber
-\;{1\over2}(d-2){1\over(p^2+m^2_D)(q^2+m^2_D)({\bf q}-{\bf r})^2({\bf r}-{\bf p})^2}
\right. \\&&\left. \nonumber 
+\;{1\over2}(3-d){(r^2+m^2_D)\over(p^2+m^2_D)(q^2+m^2_D)({\bf p}-{\bf q})^2({\bf q}-{\bf r})^2({\bf r}-{\bf p})^2}
\right. \\&&\left. \nonumber 
-\;{1\over2}(d-2){(r^2+m^2_D)^2\over(p^2+m^2_D)(q^2+m^2_D)({\bf p}-{\bf q})^4({\bf q}-{\bf r})^2({\bf r}-{\bf p})^2}
\right. \\&&\left. \nonumber 
+\;{4m_D^2\over(p^2+m^2_D)(q^2+m^2_D)(r^2+m^2_D)({\bf q}-{\bf r})^2({\bf r}-{\bf p})^2}
\right. \\&&\left. \nonumber
+\;{2m_D^2\over(p^2+m^2_D)(q^2+m^2_D)({\bf p}-{\bf q})^2({\bf q}-{\bf r})^2({\bf r}-{\bf p})^2}
\right. \\&&\left. \nonumber 
-\;{4m_D^4\over(p^2+m^2_D)(q^2+m^2_D)(r^2+m^2_D)^2({\bf q}-{\bf r})^2({\bf r}-{\bf p})^2}
\right. \\&&\left. \nonumber
-\;{3\over8}{1\over(p^2+m^2_D)(q^2+m^2_D)[({\bf q}-{\bf r})^2+m^2_D][({\bf r}-{\bf p})^2+m^2_D)]}
\right. \\&&\left. \nonumber
-\;{1\over2}{({\bf p}-{\bf q})^2\over(p^2+m^2_D)(q^2+m^2_D)[({\bf q}-{\bf r})^2+m^2_D][({\bf r}-{\bf p})^2+m^2_D]r^2}
\right. \\&&\left. \nonumber
-\;{1\over4}{({\bf p}-{\bf q})^4\over(p^2+m^2_D)(q^2+m^2_D)[({\bf q}-{\bf r})^2+m^2_D][({\bf r}-{\bf p})^2+m^2_D]r^4}
\right. \\&&\left. \nonumber
-\;{2m_D^2\over(p^2+m^2_D)(q^2+m^2_D)[({\bf q}-{\bf r})^2+m^2_D][({\bf r}-{\bf p})^2+m^2_D)]r^2}
\right. \\&&\left. \nonumber
-\;{m_D^2({\bf p}-{\bf q})^2\over(p^2+m^2_D)(q^2+m^2_D)[({\bf q}-{\bf r})^2+m^2_D][({\bf r}-{\bf p})^2+m^2_D]r^4}
\right. \\&&\left. \nonumber
-\;{m_D^4\over(p^2+m^2_D)(q^2+m^2_D)[({\bf q}-{\bf r})^2+m^2_D][({\bf r}-{\bf p})^2+m^2_D]r^2({\bf p}-{\bf q})^2}
\right. \\&&\left. \nonumber
-\;{m_D^4\over(p^2+m^2_D)(q^2+m^2_D)[({\bf q}-{\bf r})^2+m^2_D][({\bf r}-{\bf p})^2+m^2_D]r^4}
\right. \\&&\left. 
-\;{1\over4}{(q^2+m^2_D)\over(p^2+m^2_D)[({\bf r}-{\bf p})^2+m^2_D][({\bf q}-{\bf r})^2+m^2_D]r^2({\bf p}-{\bf q})^2}
\right\}\;.
\eqa
The expression for the integrals are given in Appendix~\ref{app:int}. Adding
Eqs.~(\ref{sssfirst})--~(\ref{ssslast}), the final result
is
\bqa
{\cal F}^{(sss)}_{\rm 3a-3g}&=&{9\pi\over4}\left({N_c\alpha_s\over3\pi}\right)^2
\left[{89\over24}-{11\over6}\log2+{1\over6}\pi^2\right]m_DT^3\;.
\eqa
Note that all the poles in $\epsilon$ cancel.

\section{Thermodynamic potentials}

In this section we present the final renormalized thermodynamic potential explicitly through order $\delta^2$, aka NNLO. The final NNLO expression is completely analytic; however, there are some numerically determined constants which remain in the final expressions at NLO.

\subsection{Leading order}

The leading order thermodynamic potential is given by the contribution from the diagrams (1a) and (1b)

\bqa\nonumber
\Omega_{\rm 1-loop}\!\!&=&\!\! 
{\cal F}_{\rm ideal}\left\{ 1 - {15 \over 2} \hat m_D^2 + 30 \hat m_D^3+ {45 \over 8}
\left({1\over\epsilon} +2\log {\hat \mu \over 2} - 7 + 2\gamma + {2\pi^2\over 3} \right)\hat m_D^4\right\} \;, \\ 
\label{Omega-1-YM}
\eqa
where ${\cal F}_{\rm ideal}$ is the free energy of a gas of $N_c^2-1$ massless spin-one bosons and $\hat m_D$ and $\hat \mu$ are dimensionless variables:
\bqa
{\cal F}_{\rm ideal}\!\!&=&\!\!\left(N_c^2-1\right)\left(-{\pi^2\over45}T^4\right) \;, \\
\hat m_D\!\!&=&\!\!{m_D \over 2 \pi T} \;, \\
\hat \mu\!\!&=&\!\!{\mu \over 2 \pi T} \;. 
\eqa
The complete expression for the leading order thermodynamic potential is given by adding the leading vacuum energy counterterm~(\ref{lovac-YM}) to Eq.~(\ref{Omega-1-YM}):
\bqa
\Omega_{\rm LO}\!\!&=&\!\! 
{\cal F}_{\rm ideal}\left\{ 1 - {15 \over 2} \hat m_D^2 + 30 \hat m_D^3 + {45 \over 4}
\left(\log {\hat \mu \over 2} - {7\over2} + \gamma + {\pi^2\over 3} \right)\hat m_D^4\right\} \;,
\label{Omega-1}
\eqa

\subsection{Next-to-leading order}

The renormalization contributions at first order in $\delta$ are
\bqa
\Delta\Omega_1\!\!&=&\!\!\Delta_1{\cal E}+\Delta_1m_D^2{\partial\over\partial m_D^2}\Omega_{\rm LO}\;.
\eqa
Using the results listed in Eqs.~(\ref{del111-YM}) and (\ref{del333-YM}) the complete contribution from the counterterm at first order in $\delta$ is
\bqa
\nonumber 
\Delta\Omega_1\!\!&=&\!\! 
{\cal F}_{\rm ideal}\left\{ {45\over4\epsilon} \hat m_D^4 +{165\over8}\left[{1\over\epsilon}+2\log{\hat\mu\over 2} + 2 {\zeta'(-1)\over\zeta(-1)}+2 \right]{N_c\alpha_s \over 3\pi}\hat m_D^2
\right. \\ && \left.
-\;{495\over4}\left[{1\over\epsilon}+2\log{\hat\mu\over 2} - 2 \log \hat m_D +2 \right] {N_c\alpha_s \over 3\pi}  \hat m_D^3\right\}\;. 
\label{OmegaVMct1-YM}
\eqa
Adding the NLO counterterms~(\ref{OmegaVMct1-YM}) to the contributions from the various NLO diagrams, we obtain the renormalized NLO thermodynamic potential~\cite{htl2}
\bqa
\Omega_{\rm NLO}\!\!&=&\!\!
{\cal F}_{\rm ideal}\left\{ 
	1  - 15 \hat m_D^3 
	- {45\over4}\left(\log\hat{\mu\over2}-{7\over2}+\gamma+{\pi^2\over3}
\right)\hat m_D^4
\right. \nonumber \\ && \left.
	+\;\left[ -{15\over4}
	+ 45 \hat m_D-{165\over4}
\left(\log{\hat\mu \over 2}-{36\over11}\log\hat{m}_D
-2.001\right)\hat m_D^2
\right. \right. \nonumber \\ && \left. \left.
	+\;{495\over2}\left(\log{\hat\mu \over 2}+{5\over22}
+\gamma\right)\!\!\hat m_D^3
\right]
	{N_c\alpha_s\over3\pi} 
\right\} \;.
\label{Omega-NLO-YM}
\eqa

\subsection{Next-to-next-to-leading order}

The renormalization contributions at second order in $\delta$ are
\bqa\nonumber
\Delta\Omega_2\!\!&=&\!\!\Delta_2{\cal E}_0
+\Delta_2m_D^2{\partial\over\partial m_D^2}\Omega_{\rm LO}
+\Delta_1m_D^2{\partial\over\partial m_D^2}\Omega_{\rm NLO}
\\ &&
+\;{1\over2}\left({\partial^2\over(\partial m_D^2)^2}\Omega_{\rm LO}\right)\left(\Delta_1m_D^2\right)^2
+{F_{\rm 2a-2c}\over\alpha}\Delta_1\alpha_s\;.
\eqa
Using the results listed above, we obtain
\bqa\nonumber
\Delta\Omega_2\!\!&=&\!\! 
{\cal F}_{\rm ideal}
\left\{ 
-{45\over8\epsilon}\hat{m}_D^4			
-{165\over8}{N_c\alpha_s \over 3\pi}  
\left[{1\over\epsilon}+2\log{\hat\mu\over 2} + 2 {\zeta'(-1)\over\zeta(-1)}+2 \right]\hat m_D^2
\right. \\ \nonumber && \left.
+\;{1485\over8}{N_c\alpha_s \over 3\pi}\left[{1\over\epsilon}+2\log{\hat\mu\over 2} -2\log{\hat{m}_D}+{4\over3}\right]\hat{m}_D^3
\right. \\ \nonumber && \left.
+\;\left({N_c\alpha_s\over3\pi}\right)^2\left[
{165\over16}\left({1\over\epsilon}+4\log{\hat{\mu}\over2}+2+4{\zeta^{\prime}(-1)\over\zeta(-1)}\right)
\right. \right.\\ && \left. \left.
-\;{1485\over8}\left({1\over\epsilon}+4\log{\hat{\mu}\over2}
-2\log\hat{m}_D
+{4\over3}+2{\zeta^{\prime}(-1)\over\zeta(-1)}
\right)\hat{m}_D
\right]
\right\}\;. 
\label{OmegaVMct2-YM}
\eqa
Adding the NNLO counterterms (\ref{OmegaVMct2-YM}) to the contributions from the various NNLO diagrams we obtain the renormalized NNLO thermodynamic potential.  We note that at NNLO all numerically determined coefficients of order $\epsilon^0$ drop out and we are left with a final result which is completely analytic. The resulting NNLO thermodynamic potential is
\begin{eqnarray}\nonumber
\Omega_{\rm NNLO}\!\!&=&\!\!{\cal F}_{\rm ideal}
\left\{1-{15\over4}\hat{m}_D^3
+{N_c\alpha_s\over3\pi}\left[
-{15\over4}+{45\over2}\hat{m}_D-{135\over2}\hat{m}^2_D
\right. \right. \\ \nonumber && \left. \left.
-\;{495\over4}\left(\log\hat{\mu\over2}+{5\over22}+\gamma\right)\hat{m}^3_D\right]
+\left({N_c\alpha_s\over3\pi}\right)^2\left[{45\over4\hat{m}_D}
\right. \right. \\ \nonumber && \left. \left.
-\;{165\over8}\left(\log{\hat{\mu}\over2}-{72\over11}\log{\hat{m}_D}-{84\over55}-{6\over11}\gamma
-{74\over11}{\zeta^{\prime}(-1)\over\zeta(-1)}
+{19\over11}{\zeta^{\prime}(-3)\over\zeta(-3)}\right)
\right. \right. \\ && \left. \left.
+{1485\over4}\left(\log{\hat{\mu}\over2}-{79\over44}+\gamma+\log2-{\pi^2\over11}\right)\hat{m}_D
\right]
\right\} \;.
\label{Omega-NNLO-YM}
\end{eqnarray}

We note that the coupling constant counterterm listed in Eq.~(\ref{delalpha-YM}) coincides with the known one-loop running of the QCD coupling constant
\beq
\mu \frac{d g^2}{d \mu}\;=\;-\frac{11N_cg^4}{24\pi^2} \, .
\label{runningcoupling-YM}
\eeq
Finally, note that if we use the weak-coupling value for the Debye mass $m_D^2=4\pi N_c\alpha_sT^2/3$, the NNLO HTLpt result~(\ref{Omega-NNLO-YM}) reduces to the weak-coupling result through order $g^5$ and we have checked that this is the case.

\section{Thermodynamic functions}
\label{functions}

\subsection{Mass prescriptions}

The mass parameter $m_D$ in HTLpt is completely arbitrary. To complete a calculation, it is necessary to specify $m_D$ as function of $g$ and $T$. In our case this implies that the free energy ${\cal F}$ is obtained by specifying $m_D$ as a function of $T$ and $\alpha_s$ in the thermodynamic potential $\Omega$. In this section we will discuss several prescriptions for the mass parameter.

\subsubsection{Variational Debye mass}

The variational mass is given by the solution to the variational gap equation which is defined by 
\bqa
{\partial \ \ \over \partial m_D}\Omega(T,\alpha_s,m_D,\mu,\delta=1) \;=\; 0 \;.
\label{gap-YM}
\eqa
Applying it to~(\ref{Omega-NNLO-YM}), the NNLO gap equation reads
\begin{eqnarray}
{45\over4} \hat m_D^2\!\!&=&\!\!{N_c\alpha_s\over3\pi} \left[{45\over2} - 135 \hat m_D - {1485\over4} \left( \log{\hat\mu\over2} + {5\over22} + \gamma\right)\!\!\hat m_D^2 \right]
\nonumber \\ &&
+ \left({N_c\alpha\over3\pi}\right)^2 \left[-{45\over4}{1\over\hat m_D^2} + {135\over \hat m_D} + {1485\over4}\left(\log{\hat\mu \over 2} - {79\over44} + \gamma + \log2 - {\pi^2\over11}\right)\right]\;.
\nonumber \\
\end{eqnarray}

At leading order in HTLpt, the only solution is the trivial solution, i.e. $m_D=0$. In that case, it is natural to chose the weak-coupling result for the Debye mass. This was done in Ref.~\cite{htl1}. At NLO, the resulting gap equation has a nontrivial solution, which is real for all values of the coupling~\cite{htl2}. At NNLO, the solution to the gap equation is plagued by imaginary parts for all values of the coupling. The problem with complex solutions seems to be generic since it has also been observed in screened perturbation theory~\cite{Andersen:2000yj} and QED~\cite{Andersen:2009tw}. In those cases, however, it was complex only for small values of the coupling.

\subsubsection{Perturbative Debye mass}

At leading order in the coupling constant $g$, the Debye mass is given by the static longitudinal gluon self-energy at zero three-momentum, $m_D^2=\Pi_L(0,0)$, i.e.
\bqa\nonumber
m_D^2\!\!&=&\!\!N_c(d-1)^2g^2\sumint_P{1\over P^2}\\
\!\!&=&\!\!{4\pi\over3}N_c\alpha_sT^2\;.
\eqa
The next-to-leading order correction to the Debye mass is determined by resummation of one-loop diagrams with dressed vertices. Furthermore, since it suffices to take into account static modes in the loops, the HTL-corrections to the vertices also vanish. The result, however, turns out to be logarithmically infrared divergent, which reflects the sensitivity to the nonperturbative magnetic mass scale. The result was first obtained by Rebhan~\cite{Rebhan:1993az} and reads~\footnote{In Ref.~\cite{Rebhan:1993az}, it was shown that the gauge dependent part of the static gluon self-energy $\Pi_L(0,{\bf k})$ vanishes when it is evaluated on shell, i.e. when $k^2=-m_D^2$. This is in accordance with general gauge-dependece identities~\cite{Kobes:1990dc}.}
\bqa
\delta m_D^2\;=\:m_D^2\sqrt{3N\over\pi}\alpha^{1/2}
\left[\log{2m_D\over m_{\rm mag}}-{1\over2}
\right]\;,
\eqa
where $m_{\rm mag}$ is the nonperturbative magnetic mass. We will not use this mass prescription since it involves the magnetic mass which would require input from e.g. lattice simulations.

\subsubsection{BN mass parameter $m_E^2$}

In the previous subsection, we saw that the Debye mass is sensitive to the nonperturbative magnetic mass which is of order $g^2T$. In QED, the situation is much better. The Debye mass can be calculated order by order in $e$ using resummed perturbation theory. The Debye mass then receives contributions from the scale $T$ and $eT$. Effective field theory methods and dimensional reduction can be conveniently used to calculate separately the contributions to $m_D$ from the two scales in the problem. The contributions to $m_D$ and other physical quantities from the scale $T$ can be calculated using bare propagators and vertices. The contributions from the soft scale can be calculated using an effective three-dimensional field theory called electrostatic QED. The parameters of this effective theory are obtained by a matching procedure and encode the physics from the scale $T$. The effective field theory contains a massive field $A_0$ that up to normalization can be identified with the zeroth component of the gauge field in QED. The mass parameter $m_E$ of $A_0$ gives the contribution to the Debye mass from the hard scale $T$ and can be written as a power series in $e^2$. For non-Abelian gauge theories, the corresponding effective three-dimensional theory is called electrostatic QCD. The mass parameter $m_E$ for the field $A_0^a$ (which lives in the adjoint representation) can also be calculated as a power series in $g^2$. It has been determined to order $g^4$ by Braaten and Nieto~\cite{BN-96}. For pure-glue QCD, it reads
\bqa
m_E^2\;=\;
{4\pi\over3}N_c\alpha_sT^2\left[
1+{N_c\alpha_s\over3\pi}\left(
{5\over4}+{11\over2}\gamma
+{11\over2}\log{\mu\over4\pi T}
\right)
\right] \;.
\label{BN}
\eqa
We will use the mass parameter $m_E$ as another prescription for the Debye mass and denote it by the Braaten Nieto (BN) mass prescription.

\subsection{Pressure}

In this subsection, we present our results for the pressure using the variational mass prescription and the BN mass prescription.

\subsubsection{Variational mass}

%%%%%%%%%%%%%%%%%%%%%%%%%%%%%%%%%%%%%%%%%%%%%%%%%%%%
\begin{figure}[t]
\begin{center}
\includegraphics[width=10cm]{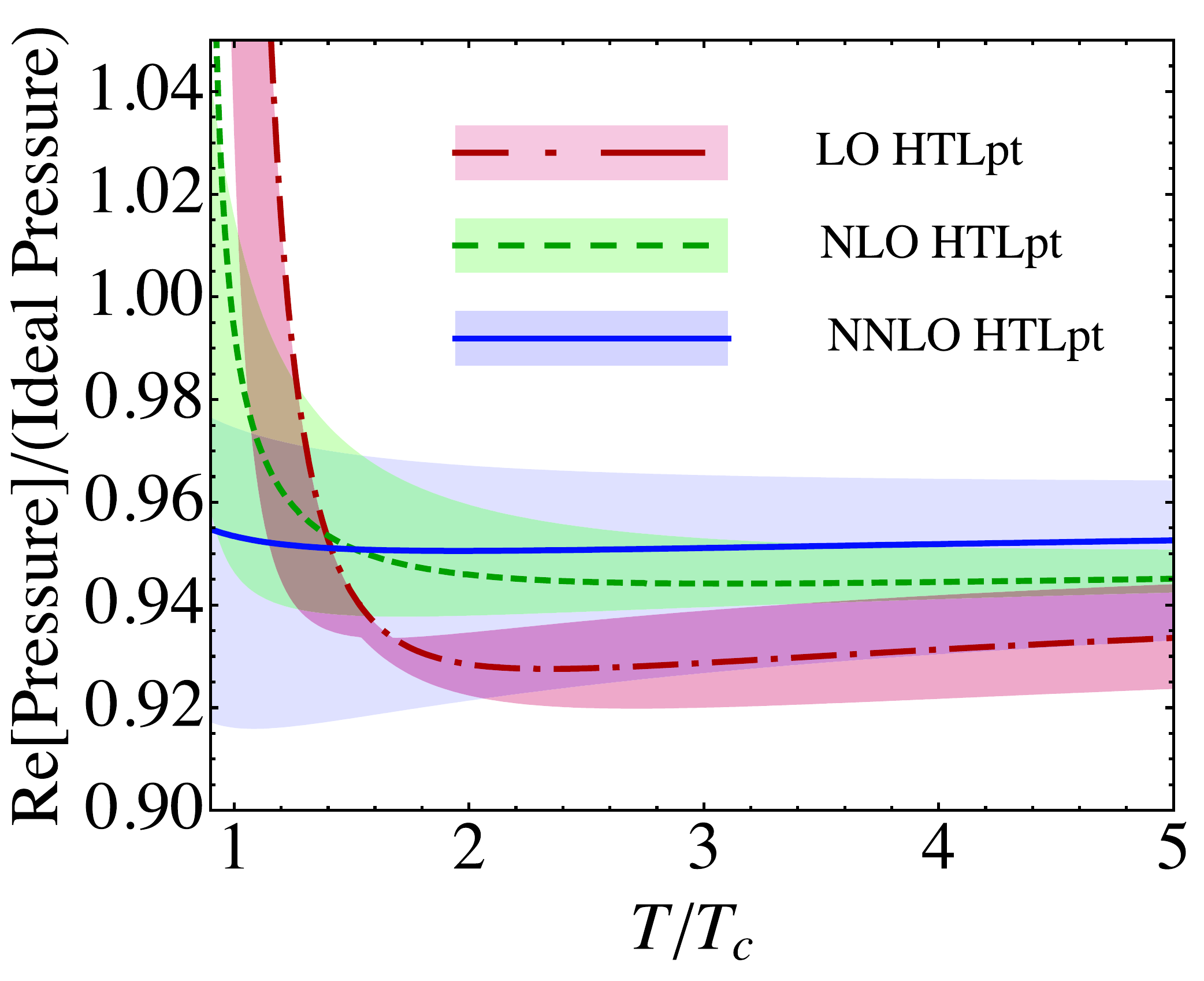}
\end{center}
\vspace{-3mm}
\caption{Comparison of LO, NLO, and NNLO predictions for the scaled real part of the pressure using the variational mass and three-loop running~\cite{Amsler:2008zzb}.  Shaded bands show the result of varying the renormalization scale $\mu$ by a factor of 2 around $\mu = 2 \pi T$.}
\label{fig:pressure}
\end{figure}
%%%%%%%%%%%%%%%%%%%%%%%%%%%%%%%%%%%%%%%%%%%%%%%%%%%%
%%%%%%%%%%%%%%%%%%%%%%%%%%%%%%%%%%%%%%%%%%%%%%%%%%%%
\begin{figure}[h]
\begin{center}
\includegraphics[width=10cm]{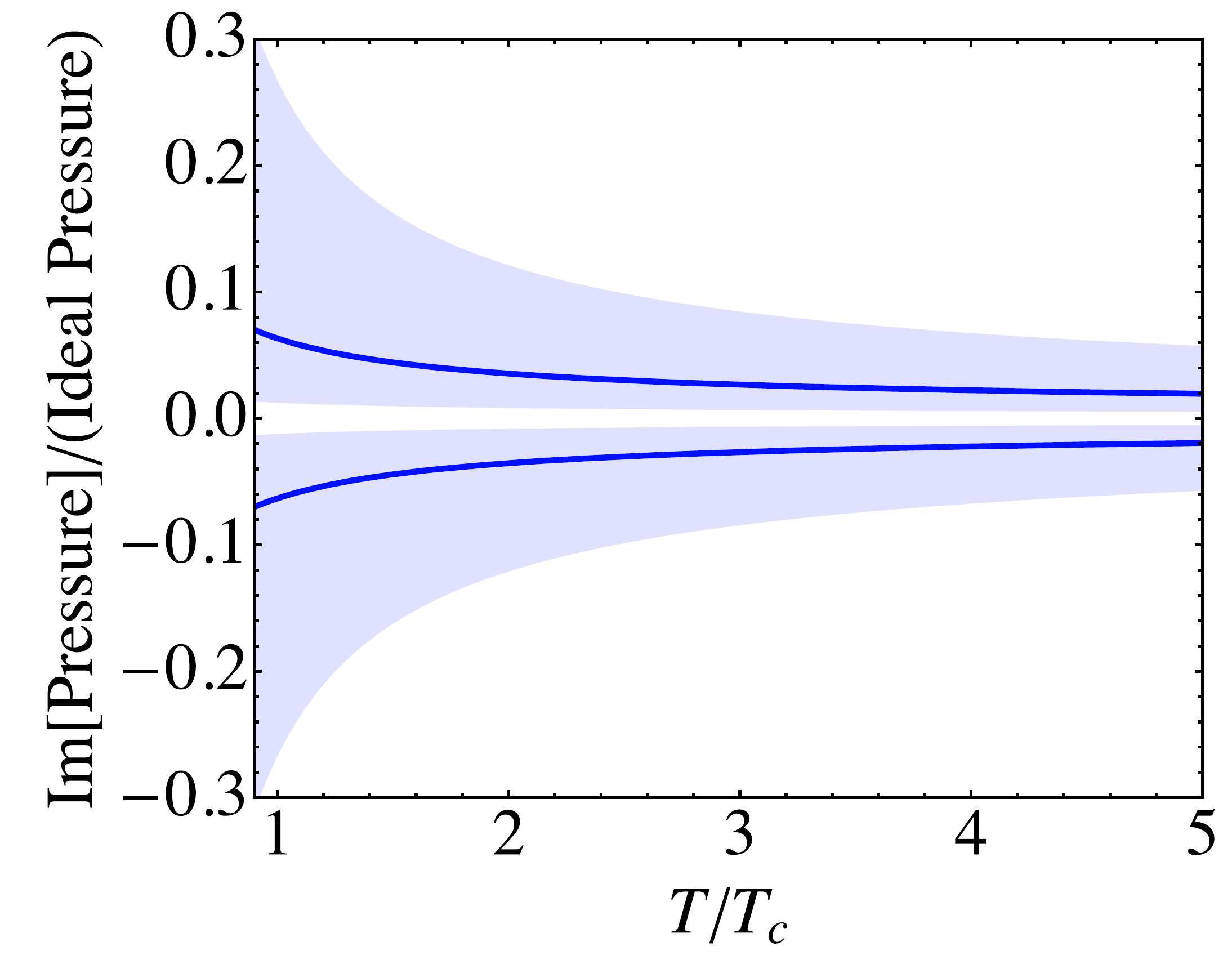}
\end{center}
\vspace{-3mm}
\caption{The NNLO result for the scaled imaginary part of the pressure using the variational mass and three-loop running~\cite{Amsler:2008zzb}. The two curves arises from the two complex conjugate solutions to the gap equations. Shaded bands show the result of varying the renormalization scale $\mu$ by a factor of 2 around $\mu = 2 \pi T$.}
\label{imagp}
\end{figure}
%%%%%%%%%%%%%%%%%%%%%%%%%%%%%%%%%%%%%%%%%%%%%%%%%%%%
In Fig.~\ref{fig:pressure}, we compare the LO, NLO, and NNLO predictions for the real part of the pressure normalized to that of an ideal gas using the variational mass and three-loop running of $\alpha_s$~\cite{Amsler:2008zzb}. Shaded bands show the result of varying the renormalization scale $\mu$ by a factor of 2 around $\mu = 2 \pi T$. 

In Fig.~\ref{imagp}, we show the NNLO result for the imaginary part of the pressure normalized by the ideal gas pressure using the variational mass and three-loop running of $\alpha_s$~\cite{Amsler:2008zzb}. The imaginary part decreases with increasing temperature and is rather small beyond $3-4\,T_c$.

Due to the imaginary parts, we abandon the variational prescription in the remainder of the chapter.

\subsubsection{BN mass}

\begin{figure}[h]
\begin{center}
\includegraphics[width=10cm]{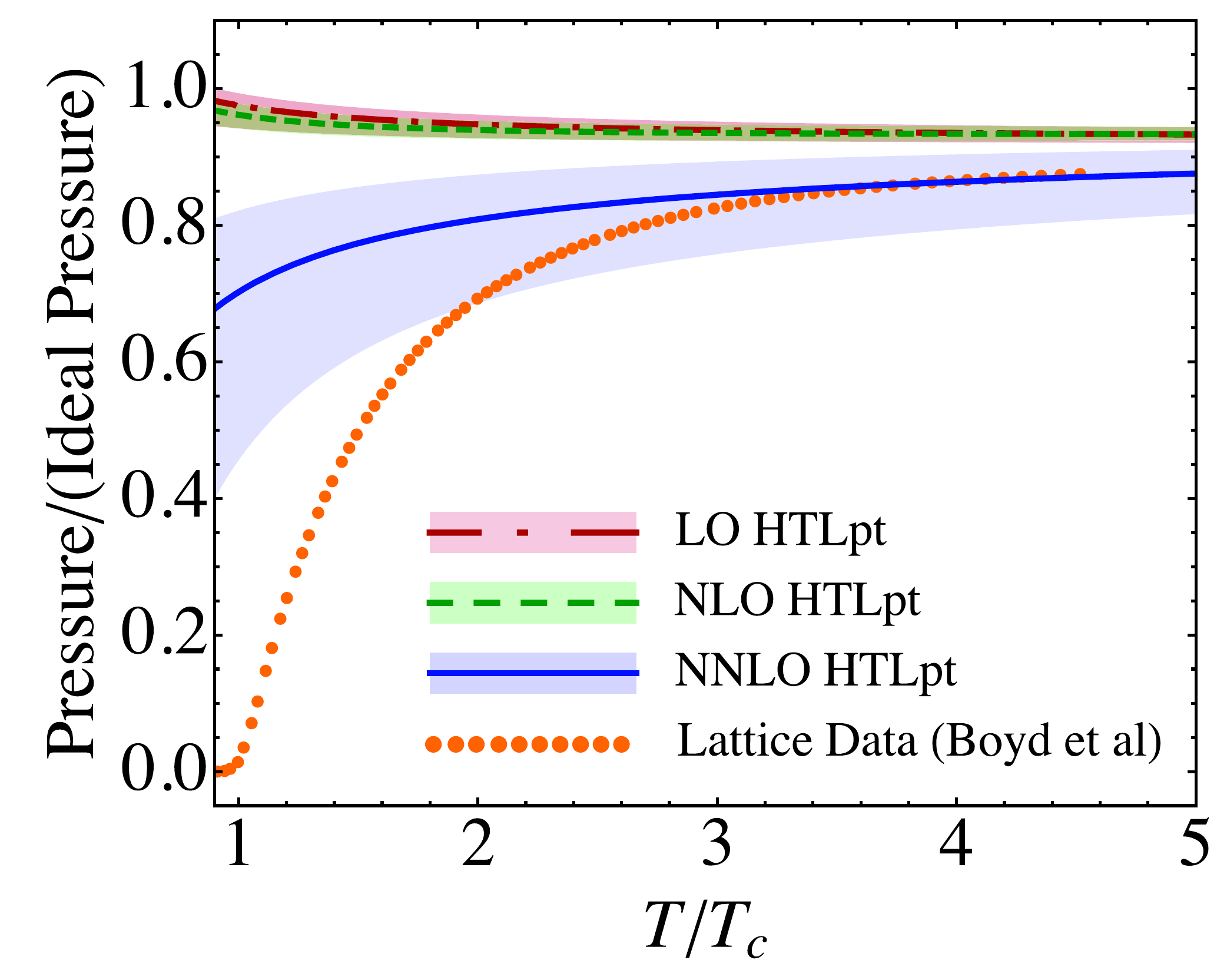}
\end{center}
\vspace{-3mm}
\caption{Comparison of LO, NLO, and NNLO predictions for the scaled pressure using the BN mass and one-loop running~(\ref{runningcoupling-YM}). The points are lattice data for pure-glue with $N_c=3$ from Boyd et al. \cite{Boyd:1996bx}. Shaded bands show the result of varying the renormalization scale $\mu$ by a factor of 2 around $\mu = 2 \pi T$.}
\label{bn1pressure}
\end{figure}

\begin{figure}[h]
\begin{center}
\includegraphics[width=10cm]{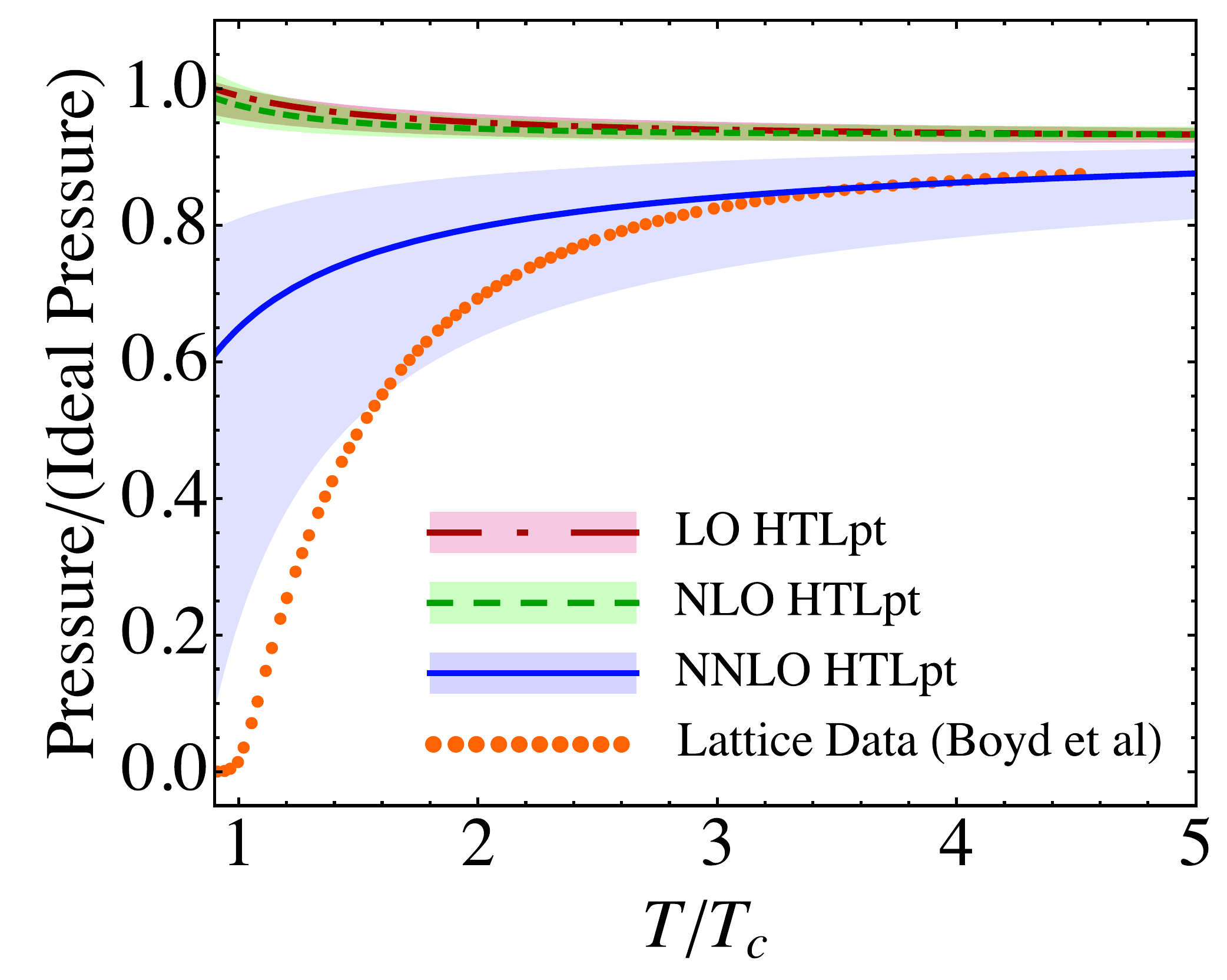}
\end{center}
\vspace{-3mm}
\caption{Comparison of LO, NLO, and NNLO predictions for the scaled pressure using the BN mass and three-loop running~\cite{Amsler:2008zzb} with SU(3) pure-glue lattice data from Boyd et al. \cite{Boyd:1996bx}. Shaded bands show the result of varying the renormalization scale $\mu$ by a factor of 2 around $\mu = 2 \pi T$.}
\label{bn3pressure}
\end{figure}

In Fig.~\ref{bn1pressure}, we show the HTLpt predictions for the pressure normalized to that of an ideal gas using the BN mass prescription and one-loop running of $\alpha_s$ in Eq.~(\ref{runningcoupling-YM}). The bands are obtained by varying the renormalization scale by a factor of 2 around $\mu=2\pi T$.
%%%%%%%%%%%%%%%%%%%%%%%%%%%%%%%%%%%%%%%%%%%%%%%%%%%%
%%%%%%%%%%%%%%%%%%%%%%%%%%%%%%%%%%%%%%%%%%%%%%%%%%%%
In Fig.~\ref{bn3pressure}, we again plot the normalized pressure, but now with three-loop running of $\alpha_s$~\cite{Amsler:2008zzb}. The agreement between the lattice data from Boyd et al. \cite{Boyd:1996bx} is very good down to temperatures of around $3\,T_c$. Comparing Figs.~\ref{bn1pressure}--\ref{bn3pressure} we see that using the three-loop running, the band becomes wider. However, the difference is significant only for low $T$, where the HTLpt results disagrees with the lattice anyway. For $T>3T_c$, the prescription for the running makes very little difference.

%%%%%%%%%%%%%%%%%%%%%%%%%%%%%%%%%%%%%%%%%%%%%%%%%%%%
%%%%%%%%%%%%%%%%%%%%%%%%%%%%%%%%%%%%%%%%%%%%%%%%%%%%

\begin{figure}[htb]
\begin{center}
\includegraphics[width=10cm]{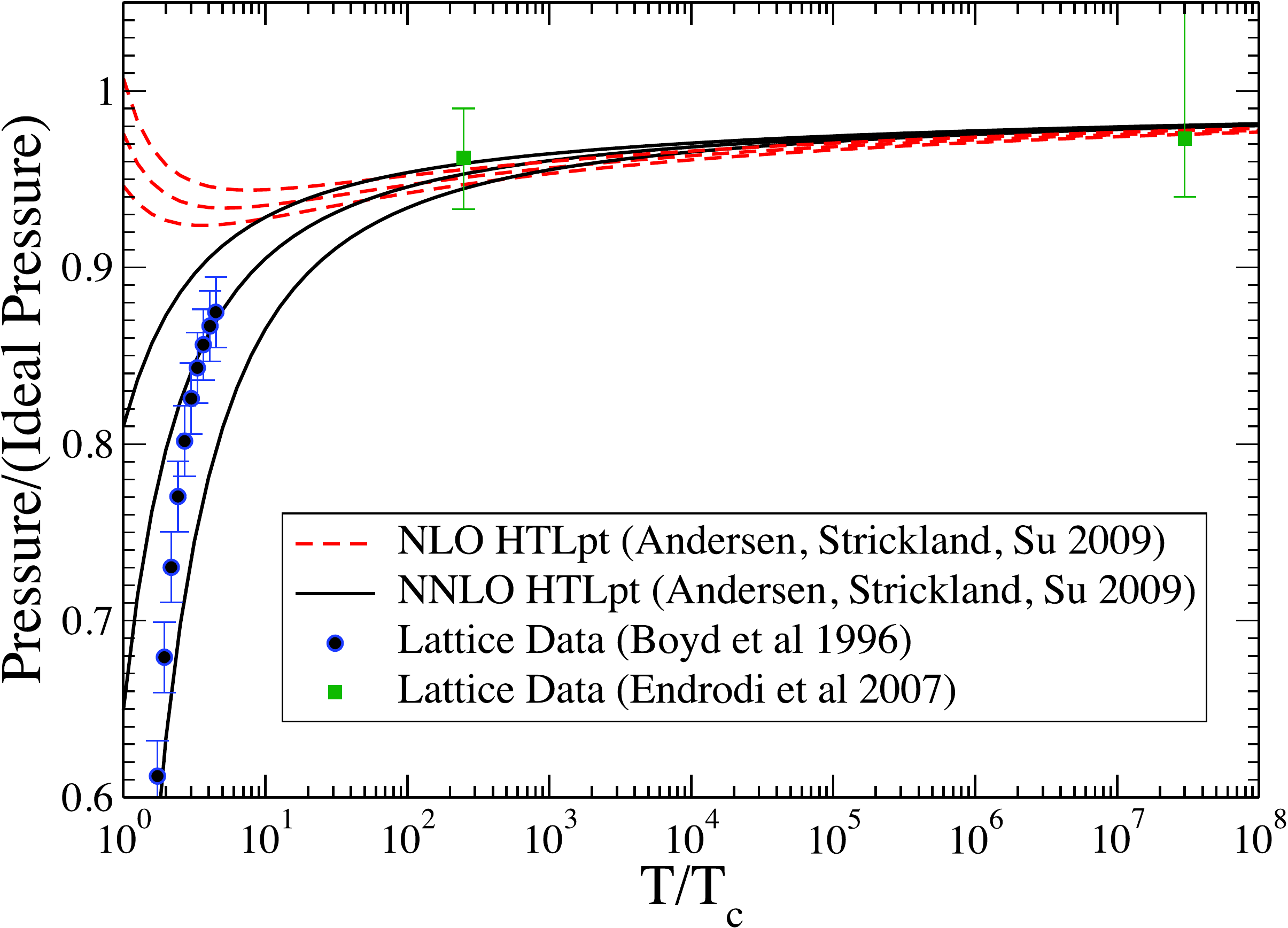}
\end{center}
\vspace{-3mm}
\caption{Comparison of NLO, and NNLO predictions for the scaled pressure using the BN mass and three-loop running~\cite{Amsler:2008zzb} with SU(3) pure-glue lattice data from Boyd et al. \cite{Boyd:1996bx} and Endrodi et al.~\cite{Endrodi:2007tq}. Shaded bands show the result of varying the renormalization scale $\mu$ by a factor of 2 around $\mu = 2 \pi T$.}
\label{highpressure}
\end{figure}

Until recently, lattice data for thermodynamic variables only existed for temperatures up to approximately $5\,T_c$. In the paper by Endrodi et al~\cite{Endrodi:2007tq}, the authors calculate the pressure on the lattice for pure-glue QCD at very large temperatures. In Fig.~\ref{highpressure}, we show the results of Endrodi et al as well as Boyd et al, together with the HTLpt NLO and NNLO predictions for the pressure using the BN mass prescription and three-loop running of $\alpha_s$~\cite{Amsler:2008zzb}. The two points from Ref.~\cite{Endrodi:2007tq} have large error bars, but data points are consistent with the HTLpt predictions.

%%%%%%%%%%%%%%%%%%%%%%%%%%%%%%%%%%%%%%%%%%%%%%%%%%%%
%%%%%%%%%%%%%%%%%%%%%%%%%%%%%%%%%%%%%%%%%%%%%%%%%%%%

It is interesting to make a comparison of convergence between HTLpt and weak-coupling expansion. Analyzing the weak-coupling result listed in Eq.~(\ref{f-qcd-w}), we find for the case of SU(3) Yang-Mills theory that in order to make the magnitude of the coefficients of each order smaller than that of the previous order, one has to require the temperature be higher than $5.36\times10^6\,T_c$. However HTLpt meets the same requirement at temperatures higher than $25.6\,T_c$, which is an improvement of five orders of magnitude.

\subsection{Pressure at large $N_c$}

%%%%%%%%%%%%%%%%%%%%%%%%%%%%%%%%%%%%%%%%%%%%%%%%%%%%

\begin{figure}[htb]
\vspace{3mm}
\begin{center}
\includegraphics[width=10cm]{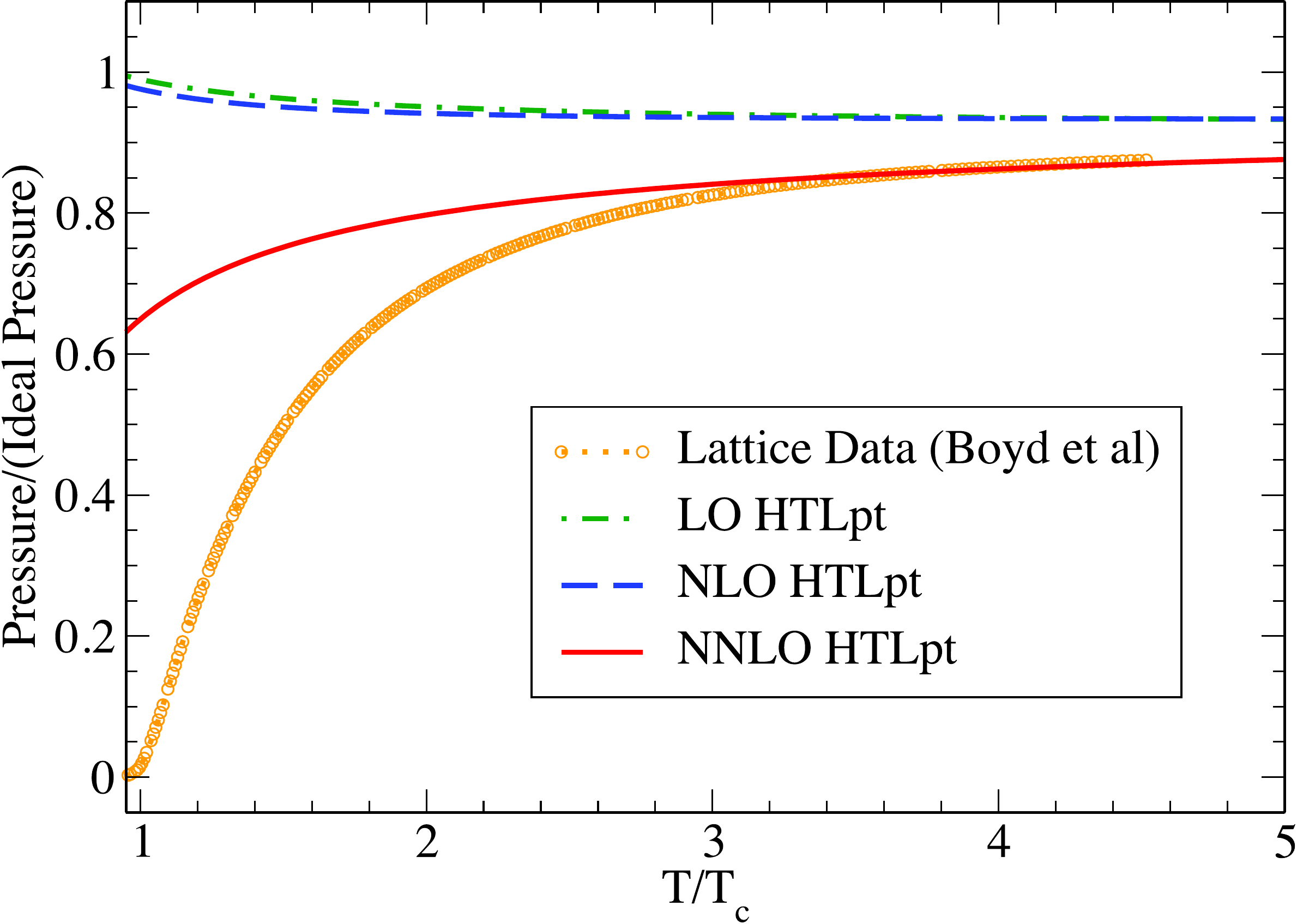}
\end{center}
\vspace{-3mm}
\caption{Comparison of LO, NLO, and NNLO predictions for the scaled pressure at large $N_c$   using the BN mass and three-loop running~\cite{Amsler:2008zzb} with SU(3) pure-glue lattice data from Boyd et al. \cite{Boyd:1996bx}. $\mu = 2 \pi T$ is taken here.}
\label{fig:N_c}
\end{figure}
%%%%%%%%%%%%%%%%%%%%%%%%%%%%%%%%%%%%%%%%%%%%%%%%%%%%

The large $N_c$ limit is achieved by taking $N_c$ to be large while holding $g^2N_c$ fixed. The large $N_c$ coupling is defined by $\lambda \equiv g^2 N_c$. As can be seen from Eqs.~(\ref{Omega-1}), (\ref{Omega-NLO-YM}) and (\ref{Omega-NNLO-YM}) that the ratios of the thermodynamic potentials over ${\cal F}_{\rm ideal}$ are solely functions of $\lambda$ which have no residual dependence on $N_c$ or $g$ after the substitution $g^2 \rightarrow \lambda / N_c$, while the same is true for the BN mass~(\ref{BN}). Therefore the scaled HTLpt thermodynamics, i.e. the thermodynamics obtained by taking ratio over ${\cal F}_{\rm ideal}$, is independent of the actual number of colors $N_c$ up to three-loop order. This is in line with a recent lattice study by Panero~\cite{Panero:2009tv} showing that the thermodynamics of SU($N$) Yang-Mills theories has a very mild dependence on $N_c$, supporting the idea that the QCD plasma could be described by models based on the large $N_c$ limit. In Fig.~(\ref{fig:N_c}) we plot the HTLpt predictions for the pressure at large $N_c$ through NNLO with three-loop running~\cite{Amsler:2008zzb} at $\mu = 2 \pi T$ together with the SU(3) prue-glue lattice data from Boyd et al.~\cite{Boyd:1996bx}. The curves in Fig.~(\ref{fig:N_c}) are exactly the same as those in Fig.~(\ref{bn3pressure}) demonstrating the independence of the scaled SU$(N_c)$ pressure with respect to $N_c$. It is unknown at this stage whether higher-order HTLpt contributions would spoil this independence, however from the comparison of the NNLO result with the lattice data, the $N_c$ dependence from the higher order corrections for $T>3T_c$ might be tiny in case there are any.

\subsection{Energy density}

\begin{figure}[htb]
\begin{center}
\includegraphics[width=10cm]{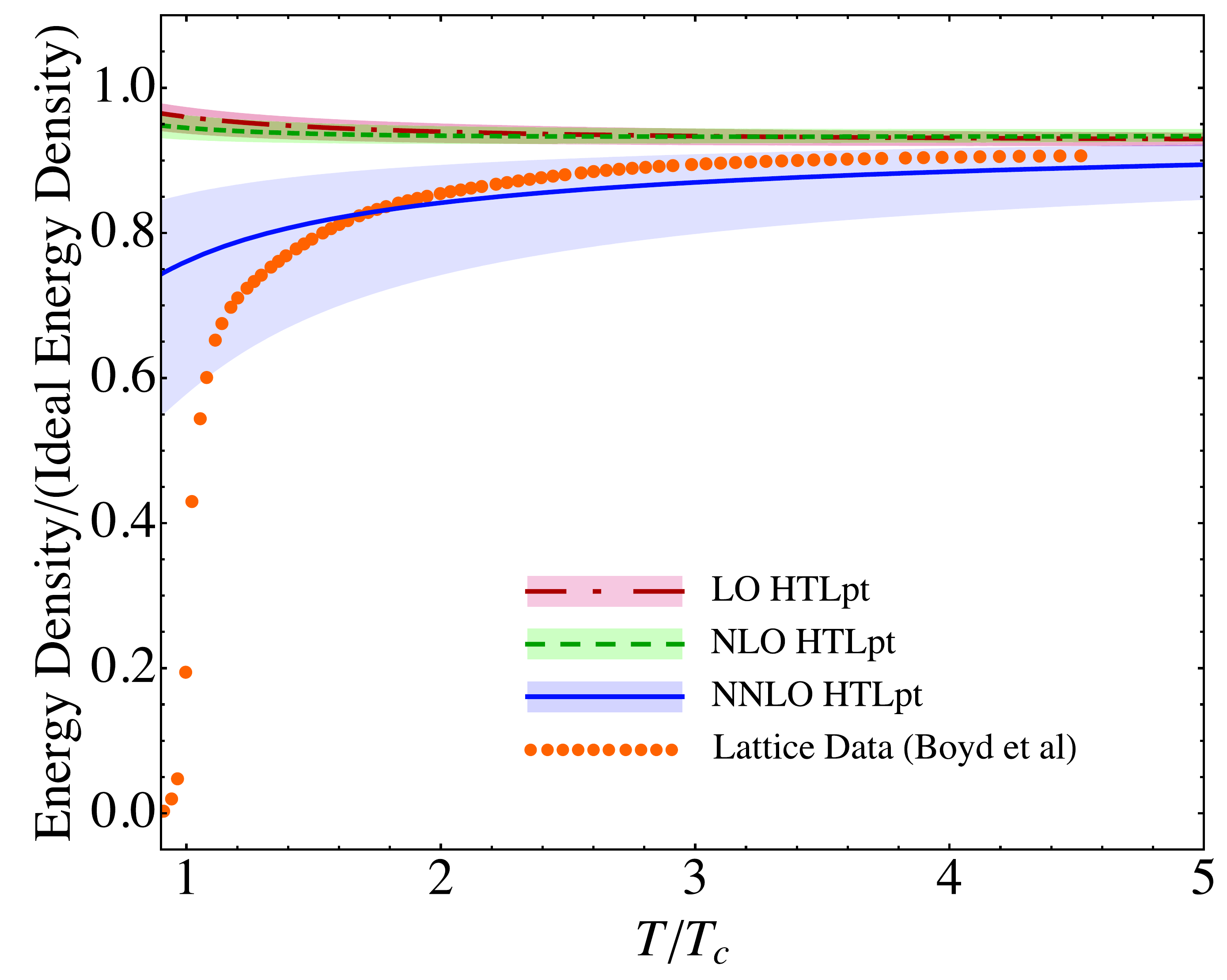}
\end{center}
\vspace{-3mm}
\caption{Comparison of LO, NLO, and NNLO predictions for the scaled energy density using the BN mass and three-loop running~\cite{Amsler:2008zzb} with SU(3) pure-glue lattice data from Boyd et al. \cite{Boyd:1996bx}. Shaded bands show the result of varying the renormalization scale $\mu$ by a factor of 2 around $\mu = 2 \pi T$.}
\label{fig:energy}
\end{figure}

The energy density $\cal E$ is defined by
\bqa
{\cal E}\;=\;{\cal F}-T{d{\cal F}\over dT}\;.
\eqa
In Fig.~\ref{fig:energy}, we show the LO, NLO, and NNLO predictions for energy density normalized to that of an ideal gas using the BN mass prescription and three-loop running of $\alpha_s$~\cite{Amsler:2008zzb}. The bands show the result of varying the renormalization scale $\mu$ by a factor of 2 around $\mu = 2 \pi T$. Our NNLO predictions are in very good agreement with the lattice data down to $T\simeq2\,T_c$.

%%%%%%%%%%%%%%%%%%%%%%%%%%%%%%%%%%%%%%%%%%%%%%%%%%%%
%%%%%%%%%%%%%%%%%%%%%%%%%%%%%%%%%%%%%%%%%%%%%%%%%%%%

\subsection{Entropy}

\begin{figure}[t]
\begin{center}
\includegraphics[width=10cm]{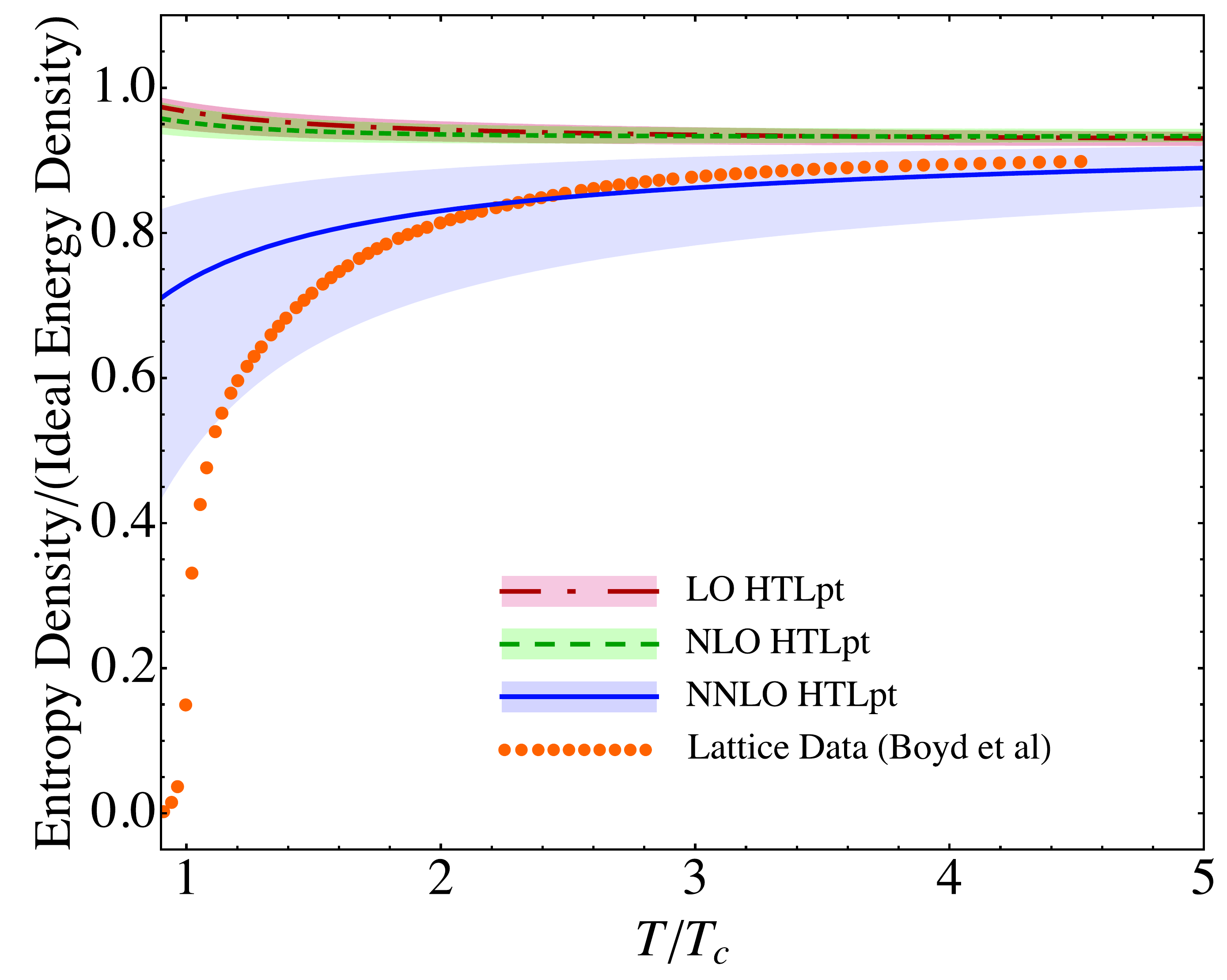}
\end{center}
\vspace{-3mm}
\caption{Comparison of LO, NLO, and NNLO predictions for the scaled entropy using the BN mass and three-loop running~\cite{Amsler:2008zzb} with SU(3) pure-glue lattice data from Boyd et al. \cite{Boyd:1996bx}. Shaded bands show the result of varying the renormalization scale $\mu$ by a factor of 2 around $\mu = 2 \pi T$.}
\label{fig:entropy}
\end{figure}

The entropy density is defined by
\bqa
{\cal S}\;=\;-{\partial{\cal F}\over\partial T} \;.
\eqa
In Fig.~\ref{fig:entropy}, we show the entropy density normalized to that of an ideal gas using the BN mass prescription and three-loop running of $\alpha_s$~\cite{Amsler:2008zzb}. The points are lattice data from Boyd et al. \cite{Boyd:1996bx}. Our NNLO predictions are in excellent agreement with the lattice data down to $T\simeq2\,T_c$.

%%%%%%%%%%%%%%%%%%%%%%%%%%%%%%%%%%%%%%%%%%%%%%%%%%%%
%%%%%%%%%%%%%%%%%%%%%%%%%%%%%%%%%%%%%%%%%%%%%%%%%%%%

\subsection{Trace anomaly}

\begin{figure}[t]
\begin{center}
\includegraphics[width=10cm]{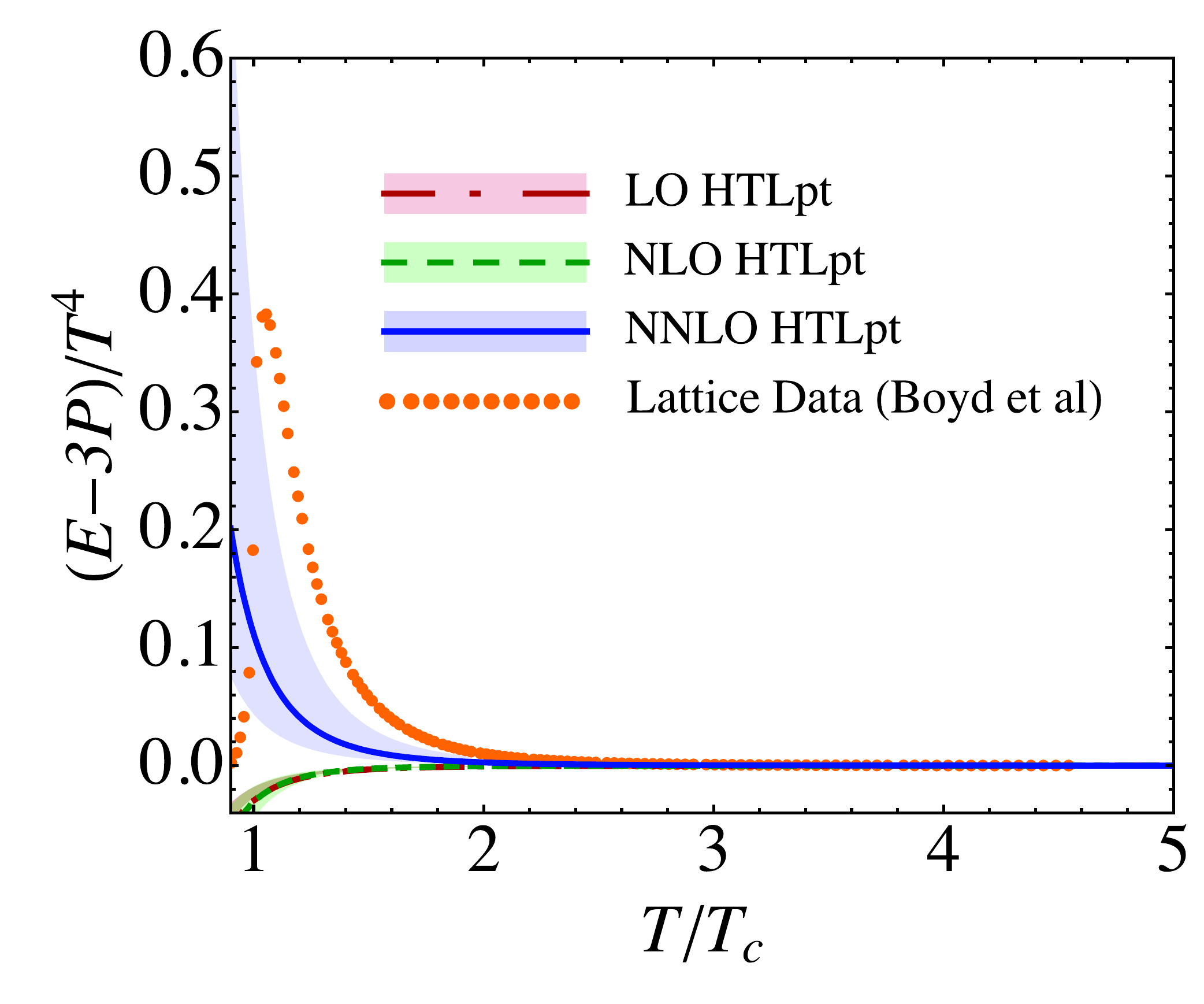}
\end{center}
\vspace{-3mm}
\caption{Comparison of LO, NLO, and NNLO predictions for the scaled trace anomaly using the BN mass and three-loop running~\cite{Amsler:2008zzb} with SU(3) pure-glue lattice data from Boyd et al. \cite{Boyd:1996bx}. Shaded bands show the result of varying the renormalization scale $\mu$ by a factor of 2 around $\mu = 2 \pi T$.}
\label{trace}
\end{figure}

In pure-glue QCD or in QCD with massless quarks, there is no mass scale in the Lagrangian and the theory is scale invariant. At the classical level, this implies that the trace of the energy-momentum tensor vanishes. At the quantum level, scale invariance is broken by renormalization effects. It is convenient to introduce the scale anomaly density ${\cal E}-3{\cal P}$, which is proportional to the trace of the energy-momentum tensor. The trace anomaly can be written as
\bqa
{\cal E}-3{\cal P}\;=\;-T^5{d\over dT}\left({{\cal F}\over T^4}\right)\;.
\eqa
In Fig.~\ref{trace}, we show the HTLpt predictions for the trace anomaly divided by $T^4$ using the BN mass prescription and three-loop running of $\alpha_s$~\cite{Amsler:2008zzb}. The points are lattice data from Boyd et al. \cite{Boyd:1996bx}. For temperatures below approximately $2\,T_c$, there is a large discrepancy between the HTLpt predictions and the lattice. At LO and NLO, the curves are even bending downwards.

At temperatures close to the phase transition it has been suggested that the discrepancy between HTLpt resummed predictions for thermodynamics functions and, in particular, the trace anomaly is due to influence of a dimension two condensate~\cite{Pisarski:2000eq,Pisarski:2002ji,Kondo:2001nq} which is related to confinement.  Phenomenological fits of lattice data which include such a condensate show that the agreement with lattice data is improved \cite{Megias:2009mp,Megias:2009ar}. Alternatively, others have constructed AdS/CFT inspired models which break conformal invariance ``by hand''~\cite{Gubser:2008ny,Gubser:2008yx,Noronha:2009ud}. These models are also able to fit the thermodynamical functions of QCD at temperatures close to the phase transition.

%%%%%%%%%%%%%%%%%%%%%%%%%%%%%%%%%%%%%%%%%%%%%%%%%%%%
%%%%%%%%%%%%%%%%%%%%%%%%%%%%%%%%%%%%%%%%%%%%%%%%%%%%

\begin{figure}[t]
\begin{center}
\includegraphics[width=10cm]{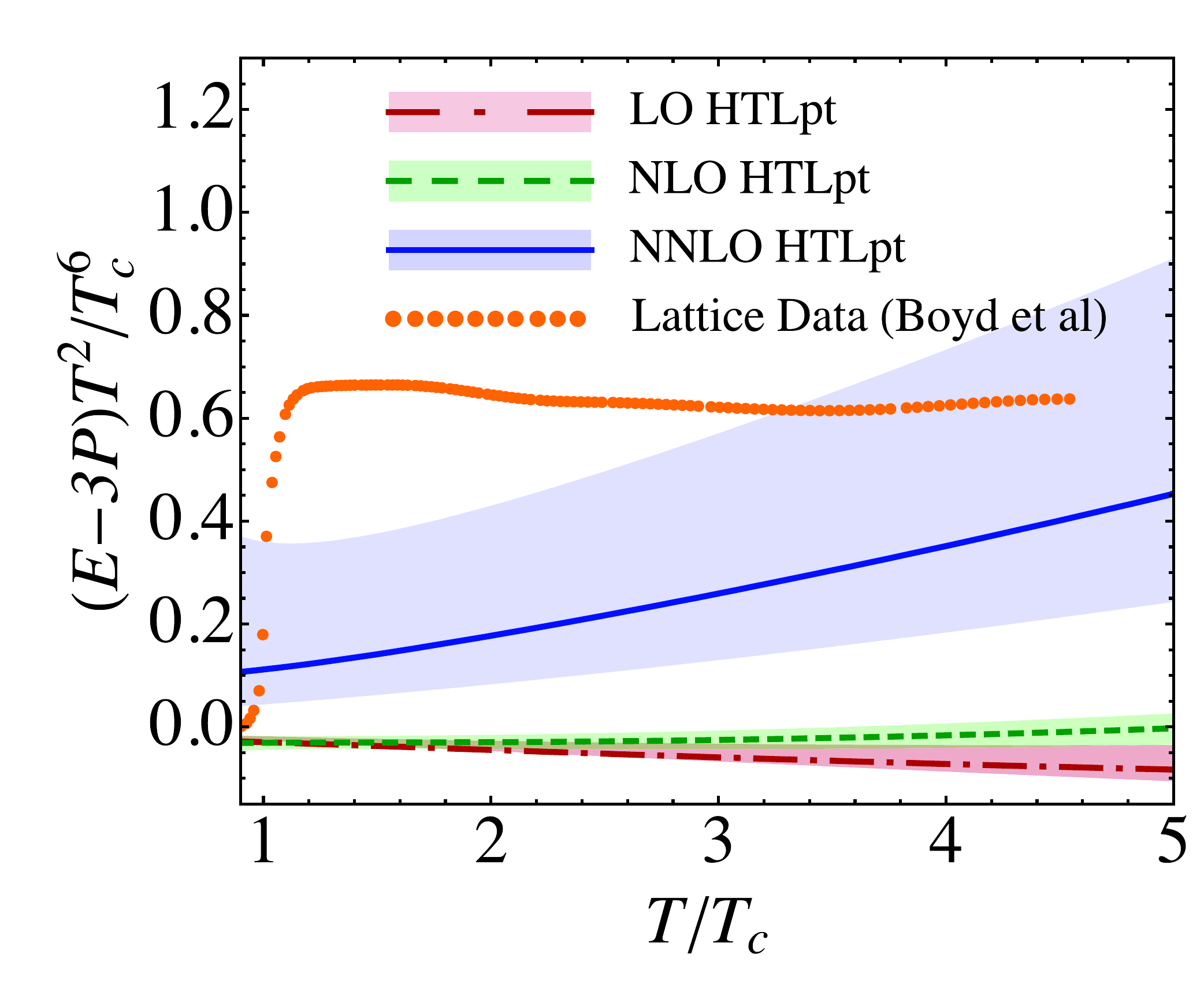}
\end{center}
\vspace{-3mm}
\caption{Comparison of LO, NLO, and NNLO predictions for the scaled trace anomaly using the BN mass and three-loop running~\cite{Amsler:2008zzb} with SU(3) pure-glue lattice data from Boyd et al. \cite{Boyd:1996bx}. Shaded bands show the result of varying the renormalization scale $\mu$ by a factor of 2 around $\mu = 2 \pi T$.}
\label{trace2}
\end{figure}

In Fig.~\ref{trace2}, we show the HTLpt predictions for the trace anomaly scaled by $T^2/T_c^6$ using the BN mass prescription and three-loop running of $\alpha_s$~\cite{Amsler:2008zzb}. The points are lattice data from Boyd et al. \cite{Boyd:1996bx}. The most remarkable feature is that lattice data are essentially constant over a very large temperature range. Clearly, HTLpt does not reproduce the scaled lattice data precisely; however, the agreement is dramatically improved when going from NLO to NNLO.
%%%%%%%%%%%%%%%%%%%%%%%%%%%%%%%%%%%%%%%%%%%%%%%%%%%%
%%%%%%%%%%%%%%%%%%%%%%%%%%%%%%%%%%%%%%%%%%%%%%%%%%%%

\section{Conclusions}

In this chapter, we have presented results for the LO, NLO, and NNLO thermodynamic functions for SU($N_c$) Yang-Mills theory using HTLpt. We compared our predictions with lattice data for $N_c=3$ and found that HTLpt is consistent with available lattice data down to approximately $T \sim 3\,T_c$ in the case of the pressure and $T \sim 2\,T_c$ in the case of the energy density and entropy. These results are in line with expectations since below $T \sim 2-3\, T_c$ a simple ``electric'' quasiparticle approximation breaks down due to nonperturbative chromomagnetic effects~\cite{Linde:1980ts,Gross:1980br}~\footnote{There have been also hints that the Z($N$) interface~\cite{Z(N)}, gauge-fixing ambiguities~\cite{GZ}, and topological objects such as quantum instantons~\cite{Andersen:2006sf} and magnetic monopoles~\cite{monopole} might play important roles on the thermodynamics at intermediate temperature.}. This is a nontrivial result since, in this temperature regime the QCD coupling constant is neither infinitesimally weak nor infinitely strong with $g \sim 2$, or equivalently $\alpha_s = g^2/(4\pi) \sim 0.3$. Therefore, we have a crucial test of the quasiparticle picture in the intermediate coupling regime.

The mass parameter $m_D$ in HTLpt is arbitrary and we employed two different prescriptions for fixing it. Unfortunately, the variational gap equation has four complex conjugate solutions, two with positive real parts. This has also been observed in scalar theory and QED. Whether this is a problem of HTLpt as such or is related to our $m_D/T$ expansion is unknown. Since it is not currently possible to evaluate the NNLO HTLpt diagrams in gauge theories exactly, it is impossible to settle the issue at this stage. On the other hand, the BN mass prescription is well defined to all orders in perturbation theory and does a reasonable job reproducing available lattice data for temperatures above $T\gtrsim3T_c$. With QED and Yang-Mills results at hand, the NNLO full QCD HTLpt thermodynamics will be a routine extension~\cite{Andersen:2010wu,Andersen:2011sf}.

%%%%%%%%%%%%%%%%%%%%%%%%%%%%%%%%%%%%%%%%%%%%%%%%%%%%%%%%%%%%%
%
%	Include File:			DON'T COMPILE !!!
%
%%%%%%%%%%%%%%%%%%%%%%%%%%%%%%%%%%%%%%%%%%%%%%%%%%%%%%%%%%%%%

\chapter{Summary and Outlook}\label{chapter:sum}

This dissertation is devoted to the study of thermodynamics for thermal gauge theories. The poor convergence of conventional perturbation theory has been the main obstacle in the practical application of thermal QCD for decades. To improve this embarrassing situation, a considerable effort has been put into reorganizing the perturbative series at phenomenologically relevant temperatures. The application of hard-thermal-loop perturbation theory to the problem carried out in this dissertation leads to laudable results for both Abelian and non-Abelian theories.

The success of HTLpt is not totally unexpected since it is essentially just a reorganization of perturbation theory which shifts the expansion from around an ideal gas of massless particles to that of massive quasiparticles which are the real degrees of freedom at high temperature. The HTL Feynman rules listed in Appendix~\ref{app:rules} show clearly that the propagators and vertices are dressed systematically by the thermal medium, as a result the interactions get screened in the medium which can be seen, for instance, from the quark-gluon three-vertex~(\ref{3qgv}) that the coupling strength gets screened by the thermal mass term explicitly. Therefore the expansion in terms of the HTL Feynman rules are self-consistently around a gas of thermal quasiparticles. The fact that the mass parameter is not arbitrary but a function of $g$ and $T$ determined variationally or perturbatively also indicates that HTLpt doesn't modify the original gauge theory but just reorganizes its perturbation series. Gauge invariance which is guaranteed by construction in HTLpt is useful both as a consistency check in calculations and as a way to simplify calculations. Although the renormalizability of HTLpt is not yet proven, the fact that it is renormalizable at NNLO using only known counterterms shows promising light along the way.

So far, thermodynamics for quantum fields has been studied intensively in the community, both perturbatively through higher orders or numerically on the lattice, however real-time dynamics is still in its very early stage of development. Transport coefficients are of great interest since they are theoretically clean and well defined non-equilibrium dynamical quantities. Along the line of perturbative approach to transport coefficients, considerable efforts have been devoted at leading order in weak-coupling expansion~\cite{trans-coeff-LO}. However the only known transport coefficients to next-to-leading order are shear viscosity in scalar $\phi^4$ theory~\cite{Moore:2007ib}, heavy quark diffusion rate in QCD and ${\cal N}=4$ supersymmetric Yang-Mills theory~\cite{CHM}, and transverse diffusion rate $\hat{q}$ in QCD~\cite{CaronHuot:2008ni}, and all of them exhibit poor convergence as bad as the case of thermodynamic quantities, such as the pressure. Since dynamical quantities are still not well described by lattice gauge theory, new resummation techniques are urgently needed in order to achieve a better understanding of transport coefficients.

Although the papers written to date have focussed on using HTLpt to compute thermodynamic observables, the goal of this work is to create a framework which can be applied to both equilibrium and non-equilibrium systems. HTLpt is formulated in Minkowski space, so its application to non-equilibrium dynamics is straightforward. With the confidence from thermodynamics, HTLpt is ready to enter the domain of real-time dynamics and this might be of great help in deepening our knowledge in the properties of the quark-gluon plasma.

abs.tex include
ack.tex include
intro.tex include
sum.tex include

%%%%%%%%%%%%%%%%%%%%%%%%%%%%%%%%%%%%%%%%%%%%%%%%%%%%%%%%%%%%%%%%%%%%%%

\cleardoublepage
\thispagestyle{empty}

\appendix
%%%%%%%%%%%%%%%%%%%%%%%%%%%%%%%%%%%%%%%%%%%%%%%%%%%%%%%%%%%%%
%
%	Include File:			DON'T COMPILE !!!
%
%%%%%%%%%%%%%%%%%%%%%%%%%%%%%%%%%%%%%%%%%%%%%%%%%%%%%%%%%%%%%

\chapter{HTL Feynman Rules}
\label{app:rules}

In this appendix, we present Feynman rules for HTL perturbation theory in QCD, from which QED Feynman rules can be obtained by simplifying the relevant color structures. We give explicit expressions for the propagators and for the quark-gluon three- and four-vertices. The Feynman rules are given in Minkowski space to facilitate future applications to real-time processes. A Minkowski momentum is denoted $p = (p_0, {\bf p})$, and the inner product is $p \cdot q = p_0 q_0 - {\bf p} \cdot {\bf q}$. The vector that specifies the thermal rest frame is $n = (1,{\bf 0})$.

\section{Gluon self-energy}

The HTL gluon self-energy tensor for a gluon of momentum $p$ is
\bqa
\label{a1}
\Pi^{\mu\nu}(p)\;=\;m_D^2\left[{\cal T}^{\mu\nu}(p,-p)-n^{\mu}n^{\nu}\right]\;.
\eqa
The tensor ${\cal T}^{\mu\nu}(p,q)$, which is defined only for momenta that satisfy $p+q=0$, is
\bqa
{\cal T}^{\mu\nu}(p,-p)\;=\;\left \langle y^{\mu}y^{\nu}{p\!\cdot\!n\over p\!\cdot\!y}\right\rangle_{\bf\hat{y}} \;.
\label{T2-def}
\eqa
The angular brackets indicate averaging over the spatial directions of the light-like vector $y=(1,\hat{\bf y})$. The tensor ${\cal T}^{\mu\nu}$ is symmetric in $\mu$ and $\nu$ and satisfies the ``Ward identity''
\bqa
p_{\mu}{\cal T}^{\mu\nu}(p,-p)\;=\;p\!\cdot\!n\;n^{\nu}\;.
\label{ward-t2}
\eqa
The self-energy tensor $\Pi^{\mu\nu}$ is therefore also symmetric in $\mu$ and $\nu$ and satisfies
\bqa
p_{\mu}\Pi^{\mu\nu}(p)\!\!&=&\!\!0\;,\\
\label{contr}
g_{\mu\nu}\Pi^{\mu\nu}(p)\!\!&=&\!\!-m_D^2\;.
\eqa

The gluon self-energy tensor can be expressed in terms of two scalar functions, the transverse and longitudinal self-energies $\Pi_T$ and $\Pi_L$, defined by
\bqa
\label{pit2}
\Pi_T(p)\!\!&=&\!\!{1\over d-1}\left(\delta^{ij}-\hat{p}^i\hat{p}^j\right)\Pi^{ij}(p)\;, \\
\label{pil2}
\Pi_L(p)\!\!&=&\!\!-\Pi^{00}(p)\;,
\eqa
where ${\bf \hat p}$ is the unit vector in the direction of ${\bf p}$. In terms of these functions, the self-energy tensor is
\bqa
\label{pi-def}
\Pi^{\mu\nu}(p) \;=\; - \Pi_T(p) T_p^{\mu\nu} - {1\over n_p^2} \Pi_L(p) L_p^{\mu\nu}\;,
\eqa
where the tensors $T_p$ and $L_p$ are
\bqa
T_p^{\mu\nu}\!\!&=&\!\!g^{\mu\nu} - {p^{\mu}p^{\nu} \over p^2}-{n_p^{\mu}n_p^{\nu}\over n_p^2}\;,\\
L_p^{\mu\nu}\!\!&=&\!\!{n_p^{\mu}n_p^{\nu} \over n_p^2}\;.
\eqa
The four-vector $n_p^{\mu}$ is
\bqa
n_p^{\mu} \;=\; n^{\mu} - {n\!\cdot\!p\over p^2} p^{\mu}
\eqa
and satisfies $p\!\cdot\!n_p=0$ and $n^2_p = 1 - (n\!\cdot\!p)^2/p^2$. (\ref{contr}) reduces to the identity
\bqa
(d-1)\Pi_T(p)+{1\over n^2_p}\Pi_L(p) \;=\; m_D^2 \;.
\label{PiTL-id}
\eqa
We can express both self-energy functions in terms of the function ${\cal T}^{00}$ defined by (\ref{T2-def}):
\bqa
\Pi_T(p)\!\!&=&\!\!{m_D^2 \over (d-1) n_p^2}\left[ {\cal T}^{00}(p,-p) - 1 + n_p^2  \right] \;,
\label{PiT-T}
\\
\Pi_L(p)\!\!&=&\!\!m_D^2\left[ 1- {\cal T}^{00}(p,-p) \right]\;,
\label{PiT-L}
\eqa

In the tensor ${\cal T}^{\mu \nu}(p,-p)$ defined in~(\ref{T2-def}), the angular brackets indicate the angular average over the unit vector $\hat{\bf y}$. In almost all previous work, the angular average in~(\ref{T2-def}) has been taken in $d=3$ dimensions. For consistency of higher order corrections, it is essential to take the angular average in $d=3-2\epsilon$ dimensions and analytically continue to $d=3$ only after all poles in $\epsilon$ have been cancelled. Expressing the angular average as an integral over the cosine of an angle, the expression for the $00$ component of the tensor is
\bqa
{\cal T}^{00}(p,-p) \!\!&=&\!\! {w(\epsilon)\over2}\int_{-1}^1dc\;(1-c^2)^{-\epsilon}{p_0\over p_0-|{\bf p}|c} \;,
\label{T00-int}
\eqa
where the weight function $w(\epsilon)$ is
\bqa
w(\epsilon)\;=\;{\Gamma(2-2\epsilon)\over\Gamma^2(1-\epsilon)}\;2^{2\epsilon}
                    \;=\; {\Gamma({3\over2}-\epsilon)\over \Gamma({3\over2}) \Gamma(1-\epsilon)} \;.
\label{weight}
\eqa
The integral in (\ref{T00-int}) must be defined so that it is analytic at $p_0=\infty$. It then has a branch cut running from $p_0=-|{\bf p}|$ to $p_0=+|{\bf p}|$. If we take the limit $\epsilon\rightarrow 0$, it reduces to
\begin{eqnarray}
{\cal T}^{00}(p,-p)\;=\;{p_0 \over 2|{\bf p}|}\log {p_0 +|{\bf p}| \over p_0-|{\bf p}|}\;,
\end{eqnarray}
which is the expression that appears in the usual HTL self-energy functions.

\section{Gluon propagator}
\label{app:prop}

The Feynman rule for the gluon propagator is
\bqa
i \delta^{a b} \Delta_{\mu\nu}(p) \;,
\eqa
where the gluon propagator tensor $\Delta_{\mu\nu}$ depends on the choice of gauge fixing. We consider two possibilities that introduce an arbitrary gauge parameter $\xi$:  general covariant gauge and general Coulomb gauge. In both cases, the inverse propagator reduces in the limit $\xi\rightarrow\infty$ to
\bqa
\Delta^{-1}_{\infty}(p)^{\mu\nu}\;=\;-p^2 g^{\mu \nu} + p^\mu p^\nu - \Pi^{\mu\nu}(p)\;.
\label{delta-inv:inf0}
\eqa
This can also be written
\bqa
\Delta^{-1}_{\infty}(p)^{\mu\nu} \;=\; - {1 \over \Delta_T(p)} T_p^{\mu\nu} + {1 \over n_p^2 \Delta_L(p)} L_p^{\mu\nu}\;,
\label{delta-inv:inf}
\eqa
where $\Delta_T$ and $\Delta_L$ are the transverse and longitudinal propagators:
\bqa
\Delta_T(p)\!\!&=&\!\!{1 \over p^2-\Pi_T(p)}\;,
\label{Delta-T:M} \\
\Delta_L(p)\!\!&=&\!\!{1 \over - n_p^2 p^2+\Pi_L(p)}\;.
\label{Delta-L:M}
\eqa
The inverse propagator for general $\xi$ is
\bqa
\Delta^{-1}(p)^{\mu\nu}\!\!&=&\!\!\Delta^{-1}_{\infty}(p)^{\mu\nu}-{1\over\xi}p^{\mu}p^{\nu}\hspace{4.4cm}\mbox{covariant}\;,
\label{Delinv:cov} \\
\!\!&=&\!\!\Delta^{-1}_{\infty}(p)^{\mu\nu}-{1\over\xi}\left(p^{\mu}-p\!\cdot\!n\;n^{\mu}\right)\left(p^{\nu}-p\!\cdot\!n\;n^{\nu}\right)\hspace{0.4cm}\mbox{Coulomb} \;.
\label{Delinv:C}
\eqa
The propagators obtained by inverting the tensors in~(\ref{Delinv:C}) and~(\ref{Delinv:cov}) are
\bqa
\Delta^{\mu\nu}(p)\!\!&=&\!\!-\Delta_T(p)T_p^{\mu\nu} +\Delta_L(p)n_p^{\mu}n_p^{\nu} - \xi {p^{\mu}p^{\nu} \over (p^2)^2}\hspace{2cm}\mbox{covariant}\;,
\label{D-cov}
\\
\!\!&=&\!\!-\Delta_T(p)T_p^{\mu\nu}+\Delta_L(p)n^{\mu}n^{\nu}-\xi{p^{\mu}p^{\nu}\over(n_p^2p^2)^2}\hspace{1.6cm}\mbox{Coulomb} \;.
\label{D-C}
\eqa

It is convenient to define the following combination of propagators:
\bqa
\Delta_X(p) \;=\; \Delta_L(p)+{1\over n_p^2}\Delta_T(p) \;.
\label{Delta-X}
\eqa
Using (\ref{PiTL-id}), (\ref{Delta-T:M}), and (\ref{Delta-L:M}), it can be expressed in the alternative form
\bqa
\Delta_X(p) \;=\;\left[ m_D^2 - d \, \Pi_T(p) \right] \Delta_L(p) \Delta_T(p) \;,
\label{Delta-X:2}
\eqa
which shows that it vanishes in the limit $m_D \to 0$. In the covariant gauge, the propagator tensor can be written
\bqa\nonumber
\Delta^{\mu\nu}(p)\!\!&=&\!\!\left[ - \Delta_T(p) g^{\mu \nu} + \Delta_X(p) n^\mu n^\nu \right] - {n \!\cdot\! p \over p^2} \Delta_X(p) \left( p^\mu n^\nu  + n^\mu p^\nu \right)
\\
&&
+ \left[ \Delta_T(p) + {(n \!\cdot\! p)^2 \over p^2} \Delta_X(p)
        - {\xi \over p^2} \right] {p^\mu p^\nu \over p^2} \;.
\label{gprop-TC}
\eqa
This decomposition of the propagator into three terms has proved to be particularly convenient for explicit calculations. For example, the first term satisfies the identity
\bqa
\left[- \Delta_T(p) g_{\mu \nu} + \Delta_X(p) n_\mu n_\nu \right] \Delta^{-1}_{\infty}(p)^{\nu\lambda} \;=\;
g_\mu^\lambda - {p_\mu p^\lambda \over p^2}
+ {n \!\cdot\! p \over n_p^2 p^2} {\Delta_X(p) \over \Delta_L(p)}
        p_\mu n_p^\lambda \;.
\label{propid:2}
\eqa

\section{Three-gluon vertex}
\label{app:3gluon}

The three-gluon vertex for gluons with outgoing momenta $p$, $q$, and $r$, Lorentz indices $\mu$, $\nu$, and $\lambda$, and color indices $a$, $b$, and $c$ is
\bqa
i\Gamma_{abc}^{\mu\nu\lambda}(p,q,r)\;=\;-gf_{abc}\Gamma^{\mu\nu\lambda}(p,q,r)\;,
\eqa
where $f^{abc}$ are the structure constants and the three-gluon vertex tensor is
\bqa\nonumber
\Gamma^{\mu\nu\lambda}(p,q,r)\!\!&=&\!\!
g^{\mu\nu}(p-q)^{\lambda}+g^{\nu\lambda}(q-r)^{\mu}+g^{\lambda\mu}(r-p)^{\nu}-m_D^2{\cal T}^{\mu\nu\lambda}(p,q,r)\;.\\ 
&&
\label{Gam3}
\eqa
The tensor ${\cal T}^{\mu\nu\lambda}$ in the HTL correction term is defined only for $p+q+r=0$:
\bqa
{\cal T}^{\mu\nu\lambda}(p,q,r) \;=\;
 - \Bigg\langle y^{\mu} y^{\nu} y^{\lambda}
\left( {p\!\cdot\!n\over p\!\cdot\!y\;q\!\cdot\!y}
	- {r\!\cdot\!n\over\!r\cdot\!y\;q\!\cdot\!y} \right)
	\Bigg\rangle\;.
\label{T3-def}
\eqa
This tensor is totally symmetric in its three indices and traceless in any pair of indices: $g_{\mu\nu}{\cal T}^{\mu\nu\lambda}=0$. It is odd (even) under odd (even) permutations of the momenta $p$, $q$, and $r$. It satisfies the ``Ward identity''
\bqa
q_{\mu}{\cal T}^{\mu\nu\lambda}(p,q,r) \;=\;
{\cal T}^{\nu\lambda}(p+q,r)-
{\cal T}^{\nu\lambda}(p,r+q)\;.
\label{ward-t3}
\eqa
The three-gluon vertex tensor therefore satisfies the Ward identity
\bqa
p_{\mu}\Gamma^{\mu\nu\lambda}(p,q,r) \;=\;
\Delta_{\infty}^{-1}(q)^{\nu\lambda}-\Delta_{\infty}^{-1}(r)^{\nu\lambda}\;.
\label{ward-3}
\eqa

\section{Four-gluon vertex}
\label{app:4gluon}

The four-gluon vertex for gluons with outgoing momenta $p$, $q$, $r$, and $s$, Lorentz indices $\mu$, $\nu$, $\lambda$, and $\sigma$, and color indices $a$, $b$, $c$, and $d$ is
\bqa
i\Gamma^{\mu\nu\lambda\sigma}_{abcd}(p,q,r,s) \!\!&=&\!\!
- ig^2\big\{ f_{abx}f_{xcd} \left(g^{\mu\lambda}g^{\nu\sigma}
				-g^{\mu\sigma}g^{\nu\lambda}\right)
\nonumber
\\
&&
+\;2m_D^2\mbox{tr}\left[T^a\left(T^bT^cT^d+T^dT^cT^b
\right)\right]{\cal T}^{\mu\nu\lambda\sigma}(p,q,r,s)
\big\}
\nonumber
\\
&&
+ \; 2 \; \mbox{cyclic permutations}\;,
\eqa
where the cyclic permutations are of $(q,\nu,b)$, $(r,\lambda,c)$, and $(s,\sigma,d)$. The matrices $T^a$ are in the fundamental representation of the SU$(N_c)$ algebra with the standard normalization ${\rm tr}(T^a T^b) = {1 \over 2} \delta^{ab}$. The tensor ${\cal T}^{\mu\nu\lambda\sigma}$ in the HTL correction term is defined only for $p+q+r+s=0$:
\bqa
{\cal T}^{\mu\nu\lambda\sigma}(p,q,r,s) \!\!&=&\!\!
\Bigg\langle y^{\mu} y^{\nu} y^{\lambda} y^{\sigma}
\left( {p\!\cdot\!n \over p\!\cdot\!y \; q\!\cdot\!y \; (q+r)\!\cdot\!y}
\right.
\nonumber
\\
&&
\left.
+\;{(p+q)\!\cdot\!n\over q\!\cdot\!y\;r\!\cdot\!y\;(r+s)\!\cdot\!y}
+{(p+q+r)\!\cdot\!n\over r\!\cdot\!y\;s\!\cdot\!y\;(s+p)\!\cdot\!y}\right)
\Bigg\rangle\;.
\label{T4-def}
\eqa
This tensor is totally symmetric in its four indices and traceless in any pair of indices: $g_{\mu\nu}{\cal T}^{\mu\nu\lambda\sigma}=0$. It is even under cyclic or anti-cyclic permutations of the momenta $p$, $q$, $r$, and $s$. It satisfies the ``Ward identity''
\bqa
q_{\mu}{\cal T}^{\mu\nu\lambda\sigma}(p,q,r,s)\;=\;{\cal T}^{\nu\lambda\sigma}(p+q,r,s)-{\cal T}^{\nu\lambda\sigma}(p,r+q,s)
\label{ward-t4}
\eqa
and the ``Bianchi identity''
\bqa
{\cal T}^{\mu\nu\lambda\sigma}(p,q,r,s)
+ {\cal T}^{\mu\nu\lambda\sigma}(p,r,s,q)+
{\cal T}^{\mu\nu\lambda\sigma}(p,s,q,r)\;=\;0\;.
\label{Bianchi}
\eqa

When its color indices are traced in pairs, the four-gluon vertex becomes particularly simple:
\bqa
\delta^{ab} \delta^{cd} i \Gamma_{abcd}^{\mu\nu\lambda\sigma}(p,q,r,s)
\;=\; -i g^2 N_c (N_c^2-1) \Gamma^{\mu\nu,\lambda\sigma}(p,q,r,s) \;,
\eqa
where the color-traced four-gluon vertex tensor is
\bqa
\Gamma^{\mu\nu,\lambda\sigma}(p,q,r,s)\;=\;
2g^{\mu\nu}g^{\lambda\sigma}
-g^{\mu\lambda}g^{\nu\sigma}
-g^{\mu\sigma}g^{\nu\lambda}
-m_D^2{\cal T}^{\mu\nu\lambda\sigma}(p,s,q,r)\;.
\label{Gam4}
\eqa
Note the ordering of the momenta in the arguments of the tensor ${\cal T}^{\mu\nu\lambda\sigma}$, which comes from the use of the Bianchi identity (\ref{Bianchi}). The tensor (\ref{Gam4}) is symmetric under the interchange of $\mu$ and $\nu$, under the interchange of $\lambda$ and $\sigma$, and under the interchange of $(\mu,\nu)$ and $(\lambda,\sigma)$. It is also symmetric under the interchange of $p$ and $q$, under the interchange of $r$ and $s$, and under the interchange of $(p,q)$ and $(r,s)$. It satisfies the Ward identity
\bqa
p_{\mu}\Gamma^{\mu\nu,\lambda\sigma}(p,q,r,s)
\;=\;\Gamma^{\nu\lambda\sigma}(q,r+p,s) - \Gamma^{\nu\lambda\sigma}(q,r,s+p)\;.
\label{ward-4}
\eqa

\section{HTL gluon counterterm}
\label{app:HTLct}

The Feynman rule for the insertion of an HTL counterterm into a gluon propagator is
\bqa
-i\delta^{ab}\Pi^{\mu\nu}(p)\;,
\eqa
where $\Pi^{\mu\nu}(p)$ is the HTL gluon self-energy tensor given in~(\ref{pi-def}).

\section{Quark self-energy}

The HTL self-energy of a quark with momentum $p$ is given by
\bqa
\label{selfq}
\Sigma(P)\;=\;m_q^2/\!\!\!\!{\cal T}(p) \;,
\eqa
where
\bqa
\label{deftf}
{\cal T}^{\mu}(p)\;=\;\left\langle{y^{\mu}\over p\cdot y}\right\rangle_{\hat{\bf y}} \;.
\eqa
Expressing the angular average as an integral over the cosine of an angle, it becomes
\bqa 
\label{def-tf}
{\cal T}^{\mu}(p)\;=\;{w(\epsilon)\over2}
\int_{-1}^1dc\;(1-c^2)^{-\epsilon}{y^{\mu}\over p_0-|{\bf p}|c}\;,
\eqa
The integral in (\ref{def-tf}) must be defined so that it is analytic at $p_0=\infty$. It then has a branch cut running from $p_0=-|{\bf p}|$ to $p_0=+|{\bf p}|$. In three dimensions, this reduces to
\bqa
\Sigma(P)\;=\;
{m_q^2\over 2|{\bf p}|}\gamma_0\log{p_0+|{\bf p}|\over p_0-|{\bf p|}}
+{m_q^2\over |{\bf p}|}\gamma\cdot \hat{\bf p}
\left(1-{p_0\over 2|{\bf p}|}\log{p_0+|{\bf p}|\over p_0-|{\bf p|}}\right)\;.
\eqa

\section{Quark propagator}

The Feynman rule for the quark propagator is 
\bqa
i\delta^{ab}S(p)\;.
\eqa
The quark propagator can be written as
\bqa
\label{qprop}
S(p)\;=\;{1\over/\!\!\!\!p-\Sigma(p)}\;,
\eqa
where the quark self-energy is given by~(\ref{selfq}). The inverse quark propagator can be written as
\bqa
S^{-1}(p)\;=\;/\!\!\!\!p-\Sigma(p)\;.
\eqa
This can be written as
\bqa
S^{-1}(p)\;=\;/\!\!\!\!\!{\cal A}(p)\;,
\eqa
where we have organized $A_0(p)$ and $A_S(p)$ into:
\bqa
\label{qself}
A_{\mu}(p)\;=\;(A_0(p),A_S(p)\hat{\bf p})\;.
\eqa
The functions $A_0(p)$ and $A_S(p)$ are defined as
\bqa
\label{aodef}
A_0(p)\!\!&=&\!\!p_0-{m_q^2\over p_0}{\cal T}_p\;,\\
A_S(p)\!\!&=&\!\! |{\bf p}|+{m_q^2\over |{\bf p}|}\left[1-{\cal T}_p\right]\;.
\label{asdef}
\eqa

\section{Quark-gluon three-vertex}

The quark-gluon three-vertex with outgoing gluon momentum $p$, incoming fermion momentum $q$, and outgoing quark momentum $r$, Lorentz index $\mu$ and color index $a$ is
\bqa
\label{3qgv}
\Gamma^{\mu}_a(p,q,r)
\;=\;gt_a\left(\gamma^{\mu}-m_q^2\tilde{{\cal T}}^{\mu}(p,q,r)\right)\;.
\eqa
The tensor in the HTL correction term is only defined for $p-q+r=0$:
\bqa
\tilde{{\cal T}}^{\mu}(p,q,r)
\;=\;\left\langle
y^{\mu}\left({y\!\!\!/\over q\!\cdot\!y\;\;r\!\cdot\!y}\right)
\right\rangle_{\hat{\bf y}}\;.
\label{T3-def}
\eqa
This tensor is even under the permutation of $q$ and $r$. It satisfies the ``Ward identity''
\bqa
p_{\mu}\tilde{\cal T}^{\mu}(p,q,r)\;=\;
\tilde{\cal T}^{\mu}(q)-\tilde{\cal T}^{\mu}(r)\;.
\eqa
The quark-gluon three-vertex therefore satisfies the Ward identity
\bqa
p_{\mu}\Gamma^{\mu}(p,q,r)\;=\;S^{-1}(q)-S^{-1}(r)\;.
\label{qward1}
\eqa

\section{Quark-gluon four-vertex}
We define the quark-gluon four-point vertex with outgoing gluon momenta $p$ and $q$, incoming fermion momentum $r$, and outgoing fermion momentum $s$. Generally this vertex has both adjoint and fundamental indices, however, for this calculation we will only need the quark-gluon four-point vertex traced over the adjoint color indices. In this case
\bqa\nonumber
\delta^{ab} \Gamma^{\mu\nu}_{abij}(p,q,r,s) \!\!&=&\!\! 
    - g^2 m_q^2 c_F \delta_{ij} \tilde{\cal T}^{\mu\nu}(p,q,r,s)  \\
\!\!&\equiv&\!\!g^2 c_F \delta_{ij} \Gamma^{\mu\nu}  \, ,
\label{4qgv}
\eqa
where $c_F = (N_c^2-1)/(2 N_c)$. There is no tree-level term. The tensor in the  HTL correction term is only defined for $p+q-r+s=0$
\bqa
\tilde{{\cal T}}^{\mu\nu}(p,q,r,s)
\;=\;\left\langle y^{\mu}y^{\nu}\left({1\over r\!\cdot\!y}+{1\over s\!\cdot\!y}\right){y\!\!\!/\over[(r-p)\!\cdot\!y]\;[(s+p)\!\cdot\!y]}\right\rangle\;.
\label{T4-def}
\eqa
This tensor is symmetric in $\mu$ and $\nu$ and is traceless. It satisfies the Ward identity:
\bqa
p_{\mu}\Gamma^{\mu\nu}(p,q,r,s)\;=\;\Gamma^{\nu}(q,r-p,s)-\Gamma^{\nu}(q,r,s+p)\;.
\label{qward2}
\eqa

\section{HTL quark counterterm}

The Feynman rule for the insertion of an HTL quark counterterm into a quark propagator is
\bqa
i\delta^{ab}\Sigma(p)\;,
\eqa
where $\Sigma(p)$ is the HTL quark self-energy given in~(\ref{selfq}).

\section{Ghost propagator and vertex}
\label{app:ghost}

The ghost propagator and the ghost-gluon vertex depend on the gauge. The Feynman rule for the ghost propagator is
\bqa
&&{i\over p^2}\delta^{ab} \hspace{3.4cm}\mbox{covariant}\;,
\\
&&{i\over n_p^2 p^2}\delta^{ab} \hspace{3cm}\mbox{Coulomb}\;.
\eqa
The Feynman rule for the vertex in which a gluon with indices $\mu$ and $a$ interacts with an outgoing ghost with outgoing momentum $r$ and color index $c$ is
\bqa
&&-gf^{abc}r^{\mu} \hspace{3cm}\mbox{covariant}\;,
\\
&&-gf^{abc}\left(r^{\mu}-r\!\cdot\!n\,n^{\mu}\right)
\hspace{1.1cm}\mbox{Coulomb}\;.
\eqa
Every closed ghost loop requires a multiplicative factor of $-1$.

\section{Imaginary-time formalism}
\label{app:ITF}

In the imaginary-time formalism, Minkoswski energies have discrete imaginary values $p_0 = i \, 2 n \pi T$ for bosons and $p_0 = i \, (2 n+1) \pi T$ for fermions, and integrals over Minkowski space are replaced by sum-integrals over Euclidean vectors $(2 n \pi T, {\bf p})$ or $((2 n+1) \pi T, {\bf p})$, respectively. We will use the notation $P=(P_0,{\bf p})$ for Euclidean momenta. The magnitude of the spatial momentum will be denoted $p = |{\bf p}|$, and should not be confused with a Minkowski vector. The inner product of two Euclidean vectors is $P \cdot Q = P_0 Q_0 + {\bf p} \cdot {\bf q}$. The vector that specifies the thermal rest frame remains $n = (1,{\bf 0})$.

The Feynman rules for Minkowski space given above can be easily adapted to Euclidean space. The Euclidean tensor in a given Feynman rule is obtained from the corresponding Minkowski tensor with raised indices by replacing each Minkowski energy $p_0$ by $iP_0$, where $P_0$ is the corresponding Euclidean energy, and multipying by $-i$ for every $0$ index. This prescription transforms $p=(p_0,{\bf p})$ into $P=(P_0,{\bf p})$, $g^{\mu \nu}$ into $- \delta^{\mu \nu}$, and $p\!\cdot\!q$ into $-P\!\cdot\!Q$. The effect on the HTL tensors defined in (\ref{T2-def}), (\ref{T3-def}), and (\ref{T4-def}) is equivalent to substituting $p\!\cdot\!n \to - P\!\cdot\!N$ where $N = (-i,{\bf 0})$, $p\!\cdot\!y \to -P\!\cdot\!Y$ where $Y = (-i,{\bf \hat y})$, and $y^\mu \to Y^\mu$. For example, the Euclidean tensor corresponding to (\ref{T2-def}) is
\bqa
{\cal T}^{\mu\nu}(P,-P)\;=\;\left \langle Y^{\mu}Y^{\nu}{P\!\cdot\!N \over P\!\cdot\!Y}\right\rangle \;.
\label{T2E-def}
\eqa
The average is taken over the directions of the unit vector ${\bf \hat y}$.

Alternatively, one can calculate a diagram by using the Feynman rules for Minkowski momenta, reducing the expressions for diagrams to scalars, and then make the appropriate substitutions, such as $p^2 \to -P^2$, $p \cdot q \to - P \cdot Q$, and $n \cdot p \to i n \cdot P$. For example, the propagator functions (\ref{Delta-T:M}) and (\ref{Delta-L:M}) become
\bqa
\Delta_T(P)\!\!&=&\!\!{-1 \over P^2 + \Pi_T(P)}\;,
\label{Delta-T}
\\
\Delta_L(P)\!\!&=&\!\!{1 \over p^2+\Pi_L(P)}\;.
\label{Delta-L}
\eqa
The expressions for the HTL self-energy functions $\Pi_T(P)$ and $\Pi_L(P)$ are given by (\ref{PiT-T}) and (\ref{PiT-L}) with $n_p^2$ replaced by $n_P^2 = p^2/P^2$ and ${\cal T}^{00}(p,-p)$ replaced by
\bqa
{\cal T}_P \;=\; {w(\epsilon)\over2}
        \int_{-1}^1dc\;(1-c^2)^{-\epsilon}{iP_0\over iP_0-pc} \;.
\label{TP-def}
\eqa
Note that this function differs by a sign from the 00 component of the Euclidean tensor corresponding to~(\ref{T2-def}):
\bqa
{\cal T}^{00}(P,-P) \;=\; - {\cal T}^{00}(p,-p)\bigg|_{p_0 \to iP_0}
                   \;=\; - {\cal T}_P \;.
\eqa
A more convenient form for calculating sum-integrals that involve the function ${\cal T}_P$ is
\bqa
{\cal T}_P \;=\;
        \left\langle {P_0^2 \over P_0^2 + p^2c^2} \right\ranglec \, ,
\label{TP-int}
\eqa
where the angular brackets represent an average over $c$ defined by
\begin{equation}
\left\langle f(c) \right\rangle_{\!c} \;\equiv\; w(\epsilon) \int_0^1 dc \,
(1-c^2)^{-\epsilon} f(c)
\label{c-average}
\end{equation}
and $w(\epsilon)$ is given in~(\ref{weight}).

%%%%%%%%%%%%%%%%%%%%%%%%%%%%%%%%%%%%%%%%%%%%%%%%%%%%%%%%%%%%%
%
%	Include File:			DON'T COMPILE !!!
%
%%%%%%%%%%%%%%%%%%%%%%%%%%%%%%%%%%%%%%%%%%%%%%%%%%%%%%%%%%%%%

\chapter{Four-Dimensional Sum-Integrals}
\label{app:sumint}

In the imaginary-time formalism for thermal field theory, the four-momentum $P=(P_0,{\bf p})$ is Euclidean with $P^2=P_0^2+{\bf p}^2$. The Euclidean energy $P_0$ has discrete values: $P_0=2n\pi T$ for bosons and $P_0=(2n+1)\pi T$ for fermions, where $n$ is an integer. Loop diagrams involve sums over $P_0$ and integrals over ${\bf p}$. With dimensional regularization, the integral is generalized to $d = 3-2 \epsilon$ spatial dimensions. We define the dimensionally regularized sum-integrals by
\bqa
  \hbox{$\sum$}\!\!\!\!\!\!\int_{P}\!\!& \equiv &\!\!
  \left(\frac{e^\gamma\mu^2}{4\pi}\right)^\epsilon\;
  T\sum_{P_0=2n\pi T}\:\int {d^{3-2\epsilon}p \over (2 \pi)^{3-2\epsilon}} 
  \hspace{2cm}\mbox{bosons}\;, 
\label{sumint-def} \\ 
  \hbox{$\sum$}\!\!\!\!\!\!\int_{\{P\}}\!\!& \equiv &\!\!
  \left(\frac{e^\gamma\mu^2}{4\pi}\right)^\epsilon\;
  T\sum_{P_0=(2n+1)\pi T}\:\int {d^{3-2\epsilon}p \over (2 \pi)^{3-2\epsilon}} 
  \hspace{1.2cm}\mbox{fermions}\;,
\label{sumint-def1}
\eqa
where $3-2\epsilon$ is the dimension of space and $\mu$ is an arbitrary momentum scale. The factor $(e^\gamma/4\pi)^\epsilon$ is introduced so that, after minimal subtraction of the poles in $\epsilon$ due to ultraviolet divergences, $\mu$ coincides with the renormalization scale of the $\overline{\rm MS}$ renormalization scheme.

\section{One-loop sum-integrals}

The simple one-loop sum-integrals required in our calculations can be derived from the formulas
\bqa\nonumber
\sumint_{P}{p^{2m}\over(P^2)^n}
\!\!&=&\!\!
\left({\mu\over4\pi T}\right)^{2\epsilon}
{2\Gamma({3\over2}+m-\epsilon)\Gamma(n-{3\over2}-m+\epsilon)
\over\Gamma(n)\Gamma(2-2\epsilon)}\Gamma(1-\epsilon)e^{\epsilon\gamma} \\
&&\times\;
\zeta(2n-2m-3+2\epsilon)
T^{4+2m-2n}(2\pi)^{1+2m-2n}\;,
\\
\sumint_{\{P\}}{p^{2m}\over(P^2)^n}\!\!&=&\!\!
(2^{2n-2m-d}-1)\sumint_{P}{p^{2m}\over(P^2)^n}\;. 
\eqa

The specific bosonic one-loop sum-integrals needed are
\bqa
\sumint_{P}\log P^2\!\!&=&\!\!-{\pi^2\over45}T^4 \;,
\\
\sumint_{P}{1\over P^2}
\!\!&=&\!\!{T^2\over12}
\left({\mu\over4\pi T}\right)^{2\epsilon}
\left[1+\left(2+2{\zeta^{\prime}(-1)\over\zeta(-1)}\right)\epsilon+{\cal O}(\epsilon^2)
\right] \;,
\label{sumint:2}
\\
\sumint_P {1 \over (P^2)^2}\!\!&=&\!\!
{1 \over (4\pi)^2} \left({\mu\over4\pi T}\right)^{2\epsilon} 
\left[ {1 \over \epsilon} + 2 \gamma + {\cal O}(\epsilon) \right] \;,
\\
\sumint_P {1 \over p^2 P^2}\!\!&=&\!\!
{2 \over (4\pi)^2} \left({\mu\over4\pi T}\right)^{2\epsilon} 
\left[ {1\over\epsilon} + 2 \gamma + 2 + {\cal O}(\epsilon) \right] \;.
 \label{ex1}
\eqa

The specific fermionic one-loop sum-integrals needed are
\bqa
\sumint_{\{P\}}\log P^2\!\!&=&\!\!{7\pi^2\over360}T^4 \;,
\\
\sumint_{\{P\}}{1\over P^2}\!\!&=&\!\!
-{T^2\over24}\left({\mu\over4\pi T}\right)^{2\epsilon}
\left[1+\left(2-2\log2+2{\zeta^{\prime}(-1)\over\zeta(-1)}\right)\epsilon + {\cal O}(\epsilon^2)\right] \;,
\label{simple1}
\nonumber \\ &&
\\
\sumint_{\{P\}}{1\over(P^2)^2}\!\!&=&\!\!
{1\over(4\pi)^2}\left({\mu\over4\pi T}\right)^{2\epsilon}
\left[ {1 \over \epsilon} + 2 \gamma + 4\log2 + {\cal O}(\epsilon) \right] \;,
\\
\sumint_{\{P\}}{p^2\over(P^2)^2}\!\!&=&\!\!
-{T^2\over16}\left({\mu\over4\pi T}\right)^{2\epsilon}
\left[1 + \left({4\over3} - 2\log2 + 2{\zeta^{\prime}(-1)\over\zeta(-1)}\right)\epsilon + {\cal O}(\epsilon^2)
\right] \;, \nonumber \\ &&
\\ 
\sumint_{\{P\}}{p^2\over(P^2)^3}\!\!&=&\!\!
{3\over4}{1\over(4\pi)^2}\left({\mu\over4\pi T}\right)^{2\epsilon}
\left[{1\over\epsilon} + 2\gamma - {2\over3} + 4\log2 + {\cal O}(\epsilon) \right] \;,
\\ 
\sumint_{\{P\}}{p^4\over(P^2)^4}\!\!&=&\!\!
{5\over8}{1\over(4\pi)^2}\left({\mu\over4\pi T}\right)^{2\epsilon}
\left[{1\over\epsilon} + 2\gamma - {16\over15} + 4\log2 + {\cal O}(\epsilon) \right] \;,
\\ 
\sumint_{\{P\}}{1\over p^2P^2}\!\!&=&\!\!
{2\over(4\pi)^2}\left({\mu\over4\pi T}\right)^{2\epsilon}
\left[{1\over\epsilon} + 2 + 2\gamma + 4\log2 + {\cal O}(\epsilon)\right] \;.
\eqa

The number $\gamma_1$ is the first Stieltjes gamma constant defined by the equation
\begin{equation}
\label{zeta}
\zeta(1+z) \;=\; {1 \over z} + \gamma - \gamma_1 z + O(z^2)\;.
\end{equation}

\section{One-loop HTL sum-integrals}

We also need some more difficult one-loop sum-integrals that involve the HTL function defined in (\ref{TP-def}).

The specific bosonic sum-integrals needed are
\bqa
\sumint_P {1 \over P^2} {\cal T}_P \!\!&=&\!\!
- {T^2 \over 24} \left({\mu\over4\pi T}\right)^{2\epsilon}
\left[ {1 \over \epsilon} + 2 {\zeta'(-1) \over \zeta(-1)} + {\cal O}(\epsilon) \right] \;,
\label{sumint-T:1}
\\
\sumint_P {1 \over p^4} {\cal T}_P \!\!&=&\!\!
-{1 \over (4\pi)^2} \left({\mu\over4\pi T}\right)^{2\epsilon}
\left[ {1 \over \epsilon} + 2 \gamma + 2\log2 + {\cal O}(\epsilon) \right] \;,
\label{sumint-T:2}
\\ 
\sumint_P {1 \over p^2 P^2} {\cal T}_P \!\!&=&\!\!
{1 \over (4\pi)^2} \left({\mu\over4\pi T}\right)^{2\epsilon}
\bigg[ 2 \log2 \left({1 \over \epsilon} + 2 \gamma \right) + 2 \log^2 2
\nonumber \\ && \hspace{2.9cm}
+\;{\pi^2 \over 3} + {\cal O}(\epsilon) \bigg] \;,
\\
\sumint_P {1 \over p^4} ({\cal T}_P)^2 \!\!&=&\!\!
- {2 \over 3}{1 \over (4\pi)^2} \left({\mu\over4\pi T}\right)^{2\epsilon}
\bigg[ (1+ 2 \log 2) \left( {1 \over \epsilon} + 2 \gamma \right) - {4 \over 3} + {22 \over 3} \log 2
\nonumber \\ && \hspace{3.4cm}
 +\;2 \log^2 2 + {\cal O}(\epsilon) \bigg] \;.
\label{sumint-T:5}
\eqa

The specific fermionic sum-integrals needed are
\bqa
\sumint_{\{P\}} {1 \over (P^2)^2} {\cal T}_P \!\!&=&\!\!
{1 \over 2}{1 \over (4\pi)^2} \left({\mu\over4\pi T}\right)^{2\epsilon}
\left[ {1 \over \epsilon} + 2 \gamma + 1 + 4\log2 +{\cal O}(\epsilon)\right] \;,
\label{ht1} 
\\
\sumint_{\{P\}}{1\over p^2P^2}{\cal T}_P \!\!&=&\!\!
{1\over(4\pi)^2}\left({\mu\over4\pi T}\right)^{2\epsilon}
\bigg[ 2\log2 \left({1 \over \epsilon} + 2 \gamma \right) + 10 \log^22 
\nonumber \\ && \hspace{2.9cm}
+\;{\pi^2 \over 3} + {\cal O}(\epsilon) \bigg] \;,
\\
\sumint_{\{P\}}{1\over P^2P_0^2}{\cal T}_P \!\!&=&\!\!
{1\over(4\pi)^2}\left({\mu\over4\pi T}\right)^{2\epsilon}
\bigg[ {1\over \epsilon^2} + 2(\gamma+2\log2){1\over \epsilon} + {\pi^2\over4} + 4\log^22
\nonumber \\ && \hspace{2.9cm}
+\;8\gamma\log2 - 4\gamma_1 + {\cal O}(\epsilon) \bigg] \;,
\\
\sumint_{\{P\}}{1\over p^2P_0^2}\left({\cal T}_P\right)^2 \!\!&=&\!\!
{4\over(4\pi)^2}\left({\mu\over4\pi T}\right)^{2\epsilon}
\left[ \log2 \left({1\over\epsilon} + 2\gamma\right) +5\log^2 2 + {\cal O}(\epsilon) \right] \;, \nonumber \\ &&
\label{htlf} 
\\
\sumint_{\{P\}}{1\over P^2}
\bigg\langle {1\over(P\!\cdot\!Y)^2} \bigg\rangle_{\hat{\bf y}} \!\!&=&\!\!
-{1\over(4\pi)^2}\left({\mu\over4\pi T}\right)^{2\epsilon} 
\left[{1\over\epsilon}-1+2\gamma + 4\log2 + {\cal O}(\epsilon)\right]\;.
\label{cp1ly} 
\\ \nonumber
\eqa

It is straightforward to calculate the sum-integrals (\ref{sumint-T:1})--(\ref{htlf}) using the representation (\ref{TP-int}) of the function ${\cal T}_P$. For example, the sum-integral (\ref{sumint-T:2}) can be written
\bqa
\sumint_P {1 \over p^4} {\cal T}_P \;=\;
\sumint_P {1 \over p^4}
        \left\langle {P_0^2\over P_0^2  + p^2c^2} \right\ranglec \;,
\label{example}
\eqa
where the angular brackets denote an average over $c$ as defined in (\ref{c-average}). Using the factor $P_0^2$ in the numerator to cancel denominators, this becomes
\bqa
\sumint_P {1 \over p^4} {\cal T}_P \;=\; \sumint_{P}{1\over p^4}
\left[1-\left\langle{p^2c^2\over P_0^2+p^2c^2}\right\rangle_c\right] \;.
\eqa
The first term in the square brackets vanishes with dimensional regularization, while after rescaling the momentum by ${\bf p}\rightarrow{\bf p}/c$, the second term reads
\bqa
\sumint_P {1 \over p^4} {\cal T}_P \;=\; 
- \left\langle c^{1+2\epsilon}\right\rangle_c \sumint_{P}{1\over p^2P^2} \;.
\eqa
Evaluating the average over $c$, using the expression (\ref{ex1}) for the sum-integral, and expanding in powers of $\epsilon$, we obtain the result (\ref{sumint-T:2}). Following the same strategy, all the sum-integrals (\ref{sumint-T:1})--(\ref{cp1ly}) can be reduced to linear combinations of simple sum-integrals with coefficients that are averages over $c$. The only difficult integrals are the double average over $c$ that arises from (\ref{htlf}) and (\ref{sumint-T:5}):
\bqa
\left\langle{c_1^{1+2\epsilon}-c_2^{1+2\epsilon}\over c_1^2-c_2^2}\right\rangle_{c_1,c_2}
\!\!&=&\!\!2\log2-2\log2\left(2-\log2\right)\epsilon+{\cal O}(\epsilon^2) \;,
\\ 
\left\langle{c_1^{3+2\epsilon}-c_2^{3+2\epsilon}\over c_1^2-c_2^2}\right\rangle_{c_1,c_2}
\!\!&=&\!\!{1+2\log2\over3}-{2\over3}\left({5\over3}-{5\over3}\log2-\log^22\right)\epsilon+{\cal O}(\epsilon^2) \;. \nonumber \\
\eqa

\section{Two-loop sum-integrals}

The simple bosonic two-loop sum-integrals that are needed are
\bqa
\sumint_{PQ} {1\over P^2 Q^2 R^2} \!\!&=&\!\! {\cal O}(\epsilon) \;,
\\
\sumint_{PQ} {1 \over P^2 Q^2 r^2} \!\!&=&\!\!
{1 \over 12} {T^2 \over (4 \pi)^2} \left({\mu\over4\pi T}\right)^{4\epsilon}
\left[ {1 \over \epsilon} + 10 - 12 \log 2
	+ 4 {\zeta'(-1) \over \zeta(-1)} + {\cal O}(\epsilon) \right] \;,
\label{sumint2:2}
\\
\sumint_{PQ} {q^2 \over P^2 Q^2 r^4} \!\!&=&\!\!
{1 \over 6} {T^2 \over (4 \pi)^2} \left({\mu\over4\pi T}\right)^{4\epsilon}
\left[ {1 \over \epsilon} + {8 \over 3} + 2 \gamma
	+ 2 {\zeta'(-1) \over \zeta(-1)} + {\cal O}(\epsilon) \right] \;,
\label{sumint2:3}
\\
\sumint_{PQ} {q^2 \over P^2 Q^2 r^2 R^2} \!\!&=&\!\!
{1 \over 9} {T^2 \over (4 \pi)^2} \left({\mu\over4\pi T}\right)^{4\epsilon}
\left[ {1 \over \epsilon} + 7.521 + {\cal O}(\epsilon) \right] \;,
\label{sumint2:4}
\\
\sumint_{PQ} {P\!\cdot\!Q \over P^2 Q^2 r^4} \!\!&=&\!\!
-{1 \over 8} {T^2 \over (4 \pi)^2} \left({\mu\over4\pi T}\right)^{4\epsilon}
\left[ {1 \over \epsilon} + {2 \over 9} + 4 \log 2 + {8\over3} \gamma
	+ {4\over3} {\zeta'(-1) \over \zeta(-1)} + {\cal O}(\epsilon) \right] \;, \nonumber \\
\label{sumint2:5}
\eqa
where $R=-(P+Q)$ and $r=|{\bf p}+{\bf q}|$.

The simple fermionic two-loop sum-integrals that are needed are
\bqa
\sumint_{\{PQ\}}{1\over P^2Q^2R^2} \!\!&=&\!\! {\cal O}(\epsilon) \;, 
\\
\sumint_{\{PQ\}}{1\over P^2Q^2r^2} \!\!&=&\!\!
-{1\over6} {T^2 \over (4 \pi)^2} \left({\mu\over4\pi T}\right)^{4\epsilon}
\left[ {1\over\epsilon} + 4
        - 2\log2 + 4{\zeta^{\prime}(-1)\over\zeta(-1)} + {\cal O}(\epsilon) \right] \;,
\label{two1}
\nonumber \\ && 
\\
\sumint_{\{PQ\}}{q^2\over P^2Q^2r^4} \!\!&=&\!\!
-{1\over12} {T^2 \over (4 \pi)^2} \left({\mu\over4\pi T}\right)^{4\epsilon} 
\bigg[ {1\over\epsilon} + {11\over3} + 2\gamma - 2\log2 
       + 2{\zeta^{\prime}(-1)\over\zeta(-1)} \nonumber \\
&& \hspace{4cm} +\;{\cal O}(\epsilon) \bigg] \;, 
\label{two2}
\\
\sumint_{\{PQ\}}{q^2\over P^2Q^2r^2R^2} \!\!&=&\!\!
-{1\over72} {T^2 \over (4 \pi)^2} \left({\mu\over4\pi T}\right)^{4\epsilon}
\left[ {1\over\epsilon} - 7.00164 + {\cal O}(\epsilon)\right] \;,
\label{two3}
\\
\sumint_{\{PQ\}}{P\cdot Q\over P^2Q^2r^4} \!\!&=&\!\!
-{1\over36} {T^2 \over (4 \pi)^2} \left({\mu\over4\pi T}\right)^{4\epsilon}
\left[ 1 - 6\gamma + 6{\zeta^{\prime}(-1)\over\zeta(-1)} + {\cal O}(\epsilon) \right] \;,
\label{twolast}
\\
\sumint_{\{PQ\}}{p^2\over q^2P^2Q^2R^2} \!\!&=&\!\!
{5\over72} {T^2\over(4\pi)^2}\left({\mu\over4\pi T}\right)^{4\epsilon}
\left[ {1\over\epsilon} + 9.55216 + {\cal O}(\epsilon) \right] \;,
\label{ntwo1}
\\
\sumint_{\{PQ\}}{r^2\over q^2P^2Q^2R^2} \!\!&=&\!\!
-{1\over18} {T^2\over(4\pi)^2}\left({\mu\over4\pi T}\right)^{4\epsilon}
\left[ {1\over\epsilon} +8.14234 + {\cal O}(\epsilon) \right] \;,
\label{ntwo2}
\eqa
where $R=-(P+Q)$ and $r=|{\bf p}+{\bf q}|$.

To motivate the integration formula we will use to evaluate the two-loop sum-integrals, we first present the analogous integration formula for one-loop sum-integrals. In a one-loop sum-integral, the sum over $P_0$ can be replaced by a contour integral in $p_0 = i P_0$:
\begin{eqnarray}
\sumint_P F(P) \;=\; \lim_{\eta \to 0^+} \int {d p_0 \over 2 \pi i} \int_{\bf p}
\left[ F(-i p_0,{\bf p}) - F(0,{\bf p}) \right] e^{\eta p_0} \, n_B(p_0) \;,
\end{eqnarray}
where $n_B(p_0) = 1/(e^{\beta p_0} - 1)$ is the Bose-Einstein thermal distribution and the contour runs from $-\infty$ to $+\infty$ above the real axis and from  $+\infty$ to $-\infty$ below the real axis. This formula can be expressed in a more convenient form by collapsing the contour onto the real axis and separating out those terms  with the exponential convergence factor $n_B(|p_0|)$. The remaining terms run along contours from $-\infty \pm i \varepsilon$ to 0 and have the convergence factor $e^{\eta p_0}$. This allows the contours to be deformed so that they run from 0 to $\pm i \infty$ along the imaginary $p_0$ axis, which corresponds to real values of $P_0 = -i p_0$. Assuming that $F(-i p_0,{\bf p})$ is a real function of $p_0$, i.e. that it satisfies $F(-i p_0^*,{\bf p})= F^*(-i p_0,{\bf p})$, the resulting formula for the sum-integral is
\begin{eqnarray}
\sumint_P F(P) \;=\; \int_P F(P) + \int_p \epsilon(p_0) n_B(|p_0|) \,
2 {\rm Im} F(-i p_0+ \varepsilon,{\bf p}) \;,
\label{int-1loop}
\end{eqnarray}
where $\epsilon(p_0)$ is the sign of $p_0$. The first integral on the right side is over the $(d+1)$-dimensional Euclidean vector $P = (P_0,{\bf p})$ and the second is over the $(d+1)$-dimensional
Minkowskian vector $p = (p_0,{\bf p})$.

The two-loop bosonic sum-integrals can be evaluated by using a generalization of the one-loop formula (\ref{int-1loop}):
\begin{eqnarray}
&& \sumint_{PQ} F(P) G(Q) H(R) \;=\;
{1 \over 3} \int_{PQ} F(P) G(Q) H(R)
\nonumber
\\
&& \hspace{1cm}
+\; \int_p \epsilon(p_0) n_B(|p_0|) \,
	2 {\rm Im} F(-i p_0+ \varepsilon,{\bf p}) \,
	{\rm Re} \int_Q G(Q) H(R)\bigg|_{P_0 = -ip_0 + \varepsilon}
\nonumber
\\
&&  \hspace{1cm}
+\; \int_p \epsilon(p_0) n_B(|p_0|) \,
		2 {\rm Im} F(-i p_0+ \varepsilon,{\bf p}) \,
	\int_q \epsilon(q_0) n_B(|q_0|) \,
		2 {\rm Im} G(-i q_0+ \varepsilon,{\bf q}) \,
\nonumber
\\
&&  \hspace{3cm}
	\times {\rm Re} H(R)\bigg|_{R_0 = i (p_0 + q_0)+ \varepsilon}
\nonumber
\\
&&  \hspace{1cm}
+\; ({\rm cyclic \; permutations \; of \;} F, \, G, \, H) \;.
\label{int-2loop}
\end{eqnarray}
The sum over cyclic permutations multiplies the first term on the right side by a factor of 3, so there are a total of seven terms. This formula can be derived in three steps. First, express the sum over $P_0$ as the sum of two contour integrals over $p_0$, one that encloses the real axis ${\rm Im}\, p_0 = 0$ and another that encloses the line ${\rm Im} \, p_0= - {\rm Im} \, q_0$. Second, express the sum over $q_0$ as a contour integral that encloses the real-$q_0$ axis. Third, symmetrize the resulting expression under the six permutations of $F$, $G$, and $H$. The resulting terms can be combined into the expression (\ref{int-2loop}).

The fermionic generalization of (\ref{int-2loop}) reads
\begin{eqnarray} \nonumber
&& \sumint_{\{PQ\}} F(P) G(Q) H(R) \;=\;
\int_{PQ} F(P) G(Q) H(R)
\\ && \nonumber
\hspace{1cm}
-\; \int_p \epsilon(p_0) n_F(|p_0|) \, 2 \, {\rm Im} F(-i p_0+ \varepsilon,{\bf p}) 
	\, {\rm Re} \int_Q G(Q) H(R)\bigg|_{P_0 = -ip_0 + \varepsilon}
\\ && \nonumber
\hspace{1cm}
-\; \int_p \epsilon(p_0) n_F(|p_0|) \, 2 \, {\rm Im} G(-i p_0+ \varepsilon,{\bf p}) 
	\, {\rm Re} \int_Q H(Q) F(R)\bigg|_{P_0 = -ip_0 + \varepsilon}
\\ && \nonumber
\hspace{1cm}
+\; \int_p \epsilon(p_0) n_B(|p_0|) \, 2 \, {\rm Im} H(-i p_0+ \varepsilon,{\bf p}) 
	\, {\rm Re} \int_Q F(Q) G(R)\bigg|_{P_0 = -ip_0 + \varepsilon}
\\ && \nonumber 
\hspace{1cm}
+\; \int_p \epsilon(p_0) n_F(|p_0|) \, 2 \, {\rm Im} F(-i p_0+ \varepsilon,{\bf p}) \,
        \int_q \epsilon(q_0) n_F(|q_0|) \, 2 \, {\rm Im} G(-i q_0+ \varepsilon,{\bf q}) 
	\\ \nonumber
&&  \hspace{3cm}
	\times{\rm Re} H(R)\bigg|_{R_0 = i (p_0 + q_0)+ \varepsilon}
\\ && \nonumber
\hspace{1cm}
-\; \int_p \epsilon(p_0) n_F(|p_0|) \, 2 \, {\rm Im} G(-i p_0+ \varepsilon,{\bf p}) \,
        \int_q \epsilon(q_0) n_B(|q_0|) \, 2 \, {\rm Im} H(-i q_0+ \varepsilon,{\bf q})
\\ \nonumber
&&  \hspace{3cm}
         \times{\rm Re} F(R)\bigg|_{R_0 = i (p_0 + q_0)+ \varepsilon}
\\ && \nonumber
\hspace{1cm}
-\; \int_p \epsilon(p_0) n_B(|p_0|) \, 2 \, {\rm Im} H(-i p_0+ \varepsilon,{\bf p}) \,
        \int_q \epsilon(q_0) n_F(|q_0|) \, 2 \, {\rm Im} F(-i q_0+ \varepsilon,{\bf q}) 
	\\
&&  \hspace{3cm}
         \times{\rm Re} G(R)\bigg|_{R_0 = i (p_0 + q_0)+ \varepsilon} \;,
\label{int-2loop-f}
\end{eqnarray}
where $n_F(p_0) = 1/(e^{\beta p_0} + 1)$ is the Fermi-Dirac thermal distribution. The integrals of the imaginary parts that enter into our calculation can be reduced to
\begin{eqnarray}
&&\int_p \epsilon(p_0) n(|p_0|) \, 2 {\rm Im}
	{1 \over P^2}\bigg|_{P_0=-i p_0+ \varepsilon}
	f(-i p_0 + \varepsilon,{\bf p})
	\;=\; \int_{\bf p} {n(p)\over p} {1 \over 2} \sum_{\pm}
	f(\pm i p + \varepsilon,{\bf p}) \;, \nonumber \\ &&
\label {impart1}
\\
&&\int_p \epsilon(p_0) n(|p_0|) \,  2 {\rm Im}
	{\cal T}_P\bigg|_{P_0=-i p_0+ \varepsilon}
	f(-i p_0 + \varepsilon,{\bf p}) \nonumber \\
	&& \hspace {6cm} \;=\;  - \int_{\bf p} p\,n(p) {1 \over 2} \sum_{\pm}
	\left\langle c^{-3+2\epsilon}
	f(\pm i p + \varepsilon,{\bf p}/c) \right\ranglec \;. \nonumber \\ &&
\end{eqnarray}
The latter equation is obtained by inserting the expression (\ref{TP-int}) for ${\cal T}_P$, using (\ref{impart1}), and then making the change of variable ${\bf p} \rightarrow {\bf p}/c$ to put the thermal integral into a standard form.

As an illustration for calculating the simple two-loop bosonic sum-integrals, we apply the formula (\ref{int-2loop}) to the sum-integral (\ref{sumint2:2}). The nonvanishing terms are
\begin{eqnarray}\hspace{-8mm}
\sumint_{PQ} {1 \over P^2 Q^2 r^2} \!\!&=&\!\!
2 \int_p n_B(|p_0|) \, 2 \pi \delta(p_0^2 - p^2) \int_Q {1 \over Q^2 r^2}
\nonumber
\\
&& +\; \int_p n_B(|p_0|) \, 2 \pi \delta(p_0^2 - p^2)
	  \int_q n_B(|q_0|) \, 2 \pi \delta(q_0^2 - q^2) {1 \over r^2} \;.
\end{eqnarray}
The delta functions can be used to evaluate the integrals over $p_0$ and $q_0$. The integral over $Q$ is given in (\ref{int4:1}). This reduces the sum-integral to
\begin{eqnarray}
\sumint_{PQ} {1 \over P^2 Q^2 r^2} \!\!&=&\!\!
{4 \over (4 \pi)^2} \left[ {1 \over \epsilon} + 4 - 2 \log 2 \right]
\mu^{2 \epsilon} \int_{\bf p} {n_B(p) \over p} p^{-2 \epsilon}
\nonumber
\\
&& +\; \int_{\bf p q} {n_B(p) n_B(q) \over p q} {1 \over r^2} \;.
\end{eqnarray}
The momentum integrals are evaluated in (\ref{int-th:1}) and (\ref{int-th:2}).  Keeping all terms that contribute through order $\epsilon^0$, we get the result (\ref{sumint2:2}). The sum-integral (\ref{sumint2:3}) can be evaluated in the same way:
\begin{eqnarray}\hspace{-4mm}
\sumint_{PQ} {q^2 \over P^2 Q^2 r^4} \;=\;
{2 \over (4 \pi)^2 }
\left[ {1 \over \epsilon} - 2 \log 2 \right]
\mu^{2 \epsilon} \int_{\bf p} {n_B(p) \over p} p^{-2 \epsilon}
+ \int_{\bf p q} {n_B(p) n_B(q) \over p q} {q^2 \over r^4} \;.
\end{eqnarray}
The sum-integral (\ref{sumint2:5}) can be reduced to a linear combination of (\ref{sumint2:2}) and (\ref{sumint2:3}) by expressing the numerator in the form $P\!\cdot\!Q = P_0 Q_0 + (r^2 - p^2 - q^2)/2$ and noting that the $P_0 Q_0$ term vanishes upon summing over $P_0$ or $Q_0$. The sum-integral (\ref{sumint2:4}) is a little more difficult. After applying the formula (\ref{int-2loop}) and using the delta functions to integrate over $p_0$, $q_0$, and $r_0$, it can be reduced to
\begin{eqnarray}
\sumint_{PQ} {q^2\over P^2 Q^2 r^2 R^2} \!\!&=&\!\!
\int_{\bf p} {n_B(p) \over p} \int_Q {1 \over Q^2 R^2}
	\left( {p^2 \over r^2} + {q^2 \over r^2} + {q^2 \over p^2} \right)
	\bigg|_{P_0 = -i p}
\nonumber
\\
&& +\; \int_{\bf p q} {n_B(p) n_B(q) \over p q}
	\left( {p^2 \over r^2} + {p^2 \over q^2} + {r^2 \over q^2} \right)
	{r^2 - p^2 - q^2 \over \Delta(p,q,r)} \;,
\label{sumint:q/PQrR}
\end{eqnarray}
where $\Delta(p,q,r)$ is the triangle function that is negative when $p$, $q$, and $r$ are the lengths of 3 sides of a triangle:
\begin{equation}
\Delta(p,q,r) \;=\; p^4 + q^4 + r^4 - 2 (p^2 q^2 + q^2 r^2 + r^2 p^2) \;.
\label{triangle}
\end{equation}
After using (\ref{int4:4})--(\ref{int4:6}) to integrate over $Q$, the first term on the right side of (\ref{sumint:q/PQrR}) is evaluated using (\ref{int-th:1}). The two-loop thermal integrals on the right side of (\ref{sumint:q/PQrR}) are given in (\ref{int-thT:1})--(\ref{int-thT:4}). Adding together all the terms, we get the final result (\ref{sumint2:4}).

As an illustration for calculating the simple two-loop fermionic sum-integrals, we apply the formula (\ref{int-2loop-f}) to the sum-integral (\ref{two1}). The nonvanishing terms are
\begin{eqnarray}\hspace{-8mm}
\sumint_{\{PQ\}} {1 \over P^2 Q^2 r^2} \!\!&=&\!\!
-2 \int_p n_F(|p_0|) \, 2 \pi \delta(p_0^2 - p^2) \int_Q {1 \over Q^2 r^2}
\nonumber
\\ 
&& +\; \int_p n_F(|p_0|) \, 2 \pi \delta(p_0^2 - p^2)
            \int_q n_F(|q_0|) \, 2 \pi \delta(q_0^2 - q^2) {1 \over r^2} \;.
\end{eqnarray}
The delta functions can be used to evaluate the integrals over $p_0$ and $q_0$. The integral over $Q$ is given in (\ref{int4:1}). This reduces the sum-integral to
\begin{eqnarray}
\sumint_{\{PQ\}} {1 \over P^2 Q^2 r^2} \!\!&=&\!\!
-{4 \over (4 \pi)^2} \left[ {1 \over \epsilon} + 4 - 2 \log 2 \right]
\mu^{2 \epsilon} \int_{\bf p} {n_F(p) \over p} p^{-2 \epsilon} \nonumber
\\ &&
+\; \int_{\bf p q} {n_F(p) n_F(q) \over p q} {1 \over r^2} \;.
\end{eqnarray}
The momentum integrals are evaluated in (\ref{int-th:1}) and (\ref{int-th:2}).  Keeping all terms that contribute through order $\epsilon^0$, we get the result (\ref{two1}). The sum-integral (\ref{two2}) can be evaluated in the same way:
\begin{eqnarray}
\sumint_{\{PQ\}} {q^2 \over P^2 Q^2 r^4} \!\!&=&\!\!
-{2 \over (4 \pi)^2 }
\left[ {1 \over \epsilon} - 2 \log 2 \right]
\mu^{2 \epsilon} \int_{\bf p} {n_F(p) \over p} p^{-2 \epsilon}
\nonumber
\\
&& +\; \int_{\bf p q} {n_F(p) n_F(q) \over p q} {q^2 \over r^4} \;.
\end{eqnarray}
The sum-integral (\ref{twolast}) can be reduced to a linear combination of (\ref{two1}) and (\ref{two2}) by expressing the numerator in the form $P\!\cdot\!Q = P_0 Q_0 + (r^2 - p^2 - q^2)/2$ and noting that the $P_0 Q_0$ term vanishes upon summing over $P_0$ or $Q_0$. The sum-integral (\ref{two3}) is a little more difficult. After applying the formula (\ref{int-2loop-f}) and using the delta functions to integrate over $p_0$, $q_0$, and $r_0$, it can be reduced to
\begin{eqnarray}\nonumber
\sumint_{\{PQ\}} {q^2\over P^2 Q^2 r^2 R^2} \!\!&=&\!\!
\int_{\bf p}{n_B(p)\over p}\int_{Q}{q^2\over p^2Q^2R^2}\bigg|_{P =-i p}
- \int_{\bf p}{n_F(p)\over p}\int_{Q}{1\over Q^2R^2}
\left({q^2\over r^2}+{p^2\over q^2}\right)\bigg|_{P =-i p}
\\ && \nonumber
+\; \int_{\bf pq}{n_F(p)n_F(q)\over pq}{p^2\over r^2}
{r^2-p^2-q^2\over\Delta(p,q,r)}
\\ &&
-\; \int_{\bf pq}{n_F(p)n_B(q)\over pq}\left({p^2\over q^2}+{r^2\over q^2}\right) 
{r^2-p^2-q^2\over\Delta(p,q,r)} \;.
\label{sumint:q/PQrR-f}
\end{eqnarray}
After using (\ref{int4:4})--(\ref{int4:6}) to integrate over $Q$, the first term on the right side of (\ref{sumint:q/PQrR-f}) is evaluated using (\ref{int-th:1}). The two-loop thermal integrals on the right side of (\ref{sumint:q/PQrR-f}) are given in (\ref{int-thT:1})--(\ref{int-thT:4}). Adding together all the terms, we get the final result (\ref{two3}).

\section{Two-loop HTL sum-integrals}

We also need some more difficult two-loop sum-integrals that involve the functions ${\cal T}_P$ defined in (\ref{TP-def}).

The specific bosonic sum-integrals needed are
\begin{eqnarray}
\sumint_{PQ} {1 \over P^2 Q^2 r^2} {\cal T}_R \!\!&=&\!\!
- {1\over 48} {T^2 \over (4 \pi)^2} \left({\mu\over4\pi T}\right)^{4\epsilon}
\bigg[ {1 \over \epsilon^2}
+ \left( 2 - 12 \log2 + 4 {\zeta'(-1) \over \zeta(-1)} \right){1 \over \epsilon}
\nonumber \\
&& \hspace{4.1cm}
 -\; 19.83 + {\cal O}(\epsilon) \bigg] \;,
\label{sumint2:6}
\\
\sumint_{PQ} {q^2 \over P^2 Q^2 r^4} {\cal T}_R \!\!&=&\!\!
- {1\over 576} {T^2 \over (4 \pi)^2} \left({\mu\over4\pi T}\right)^{4\epsilon}
\bigg[ {1 \over \epsilon^2} 
+ \left( {26\over3} - {24 \over \pi^2} - 92 \log2 + 4 {\zeta'(-1) \over \zeta(-1)} \right) {1 \over \epsilon}
\nonumber \\
&& \hspace{4.3cm}
-\; 477.7 +{\cal O}(\epsilon) \bigg] \;,
\label{sumint2:7}
\\
\sumint_{PQ} {P\!\cdot\!Q \over P^2 Q^2 r^4} {\cal T}_R \!\!&=&\!\!
- {1\over 96} {T^2 \over (4 \pi)^2} \left({\mu\over4\pi T}\right)^{4\epsilon}
\bigg[ {1 \over \epsilon^2}
+ \left( {8 \over \pi^2} + 4 \log2 + 4 {\zeta'(-1) \over \zeta(-1)} \right){1 \over \epsilon}
\nonumber \\
&& \hspace{4.1cm}
+\; 59.66 + {\cal O}(\epsilon) \bigg] \;.
\label{sumint2:8}
\end{eqnarray}
%
%[[ DELETED 2 DIGITS (OLD: $- 19.8311$, $- 477.735$, $+ 59.6639$) ]]

The specific fermionic sum-integrals needed are
\bqa
\sumint_{\{PQ\}}{1\over P^2Q^2r^2}{\cal T}_R \!\!&=&\!\!
-{1\over48} {T^2 \over (4 \pi)^2} \left({\mu\over4\pi T}\right)^{4\epsilon}
	\bigg[{1\over\epsilon^2} + \left( 2 +12\log 2 + 4 {\zeta'(-1) \over \zeta(-1)} \right)
	{1\over\epsilon} 
\nonumber \\
&& \hspace{4.1cm}
	+\; 136.362 + {\cal O}(\epsilon) \bigg] \;, 
\label{htlf1}
\\ 
\sumint_{\{PQ\}} {q^2\over P^2Q^2r^4}{\cal T}_R \!\!&=&\!\!
-{1\over576} {T^2 \over (4 \pi)^2} \left({\mu\over4\pi T}\right)^{4\epsilon}
	\bigg[{1\over\epsilon^2} 
	+\left({26\over3}+52\log2+4{\zeta'(-1) \over \zeta(-1)}\right){1\over\epsilon}
\nonumber \\
&& \hspace{4.2cm}
	+\; 446.412 + {\cal O}(\epsilon) \bigg] \;,
\label{htlf2}
\\ 
\sumint_{\{PQ\}}{P\!\cdot\!Q\over P^2Q^2r^4}{\cal T}_R \!\!&=&\!\!
-{1\over96} {T^2 \over (4 \pi)^2} \left({\mu\over4\pi T}\right)^{4\epsilon}
	\bigg[{1\over\epsilon^2} +\left(4\log2+4{\zeta'(-1)\over\zeta(-1)} \right) 
	{1\over\epsilon} 
\nonumber \\
&& \hspace{4.1cm}
	+\; 69.174 + {\cal O}(\epsilon) \bigg] \;, 
\label{htl3}
\\ 
\sumint_{\{PQ\}}{r^2-p^2\over P^2q^2Q^2_0R^2}{\cal T}_Q \!\!&=&\!\!
-{1\over8} {T^2\over(4\pi)^2}\left({\mu\over4\pi T}\right)^{4\epsilon}
	\bigg[{1\over\epsilon^2} +\left(2+2\gamma +{10\over3}\log2
	+2{\zeta^{\prime}(-1)\over\zeta(-1)}\right){1\over\epsilon}
\nonumber \\
&& \hspace{3.9cm}	
	+\; 46.8757 + {\cal O}(\epsilon) \bigg] \;.
\label{com2l}
\eqa

To calculate the sum-integral (\ref{sumint2:6}), we begin by using the representation (\ref{TP-int}) of the function ${\cal T}_R$:
\begin{equation}
\sumint_{PQ} {1 \over P^2 Q^2 r^2} {\cal T}_R \;=\;
	\sumint_{PQ} {1 \over P^2 Q^2 r^2}
	- \sumint_{PQ} {1 \over P^2 Q^2 } \left\langle
	{c^2 \over R_0^2 + r^2 c^2} \right\ranglec \;.
\label{sumint2:6a}
\end{equation}
The first sum-integral on the right side is given by (\ref{sumint2:2}). To evaluate the second sum-integral, we apply the sum-integral formula (\ref{int-2loop}):
\begin{eqnarray}
&& \hspace{-1.5cm}
\sumint_{PQ} {1 \over P^2 Q^2 (R_0^2 + r^2 c^2)}
\nonumber
\\
\!\!&=&\!\! \int_{\bf p} {n_B(p) \over p}
	\left(2 {\rm Re} \int_Q {1 \over Q^2 (R_0^2+r^2c^2)}
	\bigg|_{P_0 = - i p + \varepsilon}
	+ c^{-3+2\epsilon}
	\int_Q {1 \over Q^2 R^2} \bigg|_{P \to (-i p,{\bf p}/c)}
	\, \right)
\nonumber
\\
&&
+\; \int_{\bf pq} {n_B(p) n_B(q) \over p q}
	\left( {\rm Re} {r^2 c^2 - p^2 - q^2 \over
	\Delta(p+i\varepsilon,q,r c)}
	+ 2 c^{-3+2\epsilon} \, {\rm Re}
	{ r_c^2 - p^2 - q^2 \over \Delta(p+i\varepsilon,q,r_c)} \,
	\right) \;,
\label{sumint2:6b}
\end{eqnarray}
where $r_c = |{\bf p} + {\bf q}/c|$. In the terms on the right side with a single thermal integral, the appropriate averages over $c$ of the integrals over $Q$ are given in (\ref{int4HTL:1}) and (\ref{int4:8.1}).
\begin{eqnarray}
&& \hspace{-1.5cm}
\left\langle c^2 \left(2 {\rm Re}
\int_Q {1 \over Q^2 (R_0^2+r^2c^2)} \bigg|_{P_0 = - i p + \varepsilon}
+ c^{-3+2\epsilon}
	\int_Q {1 \over Q^2 R^2} \bigg|_{P \to (-i p,{\bf p}/c)}
	\, \right) \right\ranglec
\nonumber
\\ 
&=&\!\! {1 \over (4 \pi)^2} \mu^{2\epsilon} p^{-2 \epsilon}
\left[ {1\over4\epsilon^2}+\left( 4-{7\over2}\log 2\right){1\over\epsilon}
+16-{13\pi^2\over16}-8\log 2+{17\over2}\log^2 2 \right] \,.
\nonumber
\\
\label{intave:1a}
\end{eqnarray}
The subsequent integral over ${\bf p}$ is a special case of (\ref{int-th:1}):
\begin{equation}
\int_{\bf p} n_B(p) \, p^{-1-2\epsilon} \;=\;
2^{8 \epsilon}
{(1)_{-4\epsilon} ({1\over2})_{2\epsilon}
	\over (1)_{-2\epsilon} ({3\over2})_{-\epsilon}} \,
{\zeta(-1+4\epsilon) \over \zeta(-1)}
(e^\gamma \mu^2)^\epsilon (4 \pi T)^{-4\epsilon} \, {T^2 \over 12} \;,
\label{int-th:-1}
\end{equation}
where $(a)_b=\Gamma(a+b)/\Gamma(a)$ is Pochhammer's symbol. Combining this with (\ref{intave:1a}), we obtain
\begin{eqnarray}
&& \hspace{-1cm}
 \int_{\bf p} {n_B(p) \over p}
\left( 2 \, {\rm Re} \int_Q {1 \over Q^2}
	\left\langle {c^2 \over R_0^2 + r^2 c^2} \right\ranglec
		\bigg|_{P_0 = - i p + \varepsilon}
+ \left\langle c^{-1+2\epsilon}
	\int_Q {1 \over Q^2 R^2} \bigg|_{P \to (-i p,{\bf p}/c)}
	\right\ranglec\, \right)
\nonumber
\\
&=&\!\! {T^2 \over (4 \pi)^2} \left({\mu\over4\pi T}\right)^{4\epsilon}
{1 \over 48}
\left[ {1 \over \epsilon^2}
+ \left( 18 - 12 \log 2 + 4 {\zeta'(-1) \over \zeta(-1)} \right)
	{1 \over \epsilon}
+ 173.30233 \right] \, .
\label{intave:1}
\end{eqnarray}
For the two terms in (\ref{sumint2:6b}) with a double thermal integral, the averages weighted by $c^2$ are given in (\ref{intHTL:1}) and (\ref{intHTL:4}). Adding them to (\ref{intave:1}), the final result is
\begin{eqnarray}
&& \hspace{-1.5cm}
 \sumint_{PQ} {1 \over P^2 Q^2}
	\left\langle {c^2 \over R_0^2 + r^2 c^2} \right\ranglec \nonumber \\
&=&\!\!  {T^2 \over (4 \pi)^2} \left({\mu\over4\pi T}\right)^{4\epsilon}
{1\over 48}
\left[ {1 \over \epsilon^2} + \left( 6 - 12 \log2 + 4 {\zeta'(-1) \over \zeta(-1)} \right)
	{1 \over \epsilon} + 18.66 \right] \, .
\label{sumintave:2}
\end{eqnarray}
% 
% [[ FEWER DIGITS, OLD: $+ 18.6587$]]
Inserting this into (\ref{sumint2:6a}), we obtain the final result (\ref{sumint2:6}).

The sum-integral (\ref{sumint2:7}) is evaluated in a similar way to (\ref{sumint2:6}). Using the representation (\ref{TP-int}) for ${\cal T}_R$, we get
\begin{equation}
\sumint_{PQ} {q^2 \over P^2 Q^2 r^4} {\cal T}_R \;=\;
	\sumint_{PQ} {q^2 \over P^2 Q^2 r^4}
	- \sumint_{PQ} {q^2 \over P^2 Q^2 r^2} \left\langle
	{c^2 \over R_0^2 + r^2 c^2} \right\ranglec \, .
\label{sumint2:7a}
\end{equation}
The first sum-integral on the right hand side is given by (\ref{sumint2:3}). To evaluate the second sum-integral, we apply the sum-integral formula (\ref{int-2loop}):
\begin{eqnarray}
&& \hspace{-1cm}
\sumint_{PQ} {q^2 \over P^2 Q^2 r^2 (R_0^2+r^2c^2)}
\nonumber
\\
&=&\!\! \int_{\bf p} {n_B(p) \over p} \left( {\rm Re}
	\int_Q {p^2+q^2 \over Q^2 r^2 (R_0^2+r^2c^2)}
	\bigg|_{P_0 = -i p + \varepsilon}
	+ {1 \over p^2} c^{-1+2\epsilon}
	\int_Q {q^2 \over Q^2 R^2}\bigg|_{P \to (-i p,{\bf p}/c)} \right)
\nonumber
\\
&&
+\; \int_{\bf pq} {n_B(p) n_B(q) \over p q}
	\left({q^2 \over r^2} \, {\rm Re}
	{r^2 c^2 - p^2 - q^2 \over \Delta(p+i\varepsilon,q,r c)}
	+ c^{-1 + 2\epsilon} {p^2 + r_c^2 \over q^2} \,
	{\rm Re} { r_c^2-p^2-q^2
	\over \Delta(p+i\varepsilon,q,r_c)} \right) \;.
\nonumber
\\
\label{sumint3:6b}
\end{eqnarray}
In the terms on the right side with a single thermal integral, the weighted averages over $c$ of the integrals over $Q$ are given in (\ref{int4HTL:3}), (\ref{int4HTL:4}), and (\ref{int4:7a}):
\begin{eqnarray}
&& \hspace{-1.5cm}
\left\langle c^2 \left(
{\rm Re} \int_Q {p^2+q^2 \over Q^2 r^2 (R_0^2+r^2c^2)}
	\bigg|_{P_0 = -i p + \varepsilon}
+ {1 \over p^2} c^{-1+2\epsilon} \int_Q {q^2 \over Q^2 R^2}
	\bigg|_{P \to (-i p,{\bf p}/c)} \right) \right\ranglec
\nonumber
\\
&=&\!\!  {1 \over (4 \pi)^2} \mu^{2\epsilon} p^{-2 \epsilon}
\left[ {1\over 48\epsilon^2}+\left({35\over36}-{31\over24}\log 2 \right){1\over\epsilon}
\nonumber \right.
\\
&& \hspace{3.3cm}
\left. +\; {313\over108}-{247\pi^2\over576}-{17\over18}\log 2
+{65\over24}\log^2 2\right] \;,
\nonumber
\\
\end{eqnarray}
After using (\ref{int-th:-1}) to evaluate the thermal integral, we obtain
\begin{eqnarray}
&& \hspace{-1cm} 
\int_{\bf p} {n_B(p) \over p}
\left( {\rm Re} \int_Q {p^2+q^2 \over Q^2 r^2}
	\left\langle {c^2 \over R_0^2+r^2c^2} \right\ranglec
	\bigg|_{P_0 = -i p + \varepsilon}
+ {1 \over p^2} \left\langle c^{1+2\epsilon}
	\int_Q {q^2 \over Q^2 R^2} \bigg|_{P \to (-i p,{\bf p}/c)}
	\right\ranglec \right)
\nonumber
\\
&=&\!\!   {T^2 \over (4 \pi)^2} \left({\mu\over4\pi T}\right)^{4\epsilon}
{1\over576}\left[ {1\over\epsilon^2}+\left({146\over3}-60\log 2+
4 {\zeta'(-1) \over \zeta(-1)}\right){1\over\epsilon}+84.72308\right]\, ,
\label{intave:2}
\end{eqnarray}
For the two terms in (\ref{sumint3:6b}) with a double thermal integral, the averages weighted by $c^2$ are given in (\ref{intHTL:3}), (\ref{intHTL:6}), and (\ref{intHTL:7}). Adding them to (\ref{intave:2}), the final result is
\begin{eqnarray}
&& \hspace{-1.5cm}
\sumint_{PQ} {q^2 \over P^2 Q^2 r^2}
	\left\langle {c^2 \over R_0^2 + r^2 c^2} \right\ranglec 
\nonumber
\\
&=&\!\! {T^2 \over (4 \pi)^2} \left({\mu\over4\pi T}\right)^{4\epsilon}
{1\over 576}
\left[ {1 \over \epsilon^2} + \left( {314\over3} -{24 \over \pi^2} - 92 \log2
	+ 4 {\zeta'(-1) \over \zeta(-1)} \right) {1 \over \epsilon} + 270.2 \right] \;.
\nonumber
\\
\label{sumintHTL:c2}
\end{eqnarray}
%
%[[ DELETED 2 DIGITS (OLD: $+ 270.221$) ]]
Inserting this into (\ref{sumint2:7a}), we obtain the final result (\ref{sumint2:7}).

To evaluate (\ref{sumint2:8}), we use the expression (\ref{TP-int}) for ${\cal T}_R$ and the identity $P\!\cdot\!Q = (R^2-P^2-Q^2)/2$ to write it in the form
\begin{eqnarray}\hspace{-1.5cm}
\sumint_{PQ} {P\!\cdot\!Q \over P^2 Q^2 r^4} {\cal T}_R \!\!&=&\!\!
\sumint_{PQ} {P\!\cdot\!Q \over P^2 Q^2 r^4}
- \sumint_P {1\over P^2} \sumint_R {1 \over r^4} {\cal T}_R
\nonumber
\\
&&
-\; {1\over2} \langle c^2 \rangle_c \sumint_{PQ} {1 \over P^2 Q^2 r^2}
- {1\over2} \sumint_{PQ} {1 \over P^2 Q^2}
	\left\langle {c^2(1-c^2) \over R_0^2+r^2c^2} \right\ranglec \;.
\label{sumint2:8:2}
\end{eqnarray}
The sum-integrals in the first three terms on the right side of (\ref{sumint2:8:2}) are given in (\ref{sumint:2}), (\ref{sumint-T:2}), (\ref{sumint2:2}), and (\ref{sumint2:5}). The last sum-integral before the average weighted by $c$ is given in (\ref{sumint2:6a}). The average weighted by $c^2$ is given in (\ref{sumintave:2}). The average weighted by $c^4$ can be computed in the same way. In the integrand of the single thermal integral, the weighted averages over $c$ of the integrals over $Q$ are given in (\ref{int4HTL:2}) and (\ref{int4:8.2}):
\begin{eqnarray}
&& \hspace{-1.5cm}
\left\langle c^4 \left(
2 {\rm Re} \int_Q {1 \over Q^2 (R_0^2+r^2c^2)}
	\bigg|_{P_0 = - i p + \varepsilon}
+ c^{-3+2\epsilon} \int_Q {1 \over Q^2 R^2}
	\bigg|_{P \to (-i p,{\bf p}/c)} \right) \right\ranglec
\nonumber
\\
&& \hspace{-1.5cm}
=\;  {1 \over (4 \pi)^2} \mu^{2\epsilon} p^{-2 \epsilon}
\left[ \left({23\over6}-4\log 2\right){1\over\epsilon}+{104\over9}-\pi^2
-3\log 2+8\log^2 2 \right] \;,
\end{eqnarray}
After using (\ref{int-th:-1}) to evaluate the thermal integral, we obtain
\begin{eqnarray}
&& \hspace{-1cm}
\int_{\bf p} {n_B(p) \over p}
\left( 2 {\rm Re} \int_Q {1 \over Q^2}
	\left\langle {c^4 \over R_0^2 + r^2 c^2} \right\ranglec
		\bigg|_{P_0 = - i p + \varepsilon}
+ \left\langle c^{1+2\epsilon}
	\int_Q {1 \over Q^2 R^2} \bigg|_{P \to (-i p,{\bf p}/c)}
	\right\ranglec \right)
\nonumber
\\
&=&\!\!  {T^2 \over (4 \pi)^2} \left({\mu\over4\pi T}\right)^{4\epsilon}
\left[\left({23\over72}-{1\over3}\log 2\right){1\over\epsilon}+1.28872
\right] \;.
\label{intave:3}
\end{eqnarray}
For the two terms with a double thermal integral, the averages weighted by $c^4$ are given in (\ref{intHTL:2}) and (\ref{intHTL:5}). Adding them to (\ref{intave:3}), we obtain
\begin{eqnarray}
\sumint_{PQ} {1 \over P^2 Q^2}
\left\langle {c^4 \over R_0^2 + r^2 c^2} \right\ranglec
\!\!&=&\!\!
{T^2 \over (4 \pi)^2} \left({\mu\over4\pi T}\right)^{4\epsilon}
\left[ \left( {17\over 72} - {1 \over 6 \pi^2} - {1\over 3} \log2 \right)
	{1 \over \epsilon}
	- 0.1917 \right] \;.
\nonumber
\\
\label{sumintHTL:c4}
\end{eqnarray}
%
%{\bf [pjm]} [[ DELETED 1 DIGIT (OLD: $- 0.19174$) ]]
Inserting this into (\ref{sumint2:8:2}) along with (\ref{sumintave:2}), we get the final result (\ref{sumint2:8}).

To calculate the sum-integral~(\ref{htlf1}), we begin by using the representation~(\ref{TP-int}) of the function ${\cal T}_R$:
\bqa
\sumint_{\{PQ\}}{1\over P^2Q^2r^2}{\cal T}_R \;=\;
\sumint_{\{PQ\}}{1\over P^2Q^2r^2}
-\sumint_{\{PQ\}}{1\over P^2Q^2} \left\langle{c^2\over R_0^2+r^2c^2}\right\rangle_c \;.
\label{sumint2:6bb}
\eqa
The first sum-integral on the right hand side is given by~(\ref{two1}). To evaluate the second sum-integral, we apply the sum-integral formula~(\ref{int-2loop-f}):
\bqa \nonumber
\sumint_{\{PQ\}} {1 \over P^2 Q^2 (R_0^2 + r^2 c^2)} &&
\\\nonumber 
        && \hspace{-4.5cm} =\; -\int_{\bf p} {n_F(p) \over p}
        2 {\rm Re} \int_Q {1 \over Q^2 (R_0^2+r^2c^2)}
        \bigg|_{P_0 = - i p + \varepsilon}
        \,+\, c^{-3+2\epsilon}
           \int_{\bf p} {n_B(p) \over p}
           \int_Q {1 \over Q^2 R^2} \bigg|_{P \to (-i p,{\bf p}/c)}
\\
&&\nonumber \hspace{-4cm}
  + \int_{\bf pq} {n_F(p) n_F(q) \over p q}
        {\rm Re} {r^2 c^2 - p^2 - q^2 \over
        \Delta(p+i\varepsilon,q,r c)}
       \,-\, 2 c^{-3+2\epsilon}
          \int_{\bf pq} {n_F(p) n_B(q) \over p q}
          {\rm Re}{ r_c^2 - p^2 - q^2 \over \Delta(p+i\varepsilon,q,r_c)} 
\;, \\
\label{sumint2:6b}
\eqa
where $r_c = |{\bf p} + {\bf q}/c|$. In the terms on the right side with a single thermal integral, the appropriate averages over $c$ of the integrals over $Q$ are given in~(\ref{int4:8.1}) and~(\ref{int4HTL:1}). 

The subsequent integrals over ${\bf p}$ are special cases of (\ref{int-th:1}) and (\ref{int-th:2}):
\bqa\nonumber
\int_{\bf p} n_B(p) \, p^{-1-2\epsilon} \!\!&=&\!\!
2^{8 \epsilon}
{(1)_{-4\epsilon} ({1\over2})_{2\epsilon} \over (1)_{-2\epsilon} ({3\over2})_{-\epsilon}} \,
{\zeta(-1+4\epsilon) \over \zeta(-1)}
(e^\gamma \mu^2)^\epsilon (4 \pi T)^{-4\epsilon} \, {T^2 \over 12} \;, 
\\ 
\int_{\bf p} n_F(p) \, p^{-1-2\epsilon} \!\!&=&\!\!
\left[1-2^{-1+4\epsilon}\right] \int_{\bf p} n_B(p) \, p^{-1-2\epsilon} \;.
\label{int-th:-2}
\eqa
This yields
\bqa\nonumber
&& \hspace{-1.5cm} 
-2\int_{\bf p} {n_F(p) \over p}
        \, {\rm Re} \int_Q {1 \over Q^2}
        \left\langle {c^2 \over R_0^2 + r^2 c^2} \right\ranglec
                \bigg|_{P_0 = - i p + \varepsilon}
+ \int_{\bf p} {n_B(p)\over p} \left\langle c^{-1+2\epsilon}
        \int_Q {1 \over Q^2 R^2} \bigg|_{P \to (-i p,{\bf p}/c)}
        \right\ranglec\, 
\nonumber
\\
&=&\!\! {T^2 \over (4 \pi)^2} \left({\mu\over4\pi T}\right)^{4\epsilon} {1\over 48}
\left[{1 \over \epsilon^2}
          - \left( 6 - 12 \log 2 - 4 {\zeta'(-1) \over \zeta(-1)} \right) {1 \over \epsilon}
          + 70.122 \right] \;.
\label{intave:1}
\eqa

For the two terms in (\ref{sumint2:6bb}) with a double thermal integral, the averages weighted by $c^2$ are given in~(\ref{f1}) and~(\ref{intHTL:4x}). Adding them to (\ref{intave:1}), the final result is
\bqa\nonumber
\hspace{-1.5cm}
\sumint_{\{PQ\}} {1 \over P^2 Q^2}
        \left\langle {c^2 \over R_0^2 + r^2 c^2} \right\ranglec && \\
&& \hspace{-4.5cm}
=\; {T^2 \over (4 \pi)^2} \left({\mu\over4\pi T}\right)^{4\epsilon} \left({1\over 48}\right)
       \left[ {1 \over \epsilon^2}
                 - \left( 6 - 12 \log 2 - 4 {\zeta'(-1) \over \zeta(-1)} \right){1 \over \epsilon}
                 + 51.9306 \right] \; .
\label{sumintave:2}
\eqa
Inserting this into (\ref{sumint2:6bb}), we obtain the final result (\ref{htlf1}).

The sum-integral (\ref{htlf2}) is evaluated in a similar way to (\ref{htlf1}). Using the representation (\ref{TP-int}) for ${\cal T}_R$, we get
\bqa
\sumint_{\{PQ\}} {q^2 \over P^2 Q^2 r^4} {\cal T}_R \;=\;
        \sumint_{\{PQ\}} {q^2 \over P^2 Q^2 r^4}
        \,-\, \sumint_{\{PQ\}} {q^2 \over P^2 Q^2 r^2} \left\langle
             {c^2 \over R_0^2 + r^2 c^2} \right\ranglec \;.
\label{sumint2:7a}
\eqa
The first sum-integral on the right hand side is given by~(\ref{two2}). To evaluate the second sum-integral, we apply the sum-integral formula~(\ref{int-2loop-f}):
\bqa\nonumber\hspace{-1cm}
\sumint_{\{PQ\}} {q^2 \over P^2 Q^2 r^2 (R_0^2+r^2c^2)} &&
\\\nonumber
&& \hspace{-4cm}
=\; -\int_{\bf p} {n_F(p) \over p} {\rm Re}
       \int_Q {p^2+q^2 \over Q^2 r^2 (R_0^2+r^2c^2)}\bigg|_{P_0 = -i p + \varepsilon}
\\ && \hspace{-3.5cm} \nonumber
+\; c^{-1+2\epsilon}\int_{\bf p} {n_B(p) \over p}p^{-2}\;
                                   \int_Q {q^2 \over Q^2 R^2}\bigg|_{P \to (-i p,{\bf p}/c)} 
\nonumber
\\ \nonumber
&& \hspace{-3.5cm}
+ \int_{\bf pq} {n_F(p) n_F(q) \over p q}
                         {q^2 \over r^2} \, {\rm Re}
                         {r^2 c^2 - p^2 - q^2 \over \Delta(p+i\varepsilon,q,r c)}
\\ && \hspace{-3.5cm} 
  -\; c^{-1 + 2\epsilon} \int_{\bf pq} {n_F(p) n_B(q) \over p q}
                                                            {p^2 + r_c^2 \over q^2} \,
                                                            {\rm Re} { r_c^2-p^2-q^2 \over \Delta(p+i\varepsilon,q,r_c)} \;.
\label{sumint3:6b}
\eqa

In the terms on the right side with a single thermal integral, the weighted averages over $c$ of the integrals over $Q$ are given in~(\ref{int4:7a}),~(\ref{int4HTL:3}), and~(\ref{int4HTL:4}). After using (\ref{int-th:-2}) to evaluate the thermal integral, we obtain
\bqa\nonumber
&& \hspace{-1cm}
 - \int_{\bf p} {n_F(p) \over p}{\rm Re} \int_Q {p^2+q^2 \over Q^2 r^2}
        \left\langle {c^2 \over R_0^2+r^2c^2} \right\ranglec
        \bigg|_{P_0 = -i p + \varepsilon}
\\ \nonumber
&& \hspace{-1cm}
+ \int_{\bf p} {n_B(p) \over p}
{1 \over p^2} \left\langle c^{1+2\epsilon}
        \int_Q {q^2 \over Q^2 R^2} \bigg|_{P \to (-i p,{\bf p}/c)}
        \right\ranglec 
\\ \nonumber
&=&\!\! {T^2 \over (4 \pi)^2} \left({\mu\over4\pi T}\right)^{4\epsilon}\left({1\over576}\right)
\left[ {1\over\epsilon^2} - \left({34\over3}-36\log 2 -
          4{\zeta'(-1) \over \zeta(-1)}\right){1\over\epsilon} + 229.354 \right] \;.
\\
\label{intave:2}
\eqa

For the two terms in (\ref{sumint3:6b}) with a double thermal integral, the averages weighted by $c^2$ are given in (\ref{f3}), (\ref{intHTL:6x}), and (\ref{intHTL:7x}). Adding them to (\ref{intave:2}), the final result is
\bqa\nonumber
\sumint_{\{PQ\}} {q^2 \over P^2 Q^2 r^2}
        \left\langle {c^2 \over R_0^2 + r^2 c^2} \right\ranglec && \\
&& \hspace{-4.5cm} \nonumber
=\; {T^2 \over (4 \pi)^2} \left({\mu\over4\pi T}\right)^{4\epsilon} \left({1\over 576}\right)
        \left[ {1 \over \epsilon^2}
         - \left({118\over3} - 52 \log2 - 4 {\zeta^{\prime}(-1)\over\zeta(-1)} \right) {1\over\epsilon} 
         + 90.9762 \right] \;.
\\ 
\label{sumintHTL:c2}
\eqa

To evaluate (\ref{htl3}), we use the expression (\ref{TP-int}) for ${\cal T}_R$ and the identity $P\!\cdot\!Q = (R^2-P^2-Q^2)/2$ to write it in the form
\bqa
\sumint_{\{PQ\}} {P\!\cdot\!Q \over P^2 Q^2 r^4} {\cal T}_R \!\!&=&\!\!
\sumint_{\{PQ\}} {P\!\cdot\!Q \over P^2 Q^2 r^4}
\,-\, \sumint_{\{P\}} {1\over P^2} \sumint_R {1 \over r^4} {\cal T}_R
\,-\, {1\over2} \langle c^2 \rangle_c \sumint_{\{PQ\}} {1 \over P^2 Q^2 r^2}
\nonumber \\ &&
-\; {1\over2} \sumint_{\{PQ\}} {1 \over P^2 Q^2}
        \left\langle {c^2(1-c^2) \over R_0^2+r^2c^2} \right\ranglec \,.
\label{sumint2:8:2}
\eqa
The sum-integrals in the first three terms on the right side of (\ref{sumint2:8:2}) are given in~(\ref{simple1}), (\ref{sumint-T:2}), (\ref{two1}), and (\ref{twolast}). The last sum-integral before the average weighted by $c$ is given in (\ref{sumint2:6b}). The average weighted by $c^2$ is given in (\ref{sumintave:2}). The average weighted by $c^4$ can be computed in the same way. In the integrand of the single thermal integral, the weighted averages over $c$ of the integrals over $Q$ are given in~(\ref{int4:8.2}) and~(\ref{int4HTL:2}). After using (\ref{int-th:-2}) to evaluate the thermal integral, we obtain
\bqa\nonumber
&& \hspace{-1.2cm}
-2\int_{\bf p} {n_F(p) \over p}
{\rm Re} \int_Q {1 \over Q^2}
        \left\langle {c^4 \over R_0^2 + r^2 c^2} \right\ranglec
                \bigg|_{P_0 = - i p + \varepsilon}
+ \int_{\bf p}{n_B(p)\over p}\left\langle c^{1+2\epsilon}
        \int_Q {1 \over Q^2 R^2} \bigg|_{P \to (-i p,{\bf p}/c)}
        \right\ranglec 
\nonumber
\\
&=&\!\! {T^2 \over (4 \pi)^2} \left({\mu\over4\pi T}\right)^{4\epsilon}
\left[ -\left( {7\over72} - {1\over6}\log 2 \right){1\over\epsilon} + 0.2150 \right] \;.
\label{intave:3}
\eqa
For the two terms with a double thermal integral, the averages weighted by $c^4$ are given in (\ref{f2}) and (\ref{intHTL:5x}). Adding them to (\ref{intave:3}), we obtain
\bqa\nonumber
\sumint_{\{PQ\}} {1 \over P^2 Q^2}
\left\langle {c^4 \over R_0^2 + r^2 c^2} \right\ranglec \;=\;
{T^2 \over (4 \pi)^2} \left({\mu\over4\pi T}\right)^{4\epsilon}
\left[ -\left( {7\over72} - {1\over6}\log 2 \right){1\over\epsilon} + 0.1359 \right] \;.
\\
\eqa

We finally need to evaluate~(\ref{com2l}). Applying~(\ref{int-2loop-f}) gives
\bqa\nonumber
\sumint_{\{PQ\}}{r^2-p^2\over P^2q^2Q^2_0R^2}{\cal T}_Q
\!\!&=&\!\!
\left[\int_{\bf p}{n_B(p)\over p}\,+\,\int_{\bf p}{n_F(p)\over p}
\right]
{\rm Re}\int_{Q}\left\langle {p^2-q^2\over Q^2r^2(R_0^2+r^2c^2)}\right\ranglec\bigg|_{P_0=-i p} 
\\ \nonumber 
&&
+\int_{\bf pq}{n_F(p)n_F(q)\over pq}{\rm Re}\left\langle
{r_c^2-p^2\over q^2}{r^2_c-p^2-q^2\over\Delta(p+i\epsilon,q,r_c)}c^{-1+2\epsilon}\right\ranglec
\\ \nonumber
&&
+\int_{\bf pq}{n_B(p)n_F(q)\over pq}{\rm Re} \left\langle
{r_c^2-p^2\over q^2}{r^2_c-p^2-q^2\over\Delta(p+i\epsilon,q,r_c)}c^{-1+2\epsilon} \right\ranglec
\\
&&
+\int_{\bf pq}{n_F(p)n_B(q)\over pq}{\rm Re} \left\langle
{p^2-q^2\over r^2}{r^2c^2-p^2-q^2\over\Delta(p+i\epsilon,q,rc)} \right\ranglec \;.
\eqa

In the terms on the right side, with a single thermal factor, the weighted average is given in Eq.~(\ref{last4d}), After using Eq.~(\ref{int-th:-2}) to evaluate the thermal integral, we obtain
\bqa
\left[\int_{\bf p}{n_B(p)\over p}\,+\,\int_{\bf p}{n_F(p)\over p}\right]
\int_{Q}\left\langle{p^2-q^2\over Q^2r^2(R_0^2+r^2c^2)}\right\ranglec
\;=\; {T^2\over(4\pi)^2} \left({\pi^2\over24}\right) \;.
\label{yiha}
\eqa
The terms with two thermal factors are given in Eqs.~(\ref{ff4}),~(\ref{lll}) and~(\ref{llll}). Adding them to~(\ref{yiha}), we finally obtain~(\ref{com2l}).

\section{Three-loop sum-integrals}

The three-loop sum-integrals needed are
\bqa
\nonumber
\sumint_{PQR}{1\over P^2Q^2R^2(P+Q+R)^2} && \\
&& \hspace{-5.4cm} =\;
{1 \over 24} {T^4 \over (4\pi)^2} \left({\mu\over4\pi T}\right)^{6\epsilon}
\left[ {1\over\epsilon} + {91\over15} + 8{\zeta^{\prime}(-1)\over\zeta(-1)}
        - 2{\zeta^{\prime}(-3)\over\zeta(-3)} + {\cal O}(\epsilon)
\right] \;,\\ \nonumber && \\ \nonumber
\sumint_{PQR}{(P-Q)^4\over P^2Q^2R^4(Q-R)^2(R-P)^2} && \\
&& \hspace{-6.2cm} =\;
{11 \over 216}{T^4 \over (4\pi)^2} \left({\mu\over4\pi T}\right)^{6\epsilon}
\left[ {1\over\epsilon} + {73\over22} + {12\over11}\gamma
         + {64\over11}{\zeta^{\prime}(-1)\over\zeta(-1)}
          - {10\over11}{\zeta^{\prime}(-3)\over\zeta(-3)} + {\cal O}(\epsilon)
\right] \;, \\
\nonumber && \\ \nonumber
\sumint_{\{PQR\}}{1\over P^2Q^2R^2(P+Q+R)^2} && \\
&& \nonumber \hspace{-5.7cm} =\;
{1\over96} {T^4\over(4\pi)^2} \left({\mu\over4\pi T}\right)^{6\epsilon}
\left[ {1 \over \epsilon} + {173 \over 30} - {42 \over 5} \log2 
          + 8{\zeta^{\prime}(-1) \over \zeta(-1)}
           - 2{\zeta^{\prime}(-3)\over\zeta(-3)} + {\cal O}(\epsilon)
\right] \;,\\ && \\
\nonumber
\sumint_{PQ\{R\}}{1\over P^2Q^2R^2(P+Q+R)^2} && \\
&& \nonumber \hspace{-5.7cm} =\;
-{1\over192} {T^4\over(4\pi)^2} \left({\mu\over4\pi T}\right)^{6\epsilon}
\left[ {1 \over \epsilon} + {179 \over 30} - {34 \over 5} \log2
        + 8{\zeta^{\prime}(-1)\over\zeta(-1)} 
         - 2{\zeta^{\prime}(-3)\over\zeta(-3)} + {\cal O}(\epsilon)
\right] \;,\\ && \\
\nonumber
\sumint_{\{P\}QR}{Q\!\cdot\!R\over P^2Q^2R^2(P+Q)^2(P+R)^2} && \\
&& \nonumber \hspace{-6.5cm} =\;
{1\over384} {T^4\over(4\pi)^2} \left({\mu\over4\pi T}\right)^{6\epsilon}
\left[ {1 \over \epsilon} + {361 \over 60} + 6 \gamma + {76 \over 5} \log2
           - 4{\zeta^{\prime}(-1)\over\zeta(-1)}
          + 4{\zeta^{\prime}(-3)\over\zeta(-3)} + {\cal O}(\epsilon)
\right] \;, \\ && \\
\nonumber
\sumint_{P\{QR\}}{(Q\!\cdot\!R)^2\over P^2Q^2R^2(P+Q)^2(P+R)^2} && \\
&& \nonumber \hspace{-6.5cm} =\;
{5\over3456} {T^4\over(4\pi)^2} \left({\mu\over4\pi T}\right)^{6\epsilon}
\left[ {1 \over \epsilon} + {23 \over 5} + {6 \over 5} \gamma - {192 \over 25} \log2
         + {28\over5}{\zeta^{\prime}(-1)\over\zeta(-1)}
          - {4\over5}{\zeta^{\prime}(-3)\over\zeta(-3)} + {\cal O}(\epsilon)
\right] \;.
\\
\eqa
The three-loop sum-integrals were first calculated by Arnold and Zhai and calculational details can be found in Ref.~\cite{AZ-95}.

%%%%%%%%%%%%%%%%%%%%%%%%%%%%%%%%%%%%%%%%%%%%%%%%%%%%%%%%%%%%%
%
%	Include File:			DON'T COMPILE !!!
%
%%%%%%%%%%%%%%%%%%%%%%%%%%%%%%%%%%%%%%%%%%%%%%%%%%%%%%%%%%%%%

\chapter{Three-Dimensional Integrals}
\label{app:int}

Dimensional regularization can be used to regularize both the ultraviolet divergences and infrared divergences in three-dimensional integrals over momenta. The spacial dimension is generalized to  $d = 3-2\epsilon$ dimensions. Integrals are evaluated at a value of $d$ for which they converge and then analytically continued to $d=3$. We use the integration measure
\begin{equation}
  \int_{\bf p} \;\equiv\;
  \left(\frac{e^\gamma\mu^2}{4\pi}\right)^\epsilon\,
  \int {d^{3-2\epsilon}p \over (2 \pi)^{3-2\epsilon}} \;.
\label{3d-int-def}
\end{equation}
We require one integral that does not involve the thermal distribution function. The momentum scale in these integrals is set by the mass $m=m_D$.

\section{One-loop integrals}

The one-loop integral is given by
\bqa\nonumber
I_n \!\!&\equiv&\!\! \int_{\bf p}{1\over(p^2+m^2)^n}\\
&=&\!\! {1\over8\pi}(e^{\gamma}\mu^2)^{\epsilon}
{\Gamma(n-\mbox{$3\over2$}+\epsilon)
\over\Gamma(\mbox{$1\over2$})
\Gamma(n)}m^{3-2n-2\epsilon}
\;.
\eqa
Specifically, we need
\bqa\nonumber
I_0^{\prime} \!\!&\equiv&\!\! \int_{\bf p}\log(p^2+m^2)
\\
&=&\!\! -{m^3\over6\pi}\left({\mu\over2m}\right)^{2\epsilon}
\left[ 1 + {8\over3}\epsilon + {\cal O}(\epsilon^2) \right] \;, 
\\
I_1 \!\!&=&\!\! -{m\over4\pi}\left({\mu\over2m}\right)^{2\epsilon}
\left[ 1 + 2\epsilon + {\cal O}(\epsilon^2)
\right] \;,
\\
\label{i2}
I_2 \!\!&=&\!\! {1\over8\pi m}\left({\mu\over2m}\right)^{2\epsilon}
\left[1 + {\cal O}(\epsilon) \right]
\;.
\eqa

\section{Two-loop integrals}

We also need a few two-loop integrals on the form
\bqa
J_n \!\!&=&\!\! \int_{\bf pq}{1\over p^2+m^2}{1\over(q^2+m^2)^n}
{1\over({\bf p}+{\bf q})^2} \;,
\\
K_n \!\!&=&\!\! \int_{\bf pq}{1\over p^2+m^2}{1\over(q^2+m^2)}
{1\over[({\bf p}+{\bf q})^2]^n} \;.
\eqa
Specifically, we need $J_1$, $J_2$, and $K_1$ which were calculated in Refs.~\cite{Zhai:1995ac,BN-96}:
\bqa
J_1 \!\!&=&\!\!
{1\over4(4\pi)^2}\left({\mu\over2m}\right)^{4\epsilon}
\left[ {1\over\epsilon} + 2 + {\cal O}(\epsilon) \right] \;,
\\
J_2 \!\!&=&\!\!
{1\over4(4\pi)^2m^2}\left({\mu\over2m}\right)^{4\epsilon}
\left[ 1 + {\cal O}(\epsilon) \right] \;,
\\
K_2 \!\!&=&\!\!
-{1\over8m^2(4\pi)^2}\left({\mu\over2m}\right)^{4\epsilon}
\left[ 1 + {\cal O}(\epsilon) \right] \;.
\eqa

\section{Three-loop integrals}

We also need a number of three-loop integrals. The specific integrals we need are listed below and were calculated in Refs.~\cite{Zhai:1995ac,BN-96}. They are special cases of more general integrals defined in Ref.~\cite{Broadhurst:1991fi}.
\bqa
\label{sssfirst}\nonumber
&& 
\int_{\bf pqr}{1\over(p^2+m^2)(q^2+m^2)}{1\over r^2({\bf p}+{\bf q}+{\bf r})^2}
\\
&& \hspace{-0cm}
=\; -{m\over2(4\pi)^3}\left({\mu\over2m}\right)^{6\epsilon} 
\left[ {1\over\epsilon} + 8 + {\cal O}(\epsilon) \right] \;, 
\\ \nonumber
&& 
\int_{\bf pqr}{(r^2+m^2)\over(p^2+m^2)(q^2+m^2)}
{1\over({\bf p}-{\bf q})^2({\bf q}-{\bf r})^2({\bf r}-{\bf p})^2}
\\
&& \hspace{-0cm}
=\; {m\over4(4\pi)^3}\left({\mu\over2m}\right)^{6\epsilon}
\left[ {1\over\epsilon} + 8 + {\cal O}(\epsilon) \right] \;,
\\ \nonumber
&&
\int_{\bf pqr}{(r^2+m^2)^2\over(p^2+m^2)(q^2+m^2)}
{1\over({\bf p}-{\bf q})^4({\bf q}-{\bf r})^2({\bf r}-{\bf p})^2}
\\
&& \hspace{-0cm}
=\; -{m\over4(4\pi)^3}\left({\mu\over2m}\right)^{6\epsilon}
\left[ {1\over\epsilon} + 6 + {\cal O}(\epsilon) \right] \;,
\\ \nonumber
&&
\int_{\bf pqr}{1\over(p^2+m^2)(q^2+m^2)(r^2+m^2)}
{1\over({\bf q}-{\bf r})^2({\bf r}-{\bf p})^2}
\\
&& \hspace{-0cm}
=\; {1\over m(4\pi)^3}\left({\mu\over2m}\right)^{6\epsilon}
\left[ {\pi^2\over12} + {\cal O}(\epsilon) \right] \;,
\\ \nonumber
&&
\int_{\bf pqr}{1\over(p^2+m^2)(q^2+m^2)}
{1\over({\bf p}-{\bf q})^2({\bf q}-{\bf r})^2({\bf r}-{\bf p})^2}
\\
&& \hspace{-0cm}
=\; -{1\over8m(4\pi)^3}\left({\mu\over2m}\right)^{6\epsilon}
\left[ {1\over\epsilon} - 2 + {\cal O}(\epsilon) \right] \;,
\\ \nonumber
&&
\int_{\bf pqr}{1\over(p^2+m^2)(q^2+m^2)(r^2+m^2)^2}
{1\over({\bf q}-{\bf r})^2({\bf r}-{\bf p})^2}
\\
&& \hspace{-0cm}
=\; -{1\over4m^3(4\pi)^3}\left({\mu\over2m}\right)^{6\epsilon}
\left[ 1 - {\pi^2\over6} + {\cal O}(\epsilon) \right] \;,
\\ \nonumber
&&
\int_{\bf pqr}{1\over(p^2+m^2)(q^2+m^2)[({\bf q}-{\bf r})^2+m^2]
[({\bf r}-{\bf p})^2+m^2]}
\\
&& \hspace{-0cm}
=\;-{m\over(4\pi)^3}\left({\mu\over2m}\right)^{6\epsilon}
\left[ {1\over\epsilon} + 8 - 4\log2 + {\cal O}(\epsilon) \right] \;,
\\ \nonumber
&&
\int_{\bf pqr}{1\over(p^2+m^2)(q^2+m^2)[({\bf q}-{\bf r})^2+m^2]
[({\bf r}-{\bf p})^2+m^2]}{({\bf p}-{\bf q})^2\over r^2}
\\
&& \hspace{-0cm}
=\; {2m\over(4\pi)^3}\left({\mu\over2m}\right)^{6\epsilon}
\left[ 1 - 2\log2 +{\cal O}(\epsilon) \right] \;,
\\ \nonumber
&&
\int_{\bf pqr}{1\over(p^2+m^2)(q^2+m^2)[({\bf q}-{\bf r})^2+m^2]
[({\bf r}-{\bf p})^2+m^2]}{({\bf p}-{\bf q})^4\over r^4}
\\
&& \hspace{-0cm}
=\; -{3m\over(4\pi)^3}\left({\mu\over2m}\right)^{6\epsilon}
\left[ 1 - {4\over3}\log2 +{\cal O}(\epsilon) \right] \;,
\\ \nonumber
&&
\int_{\bf pqr}{1\over(p^2+m^2)(q^2+m^2)[({\bf q}-{\bf r})^2+m^2]
[({\bf r}-{\bf p})^2+m^2]}{1\over r^2}
\\ 
&& \hspace{-0cm}
=\; {1\over m(4\pi)^3}\left({\mu\over2m}\right)^{6\epsilon}
\left[ \log2 + {\cal O}(\epsilon) \right] \;,
\\ \nonumber
&&
\int_{\bf pqr}{1\over(p^2+m^2)(q^2+m^2)[({\bf q}-{\bf r})^2+m^2]
[({\bf r}-{\bf p})^2+m^2]}{({\bf p}-{\bf q})^2\over r^4}
\\
&& \hspace{-0cm}
=\; {1\over3m(4\pi)^3}\left({\mu\over2m}\right)^{6\epsilon}
\left[ 1 - \log2 + {\cal O}(\epsilon) \right] \;,
\\ \nonumber
&&
\int_{\bf pqr}{1\over(p^2+m^2)(q^2+m^2)[({\bf q}-{\bf r})^2+m^2]
[({\bf r}-{\bf p})^2+m^2]}{1\over r^2({\bf p}-{\bf q})^2}
\\
&& \hspace{-0cm}
=\; {1\over4m^3(4\pi)^3}\left({\mu\over2m}\right)^{6\epsilon}
\left[ 1 - \log2 + {\cal O}(\epsilon) \right] \;, 
\\ \nonumber
&&
\int_{\bf pqr}{1\over(p^2+m^2)(q^2+m^2)[({\bf q}-{\bf r})^2+m^2]
[({\bf r}-{\bf p})^2+m^2]}{1\over r^4}
\\ 
&& \hspace{-0cm}
=\; -{1\over24m^3(4\pi)^3}\left({\mu\over2m}\right)^{6\epsilon}
\left[ 1 + 2\log2 +{\cal O}(\epsilon) \right] \;.
\label{ssslast}
\eqa
Finally, we need the combination 
\bqa\nonumber
\int_{\bf pqr}{1\over(p^2+m^2)(q^2+m^2)(r^2+m^2)}
{({\bf p}-{\bf q})^2\over({\bf q}-{\bf r})^2({\bf r}-{\bf p})^2} &&
\\ \nonumber
&& \hspace{-9cm}
+ \int_{\bf pqr}{(q^2+m^2)\over(p^2+m^2)[({\bf r}-{\bf p})^2+m^2]
[({\bf q}-{\bf r})^2+m^2]} {1\over r^2({\bf p}-{\bf q})^2} 
\\
&& \hspace{-7cm}
=\;
{2m\over(4\pi)^3}\left({\mu\over2m}\right)^{6\epsilon}\left[1+{\cal O}(\epsilon)\right] \;.
\eqa
%

%%%%%%%%%%%%%%%%%%%%%%%%%%%%%%%%%%%%%%%%%%%%%%%%%%%%%%%%%%%%%
%
%	Include File:			DON'T COMPILE !!!
%
%%%%%%%%%%%%%%%%%%%%%%%%%%%%%%%%%%%%%%%%%%%%%%%%%%%%%%%%%%%%%

\chapter{Three-Dimensional Thermal Integrals}

The three-dimensional thermal integrals involve the Bose-Einstein distribution $n_B(p) = 1/(e^{\beta p} - 1)$ and the Fermi-Dirac distribution $n_F(p) = 1/(e^{\beta p} + 1)$.

\section{One-loop integrals}

The one-loop integrals can all be obtained from the general formulae
\bqa
%%%%%glue%%%%%
\int_{\bf p} {n_B(p) \over p} p^{2 \alpha} \!\!&=&\!\!
{\zeta(2+2\alpha-2\epsilon) \over  4 \pi^2}
{\Gamma(2+2\alpha-2\epsilon) \Gamma({1\over2})
	\over  \Gamma({3\over2}-\epsilon)}
\left( e^\gamma \mu^2 \right)^\epsilon T^{2+2\alpha-2\epsilon} \;, 
\label{int-th:10}
\\
%%%%%quark%%%%%
\int_{\bf p}{n_F(p)\over p}p^{2\alpha} \!\!&=&\!\!
\left(1-2^{-1-2\alpha+2\epsilon}\right) \int_{\bf p} {n_B(p) \over p} p^{2 \alpha} \;.
\label{int-th:1}
\eqa

\section{Two-loop integrals}

The simple two-loop thermal integrals are
\bqa
%%%%%glue%%%%%
\int_{\bf pq} {n_B(p) n_B(q) \over p q} \, {1 \over r^2} \!\!&=&\!\!
- {1 \over 4} {T^2 \over (4\pi)^2} \left({\mu\over4\pi T}\right)^{4\epsilon}
\left[ {1\over\epsilon} + {14\over3} + 4 \log 2
	+ 4 {\zeta'(-1) \over \zeta(-1)} + {\cal O}(\epsilon) \right] \;,
\nonumber \\ &&
\label{int-th:20}
\\ \nonumber
\int_{\bf pq} {n_B(p) n_B(q) \over p q} \, {p^2 \over r^4} \!\!&=&\!\!
{T^2 \over (4\pi)^2}  \left({\mu\over4\pi T}\right)^{4\epsilon}
\left[ {1\over9} + {1\over3} \gamma
	- {1\over3} {\zeta'(-1) \over \zeta(-1)}
	- 4.855 \, \epsilon + {\cal O}(\epsilon) \right] \;,
\label{int-th:30} \\ && \\
%%%%%quark%%%%%
\label{int-th:2}
\int_{\bf p q}{n_F(p)n_F(q)\over pq} \, {1\over r^2}
\!\!&=&\!\! {T^2\over(4\pi)^2}
\left({\mu\over4\pi T}\right)^{4\epsilon}
{1\over3}\left[ 1 - \log2 + {\cal O}(\epsilon) \right] \;,
\\ \nonumber
\int_{\bf p q}{n_F(p)n_F(q)\over pq} \, {q^2\over r^4}
\!\!&=&\!\! -{1\over36} {T^2\over(4\pi)^2}\left({\mu\over4\pi T}\right)^{4\epsilon}
\\ && \times
\left[ 5 + 6\gamma + 6\log2 - 6{\zeta^{\prime}(-1)\over\zeta(-1)}
        + 3.076 \, \epsilon + {\cal O}(\epsilon) \right] \;,
\label{int-th:3} 
\\ \nonumber
\int_{\bf p q}{n_B(p)n_F(q)\over pq} \, {p^2\over r^4}
\!\!&=&\!\! -{1\over36} {T^2\over(4\pi)^2}\left({\mu\over4\pi T}\right)^{4\epsilon}
\\ && \times
\left[ 7 - 6\gamma - 18\log2 + 6{\zeta^{\prime}(-1)\over\zeta(-1)}
        + 29.509 \, \epsilon + {\cal O}(\epsilon) \right] \;, 
\label{int-th:4}
\\ \nonumber
\int_{\bf p q}{n_B(p)n_F(q)\over pq} \, {q^2\over r^4}
\!\!&=&\!\! {1\over18} {T^2\over(4\pi)^2}\left({\mu\over4\pi T}\right)^{4\epsilon}
\\ && \times
\left[ 1 - 6\gamma - 12\log2 + 6{\zeta^{\prime}(-1)\over\zeta(-1)}
        + 31.098 \, \epsilon + {\cal O}(\epsilon) \right] \;.
\label{int-th:5}
\eqa
We also need some more complicated two-loop thermal integrals that involve the triangle function defined in Eq.~(\ref{triangle}):

\bqa
%%%%%glue%%%%%
\int_{\bf pq} {n_B(p) n_B(q) \over p q} \,
	{r^4 \over q^2 \Delta(p,q,r)} \!\!&=&\!\!
{7 \over 48} {T^2 \over (4\pi)^2} \left({\mu\over4\pi T}\right)^{4\epsilon}
\bigg[ {1\over\epsilon^2} + \bigg( {22\over 7} + 2 \gamma + 2 {\zeta'(-1)\over\zeta(-1)}
\nonumber
\\
&& -\;{\zeta(3)\over 35} \bigg) {1\over\epsilon} + 40.3896 + {\cal O}(\epsilon) \bigg] \;,
\label{int-thT:10}
\\
\int_{\bf pq} {n_B(p) n_B(q) \over p q} \,
	{r^2 \over \Delta(p,q,r)} \!\!&=&\!\!
{1 \over 24} {T^2 \over (4\pi)^2} \left({\mu\over4\pi T}\right)^{4\epsilon}
\bigg[ {1\over\epsilon^2}
	+ 2\left( 1 + \gamma + {\zeta'(-1)\over\zeta(-1)} \right){1\over\epsilon}
\nonumber
\\ &&	
         +\;4 + 4 \gamma + {\pi^2 \over 2} - 4 \gamma_1
	+ 4 (1 + \gamma) {\zeta'(-1)\over\zeta(-1)}
\nonumber
\\ &&	
          +\;2 {\zeta''(-1)\over\zeta(-1)} + {\cal O}(\epsilon) \bigg] \;,
\label{int-thT:20}
\\
\int_{\bf pq} {n_B(p) n_B(q) \over p q} \,
	{p^4 \over q^2 \Delta(p,q,r)} \!\!&=&\!\!
- {\zeta(3) \over 240} {T^2 \over (4\pi)^2} \left({\mu\over4\pi T}\right)^{4\epsilon}
\bigg[ {1\over\epsilon} + 2 + 2 {\zeta'(-3)\over\zeta(-3)}
\nonumber
\\ &&
	+\;2 {\zeta'(3)\over\zeta(3)} + {\cal O}(\epsilon) \bigg] \;,
\label{int-thT:30}
\\
\int_{\bf pq} {n_B(p) n_B(q) \over p q} \,
	{p^2 (p^2+q^2) \over r^2 \Delta(p,q,r)} \!\!&=&\!\!
{1 \over 48} {T^2 \over (4\pi)^2} \left({\mu\over4\pi T}\right)^{4\epsilon}
\bigg[ {1\over\epsilon^2}
	+ \left( {14 \over 3} + 10 \gamma - 6 {\zeta'(-1)\over\zeta(-1)} \right) {1\over\epsilon}
\nonumber
\\ &&
	-\; 86.46 + {\cal O}(\epsilon) \bigg] \;,
\label{int-thT:40}
\\
%%%%%quark%%%%%
\int_{\bf p q}{n_F(p) n_F(q)\over p q} \,
          {r^4\over q^2\Delta(p,q,r)} \!\!&=&\!\!
-{7\over96} {T^2\over(4\pi)^2} \left({\mu\over4\pi T}\right)^{4\epsilon}
\bigg[ {1\over\epsilon^2} +
\bigg( {22\over7} + 2\gamma + 2\log2
\nonumber
\\ &&
+\;2{\zeta^{\prime}(-1)\over\zeta(-1)} - {7\over20}\zeta(3)
\bigg){1\over\epsilon} 
\nonumber
\\ &&
+\; 47.2406 + {\cal O}(\epsilon)
\bigg]\;, 
\label{int-thT:1}
\\ 
\int_{\bf p q}{n_F(p) n_F(q) \over p q} \,
          {r^2\over\Delta(p,q,r)} \!\!&=&\!\!
-{1\over48} {T^2\over(4\pi)^2}\left({\mu\over4\pi T}\right)^{4\epsilon}
\bigg[ {1\over\epsilon^2}
+2
\bigg(1+\gamma+\log2
\nonumber 
\\ &&
+\;{\zeta^{\prime}(-1) \over \zeta(-1)} \bigg){1 \over \epsilon} + 4
+ 4\gamma + {\pi^2 \over 2} + 4\gamma\log2 - 6\log^2 2
\nonumber 
\\ &&
+\;4\log2 - 4\gamma_1 + 4( 1 + \gamma + \log2 ){\zeta^{\prime}(-1) \over \zeta(-1)}
\nonumber 
\\ &&
+\;2{\zeta^{\prime\prime}(-1) \over \zeta(-1)} + {\cal O}(\epsilon)
\bigg] \;,
\label{int-thT:2}
\\
\int_{\bf p q}{n_F(p) n_F(q)\over p q} \,
          {p^4\over q^2\Delta(p,q,r)} \!\!&=&\!\!
{49\,\zeta(3)\over1920} {T^2\over(4\pi)^2}\left({\mu\over4\pi T}\right)^{4\epsilon}
\bigg[ {1\over\epsilon} + 2 + 2\log2
\nonumber 
\\ &&
+\;2{\zeta^{\prime}(-3)\over\zeta(-3)}
+ 2{\zeta^{\prime}(3)\over\zeta(3)} + {\cal O}(\epsilon)
\bigg] \;,
\label{int-thT:3}
\\
\int_{\bf p q}{n_F(p) n_F(q)\over p q} \,
          {p^2(p^2+q^2)\over r^2\Delta(p,q,r)} \!\!&=&\!\!
-{1\over96} {T^2\over(4\pi)^2}\left({\mu\over4\pi T}\right)^{4\epsilon}
\bigg[{1\over\epsilon^2} + \bigg( {26 \over 3} + 10\gamma + 10\log2
\nonumber
\\ && 
-\;6{\zeta^{\prime}(-1) \over \zeta(-1)} \bigg){1\over\epsilon} + 41.1580 + {\cal O}(\epsilon)
\bigg] \;, 
\label{int-thT:4}
\\
\int_{\bf pq} {n_F(p) n_F(q)\over p q} \, 
          {p^2\over\Delta(p,q,r)} \!\!&=&\!\!
-{1\over96} {T^2\over(4\pi)^2}\left({\mu\over4\pi T}\right)^{4\epsilon}
\bigg[ {1\over\epsilon^2} + 2\bigg( 1 + \gamma + \log2
\nonumber
\\ &&
+\;{\zeta^{\prime}(-1) \over \zeta(-1)} \bigg){1\over\epsilon}
+ 37.0573 + {\cal O}(\epsilon)
\bigg]\;,
\\ 
\int_{\bf pq} {n_F(p) n_B(q)\over p q} \, 
         {p^2\over\Delta(p,q,r)} \!\!&=&\!\!
{1\over96} {T^2\over(4\pi)^2}\left({\mu\over4\pi T}\right)^{4\epsilon}
\bigg[ {1\over\epsilon^2} + 2\bigg( 1 + \gamma - \log2
\nonumber
\\ &&
+\;{\zeta^{\prime}(-1)\over\zeta(-1)} \bigg){1\over\epsilon}
+ 19.2257 + {\cal O}(\epsilon)
\bigg] \;,
\\ 
\int_{\bf pq} {n_F(p) n_B(q)\over p q} \,
          {p^4\over q^2\Delta(p,q,r)} \!\!&=&\!\!
-{7\,\zeta(3)\over1920} {T^2\over(4\pi)^2}\left({\mu\over4\pi T}\right)^{4\epsilon}
\bigg[ {1\over\epsilon} + 2 - {2\over7}\log2
\nonumber
\\ &&
+\;2{\zeta^{\prime}(-3)\over\zeta(-3)}
+ 2{\zeta^{\prime}(3)\over\zeta(3)} + {\cal O}(\epsilon)
\bigg] \;,
\\
\int_{\bf pq} {n_F(p) n_B(q)\over p q} \,
          {r^4\over q^2\Delta(p,q,r)} \!\!&=&\!\!
{1\over24} {T^2\over(4\pi)^2}\left({\mu\over4\pi T}\right)^{4\epsilon}
\bigg[ {1\over\epsilon^2} + \bigg( 4 + 2\gamma - 5\log2 - {7\zeta(3)\over80}
\nonumber
\\ &&
+\;2{\zeta^{\prime}(-1)\over\zeta(-1)} \bigg){1\over\epsilon}
+18.1551 + {\cal O}(\epsilon)
\bigg] \;,
\\
\int_{\bf pq} {n_F(p) n_B(q)\over p q} \,
         {r^2\over\Delta(p,q,r)}\!\!&=&\!\!
-{1\over96} {T^2\over(4\pi)^2}\left({\mu\over4\pi T}\right)^{4\epsilon}
\bigg[ {1\over\epsilon^2} + 2\bigg( 1 + \gamma + 5\log2
\nonumber
\\ &&
+\;{\zeta^{\prime}(-1)\over\zeta(-1)} \bigg){1\over\epsilon}
+ 84.2513 + {\cal O}(\epsilon)
\bigg] \;.
\eqa

The most difficult thermal integrals to evaluate involve both the triangle function and the HTL average defined in (\ref{c-average}). There are two sets of these integrals. The first set is
\bqa
%%%%%glue%%%%%
&& \hspace{-2cm}
\int_{\bf pq} {n_B(p) n_B(q) \over p q} \,
{\rm Re} \left\langle c^2 {r^2 c^2 - p^2 - q^2 \over
	\Delta(p+i\varepsilon,q,r c)}  \right\ranglec \;=\;
{T^2 \over (4\pi)^2}  \left[ 0.138727 + {\cal O}(\epsilon) \right] \;,
\label{intHTL:1}
\\
&& \hspace{-2cm}
\int_{\bf pq} {n_B(p) n_B(q) \over p q} \,
{\rm Re} \left\langle c^4 {r^2 c^2 - p^2 - q^2 \over
	\Delta(p+i\varepsilon,q,r c)}  \right\ranglec \;=\;
- {1 \over 6\pi^2} {T^2 \over (4\pi)^2} \left({\mu\over4\pi T}\right)^{4\epsilon}
\nonumber \\ && \hspace{6cm}
\times\left[ {1\over\epsilon} +  6.8343 + {\cal O}(\epsilon) \right] \;,
\label{intHTL:2}
\\
&& \hspace{-2cm}
\int_{\bf pq} {n_B(p) n_B(q) \over p q} \, {q^2 \over r^2} \,
{\rm Re} \left\langle c^2
	{r^2 c^2 - p^2 - q^2 \over \Delta(p+i\varepsilon,q,r c)}
	\right\ranglec \;=\;
{\pi^2 -1 \over 24 \pi^2} {T^2 \over (4\pi)^2} \left({\mu\over4\pi T}\right)^{4\epsilon}
\nonumber \\ && \hspace{6cm}
\times\left[ {1\over\epsilon} + 15.3782 + {\cal O}(\epsilon) \right] \;,
\label{intHTL:3}
\\
%%%%%quark%%%%%
&& \hspace{-2cm}
\int_{\bf pq}{n_F(p)n_F(q)\over pq} \, 
{\rm Re}\left\langle c^2{r^2c^2-p^2-q^2\over\Delta(p+i\varepsilon,q,rc)} \right\rangle_c 
\;=\;
{T^2\over(4\pi)^2}\left[0.01458 + {\cal O}(\epsilon) \right] \;,
\label{f1}
\\
&& \hspace{-2cm}
\int_{\bf pq}{n_F(p)n_F(q)\over pq} \,  
{\rm Re}\left\langle c^4{r^2c^2-p^2-q^2\over\Delta(p+i\varepsilon,q,rc)} \right\rangle_c
\;=\;
{T^2\over(4\pi)^2}\left[0.017715 + {\cal O}(\epsilon) \right] \;, 
\label{f2}
\\
&& \hspace{-2cm} 
\int_{\bf pq}{n_F(p)n_F(q)\over pq} \, {q^2 \over r^2} \,
{\rm Re}\left\langle c^2{r^2c^2-p^2-q^2\over\Delta(p+i\varepsilon,q,rc)} \right\rangle_c
\;=\;
{T^2\over(4\pi)^2}\left[ -0.011578 + {\cal O}(\epsilon) \right] \;,
\label{f3}
\\
&& \hspace{-2cm}
\int_{\bf pq }{n_B(p) n_F(q)\over pq} \, {p^2-q^2\over r^2} \,
{\rm Re}\left\langle {r^2c^2-p^2-q^2\over\Delta(p+i\epsilon,q,rc)} \right\rangle
\;=\;
{T^2\over(4\pi)^2}\left[0.17811 + {\cal O}(\epsilon) \right] \;.
\label{ff4}
\eqa

The second set of these integrals involve the variable $r_c = |{\bf p} + {\bf q}/c|$:
\bqa
%%%%%glue%%%%%
&& \int_{\bf pq} {n_B(p) n_B(q) \over p q} \,
{\rm Re} \left\langle c^{-1+2\epsilon}
	{r_c^2 - p^2 - q^2 \over \Delta(p+i\varepsilon,q,r_c)} \,
	\right\ranglec 
\nonumber \\
&& \hspace{3cm}
=\; - {1 \over 8} {T^2 \over (4\pi)^2} \left({\mu\over4\pi T}\right)^{4\epsilon}
\left[ {1\over\epsilon} + 13.442 + {\cal O}(\epsilon) \right] \;,
\label{intHTL:4}
\\
&& \int_{\bf pq} {n_B(p) n_B(q) \over p q} \,
{\rm Re} \left\langle c^{1+2\epsilon}
	{r_c^2 - p^2 - q^2 \over \Delta(p+i\varepsilon,q,r_c)} \,
	\right\ranglec 
\nonumber \\
&& \hspace{3cm}	
=\; - {1 \over 24} {T^2 \over (4\pi)^2} \left({\mu\over4\pi T}\right)^{4\epsilon}
\left[ {1\over\epsilon} +  16.381 + {\cal O}(\epsilon) \right] \;,
\label{intHTL:5}
\\
&& \int_{\bf pq} {n_B(p) n_B(q) \over p q} \, {p^2 \over q^2} \,
{\rm Re} \left\langle c^{1+2\epsilon}
	{r_c^2 - p^2 - q^2 \over \Delta(p+i\varepsilon,q,r_c)} \,
	\right\ranglec 
\nonumber \\
&& \hspace{3cm}
=\; {1 \over 48} {T^2 \over (4\pi)^2} \left({\mu\over4\pi T}\right)^{4\epsilon}
\left[ {1\over\epsilon} + 6.1227 + {\cal O}(\epsilon) \right] \;,
\label{intHTL:6}
\\
&& \int_{\bf pq} {n_B(p) n_B(q) \over p q} \,
{\rm Re} \left\langle c^{1+2\epsilon} {r_c^2 \over q^2}
	{r_c^2 - p^2 - q^2\over \Delta(p+i\varepsilon,q,r_c)} \,
	\right\ranglec 
\nonumber \\
&& \hspace{2cm}
=\; {5-8\log 2 \over 144} {T^2 \over (4\pi)^2} \left({\mu\over4\pi T}\right)^{4\epsilon}
\left[ {1\over\epsilon} +  100.73 + {\cal O}(\epsilon) \right] \;,
\label{intHTL:7}
%%%%%quark%%%%%
\\ \nonumber
&& \int_{\bf pq} {n_F(p) n_B(q) \over p q} \,
{\rm Re} \left\langle c^{-1+2\epsilon}
        {r_c^2 - p^2 - q^2 \over \Delta(p+i\varepsilon,q,r_c)} \,
        \right\ranglec 
\\ && \hspace{3cm} 
=\; {T^2 \over (4\pi)^2} \left[0.19678 + {\cal O}(\epsilon) \right] \;, 
\label{intHTL:4x}
\\ \nonumber
&& \int_{\bf pq} {n_F(p) n_B(q) \over p q} \,
{\rm Re} \left\langle c^{1+2\epsilon}
        {r_c^2 - p^2 - q^2 \over \Delta(p+i\varepsilon,q,r_c)} \,
        \right\ranglec 
\\ && \hspace{3cm} 
=\; {T^2 \over (4\pi)^2} \left[4.8368\times10^{-2} + {\cal O}(\epsilon) \right] \;,
\label{intHTL:5x}
\\ \nonumber
&& \int_{\bf pq} {n_F(p) n_B(q) \over p q} \, {p^2 \over q^2} \,
{\rm Re} \left\langle c^{1+2\epsilon}
        {r_c^2 - p^2 - q^2 \over \Delta(p+i\varepsilon,q,r_c)} \,
        \right\ranglec 
\\ && \hspace{3cm} 
=\; {1 \over 96} {T^2 \over (4\pi)^2} \left({\mu\over4\pi T}\right)^{4\epsilon}
\left[{1\over\epsilon} +7.77235 + {\cal O}(\epsilon)\right] \;,
\label{intHTL:6x}
\\ \nonumber
&& \int_{\bf pq} {n_F(p) n_B(q) \over p q} \,
{\rm Re} \left\langle c^{1+2\epsilon} {r_c^2 \over q^2}
        {r_c^2 - p^2 - q^2\over \Delta(p+i\varepsilon,q,r_c)} \,
        \right\ranglec 
\\ && \hspace{1cm} 
=\; {11-8\log2\over288} {T^2 \over(4\pi)^2} \left({\mu\over4\pi T}\right)^{4\epsilon}
\left[ {1\over\epsilon}+7.79813 + {\cal O}(\epsilon) \right] \;,
\label{intHTL:7x}
\\ \nonumber
&&\int_{\bf pq}{n_F(p) n_F(p)\over p q} \,
{\rm Re}\left\langle
{r_c^2-p^2\over q^2}{r^2_c-p^2-q^2\over\Delta(p+i\epsilon,q,r_c)}
c^{-1+2\epsilon}
\right\ranglec
\\ \nonumber
&& \hspace{1cm}
=\; -{1\over24}{T^2\over(4\pi)^2}\left({\mu\over4\pi T}\right)^{4\epsilon}
\bigg[ {1\over\epsilon^2} 
+ \left( 2 + 2\gamma + 2\log2 + 2{\zeta^{\prime}(-1)\over\zeta(-1)} \right){1\over\epsilon}
\\ && \hspace{5.4cm}
+\;40.316 + {\cal O}(\epsilon) \bigg] \;,
\label{lll}
\\ \nonumber
&& \int_{\bf pq}{n_B(p) n_F(p)\over p q} \,
{\rm Re}\left\langle
{r_c^2-p^2\over q^2}{r^2_c-p^2-q^2\over\Delta(p+i\epsilon,q,r_c)}
c^{-1+2\epsilon}
\right\ranglec
\\ \nonumber&&
\hspace{1cm}
=\; - {1\over12} {T^2\over(4\pi)^2}\left({\mu\over4\pi T}\right)^{4\epsilon}
\bigg[ {1\over\epsilon^2}
+ \left(2 + 2\gamma + 4\log2 + 2{\zeta^{\prime}(-1)\over\zeta(-1)} \right){1\over\epsilon}
\\ && \hspace{5.4cm}
+\;52.953 + {\cal O}(\epsilon) \bigg] \;.
\label{llll}
\eqa

The simplest way to evaluate integrals like (\ref{int-th:20})--(\ref{int-th:5}) whose integrands factor into separate functions of $p$, $q$, and $r$  is to Fourier transform to coordinate space where they reduce to an integral over a single coordinate ${\bf R}$:
\bqa
\int_{\bf pq} f(p) \, g(q) \, h(r) \;=\;
\int_{\bf R} \tilde f(R) \, \tilde g(R) \, \tilde h(R) \;.
\label{int-fgh}
\eqa
The Fourier transform is
\bqa
\tilde{f}(R) \;=\; \int_{\bf p}e^{i{\bf p}\cdot {\bf R}}f(p) \;,
\eqa
and the dimensionally regularized coordinate integral is
\bqa
\int_{\bf R} \;=\;
\left( {e^\gamma \mu^2 \over 4 \pi} \right)^{-\epsilon}
\int d^{3-2 \epsilon}R \;.
\eqa
The Fourier transforms we need are
\bqa
\int_{\bf p} p^{2 \alpha} \,  e^{i {\bf p} \cdot {\bf R}} \!\!&=&\!\!
{1 \over 8\pi} 
{\Gamma({3\over2} + \alpha - \epsilon)
        \over \Gamma({1\over2}) \Gamma(-\alpha)}
\left( e^\gamma \mu^2 \right)^\epsilon 
\left( {2 \over R} \right)^{3 + 2 \alpha - 2\epsilon} \;, 
\\
\int_{\bf p} {n(p) \over p} \,  p^{2 \alpha} \,
        e^{i {\bf p} \cdot {\bf R}} \!\!&=&\!\!
{1 \over 4\pi} {1 \over \Gamma({1\over2})}
\left( e^\gamma \mu^2 \right)^\epsilon 
\left( {2 \over R} \right)^{{1\over2} - \epsilon}
\int_0^\infty dp \, p^{2 \alpha + {1\over2} - \epsilon} n(p)
        J_{{1\over2}-\epsilon}(pR) \;.
\nonumber \\
\label{fourier-n}
\eqa
If $\alpha$ is an even integer, the Fourier transform (\ref{fourier-n}) is particularly simple in the limit $d \to 3$:
\bqa
\int_{\bf p} {n_B(p) \over p} \, e^{i {\bf p} \cdot {\bf R}} \!\!&=&\!\!
{T \over 4 \pi R}
\left( \coth x - {1 \over x} \right) \;,
\\
\int_{\bf p} {n_B(p) \over p} \, p^2 \,  e^{i {\bf p} \cdot {\bf R}} \!\!&=&\!\!
- {\pi T^3 \over 2 R}
\left( \coth^3x - \coth x - {1 \over x^3} \right) \;,
\\
\int_{\bf p} {n_F(p) \over p} \, e^{i {\bf p} \cdot {\bf R}} \!\!&=&\!\!
{T \over 4\pi R}
\left( {1\over x} - {\rm csch} x \right) \;,
\\
\int_{\bf p} {n_F(p) \over p} \, p^2 \, e^{i {\bf p} \cdot {\bf R}} \!\!&=&\!\!
{\pi T^3\over2R}
\left( {\rm csch}^3x + {1\over2}{\rm csch}x - {1\over x^3} \right) \;,
\eqa
where $x=\pi RT$. We can use these simple expressions only if the integral over the coordinate ${\bf R}$ in (\ref{int-fgh}) converges for $d=3$. Otherwise, we must first make subtractions inside of the integrand to make the integral convergent. 

The integrals~(\ref{int-th:30})--(\ref{int-th:5}) can be evaluated directly by applying the Fourier transform formula (\ref{int-fgh}) in the limit $\epsilon \to 0$. The integral (\ref{int-th:20}) however requires subtractions. It can be written
\begin{eqnarray}
\hspace{-5mm}
\int_{\bf pq} {n_B(p) n_B(q) \over p q} \, {1 \over r^2} \;=\;
\int_{\bf pq} {n_B(p) \over p} \,
	\left( {n_B(q) \over q} - {T \over q^2} \right) \, {1 \over r^2}
\,+\, T \int_{\bf p} {n_B(p) \over p} \int_{\bf q} {1 \over q^2 r^2} \;.
\end{eqnarray}
In the second term on the right side, the integral over ${\bf q}$ is proportional to $p^{-1-2 \epsilon}$, so the integral over ${\bf p}$ can be evaluated using (\ref{int-th:10}). This first term on the right side is convergent for $d=3$ so it can be evaluated easily using the Fourier transform formula (\ref{int-fgh}). The integral over ${\bf R}$ reduces to a sum of integrals of the form $\int_0^\infty dx \, x^m \coth ^n x$. Although the sum of the integrals converges, each of the individual integrals diverges either as $x \to 0$ or as $x \to \infty$. A convenient way to evaluate these integrals is to use the strategy in Appendix C of Ref.~\cite{AZ-95}. The integrals are regularized by using the substitution
\begin{eqnarray}
\int_0^\infty dx \, x^m \coth^n x \longrightarrow
{\Gamma(1+\delta) \over 2^\delta}
\int_0^\infty dx \, x^{m+\delta} \coth^n x \;.
\end{eqnarray}
The divergences appear as poles in $\delta$ that cancel upon adding a convergent combination of these integrals.

The integrals (\ref{int-thT:10})--(\ref{int-thT:30}) and (\ref{int-thT:1})--(\ref{int-thT:3}) can be evaluated by first averaging over angles. The triangle function can be expressed as
\begin{equation}
\Delta(p,q,r) \;=\; - 4 p^2 q^2 (1 - \cos^2 \theta) \;,
\label{triangle-theta}
\end{equation}
where $\theta$ is the angle between ${\bf p}$ and ${\bf q}$. For example, the angle average for (\ref{int-thT:10}) and (\ref{int-thT:1}) is
\bqa
\left\langle {r^4 \over \Delta(p,q,r)} 
\right\rangle_{\!\!{\bf \hat p}\cdot{\bf \hat q}}
\;=\; -{w(\epsilon) \over 8} \int_{-1}^{+1} dx \, (1-x^2)^{-1-\epsilon} \,
        {(p^2 + q^2 + 2 p q x)^2 \over p^2 q^2} \;.
\label{ang-ave:1}
\eqa
After integrating over $x$ and inserting the result into (\ref{int-thT:10}) and (\ref{int-thT:1}), the integral reduces to
\bqa
\int_{\bf pq} {n(p) n(q) \over p q} \,
	{r^4 \over q^2\Delta(p,q,r)} \;=\;
\int_{\bf pq} {n(p) n(q) \over p q}
\left( {1 - 2 \epsilon \over 8 \epsilon} \, {p^2 \over q^4}
	\,+\, {7 - 6 \epsilon \over 8 \epsilon} \, {1 \over q^2} \right) \;.
\eqa
The integrals over ${\bf p}$ and ${\bf q}$ factor into separate integrals that can be evaluated using (\ref{int-th:10}) and (\ref{int-th:1}). After averaging over angles, the integrals (\ref{int-thT:20}), (\ref{int-thT:30}), (\ref{int-thT:2}) and (\ref{int-thT:3}) reduce to
\bqa
\int_{\bf pq} {n(p) n(q) \over p q} \,
	{r^2 \over \Delta(p,q,r)}
\!\!&=&\!\! {1 - 2 \epsilon \over 4 \epsilon}
\int_{\bf p} {n(p) \over p}
\int_{\bf q} {n(q) \over q} \, {1 \over  q^2} \;,
\\
\int_{\bf pq} {n(p) n(q) \over p q} \,
	{p^4 \over q^2 \Delta(p,q,r)}
\!\!&=&\!\! {1 - 2 \epsilon \over 8 \epsilon}
\int_{\bf p} {n(p) \over p} \, p^2
\int_{\bf q} {n(q) \over q} \, {1 \over  q^4} \;,
\eqa
which again can be evaluated using (\ref{int-th:10}) and (\ref{int-th:1}).
 
The integral~(\ref{int-thT:40}) and (\ref{int-thT:4}) can be evaluated by using the identity
\begin{equation}
\left\langle {p^2+q^2 \over r^2\Delta(p,q,r)} 
\right\rangle_{\!\!{\bf \hat p}\cdot{\bf \hat q}}
\;=\; {1 \over 2 \epsilon} \left\langle {1 \over r^4} 
\right\rangle_{\!\!{\bf \hat p}\cdot{\bf \hat q}}
\,+\, {1-2\epsilon \over 8\epsilon} {1 \over p^2 q^2} \;.
\label{ang-ave:2}
\end{equation}
The identity can be proved by expressing the angular averages in terms of integrals over the cosine of the angle between ${\bf p}$ and ${\bf q}$ as in (\ref{ang-ave:1}), and then integrating by parts. Inserting the identity (\ref{ang-ave:2}) into (\ref{int-thT:40}) and (\ref{int-thT:4}), the integrals reduce to 
\bqa
\nonumber
\int_{\bf pq} {n(p) n(q) \over p q} \, 
        {p^2(p^2+q^2) \over r^2 \Delta(p,q,r)} \!\!&=&\!\!
{1 \over 2 \epsilon}
\int_{\bf pq} {n(p) n(q) \over p q} \, {p^2 \over r^4} 
\\ &&
+\; {1 - 2 \epsilon \over 8 \epsilon}
\int_{\bf p} {n(p) n(q) \over p q} \, {1 \over  q^2} \;.
\eqa
The integral in the first term on the right is given in (\ref{int-th:30}) and (\ref{int-th:3}), while the second term can be evaluated using (\ref{int-th:10}) and (\ref{int-th:1}).
        
The integral~(\ref{f1}) can be evaluated directly in three dimensions by first averaging over $c$ and $x$, and then integrate the resulting functions numerically over $p$ and $q$.

To evaluate the weighted averages over $c$ of the thermal integrals in Eqs.~(\ref{intHTL:1})--(\ref{intHTL:3}) and Eqs.~(\ref{f2})--(\ref{ff4}), we first isolate the divergent parts, which come from the region $p-q \rightarrow 0$. We write the product of thermal functions in the form
\bqa
n(p) n(q) \;=\;
\left( n(p) n(q) - {s^2 n^2(s) \over p q}  \right)
+ {s^2 n^2(s) \over p q} \;,
\label{nnsub-1}
\eqa
where $s= (p+q)/2$. In the difference term, the HTL average over $c$ and the angular average over $x = \hat {\bf p} \cdot \hat {\bf q}$ can be calculated in three dimensions:
\begin{eqnarray}
&& \hspace{-2cm}
{\rm Re} \left\langle c^2 {r^2 c^2 - p^2 - q^2 \over
	\Delta(p+i\varepsilon,q,r c)}  \right\ranglecx \;=\;
{1 \over 4 p q} \log {p+q \over |p-q|}
	- {1 \over 2(p^2 - q^2)} \log{p \over q} \;,
\\ \nonumber && \\
&& \hspace{-2cm}
{\rm Re} \left\langle c^4 {r^2 c^2 - p^2 - q^2 \over
	\Delta(p+i\varepsilon,q,r c)}  \right\ranglecx \;=\;
{2(p^2 + q^2) \over 3 (p^2-q^2)^2}
+ {1 \over 12 p q} \log {p+q \over |p-q|}
\nonumber
\\
&& \hspace{3.1cm}
-\; {(3p^2 + q^2)(p^2 + 3 q^2) \over 6 (p^2 - q^2)^3} \log{p \over q} \;,
\\ \nonumber && \\
&& \hspace{-2cm}
{\rm Re} \left\langle c^2 {q^2 \over r^2} {r^2 c^2 - p^2 - q^2 \over
	\Delta(p+i\varepsilon,q,r c)}  \right\ranglecx \;=\;
{q^2 \over 3 (p^2 - q^2)^2}
\left( 2 - {1 \over 2} \log{|p^2-q^2| \over p q} \right.
\nonumber
\\
&& \hspace{3.5cm}
\left. -\; {p^2+q^2 \over 4 p q} \log {p+q \over |p-q|}
	- {p^2 + q^2 \over p^2 - q^2} \log{p \over q} \right) \;,
\\ \nonumber && \\
\nonumber
&& \hspace{-2cm}
{\rm Re}\left\langle
{p^2-q^2\over r^2}{r^2c^2-p^2-q^2\over\Delta(p+i\epsilon,q,rc)}
\right\rangle_{c,x} \;=\;
{1\over4pq(p^2-q^2)}
\left[-(p^2+q^2)\log{p+q\over|p-q|}
\right. \\ && \left. \hspace{4.1cm}
-\;2pq\log{|p^2-q^2|\over pq}
\right] \;.
\end{eqnarray}
The remaining two-dimensional integral over $p$ and $q$ can be evaluated numerically:
\begin{eqnarray}
&& \hspace{-5mm}
\int_{\bf pq}
\left( {n_B(p) n_B(q) \over p q}
	- {s^2 n_B^2(s) \over p^2 q^2}  \right)
{\rm Re} \left\langle c^2 {r^2 c^2 - p^2 - q^2 \over
	\Delta(p+i\varepsilon,q,r c)}  \right\ranglec \;=\;
{T^2 \over (4\pi)^2} \left[ 5.292 \times 10^{-3} \right] \;,
\label{intHTL:1f}
\nonumber \\ && \\
&& \hspace{-5mm}
\int_{\bf pq}
\left( {n_B(p) n_B(q) \over p q}
	- {s^2 n_B^2(s) \over p^2 q^2}  \right)
{\rm Re} \left\langle c^4 {r^2 c^2 - p^2 - q^2 \over
	\Delta(p+i\varepsilon,q,r c)}  \right\ranglec \;=\;
{T^2 \over (4\pi)^2} \left[ 3.292 \times 10^{-3} \right] \;,
\label{intHTL:2f}
\nonumber \\ && \\
&& \hspace{-5mm}
\int_{\bf pq}
\left( {n_B(p) n_B(q) \over p q}
	- {s^2 n_B^2(s) \over p^2 q^2}  \right){q^2 \over r^2} \,
{\rm Re} \left\langle c^2 {r^2 c^2 - p^2 - q^2 \over
	\Delta(p+i\varepsilon,q,r c)}  \right\ranglec \;=\;
{T^2 \over (4\pi)^2} \left[ 2.822 \times 10^{-3} \right] \;,
\label{intHTL:3f0}
\nonumber \\ && \\
&& \hspace{-5mm}
\int_{\bf pq}
\left( {n_F(p) n_F(q) \over p q}
         - {s^2 n^2_F(s) \over p^2 q^2}  \right) 
{\rm Re} \left\langle c^4 {r^2 c^2 - p^2 - q^2 \over
	\Delta(p+i\varepsilon,q,r c)}  \right\ranglec \;=\;
{T^2 \over (4\pi)^2} \left[ 8.980 \times 10^{-3} \right] \;, 
\label{xxxx}
\nonumber \\ && \\
&& \hspace{-5mm}
\int_{\bf pq}
\left( {n_F(p) n_F(q) \over p q}
         - {s^2 n^2_F(s) \over p^2 q^2}  \right){q^2 \over r^2} \,
{\rm Re} \left\langle c^2 {r^2 c^2 - p^2 - q^2 \over
        \Delta(p+i\varepsilon,q,r c)}  \right\ranglec \;=\; 
{T^2 \over (4\pi)^2} \left[ 7.792 \times 10^{-3} \right] \;,
\label{intHTL:3f}
\nonumber \\ && \\
&& \hspace{-1cm}
\int_{\bf pq }
\left( {n_B(p) n_F(q)\over p q} -{s^2n_B(s)n_F(s)\over p^2q^2} \right) 
{\rm Re} \left\langle {p^2-q^2\over r^2}{r^2c^2-p^2-q^2 \over
        \Delta(p+i\epsilon,q,rc)} \right\rangle_c \;=\;
{T^2\over(4\pi)^2} \left[0.17811\right] \;.
\nonumber \\
\label{extra}
\eqa

The integrals involving the $n^2(s)$ term in (\ref{nnsub-1}) are divergent, so the HTL average over $c$ and the angular average over $x = \hat {\bf p} \cdot \hat {\bf q}$ must be calculated in $3-2\epsilon$ dimensions. The first step in the calculation of the $n^2(s)$ term is to change variables from ${\bf p}$ and ${\bf q}$ to $s = (p+q)/2$, $\beta = 4pq/(p+q)^2$, and $x= \hat{\bf p} \cdot \hat{\bf q}$:
\begin{eqnarray} \nonumber
\int_{\bf pq} {s^2 n^2(s) \over p^2 q^2} \, f(p,q,r) \!\!&=&\!\!
{64 \over (4\pi)^4}
\left[ (e^\gamma \mu^2)^\epsilon
        {\Gamma({3\over2}) \over \Gamma({3\over2}-\epsilon)} \right]^2
        \int_0^\infty ds \, s^{1-4 \epsilon}n^2(s) s^2
\\ &&
        \times \int_0^1 d \beta \, \beta^{-2 \epsilon} (1-\beta)^{-1/2}
\Big\langle f(s_+,s_-,r) + f(s_-,s_+,r) \Big\rangle_{\!\!x} \;,
\label{int:n2f}
\nonumber \\
\end{eqnarray}
where $s_\pm = s[1 \pm \sqrt{1-\beta}]$ and $r=s [4 - 2\beta(1-x)]^{1/2}$. The two terms inside the average over $x$ come from the regions $p>q$ and $p<q$, respectively. The integral over $s$ is easily evaluated:
\bqa
\int_0^\infty ds \, s^{1-4 \epsilon}n_B^2(s) \!\!&=&\!\!
\Gamma(2-4\epsilon)
\left[ \zeta(1-4\epsilon) - \zeta(2-4\epsilon) \right] T^{2-4\epsilon} \;,
\label{intth:n20}
\\ \nonumber
\int_0^\infty ds \, s^{1-4 \epsilon}n^2_F(s) \!\!&=&\!\!
\Gamma(2-4\epsilon)
\left[-(1-2^{4\epsilon})\zeta(1-4\epsilon)
\right.\\ &&\left.
+\;(1-2^{-1+4\epsilon})\zeta(2-4\epsilon) \right] T^{2-4\epsilon} \;,
\label{intth:n2}
\\
\int_0^\infty ds \, s^{1-4 \epsilon}n_F(s)n_B(s) \!\!&=&\!\!
2^{-2+4\epsilon}
\Gamma(2-4\epsilon)
\zeta(2-4\epsilon)T^{2-4\epsilon} \;.
\label{intth:n2b}
\eqa
It remains only to evaluate the averages over $c$ and $x$ and the integral over $\beta$.

The first step in the calculation of the $n^2(s)$ term of~(\ref{intHTL:1}), (\ref{intHTL:2}) and (\ref{f2}) is to decompose the integrand into two terms:
\begin{equation}
{r^2 c^2 - p^2 - q^2 \over
	\Delta(p+i\varepsilon,q,r c)} \;=\;
-{1 \over 2} \sum_\pm {1 \over (p+i\varepsilon \pm q)^2 - r^2 c^2} \;.
\end{equation}
The weighted averages over $c$ gives a hypergeometric function:
\begin{eqnarray}
\left\langle {c^2 \over (p+i\varepsilon \pm q)^2 - r^2 c^2}
	\right\ranglec \!\!&=&\!\!
{1 \over 3 - 2 \epsilon} \,
{1 \over (p+i\varepsilon \pm q)^2} \,
F\left({{3 \over 2},1 \atop {5 \over 2} - \epsilon} \Bigg|
	{r^2 \over (p+i\varepsilon \pm q)^2} \right) \;,
\label{avec:c2xx}
\\
\left\langle {c^4 \over (p+i\varepsilon \pm q)^2 - r^2 c^2}
	\right\ranglec \!\!&=&\!\!
{3 \over (3 - 2 \epsilon)(5 - 2 \epsilon)} \,
{1 \over (p+i\varepsilon \pm q)^2} \,
F\left({{5 \over 2},1 \atop {7 \over 2} - \epsilon} \Bigg|
	{r^2 \over (p+i\varepsilon \pm q)^2} \right) \;.
\label{avec:c4xx}
\nonumber \\
\end{eqnarray}

In the $+q$ case of~(\ref{avec:c2xx}) and (\ref{avec:c4xx}), the $i\varepsilon$ prescription is unnecessary. The argument of the hypergeometric function can be written $1 - \beta y$, where $y = (1-x)/2$. After using a transformation formula to change the argument to $\beta y$, we can evaluate the angular average over $x$ to obtain hypergeometric functions with argument $\beta$. The averages over $x$ of~(\ref{avec:c2xx}) and (\ref{avec:c4xx}) are
\begin{eqnarray} 
\left\langle F\left( {{3 \over 2}, 1 \atop {5 \over 2} - \epsilon}
	\Bigg| {r^2 \over (p+q)^2} \right) \right\ranglex \!\!&=&\!\!
- {3-2\epsilon \over 2 \epsilon}
\left[ F\left( { 1-\epsilon , {3 \over 2} , 1
		\atop 2-2\epsilon , 1+\epsilon } \Bigg| \beta \right)
\right.
\nonumber
\\
&& \left. -\;
{ (1)_\epsilon (1)_{-2\epsilon} (2)_{-2 \epsilon} ({3\over2})_{-\epsilon}
	\over (1)_{-\epsilon} (2)_{-3\epsilon} }
\beta^{-\epsilon}
F\left( { 1-2\epsilon , {3\over2}-\epsilon
		\atop 2-3\epsilon } \Bigg| \beta \right)
\right] \;,
\nonumber \\ && \\
\left\langle F\left( {{5 \over 2}, 1 \atop {7 \over 2} - \epsilon}
	\Bigg| {r^2 \over (p+q)^2} \right) \right\ranglex \!\!&=&\!\!
- {5-2\epsilon \over 2 \epsilon}
\left[ F\left( { 1-\epsilon , {5 \over 2} , 1
		\atop 2-2\epsilon , 1+\epsilon } \Bigg| \beta \right)
\right.
\nonumber
\\
&& \left. -\;
{ (1)_\epsilon (1)_{-2\epsilon} (2)_{-2 \epsilon} ({5\over2})_{-\epsilon}
	\over (1)_{-\epsilon} (2)_{-3\epsilon} }
\beta^{-\epsilon}
F\left( { 1-2\epsilon , {5\over2}-\epsilon
		\atop 2-3\epsilon } \Bigg| \beta \right)
\right] \;,
\nonumber \\
\end{eqnarray}
where $(a)_b$ is Pochhammer's symbol which is defined in (\ref{Poch}). Integrating over $\beta$, we obtain hypergeometric functions with argument 1:
\begin{eqnarray}
s^2 \int_0^1 d\beta \, \beta^{-2\epsilon} (1-\beta)^{-1/2}
\left\langle {c^2 \over (p+q)^2 - r^2 c^2}
	\right\ranglecx && \nonumber
\\ && \hspace{-6cm} 
=\; - {1 \over 4\epsilon}
{ (1)_\epsilon (2)_{-2\epsilon} \over (1)_{-\epsilon} }
\left[ { (1)_{-2\epsilon} (1)_{-\epsilon}
	\over ({3\over2})_{-2\epsilon} (2)_{-2\epsilon} (1)_{\epsilon} }
F\left( { 1-2\epsilon , 1-\epsilon , {3 \over 2} , 1
	\atop {3\over2}-2\epsilon , 2-2\epsilon , 1+\epsilon }
	\Bigg| 1 \right)
\right.
\nonumber
\\ && \hspace{-5.4cm} 
\left. -\; { (1)_{-3\epsilon} (1)_{-2\epsilon} ({3\over2})_{-\epsilon}
	\over ({3\over2})_{-3\epsilon} (2)_{-3\epsilon} }
F\left( { 1-3\epsilon , 1-2\epsilon , {3\over2}-\epsilon
		\atop {3\over2}-3\epsilon,2-3\epsilon } \Bigg| 1 \right)
\right] \;,
\label{FF:1}
\\
s^2 \int_0^1 d\beta \, \beta^{-2\epsilon} (1-\beta)^{-1/2}
\left\langle {c^4 \over (p+q)^2 - r^2 c^2}
	\right\ranglecx && \nonumber
\\ && \hspace{-6cm}
=\; - {3 \over 4\epsilon(3-2\epsilon)}
\left[ { (1)_{-2\epsilon} 
	\over ({3\over2})_{-2\epsilon} }
F\left( { 1-2\epsilon , 1-\epsilon , {5 \over 2} , 1
	\atop {5\over2}-2\epsilon , 2-2\epsilon , 1+\epsilon }
	\Bigg| 1 \right)
\right.
\nonumber
\\
&& \hspace{-5.4cm} 
\left. -\; { (1)_{\epsilon} (1)_{-2\epsilon} (2)_{-2\epsilon} ({5\over2})_{-\epsilon}
	\over ({1})_{-\epsilon} (2)_{-3\epsilon} }{(1)_{-3\epsilon}\over({3\over2})_{-3\epsilon}}
F\left( { 1-3\epsilon , 1-2\epsilon , {5\over2}-\epsilon
		\atop {5\over2}-3\epsilon,2-3\epsilon } \Bigg| 1 \right)\right] \;.
\nonumber \\
\label{FF:2}
\end{eqnarray}
Expanding in powers of $\epsilon$, we obtain
\begin{eqnarray}
s^2 \int_0^1 d\beta \, \beta^{-2\epsilon} (1-\beta)^{-1/2}
\left\langle {c^2 \over (p+q)^2 - r^2 c^2}
	\right\ranglecx \!\!&=&\!\!
{\pi^2 \over 24} (1 + 3.54518 \, \epsilon) \;,
\label{intavecx:1p}
\\
s^2 \int_0^1 d\beta \, \beta^{-2\epsilon} (1-\beta)^{-1/2}
\left\langle {c^4 \over (p+q)^2 - r^2 c^2}
	\right\ranglecx \!\!&=&\!\!
{\pi^2 \over 72} (1 + 10.8408 \, \epsilon) \;.
\label{intavecx:2p}
\end{eqnarray}

In the $-q$ case of~ (\ref{avec:c2xx}) and (\ref{avec:c4xx}), the argument of the hypergeometric functions can be written $(1-\beta y)/(1-\beta \pm i \varepsilon)$, where $y=(1-x)/2$ and the prescriptions $+i \varepsilon$ and $-i \varepsilon$ correspond to the regions $p>q$ and $p<q$, respectively. These regions correspond to the two terms inside the average over $x$ in (\ref{int:n2f}). In  order to obtain an analytic result in terms of hypergeometric functions, it is necessary to integrate over $\beta$ before averaging over $x$. The integrals over $\beta$ can be evaluated by first using a transformation formula to change the argument of the hypergeometric function to $-\beta(1-y)/(1-\beta)$ and then using the integration formula (\ref{int-2F1}) to obtain hypergeometric functions with arguments $y$ or $1-y$:
\begin{eqnarray}
\int_0^1 d\beta \, \beta^{-2\epsilon} (1-\beta)^{-3/2}
F\left( { {3\over2}, 1 \atop {5\over2}-\epsilon }
	\Bigg| {1-\beta y \over 1-\beta + i \varepsilon} \right) &&
\nonumber
\\
&&  \hspace{-7cm} 
=\; {3-2\epsilon \over \epsilon} \,
{(1)_{-2\epsilon} \over ({1\over2})_{-2\epsilon}} \,
F\left( { 1-2\epsilon, 1
		\atop 1+\epsilon } \Bigg| 1-y \right)
\nonumber
\\
&& \hspace{-6.4cm}
-\; {3-2\epsilon \over \epsilon} \,
{ (1)_{\epsilon} \over ({1\over2})_{\epsilon} } \,
(1-y)^{-1/2}
F\left( { {1\over2}-2\epsilon, 1 \atop {1\over2}+\epsilon }
	\Bigg| 1-y \right)
\nonumber
\\
&& \hspace{-6.4cm}
+\; {3 \over 2\epsilon(1-3\epsilon)} e^{i\pi \epsilon} \,
(1)_{\epsilon} (\mbox{${5\over2}$})_{-\epsilon}\,
(1-y)^{-\epsilon}
F\left( { 1-3\epsilon, {3\over2}-\epsilon \atop 2-3\epsilon }
	\Bigg| y \right) \;,
	\\
\int_0^1 d\beta \, \beta^{-2\epsilon} (1-\beta)^{-5/2}
F\left( { {3\over2}, 1 \atop {7\over2}-\epsilon }
	\Bigg| {1-\beta y \over 1-\beta + i \varepsilon} \right) &&
\nonumber
\\
&&  \hspace{-7cm} 
=\; {5-2\epsilon \over \epsilon} \,
{(1)_{-2\epsilon} \over ({1\over2})_{-2\epsilon}} \,
F\left( { {5 \over 2}, 1, 1-2\epsilon
		\atop 1+\epsilon, {3 \over 2} } \Bigg| 1-y \right)
\nonumber
\\
&& \hspace{-6.4cm}
-\; {2(5-2\epsilon) \over 3\epsilon} \,
{ (1)_{\epsilon} \over ({1\over2})_{\epsilon} } \,
(1-y)^{-1/2}
F\left( {2, {1\over2}-2\epsilon \atop {1\over2}+\epsilon }
	\Bigg| 1-y \right)
\nonumber
\\
&& \hspace{-6.4cm}
+\; {4({5\over2}-\epsilon)({3\over2}-\epsilon)({1\over2}-\epsilon) \over 3\epsilon(3\epsilon-1)(3\epsilon-2)} e^{i\pi \epsilon} \,
(1)_{\epsilon} (\mbox{${1\over2}$})_{-\epsilon}\,
(1-y)^{-\epsilon}
F\left({ {5\over2}-\epsilon, 1-3\epsilon \atop 3-3\epsilon }
	\Bigg| y \right) \nonumber \;.
\\
\end{eqnarray}
After averaging over $x$, we obtain hypergeometric functions with argument 1:
\begin{eqnarray}
s^2 \int_0^1 d\beta \, \beta^{-2\epsilon} (1-\beta)^{-1/2}
\left\langle {c^2 \over (p+i\varepsilon - q)^2 - r^2 c^2}
	\right\ranglecx &&
\nonumber
\\
&&  \hspace{-9cm} 
=\; {1 \over 4\epsilon}\,
{(1)_{-2\epsilon} \over ({1\over2})_{-2\epsilon}} \,
F\left( { 1-\epsilon, 1-2\epsilon, 1
		\atop 2-2\epsilon, 1+\epsilon } \Bigg| 1 \right)
\nonumber
\\
&& \hspace{-8.4cm}
-\; {1 \over 2\epsilon}\,
{ (2)_{-2\epsilon} (1)_{\epsilon} ({1\over2})_{-\epsilon}
	\over (1)_{-\epsilon} ({1\over2})_{\epsilon}
			({3\over2})_{-2\epsilon} } \,
F\left( { {1\over2}-\epsilon,{1\over2}-2\epsilon,1
		\atop {3\over2}-2\epsilon,{1\over2}+\epsilon }
	\Bigg| 1 \right)
\nonumber
\\ \nonumber
&& \hspace{-8.4cm}
+\; {1 \over 8\epsilon(1-3\epsilon)} \, e^{i\pi \epsilon}
{ (2)_{-2\epsilon} (1)_{-2\epsilon} (1)_{\epsilon}({3\over2})_{-\epsilon}
	\over (1)_{-\epsilon} (2)_{-3\epsilon} } \,
F\left( { 1-\epsilon, 1-3\epsilon, {3\over2}-\epsilon
		\atop 2-3\epsilon, 2-3\epsilon }
	\Bigg| 1 \right) \;.
\\ && \\
s^2 \int_0^1 d\beta \, \beta^{-2\epsilon} (1-\beta)^{-1/2}
\left\langle {c^2 \over (p+i\varepsilon - q)^2 - r^2 c^2}
	\right\ranglecx &&
\nonumber
\\
&&  \hspace{-9cm} 
=\; {3 \over 4\epsilon(1-2\epsilon)(3-2\epsilon)}\,
{({1\over2})_{\epsilon} (1)_{-\epsilon}(1)_{\epsilon}({3\over2})_{-\epsilon} \over ({1\over2})_{-2\epsilon}(1)_{2\epsilon} } \,
F\left( { 1-\epsilon, 1-2\epsilon, 1, {5\over2}
		\atop 2-2\epsilon, 1+\epsilon, {3\over2} } \Bigg| 1 \right)
\nonumber
\\
&& \hspace{-8.4cm}
-\; {1 \over \epsilon(3-2\epsilon)}\,
{ ({1\over2})_{-\epsilon} (1)_{\epsilon}^2 ({3\over2})_{-\epsilon}
	\over (1)_{2\epsilon} ({3\over2})_{-2\epsilon} } \,
F\left( { {1\over2}-\epsilon,{1\over2}-2\epsilon,2
		\atop {3\over2}-2\epsilon,{1\over2}+\epsilon }
	\Bigg| 1 \right)
\nonumber
\\ \nonumber
&& \hspace{-8.4cm}
+\; {3 \over 16\epsilon(1-3\epsilon)(3-2\epsilon)} \, e^{i\pi \epsilon}
{ ({1\over2})_{\epsilon} (1)_{-2\epsilon} (1)_{\epsilon}^2 ({3\over2})_{-\epsilon} ({5\over2})_{-\epsilon}
	\over (1)_{2\epsilon} (3)_{-3\epsilon} } \,
F\left( { 1-\epsilon, 1-3\epsilon, {5\over2}-\epsilon
		\atop 3-3\epsilon, 2-3\epsilon }
	\Bigg| 1 \right) \;. \\
\end{eqnarray}
Expanding in powers of $\epsilon$ and then taking the real parts, we obtain
\begin{eqnarray}
{\rm Re} \, s^2 \int_0^1 d\beta \, \beta^{-2\epsilon} (1-\beta)^{-1/2}
\left\langle {c^2 \over (p+i\varepsilon - q)^2 - r^2 c^2}
	\right\ranglecx \!\!&=&\!\!
- {\pi^2 \over 24} (1 + 0.34275 \, \epsilon) \;,
\label{intavecx:1m}
\nonumber \\ && \\
{\rm Re} \, s^2 \int_0^1 d\beta \, \beta^{-2\epsilon} (1-\beta)^{-1/2}
\left\langle {c^4 \over (p+i\varepsilon - q)^2 - r^2 c^2}
	\right\ranglecx \!\!&=&\!\!
- {12+\pi^2 \over 72} (1 + 1.10518 \, \epsilon) \;.
\label{intavecx:2m}
\nonumber \\
\end{eqnarray}
Inserting the sum of the integrals (\ref{intavecx:1p}) and (\ref{intavecx:1m}) into the thermal integral (\ref{int:n2f}) and similarly for the integrals weighted by $c^4$, we obtain
\begin{eqnarray}
&& \hspace{-2cm}
\int_{\bf pq}
{s^2 n_B^2(s) \over p^2 q^2} \,
{\rm Re} \left\langle c^2 {r^2 c^2 - p^2 - q^2 \over
	\Delta(p+i\varepsilon,q,r c)}  \right\ranglec \;=\;
{T^2 \over (4\pi)^2} \left[ \, 0.133434 \, \right] \;, 
\label{intHTL:1d}
\\
&& \hspace{-2cm}
\int_{\bf pq}
{s^2 n_B^2(s) \over p^2 q^2} \,
{\rm Re} \left\langle c^4 {r^2 c^2 - p^2 - q^2 \over
	\Delta(p+i\varepsilon,q,r c)}  \right\ranglec \;=\;
-{1\over6 \pi^2} {T^2 \over (4\pi)^2}  \left({\mu\over4\pi T}\right)^{4\epsilon}
\left[ {1\over\epsilon} + 7.0292 \right] \;,
\label{intHTL:2d}
\\
&& \hspace{-2cm}
\int_{\bf pq}
{s^2 n_F^2(s) \over p^2 q^2} \,
{\rm Re} \left\langle c^4 {r^2 c^2 - p^2 - q^2 \over
	\Delta(p+i\varepsilon,q,r c)}  \right\ranglec \;=\;
{T^2 \over (4\pi)^2} 
\left[ \, 5.53165\times10^{-5} \, \right] \;.
\label{intHTL:2dF}
\end{eqnarray}
%
%{\bf [mj]} ({\bf [p]} mod $0.133434 \to 0.13344$)
%[[ DELETED 1 DIGIT (OLD: $7.02920$) ]]
%0.0000553165
Adding these integrals to the subtracted integrals in~(\ref{intHTL:1f}), (\ref{intHTL:2f}) and (\ref{xxxx}), we obtain the final results in (\ref{intHTL:1}), (\ref{intHTL:2}) and (\ref{f2}).

To evaluate the subtraction in the integrals~(\ref{intHTL:3f0}) and (\ref{intHTL:3f}), we use the identity $q^2 = (r^2 + q^2 - p^2 - 2 {\bf p}\cdot{\bf q})/2$. The integral with $q^2-p^2$ in the numerator is purely imaginary. Thus the real part of the integral can be expressed as
\begin{eqnarray}
\int_{\bf pq} {s^2 n^2(s) \over p^2 q^2} \, {q^2 \over r^2} \,
{\rm Re} \left\langle c^2 {r^2 c^2 - p^2 - q^2 \over
        \Delta(p+i\varepsilon,q,r c)}  \right\ranglec &&
        \nonumber
\\
&& \hspace{-5cm}
=\; \int_{\bf pq} {s^2 n^2(s) \over p^2 q^2}
        \left( {1\over2} - {{\bf p}\cdot{\bf q} \over r^2} \right)
{\rm Re} \left\langle c^2 {r^2 c^2 - p^2 - q^2 \over
        \Delta(p+i\varepsilon,q,r c)}  \right\ranglec \;.
\label{intHTL:3a}
\end{eqnarray}
The evaluation of the first term in Eq.~(\ref{intHTL:3a}) follows the same procedure as for~(\ref{intHTL:1}), but just with $n_F$ now instead of $n_B$. The result reads
\bqa
\int_{\bf pq}
{s^2 n_F^2(s) \over p^2 q^2} \,
{\rm Re} \left\langle c^2 {r^2 c^2 - p^2 - q^2 \over
	\Delta(p+i\varepsilon,q,r c)}  \right\ranglec \;=\;
{\cal O}(\epsilon) \;.
\label{intHTL:1dF}
\eqa
It remains only to evaluate the integral in Eq.~(\ref{intHTL:3a}) with ${\bf p}\cdot{\bf q}$ in the numerator. We begin by using the identity
\begin{eqnarray}
\nonumber
\left\langle c^2 \, {{\bf p}\cdot{\bf q} \over r^2} \,
{r^2 c^2 - p^2 - q^2 \over
        \Delta(p+i\varepsilon,q,r c)} \right\ranglecx
\!\!&=&\!\!
- {p^2+q^2 \over (p^2 - q^2 +i\varepsilon)^2} \langle c^2 \rangle_c
        \left\langle {{\bf p}\cdot{\bf q} \over r^2} \right\ranglex
\nonumber
\\
&&
-\; {1 \over 2} \sum_\pm  {1 \over (p+i\varepsilon \pm q)^2} \,
        \left\langle {{\bf p}\cdot{\bf q} \, c^4
                        \over (p+i\varepsilon \pm q)^2 - r^2 c^2}
                \right\ranglecx \;. 
\nonumber
\\
\label{pqrdelta}
\end{eqnarray}
In the first term on the right side, the average over $c$ is a simple multiplicative factor: $\langle  c^2\rangle_c = 1/(3-2\epsilon)$. The average over $x$ gives hypergeometric functions of argument $\beta$:
\begin{equation}
\left\langle { {\bf p}\cdot{\bf q} \over r^2 } \right\ranglex \;=\;
{1 \over 8} \beta
\left[
F\left( { 1-\epsilon, 1 \atop 3-2\epsilon } \Bigg| \beta \right)
- F\left( { 2-\epsilon, 1 \atop 3-2\epsilon } \Bigg| \beta \right)
\right] \;.
\end{equation}
The integral over $\beta$ gives hypergeometric functions of argument 1:
\begin{eqnarray}
s^2 \int_0^1 d\beta \, \beta^{-2\epsilon} (1-\beta)^{-1/2}
{p^2 + q^2 \over (p^2 - q^2)^2}
\left\langle { {\bf p}\cdot{\bf q} \over r^2 } \right\ranglex &&
\nonumber
\\
&& \hspace{-6cm} 
=\;
- {1 \over 8} \,
{ (2)_{-2\epsilon} \over ({3\over2})_{-2\epsilon} } \,
\left[
F\left( { 2-2\epsilon, 1-\epsilon,1
                \atop {3\over2}-2\epsilon, 3-2\epsilon }
        \Bigg| 1 \right)
- F\left( { 2-2\epsilon, 2-\epsilon,1
                \atop {3\over2}-2\epsilon, 3-2\epsilon }
        \Bigg| 1 \right) \right]
\nonumber \\
&& \hspace{-5.4cm} 
+\; {1 \over 12} \,
{ (3)_{-2\epsilon} \over ({5\over2})_{-2\epsilon} } \,
\left[
F\left( { 1-\epsilon, 1 \atop {5\over2}-2\epsilon } \Bigg| 1 \right)
- F\left( { 2-\epsilon, 1 \atop {5\over2}-2\epsilon } \Bigg| 1 \right)
\right] \;.
\end{eqnarray}
Expanding in powers of $\epsilon$, we obtain
\begin{equation}
s^2 \int_0^1 d\beta \, \beta^{-2\epsilon} (1-\beta)^{-1/2}
{p^2 + q^2 \over (p^2 - q^2)^2}
\left\langle { {\bf p}\cdot{\bf q} \over r^2 } \right\ranglex \;=\;
- {\pi^2 \over 16}\left[1
 - 1.02148 \, \epsilon \right] \;.
\label{intavecx:3}
\end{equation}

In the second term of (\ref{pqrdelta}), the average over $c$ is given by (\ref{avec:c4xx}). In the $+q$ term, the average over $x= \hat{\bf p} \cdot \hat{\bf q}$ is
\begin{eqnarray}
\left\langle
x F\left( { 1, {5\over2} \atop {7\over2}-\epsilon }
        \Bigg| {r^2 \over (p+q)^2} \right)
\right\ranglex
\!\!&=&\!\!{5-2\epsilon \over 4\epsilon}\,
\left[
F\left( { 2-\epsilon, 1, {5\over2}
        \atop 3-2\epsilon, 1+\epsilon } \Bigg| \beta \right)
- F\left( { 1-\epsilon, 1, {5\over2}
        \atop 3-2\epsilon, 1+\epsilon } \Bigg| \beta \right)
\right]
\nonumber
\\
&& \hspace{-5.7cm}
+\; {5 \over 4\epsilon}\,
{ (1)_{\epsilon} (1)_{-2\epsilon} (3)_{-2\epsilon} ({7\over2})_{-\epsilon}
        \over (1)_{-\epsilon} (3)_{-3\epsilon} } \,
\beta^{-\epsilon}
\left[
F\left( { 1-2\epsilon, {5\over2}-\epsilon
        \atop 3-3\epsilon } \Bigg| \beta \right)
- {1-2\epsilon \over 1-\epsilon}
F\left( { 2-2\epsilon, {5\over2}-\epsilon
                \atop 3-3\epsilon } \Bigg| \beta \right)
\right] \;. 
\nonumber \\
\end{eqnarray}
Integrating over $\beta$, we obtain hypergeometric functions of argument 1:
\begin{eqnarray}
\int_0^1 d\beta \, \beta^{-2\epsilon} (1-\beta)^{-1/2} \,
\left\langle {{\bf p}\cdot{\bf q} \, c^4 \over (p+q)^2 - r^2 c^2}
        \right\ranglecx &&
\nonumber
\\ \nonumber
&& \hspace{-8cm} 
=\; {1 \over 4\epsilon(3-2\epsilon)}\,
{(2)_{-2\epsilon} \over ({5\over2})_{-2\epsilon}} \,
\left[
F\left( { 2-2\epsilon, 2-\epsilon, 1, {5\over2}
        \atop {5\over2}-2\epsilon, 3-2\epsilon, 1+\epsilon } \Bigg| 1 \right)
- F\left( { 2-2\epsilon, 1-\epsilon, 1, {5\over2}
        \atop {5\over2}-2\epsilon, 3-2\epsilon, 1+\epsilon } \Bigg| 1 \right)
\right]
\nonumber
\\
&& \hspace{-7.4cm}
+ \; {1 \over 6\epsilon(2-3\epsilon)}\,
{ (1)_{\epsilon} (1)_{-2\epsilon} (3)_{-2\epsilon} ({3\over2})_{-\epsilon}
        \over (1)_{-\epsilon} ({5\over2})_{-3\epsilon} } \,
\nonumber
\\ \nonumber
&& \hspace{-6.9cm} \times \left[
F\left( { 2-3\epsilon, 1-2\epsilon, {5\over2}-\epsilon
                \atop {5\over2}-3\epsilon, 3-3\epsilon }
        \Bigg| 1 \right)
- {1-2\epsilon \over 1-\epsilon}
F\left( { 2-3\epsilon, 2-2\epsilon, {5\over2}-\epsilon
                \atop {5\over2}-3\epsilon, 3-3\epsilon }
        \Bigg| 1 \right)
\right] \;.
\\
\end{eqnarray}
Expanding in powers of $\epsilon$, we obtain
\begin{eqnarray}
\int_0^1 d\beta \, \beta^{-2\epsilon} (1-\beta)^{-1/2}  \,
\left\langle {{\bf p}\cdot{\bf q} \, c^4 \over (p+q)^2 - r^2 c^2}
        \right\ranglecx \;=\;
{\pi^2 - 6 \over 18}(1 \,-\, 0.0728428 \, \epsilon) \;.
\label{intavecx:4p}
\end{eqnarray}

In the $-q$ term in the integral of the second term of (\ref{pqrdelta}), we integrate over $\beta$ before averaging over $x$. The integral over $\beta$ can be expressed in terms of hypergeometric functions of type $_2F_1$:
\begin{eqnarray}
s^2 \int_0^1 d\beta \, \beta^{-2\epsilon} (1-\beta)^{-1/2}  \,
{4 {\bf p}\cdot{\bf q} \over (p-q)^2}
\left\langle {c^4 \over (p+i\varepsilon - q)^2 - r^2 c^2}
        \right\ranglec &&
\nonumber
\\\nonumber
&& \hspace{-9cm} 
=\; - {1 \over 2(3-2\epsilon)\epsilon} \,
{(2)_{-2\epsilon} \over ({1\over2})_{-2\epsilon}} \,
(1-2y) \,
F\left( { 2-2\epsilon, 1 \atop 1+\epsilon } \Bigg| 1-y \right)
\nonumber
\\\nonumber
&& \hspace{-8.4cm}
-\; {1 \over 4(3-2\epsilon)\epsilon}  \,
{(1)_{\epsilon} \over (-{1\over2})_{\epsilon}}
(1-2y) \, (1-y)^{-3/2}  \,
F\left( { {1\over2}-2\epsilon, 1 \atop -{1\over2} +\epsilon } \Bigg| 1-y \right)
\nonumber
\\\nonumber
&& \hspace{-8.4cm}
+\; {1 \over 8(2-3\epsilon)\epsilon}
e^{\mp i \pi \epsilon} (1)_\epsilon (\mbox{$3\over2$})_{-\epsilon} \,
(1-2y) \, (1-y)^{-\epsilon} \,
F\left( { 2-3\epsilon, {5\over2}-\epsilon
        \atop 3-3\epsilon } \Bigg| y \right) \;.
\\
\end{eqnarray}
The phase in the last term is $e^{-i \pi \epsilon}$ for the $f(s_+,s_-,r)$ term of (\ref{int:n2f}), which comes from the $p>q$ region of the integral, and $e^{i \pi \epsilon}$ for the $f(s_-,s_+,r)$ term, which comes from the $p<q$ region. The average over $x=\hat{\bf p} \cdot \hat{\bf q}$ can be expressed in terms of hypergeometric functions of type $_3F_2$ evaluated at 1:
\begin{eqnarray}
s^2 \int_0^1 d\beta \, \beta^{-2\epsilon} (1-\beta)^{-1/2}  \,
\left\langle {4 {\bf p} \cdot {\bf q} \over (p-q)^2} \,
        {c^4 \over (p+i\varepsilon - q)^2 - r^2 c^2}
        \right\ranglecx &&
\nonumber
\\\nonumber
&& \hspace{-10cm} 
=\;
{1 \over 4(3-2\epsilon)\epsilon} \,
{(2)_{-2\epsilon}\over ({1\over2})_{-2\epsilon}}
\left[ F\left( { 1-\epsilon,2-2\epsilon, 1
                \atop 3-2\epsilon,1+\epsilon } \Bigg| 1 \right)
         - F\left( { 2-\epsilon,2-2\epsilon, 1
                \atop 3-2\epsilon,1+\epsilon } \Bigg| 1 \right) \right]
\nonumber
\\
&& \hspace{-9.4cm}
-\; {1 \over (3-2\epsilon)\epsilon} \,
{ (1)_{\epsilon} (3)_{-2\epsilon} (-{1\over2})_{-\epsilon}
\over (1)_{-\epsilon} (-{1\over2})_{\epsilon} ({3\over2})_{-2\epsilon} }
\nonumber \\
&& \hspace{-8.9cm} 
\times
\left[ 
F\left( { -{1\over2}-\epsilon, {1\over2}-2\epsilon, 1
        \atop {3\over2}-2\epsilon, -{1\over2}+\epsilon } \Bigg| 1 \right)
+ {1+2\epsilon \over 2(1-\epsilon)}
        F\left( { {1\over2}-\epsilon, {1\over2}-2\epsilon, 1
        \atop {3\over2}-2\epsilon, -{1\over2}+\epsilon } \Bigg| 1 \right)
\right]
\nonumber
\\ \nonumber
&& \hspace{-9.4cm}
+\; {1 \over 16(2-3\epsilon)\epsilon} \, e^{\mp i \pi \epsilon} \,
{ (1)_{\epsilon} (2)_{-2\epsilon} (2)_{-2\epsilon} ({3\over2})_{-\epsilon}
        \over (1)_{-\epsilon} (3)_{-3\epsilon} }
\\
&& \hspace{-8.9cm} 
\times
\left[ 
F\left( { 1-\epsilon,  2-3\epsilon, {5\over2}-\epsilon
                \atop 3-3\epsilon, 3-3\epsilon } \Bigg| 1 \right)
- {1-\epsilon \over 1-2\epsilon}
        F\left( { 2-\epsilon,  2-3\epsilon, {5\over2}-\epsilon
                \atop 3-3\epsilon, 3-3\epsilon } \Bigg| 1 \right)
\right]  \;.
\nonumber \\
\end{eqnarray}
The expansion of the real part of the integral in powers of $\epsilon$ is
\begin{eqnarray}\nonumber
s^2 \int_0^1 d\beta \, \beta^{-2\epsilon} (1-\beta)^{-1/2} \,
{\rm Re} \left\langle {4 {\bf p} \cdot {\bf q} \over (p-q)^2} \,
        {c^4 \over (p+i\varepsilon - q)^2 - r^2 c^2}
        \right\ranglecx &&
\nonumber \\ && \hspace{-5cm}
= {9-\pi^2\over18}(1 \,-\, 0.796858\, \epsilon) \;.
\label{intavecx:4m}
\end{eqnarray}
Inserting (\ref{intavecx:3}), (\ref{intavecx:4p}), and (\ref{intavecx:4m}) into the thermal integral of (\ref{pqrdelta}), we obtain
\begin{eqnarray}
&& \hspace{-1cm}
\int_{\bf pq}
{s^2 n_B^2(s) \over p^2 q^2} \, {{\bf p}\cdot{\bf q} \over r^2} \,
{\rm Re} \left\langle c^2 {r^2 c^2 - p^2 - q^2 \over
	\Delta(p+i\varepsilon,q,r c)}  \right\ranglec 
\;=\;
{T^2 \over (4\pi)^2}  \left({\mu\over4\pi T}\right)^{4\epsilon}
{1-\pi^2\over24\pi^2} \left[ {1\over\epsilon} + 13.52098 \right] \;,
\nonumber \\ && \\
&& \hspace{-1cm}
\int_{\bf pq}
{s^2 n^2_F(s) \over p^2 q^2} \, {{\bf p}\cdot{\bf q} \over r^2} \,
{\rm Re} \left\langle c^2 {r^2 c^2 - p^2 - q^2 \over
        \Delta(p+i\varepsilon,q,r c)}  \right\ranglec 
\;=\; 
{T^2 \over (4\pi)^2}
{\pi^2-1\over6\pi^2} \left[{\pi^2\over12}-\log2 \right] \;.
\end{eqnarray}
Inserting these along with~(\ref{intHTL:1d}) and (\ref{intHTL:1dF}) into (\ref{intHTL:3a}), we obtain
\begin{eqnarray}
&& \hspace{-0.5cm}
\int_{\bf pq}
{s^2 n_B^2(s) \over p^2 r^2}
{\rm Re} \left\langle c^2 {r^2 c^2 - p^2 - q^2 \over
	\Delta(p+i\varepsilon,q,r c)}  \right\ranglec 
\;=\;
{T^2 \over (4\pi)^2}  \left({\mu\over4\pi T}\right)^{4\epsilon}
{\pi^2-1\over24\pi^2} \left[ {1\over\epsilon} + 15.302796 \right] \;,
\nonumber \\ && \\
&& \hspace{-0.5cm}
\int_{\bf pq}
{s^2 n_F^2(s) \over p^2 r^2}
{\rm Re} \left\langle c^2 {r^2 c^2 - p^2 - q^2 \over
        \Delta(p+i\varepsilon,q,r c)}  \right\ranglec 
\;=\;
{T^2 \over (4\pi)^2}
{1-\pi^2\over6\pi^2} \left[{\pi^2\over12}-\log2 \right] \;.
\label{intHTL:3d}
\end{eqnarray}
Adding this integral to the subtracted integral in~(\ref{intHTL:3f0}) and (\ref{intHTL:3f}), we obtain the final result in~(\ref{intHTL:3}) and (\ref{f3}). The subtracted integral appearing in~(\ref{extra}) vanishes due to antisymmetry of the integrand. Thus the final result~(\ref{ff4}) is given by~(\ref{extra}).

To evaluate the weighted averages over $c$ of the thermal integrals in (\ref{intHTL:4})--(\ref{llll}), we first isolate the divergent parts, which arise from the region $q \to 0$. The integrals~(\ref{intHTL:4x}) and~(\ref{intHTL:5x}) can be computed directly in three dimensions without any isolation of divergence, as described above. For the integrals (\ref{intHTL:4}) and (\ref{intHTL:5}), a single subtraction of the thermal distribution $n_B(q)$ suffices to remove the divergences:
\begin{equation}
n_B(q) \;=\; \left( n_B(q) \,-\, {T \over q} \right) \,+\, {T \over q} \;.
\label{nsub-10}
\end{equation}
For the rest, a second subtraction is also needed to remove the divergences:
\bqa
\label{nsub-1}
n_B(q) \;=\; \left(n_B(q) \,-\, {T\over q} \,+\, {1\over2}\right) \,+\, {T\over q} \,-\, {1\over2} \;.
\eqa
In the integral~(\ref{intHTL:7}) and (\ref{intHTL:7x}), it is convenient to first use the identity $r_c^2 = p^2 + 2 {\bf p} \cdot {\bf q}/c + q^2/c^2$ to expand them into three integrals, two of which are (\ref{intHTL:4}) and (\ref{intHTL:6}), and (\ref{intHTL:4x}) and (\ref{intHTL:6x}), respectively. In the third integrals, the subtraction (\ref{nsub-1}) is needed to remove the divergences.

For the convergent terms, the HTL average over $c$ and the angular average over $x=\hat{\bf p}\cdot\hat{\bf q}$ can be calculated in three dimensions:
\bqa
{\rm Re}\left\langle c^{-1}
	{r_c^2 - p^2 - q^2 \over \Delta(p+i\varepsilon,q,r_c)} \,
	\right\ranglecx \!\!&=&\!\!
{1 \over 4p^2 - q^2} \log{2p \over q}
\nonumber
\\
&& +\; {1 \over 4 p q} \left( {p+q \over 2p+q} \log{p+q \over p}
			- {p-q \over 2p-q} \log{|p-q| \over p} \right)
\;,
\nonumber \\ && \\ 
{\rm Re} \left\langle c
        {r_c^2 - p^2 - q^2 \over \Delta(p+i\varepsilon,q,r_c)} \,
        \right\ranglecx \!\!&=&\!\!
{1 \over 6(4p^2 - q^2)}
+ {q^2(4p^2+3q^2) \over 3(4p^2 - q^2)^3} \log{2p \over q}
\nonumber \\
&& \hspace{-4cm} 
+\; {(p+q)(4p^2+2p q+q^2) \over 12 p q(2p+q)^3} \log{p+q \over p}
- {(p-q)(4p^2-2p q+q^2) \over 12 p q(2p-q)^3} \log{|p-q| \over p}
\;,
\nonumber \\ && \\ 
{\rm Re} \left\langle \hat{\bf p} \cdot \hat{\bf q}
        {r_c^2 - p^2 - q^2 \over \Delta(p+i\varepsilon,q,r_c)} \,
        \right\ranglecx \!\!&=&\!\!
{1 \over 6pq}
- {q(12p^2-q^2) \over 6p (4p^2 - q^2)^2} \log{4p \over q}
\nonumber \\
&& \hspace{-4cm}
+\; {(p+q)(2p^2-2p q-q^2) \over 12 p^2q(2p+q)^2} \log{p+q \over 4p}
+ {(p-q)(2p^2+2p q-q^2) \over 12 p^2q(2p-q)^2} \log{|p-q| \over 4p} \;,
\nonumber \\ && \\ 
&& \hspace{-5cm}
{\rm Re}\left\langle
{r_c^2-p^2\over q^2}{r^2_c-p^2-q^2\over\Delta(p+i\epsilon,q,r_c)}c^{-1}
-{1\over q^2}c^{-1}+{\log2\over q^2}
\right\rangle_{c,x}
\nonumber \\
&& \hspace{-6.5mm}
=\; {1\over4pq^2}\left[q\log{p+q\over|p-q|}
+p\log{|p^2-q^2|\over p^2}
\right] \;.
\eqa
The remaining two-dimensional integral over $p$ and $q$ can be evaluated numerically:
\bqa
%%%%%%%%%% glue %%%%%%%%%%
&& \int_{\bf pq} {n_B(p)\over p}
\left( {n_B(q) \over q} - {T \over q^2} \right)
{\rm Re} \left\langle c^{-1}
	{r_c^2 - p^2 - q^2 \over \Delta(p+i\varepsilon,q,r_c)} \,
	\right\ranglec \;=\;
{T^2 \over (4 \pi)^2} \left[ - 0.5113 \right] \;,
\label{intHTL:4f}
\nonumber \\ && \\
&& \int_{\bf pq} {n_B(p)\over p}
\left( {n_B(q) \over q} - {T \over q^2} \right)
{\rm Re}\left\langle c
	{r_c^2 - p^2 - q^2 \over \Delta(p+i\varepsilon,q,r_c)} \,
	\right\ranglec \;=\;
{T^2 \over (4 \pi)^2} \left[ - 0.2651 \right] \;,
\label{intHTL:5f}
\nonumber \\ && \\
&& \hspace{-1cm} \int_{\bf pq} {n_B(p)\over p}
\left( {n_B(q) \over q} - {T \over q^2} + {1 \over 2 q} \right)
{p^2 \over q^2} \,
{\rm Re}\left\langle c
	{r_c^2 - p^2 - q^2 \over \Delta(p+i\varepsilon,q,r_c)} \,
	\right\ranglec \;=\;
{T^2 \over (4 \pi)^2} \left[ 2.085 \times 10^{-2} \right] \;,
\label{intHTL:6f0}
\nonumber \\ && \\
&& \hspace{-1cm} \int_{\bf pq} {n_B(p)\over p}
\left( {n_B(q) \over q} - {T \over q^2} + {1 \over 2 q} \right)
{{\bf p} \cdot {\bf q} \over q^2} \,
{\rm Re} \left\langle
	{r_c^2 - p^2 - q^2 \over \Delta(p+i\varepsilon,q,r_c)} \,
	\right\ranglec \;=\;
{T^2 \over (4 \pi)^2} \left[ - 3.729 \times 10^{-3} \right] \;,
\label{intHTL:8f0}
\nonumber \\ && \\
%%%%%%%%%% quark %%%%%%%%%%
&& \hspace{-1.5cm} 
\int_{\bf pq} {n_F(p)\over p}
\left( {n_B(q) \over q} - {T\over q^2} +{1\over2q}\right)
{p^2 \over q^2} \,
{\rm Re}\left\langle c^{1+2\epsilon}
        {r_c^2 - p^2 - q^2 \over \Delta(p+i\varepsilon,q,r_c)} \,
        \right\ranglec \;=\;
{T^2 \over (4 \pi)^2} \left[1.482 \times 10^{-2} \right] \;,
\label{intHTL:6f}
\nonumber \\ && \\
&& \hspace{-1.5cm} 
\int_{\bf pq} {n_F(p)\over p}
\left( {n_B(q) \over q} - {T\over q^2}+{1\over2q} \right){{\bf p} \cdot {\bf q} \over q^2} \, 
{\rm Re} \left\langle c^{2\epsilon}
        {r_c^2 - p^2 - q^2 \over \Delta(p+i\varepsilon,q,r_c)} \,
        \right\ranglec \;=\;
{T^2 \over (4 \pi)^2} \left[ -2.832\times 10^{-3} \right] \;,
\label{intHTL:8f}
\nonumber \\ && \\
&& 
\int_{\bf pq}{n_F(p)\over p}{n_F(q)\over q}
{\rm Re}\left\langle
{r_c^2-p^2\over q^2}{r^2_c-p^2-q^2\over\Delta(p+i\epsilon,q,r_c)}c^{-1+2\epsilon}
-{1\over q^2}c^{-1+2\epsilon}+{\log2\over q^2} c^{2\epsilon}
\right\ranglec
\nonumber \\
&& \hspace{7cm}
=\;{T^2\over(4\pi)^2} \left[4.134\times10^{-2} \right] \;,
\label{nnn}
\\ &&
\int_{\bf pq}{n_B(p)\over p}{n_F(q)\over q}
\mbox{Re}\left\langle
{r_c^2-p^2\over q^2}{r^2_c-p^2-q^2\over\Delta(p+i\epsilon,q,r_c)}c^{-1+2\epsilon}
-{1\over q^2}c^{-1+2\epsilon}+{\log2\over q^2} c^{2\epsilon}
\right\ranglec
\nonumber \\
&& \hspace{7cm}
=\;{T^2\over(4\pi)^2} \left[2.530\times10^{-1} \right] \;.
\label{nnnn}
\eqa
The integrals involving the terms subtracted from $n_B(q)$ in~(\ref{nsub-10}) and (\ref{nsub-1}) are divergent, so the HTL average over $c$ and the angular average over $x = \hat {\bf p} \cdot \hat {\bf q}$ must be calculated in $3-2\epsilon$ dimensions. The first step in the calculation of the subtracted terms is to replace the average over $c$ of the integral over $q$ by an average over $c$ and $x$:
\begin{eqnarray}
\nonumber
\int_{\bf q} {1 \over q^n} \,
\left\langle f(c)
{r_c^2 - p^2 - q^2 \over \Delta(p+i\varepsilon,q,r_c)} \right\ranglec
\;=\; (-1)^{n-1} {1\over 8 \pi^2 \epsilon}
{ (1)_{2 \epsilon}\nonumber
 (1)_{-2\epsilon}
        \over ({3\over2})_{-\epsilon} }
(e^\gamma \mu^2)^\epsilon (2p)^{1-n-2\epsilon} &&
\\
&& \hspace{-11cm}
\times \left\langle f(c) \, c^{3-n-2\epsilon}(1-c^2)^{n-2+2\epsilon}
\sum_\pm (x\mp c - i \varepsilon)^{1-n-2\epsilon} \right\ranglecx \! \;.
\label{intthq}
\end{eqnarray}
The integral over $p$ can now be evaluated easily using either~(\ref{int-th:-1}) and (\ref{int-th:-2}) or
\bqa
\int_{\bf p} n_B(p) \, p^{-2-2\epsilon} \!\!&=&\!\!
{1\over 2\pi^2}
{(1)_{-4\epsilon} \over ({3\over2})_{-\epsilon}}
\zeta(1-4\epsilon)
(e^\gamma \mu^2)^\epsilon T^{1-4\epsilon} \;,
\label{int-th:-11}
\\
\int_{\bf p} n_F(p) \, p^{-2-2\epsilon} \!\!&=&\!\!
(1-2^{4\epsilon}) \int_{\bf p} n_B(p) \, p^{-2-2\epsilon} \;.
\label{int-th:-21}
\eqa
It remains only to calculate the averages over $c$ and $x$. The averages over $x$ give $_2F_1$ hypergeometric functions with argument $[(1 \mp c)/2 - i \varepsilon]^{-1}$:
\begin{eqnarray}
\left\langle  (x\mp c - i \varepsilon)^{-n-2\epsilon}
        \right\ranglex \!\!&=&\!\!
(1\mp c)^{-n-2\epsilon}
F\left( { 1-\epsilon,n+2\epsilon \atop 2-2\epsilon}
        \Bigg| [(1 \mp c)/2 - i \varepsilon]^{-1} \right) \;,
\label{avex:1}
\nonumber \\ && \\
\left\langle x (x\mp c - i \varepsilon)^{-n-2\epsilon}
        \right\ranglex \!\!&=&\!\!
{1 \over 2} (1\mp c)^{-n-2\epsilon}
\left[ F\left( { 1-\epsilon,n+2\epsilon \atop 3-2\epsilon}
        \Bigg| [(1 \mp c)/2 - i \varepsilon]^{-1} \right) \right.
\nonumber
\\
&& \left. 
-\; F\left( { 2-\epsilon,n+2\epsilon \atop 3-2\epsilon}
        \Bigg| [(1 \mp c)/2 - i \varepsilon]^{-1} \right) \right] \;.
\label{avex:2}
\end{eqnarray}
Using a transformation formula, the arguments can be changed to $(1 \mp c)/2 - i \varepsilon$. If the expressions (\ref{avex:1}) and (\ref{avex:2}) are averaged over $c$ with a weight that is an even function of $c$, the $+$ and $-$ terms combine to give $_3F_2$ hypergeometric functions with argument 1. For example,
\begin{eqnarray}
\left\langle (1-c^2)^{2 \epsilon}
	\sum_\pm (x\mp c - i \varepsilon)^{-1-2\epsilon}
	\right\ranglecx &&
\nonumber \\
&& \hspace{-5.6cm}
=\; {1 \over 3\epsilon}
{ (2)_{-2 \epsilon} (1)_\epsilon ({3\over2})_{-\epsilon}
	\over (1)_{-\epsilon} (1)_{-\epsilon} }
\left\{ - e^{- i \pi \epsilon}
	{ (1)_{3\epsilon} (1)_{-2 \epsilon}
	\over (1)_{2 \epsilon} (2)_{-\epsilon} }
F\left( { 1-2\epsilon,1-\epsilon,\epsilon
	\atop 2-\epsilon,1-3\epsilon} \Bigg| 1 \right)
\right.
\nonumber
\\
&& \hspace{-5cm} \left.
+\; e^{i 2\pi \epsilon}
	{ (1)_{-3\epsilon} (1)_\epsilon
	\over  (1)_{-4\epsilon} (2)_{2\epsilon} }
F\left( { 1+\epsilon,1+2\epsilon,4\epsilon
	\atop 2+2\epsilon,1+3\epsilon} \Bigg| 1 \right)
\right\} \;.
\end{eqnarray}
Upon expanding the hypergeometric functions in powers of $\epsilon$ and taking the real parts, we obtain
\begin{eqnarray}
&& \hspace{-2cm}
{\rm Re} \left\langle (1-c^2)^{2 \epsilon}
	\sum_\pm (x\mp c - i \varepsilon)^{-1-2\epsilon}
	\right\ranglecx 
\;=\; \pi^2 \left[ - \epsilon + 2 (1-\log 2) \epsilon^2 \right] \;,
\label{avecx:1}
\\ && \hspace{-2cm}
{\rm Re} \left\langle c^2 (1-c^2)^{2 \epsilon}
	\sum_\pm (x\mp c - i \varepsilon)^{-1-2\epsilon}
	\right\ranglecx 
\;=\; \pi^2 \left[ - {1 \over 3} \epsilon
	+ {2 \over 9} (2-3\log 2) \epsilon^2 \right] \;,
\label{avecx:2}
\\ && \hspace{-2cm}
{\rm Re} \left\langle (1-c^2)^{2+2 \epsilon}
	\sum_\pm (x\mp c - i \varepsilon)^{-3-2\epsilon}
	\right\ranglecx 
\;=\; \pi^2 \left[ - {8 \over 3} \epsilon^2 \right] \;,
\label{avecx:3}
\\ && \hspace{-2cm}
{\rm Re} \left\langle x (1-c^2)^{1+2 \epsilon}
	\sum_\pm (x\mp c - i \varepsilon)^{-2-2\epsilon}
	\right\ranglecx
\;=\; \pi^2 \left[ - {2 \over 3} \epsilon
	+ {2 \over 9} (1-6\log 2) \epsilon^2 \right] \;.
\label{avecx:5}
\end{eqnarray}

If the expressions (\ref{avex:1}) and (\ref{avex:2}) are averaged over $c$ with a weight that is an odd function of $c$, they reduce to integrals of $_2F_1$ hypergeometric functions with argument $y$. For example,
\begin{eqnarray}
\left\langle c (1-c^2)^{1+2 \epsilon}
        \sum_\pm (x\mp c - i \varepsilon)^{-2-2\epsilon}
        \right\ranglecx &&
\nonumber \\ 
&& \hspace{-6.6cm} 
=\; {(2)_{-2\epsilon} ({3\over2})_{-\epsilon}
\over (1)_{-\epsilon} (1)_{-\epsilon}}
\left\{
- 2 e^{-i\pi \epsilon}
{(1)_{3\epsilon} \over (2)_{2\epsilon}}
\int_0^1 dy \, y^{-2\epsilon} (1-y)^{1+\epsilon} |1-2y|
F\left( { 1-\epsilon,\epsilon \atop -3\epsilon } \Bigg| y \right)
\right.
\nonumber
\\ \nonumber
&& \hspace{-6cm} \left.
-\; {8 \over 3(1+3\epsilon)} e^{2i\pi \epsilon}
{(1)_{-3\epsilon} \over (1)_{-4\epsilon}}
\int_0^1 dy \, y^{1+\epsilon} (1-y)^{1+\epsilon} |1-2y|
F\left( { 2+2\epsilon,1+4\epsilon \atop 2+3\epsilon } \Bigg| y \right)
\right\} \;.
\\
\end{eqnarray}

The expansions of the integrals of the hypergeometric functions in powers of $\epsilon$ are given in (\ref{Fabs-1})-(\ref{Fabs-2}). The resulting expansions for the real parts of the averages over $c$ and $x$ are
\begin{eqnarray}
&& \hspace{-2.5cm}
{\rm Re} \left\langle c (1-c^2)^{1+2 \epsilon}
        \sum_\pm (x\mp c - i \varepsilon)^{-2-2\epsilon}
        \right\ranglecx 
\;=\; -1 + {14(1-\log2) \over3} \epsilon \;,
\label{avecx:4}
\\ \nonumber
&& \hspace{-2.5cm}
{\rm Re} \left\langle x c(1-c^2)^{2 \epsilon}
        \sum_\pm (x\mp c - i \varepsilon)^{-1-2\epsilon}
        \right\ranglecx \;=\;
{2 (1 - \log 2) \over3}
\nonumber
\\
&& \hspace{4cm} +\; \left( {4\over9} + {8\over 9} \log 2
- {4\over3}\log^22 + {\pi^2 \over 18} \right) \epsilon \;.
\label{avecx:6}
\end{eqnarray}
Multiplying each of these expansions by the appropriate factors from the integral over $q$ in (\ref{intthq}) and the integral over $p$ in (\ref{int-th:-11}) and (\ref{int-th:-21}), or (\ref{int-th:-1}) and (\ref{int-th:-2}), we obtain
\bqa
%%%%%%%%%% glue %%%%%%%%%%
&& \hspace{-1cm}
\int_{\bf pq} {n_B(p)\over p} \, {1 \over q^2} \,
{\rm Re} \left\langle c^{-1+2\epsilon}
	{r_c^2 - p^2 -  q^2 \over \Delta(p+i\varepsilon,q,r_c)} \,
	\right\ranglec \;=\;
{T \over (4\pi)^2} \left({\mu\over4\pi T}\right)^{4\epsilon}
\nonumber
\\
&& \hspace{5cm}
\times \left(-{1 \over 8}\right)
\left[ {1\over\epsilon} + 2 + 4 \log(2\pi) \right] \;,
\label{intHTL:4d}
\\
&& \hspace{-1cm}
\int_{\bf pq} {n_B(p)\over p} \, {1 \over q^2} \,
{\rm Re}\left\langle c^{1+2\epsilon}
	{r_c^2 - p^2 -  q^2 \over \Delta(p+i\varepsilon,q,r_c)} \,
	\right\ranglec \;=\;
{T \over (4\pi)^2}  \left({\mu\over4\pi T}\right)^{4\epsilon}
\nonumber
\\
&& \hspace{4cm}
\times \left(-{1 \over 24}\right)
\left[ {1\over\epsilon} + {8\over3} + 4 \log(2\pi) \right] \;,
\label{intHTL:5d}
\\
&& \hspace{-1cm}
\int_{\bf pq} {n_B(p)\over p} \, {p^2 \over q^4} \,
{\rm Re}\left\langle c^{1+2\epsilon}
	{r_c^2 - p^2 -  q^2 \over \Delta(p+i\varepsilon,q,r_c)} \,
	\right\ranglec \;=\;
{T \over (4\pi)^2}  \left(-{1\over 12}\right)  \;,
\label{intHTL:6da}
\\
&& \hspace{-1cm}
\int_{\bf pq} {n_B(p)\over p} \, {{\bf p} \cdot {\bf q} \over q^4} \,
{\rm Re} \left\langle  c^{2\epsilon}
	{r_c^2 - p^2 -  q^2 \over \Delta(p+i\varepsilon,q,r_c)} \,
	\right\ranglec \;=\;
{T \over (4\pi)^2}  \left({\mu\over4\pi T}\right)^{4\epsilon}
\nonumber
\\
&& \hspace{6cm}
\times {1 \over 24}
\left[ {1\over\epsilon} +  {11\over3} + 4 \log(2\pi) \right] \;,
\label{intHTL:8da}
\\
&& \hspace{-1cm}
\int_{\bf pq} {n_B(p)\over p} \, {p^2 \over q^3} \,
{\rm Re}\left\langle c^{1+2\epsilon}
	{r_c^2 - p^2 -  q^2 \over \Delta(p+i\varepsilon,q,r_c)} \,
	\right\ranglec \;=\;
{T^2 \over (4\pi)^2}  \left({\mu\over4\pi T}\right)^{4\epsilon}
\nonumber
\\
&& \hspace{3cm}
\times \left(-{1 \over 24}\right)
\left[ {1\over\epsilon} - {2\over3} + {8\over3} \log2
	+4 {\zeta'(-1) \over \zeta(-1)} \right] \;,
\label{intHTL:6db0}
\\
&& \hspace{-1cm}
\int_{\bf pq} {n_B(p)\over p} \, {{\bf p} \cdot {\bf q} \over q^3} \,
{\rm Re} \left\langle   c^{2\epsilon}
	{r_c^2 - p^2 -  q^2 \over \Delta(p+i\varepsilon,q,r_c)} \,
	\right\ranglec \;=\;
{T^2 \over (4\pi)^2}  \left({\mu\over4\pi T}\right)^{4\epsilon}
\nonumber
\\
&& \hspace{1cm}
\times \left(-{1 \over 18}\right)
\left[ (1-\log2) \left( {1\over\epsilon} + {14\over3}
		+ 4 {\zeta'(-1) \over \zeta(-1)} \right)
	+ {\pi^2 \over 12} \right] \;,
\label{intHTL:8db0}
%%%%%%%%%% quark %%%%%%%%%%
\\
&& \hspace{-1cm} 
\int_{\bf pq} {n_F(p)\over p} \, {p^2 \over q^3} \,
{\rm Re}\left\langle c^{1+2\epsilon}
        {r_c^2 - p^2 -  q^2 \over \Delta(p+i\varepsilon,q,r_c)} \,
        \right\ranglec 
 = {T^2 \over (4\pi)^2}  \left({\mu\over4\pi T}\right)^{4\epsilon}
\nonumber
\\
&& \hspace{3cm}
\times\left(-{1 \over 48}\right)
\left[ {1\over\epsilon} - {2\over3} - {4\over3} \log2
        +4 {\zeta'(-1) \over \zeta(-1)} \right] \;,
\label{intHTL:6db} 
\\
&& \hspace{-1cm} 
\int_{\bf pq} {n_F(p)\over p} \, {p^2 \over q^4} \,
{\rm Re}\left\langle c^{1+2\epsilon}
        {r_c^2 - p^2 -  q^2 \over \Delta(p+i\varepsilon,q,r_c)} \,
        \right\ranglec 
	\; = \;  {\cal O}(\epsilon) \;,
\label{intHTL:6dbext}
\\ 
&& \hspace{-1cm} 
\int_{\bf pq} {n_F(p)\over p} \, {{\bf p} \cdot {\bf q} \over q^3} \,
{\rm Re} \left\langle   c^{2\epsilon}
        {r_c^2 - p^2 -  q^2 \over \Delta(p+i\varepsilon,q,r_c)} \,
        \right\ranglec \;=\;
	{T^2 \over (4\pi)^2}  \left({\mu\over4\pi T}\right)^{4\epsilon}
\nonumber \\ &&
\hspace{0cm} \times \left(-{1 \over 36}\right)
\left[ (1-\log2)
 \left( {1\over\epsilon} + {14\over3}-4\log2 
            + 4 {\zeta'(-1) \over \zeta(-1)} \right)
        + {\pi^2 \over 12} \right] \;,
\label{intHTL:8db}
\\
&& \hspace{-1cm} \int_{\bf pq} {n_F(p)\over p} \, {{\bf p} \cdot {\bf q} \over q^4} \,
{\rm Re} \left\langle   c^{2\epsilon}
        {r_c^2 - p^2 -  q^2 \over \Delta(p+i\varepsilon,q,r_c)} \,
        \right\ranglec 
	\;=\; {T\over (4\pi)^2} \left(-{1\over6}\log2 \right) \;.
\label{intHTL:8dbext}
\eqa

Adding these integrals to the subtracted integrals in (\ref{intHTL:4f})--(\ref{intHTL:6f0}), we obtain the final results in (\ref{intHTL:4})--(\ref{intHTL:6}). Combining (\ref{intHTL:8f}) with (\ref{intHTL:8da}) and (\ref{intHTL:8db0}), we obtain
\begin{eqnarray}
\int_{\bf pq} {n_B(p) n_B(q) \over p q} \, {{\bf p} \cdot {\bf q} \over q^2} \,
{\rm Re} \left\langle c^{2\epsilon}
	{r_c^2 - p^2 - q^2\over \Delta(p+i\varepsilon,q,r_c)} \,
	\right\ranglec &&
\nonumber \\
&& \hspace{-5.5cm}
=\; {T^2 \over (4\pi)^2} \left({\mu\over4\pi T}\right)^{4\epsilon}
{5-2\log2 \over 72} \left[ {1\over\epsilon} + 11.6689 \right] \;.
\label{intHTL:8}
\end{eqnarray}
The final integral (\ref{intHTL:7}) is obtained from (\ref{intHTL:4}), (\ref{intHTL:6}), and (\ref{intHTL:8}) by using the identity $r_c^2 = p^2 + 2 {\bf p} \cdot {\bf q}/c + q^2/c^2$.

Adding Eqs.~(\ref{intHTL:6db}) and (\ref{intHTL:6dbext}) to the subtracted integral~(\ref{intHTL:6f}) we obtain the final result in Eq.~(\ref{intHTL:6x}). Combining (\ref{intHTL:8f}) with (\ref{intHTL:8db}) and~(\ref{intHTL:8dbext}), we obtain
\begin{eqnarray}
\int_{\bf pq} {n_F(p)n_B(q)\over pq} {{\bf p} \cdot {\bf q} \over q^2}
{\rm Re} \left\langle   c^{2\epsilon}
        {r_c^2 - p^2 -  q^2 \over \Delta(p+i\varepsilon,q,r_c)} \,
        \right\ranglec &&
\nonumber \\
&& \hspace{-5.5cm} 
=\; {T^2 \over (4\pi)^2}  \left({\mu\over4\pi T}\right)^{4\epsilon} 
\left({1-\log2\over 72}\right)
\left[{1\over\epsilon} - 15.2566 \right] \;.
\label{fpq}
\end{eqnarray}
The integral~(\ref{intHTL:7x}) is obtained from~(\ref{intHTL:4x}),~(\ref{intHTL:6x}) and~(\ref{fpq}). Finally consider~(\ref{lll}) and~(\ref{llll}). In order to evaluate them we need two subtractions for each integral
\bqa
&& \nonumber \hspace{-2cm}
\int_{\bf pq}{n_F(p) n_F(p) \over p q} \, {1\over q^2} \,
\langle c^{2\epsilon}\rangle_c
\;=\; {T^2\over(4\pi)^2}\left({\mu\over4\pi T}\right)^{4\epsilon}
\\ && \hspace{2cm}
\times\left(-{1\over12}\right)
\left[
{1\over\epsilon}+2+2\log2+2\gamma
+2{\zeta^{\prime}(-1)\over\zeta(-1)}
\right] \;,
\label{ss1}
\\ \nonumber
&& \hspace{-2cm}
\int_{\bf pq}{n_F(p) n_F(q)\over p q} \, {1\over q^2} \,
\langle c^{-1+2\epsilon}\rangle_c
\;=\; {T^2\over(4\pi)^2}\left({\mu\over4\pi T}\right)^{4\epsilon}
\\ && \hspace{-0cm} 
\times \left(-{1\over24}\right)
\left[
{1\over\epsilon^2}
+\left.(2+2\gamma+4\log2 + 2{\zeta^{\prime}(-1)\over\zeta(-1)} \right){1\over\epsilon} 
+53.1065
\right] \;,
\label{ss2} 
\\ \nonumber
&& \hspace{-2cm}
\int_{\bf pq}{n_B(p) n_F(q)\over p q} \, {1\over q^2} \,
\langle c^{2\epsilon}\rangle_c
\;=\; {T^2\over(4\pi)^2}\left({\mu\over4\pi T}\right)^{4\epsilon}
\\ && \hspace{3cm} 
\times\left(-{1\over6}\right)
\left[
{1\over\epsilon}+2+4\log2+2\gamma
+2{\zeta^{\prime}(-1)\over\zeta(-1)}
\right] \;, 
\label{ss3}
\\ \nonumber
&& \hspace{-2cm}
\int_{\bf pq}{n_B(p) n_F(q)\over p q} \, {1\over q^2} \,
\langle c^{-1+2\epsilon}\rangle_c
\;=\;
{T^2\over(4\pi)^2}\left({\mu\over4\pi T}\right)^{4\epsilon}
\\ &&
\times \left(-{1\over12}\right)
\left[
{1\over\epsilon^2}
+\left(2+2\gamma+6\log2+2{\zeta^{\prime}(-1)\over\zeta(-1)} \right){1\over\epsilon}
+69.7096
\right] \;.
\label{ss4}
\eqa
The subtractions can be evaluated directly in three dimensions and the results are given in Eqs.~(\ref{nnn})--(\ref{nnnn}) The integrals~(\ref{lll}) and~(\ref{llll}) are then given by the by the sum of the difference terms~(\ref{nnn}) and~(\ref{nnnn}) and the subtraction terms~(\ref{ss1})--(\ref{ss4}).

%%%%%%%%%%%%%%%%%%%%%%%%%%%%%%%%%%%%%%%%%%%%%%%%%%%%%%%%%%%%%
%
%	Include File:			DON'T COMPILE !!!
%
%%%%%%%%%%%%%%%%%%%%%%%%%%%%%%%%%%%%%%%%%%%%%%%%%%%%%%%%%%%%%

\chapter{Four-Dimensional Integrals}

In the sum-integral formula (\ref{int-2loop}), the second term on the right side involves an integral over four-dimensional Euclidean momenta. The integrands are functions of the integration variable $Q$ and $R=-(P+Q)$. The simplest integrals to evaluate are those whose integrands are independent of $P_0$:
\bqa
\int_Q {1 \over Q^2 r^2} \!\!&=&\!\!
{1 \over (4 \pi)^2} \mu^{2 \epsilon} p^{-2\epsilon}
\; 2 \left[{1 \over \epsilon} + 4 - 2 \log 2 \right] \;,
\label{int4:1}
\\
\int_Q {q^2 \over Q^2 r^4} \!\!&=&\!\!
{1 \over (4 \pi)^2} \mu^{2 \epsilon} p^{-2\epsilon}
\;2 \left[{1 \over \epsilon} + 1 - 2 \log 2 \right] \;,
\\
\int_Q {1 \over Q^2 r^4} \!\!&=&\!\!
{1 \over (4 \pi)^2} \mu^{2 \epsilon} p^{-2-2\epsilon}
\; ( -2 ) \left[1 + (-2 - 2 \log 2) \epsilon \right] \;.
\eqa
Another simple integral that is needed depends only on $P^2=P_0^2+p^2$:
\bqa
\int_Q {1 \over Q^2 R^2}
\;=\; 
{1 \over (4 \pi)^2} (e^\gamma \mu^2)^\epsilon (P^2)^{-\epsilon} \;
{1 \over \epsilon} \,
{(1)_\epsilon (1)_{-\epsilon} (1)_{-\epsilon}
        \over (2)_{-2\epsilon}} \;,
\label{int4:8}
\eqa
where $(a)_b$ is Pochhammer's symbol which is defined in (\ref{Poch}). We need the following weighted averages over $c$ of this function evaluated at $P = (-i p,{\bf p}/c)$:
\bqa
\left\langle c^{-1+2\epsilon}
        \int_Q {1 \over Q^2 R^2} \bigg|_{P \to (-i p,{\bf p}/c)}
        \right\ranglec
\!\!&=&\!\!
{1 \over (4 \pi)^2} \mu^{2 \epsilon} p^{-2\epsilon}
{1 \over 4}
\left[ {1 \over \epsilon^2} + {2 \log 2 \over \epsilon}
        + 2 \log^2 2 + {3 \pi^2 \over 4} 
\right] \;,
\label{int4:8.1}
\nonumber \\ && \\
\left\langle c^{1+2\epsilon}
        \int_Q {1 \over Q^2 R^2} \bigg|_{P \to (-i p,{\bf p}/c)}
        \right\ranglec
\!\!&=&\!\!
{1 \over (4 \pi)^2} \mu^{2 \epsilon} p^{-2\epsilon}
{1 \over 2}
\left[ {1 \over \epsilon} + 2 \log 2  
\right] \;.
\label{int4:8.2}
\eqa
The remaining integrals are functions of $P_0$ that must be analytically continued to the point $P_0 = -i p + \varepsilon$. Several of these integrals are straightforward to evaluate:
\bqa
\int_Q {q^2 \over Q^2 R^2}
        \bigg|_{P_0 = -i p} \!\!&=&\!\! 0 \;,
\label{int4:4}
\\
\int_Q {q^2 \over Q^2 r^2 R^2}
        \bigg|_{P_0 = -i p} \!\!&=&\!\!
{1 \over (4 \pi)^2} \mu^{2 \epsilon} p^{-2 \epsilon}
(-1) \bigg[ {1 \over \epsilon^2} + {1 - 2 \log 2 \over \epsilon} 
\nonumber \\ &&
+\; 10 - 2 \log 2 
	+ 2 \log^2 2 - {7 \pi^2 \over 12} 
\bigg] \;,
\label{int4:5}
\\
\int_Q {1 \over Q^2 r^2 R^2}
        \bigg|_{P_0 = -i p} \!\!&=&\!\!
{1 \over (4 \pi)^2} \mu^{2 \epsilon} p^{-2 -2 \epsilon}
\; \left[ {1 \over \epsilon} - 2 - 2 \log 2 
\right] \;.
\label{int4:6}
\eqa
We also need two weighted average over $c$ of the integral in (\ref{int4:4}) evaluated at $P = (-i p, {\bf p}/c)$. The integral itself is
\vspace{-1mm}
\bqa
\int_Q {q^2 \over Q^2 R^2}
        \bigg|_{P \to (-i p, {\bf p}/c)} \!\!&=&\!\!
{1 \over (4 \pi)^2} (e^\gamma \mu^2)^\epsilon p^{2-2\epsilon}
{(1)_\epsilon \over \epsilon} {1\over 4} 
{(1)_{-\epsilon} (1)_{-\epsilon}
        \over (2)_{-2\epsilon}}
\nonumber
\\
&& 
\times
\left( {1 \over 3 - 2 \epsilon} + c^2 \right)
c^{-2 + 2\epsilon} (1-c^2)^{-\epsilon} \;.
\label{int4:7}
\eqa
The weighted average is
\bqa\nonumber
\left\langle c^{1+2\epsilon}
        \int_Q {q^2 \over Q^2 R^2} \bigg|_{P \to (-i p,{\bf p}/c)}
        \right\ranglec
\!\!&=&\!\!
{1 \over (4 \pi)^2} \mu^{2 \epsilon} p^{2-2\epsilon} {1 \over 48}
\bigg[ {1 \over \epsilon^2} + {2 (10+3\log 2) \over 3\epsilon}
\\ && 
+\; {4\over 9} + {40\over 3}\log 2 + 2 \log^2 2
        + {3 \pi^2 \over 4} 
\bigg]
\;.
\label{int4:7a}
%\\\nonumber
%&& \hspace{-12mm}
%\left\langle c^{-1+2\epsilon}
%        \int_Q {q^2 \over Q^2 R^2} \bigg|_{P \to (-i p,{\bf p}/c)}
%        \right\ranglec =
%	{1 \over (4 \pi)^2} \mu^{2 \epsilon} p^{2-2\epsilon} \,
%\\ \nonumber&& \hspace{4mm}
%\times {1 \over 16}
%\left[ {1 \over \epsilon^2} + {2\log 2 \over \epsilon}
%        + 2\log^22 + {3 \pi^2 \over 4} 
%\right.\\ &&\left.
%+\left(
%{3\over2}\pi^2\log2
%+{4\over3}\log^32
%-{1\over3}\zeta(3)
%\right)\epsilon
%\right]
%\;.
%\label{int4:7aa}
\eqa
%
%{\bf [jm] p mod last}
%{the 2nd one seems unnecessary? n}

The most difficult four-dimensional integrals to evaluate involve an HTL average of an integral with denominator $R_0^2 + r^2 c^2$:
\bqa\nonumber
{\rm Re} \int_Q {1 \over Q^2}
\left\langle {c^2 \over R_0^2 + r^2 c^2} \right\ranglec
\!\!&=&\!\!
{1 \over (4 \pi)^2} \mu^{2\epsilon} p^{-2 \epsilon} 
\bigg[ {2 - 2\log 2 \over \epsilon}
\\ &&
+\; 8 - 4 \log 2 + 4 \log^2 2
        - {\pi^2 \over 2} \bigg] \;,
\label{int4HTL:1}
\\ 
{\rm Re} \int_Q {1 \over Q^2}
\left\langle {c^2(1-c^2) \over R_0^2 + r^2 c^2} \right\ranglec
\!\!&=&\!\!
{1 \over 3} {1 \over (4 \pi)^2} \mu^{2\epsilon} p^{-2 \epsilon}
\left[ {1 \over \epsilon} + {20\over3} - 6 \log 2 \right] \;,
\\ 
\nonumber
{\rm Re} \int_Q {1 \over Q^2}
\left\langle {c^4 \over R_0^2 + r^2 c^2} \right\ranglec
\!\!&=&\!\!
{1 \over (4 \pi)^2} \mu^{2\epsilon} p^{-2 \epsilon} 
\bigg[ {5 - 6\log 2 \over 3 \epsilon}
\\ &&
+\; {52 \over 9} - 2 \log 2 + 4 \log^2 2
        - {\pi^2 \over 2} \bigg] \;,
\label{int4HTL:2}
\\ 
{\rm Re} \int_Q {1 \over Q^2 r^2}
\left\langle {c^2 \over R_0^2 + r^2 c^2} \right\ranglec
\!\!&=&\!\!
- {1 \over 4} {1 \over (4 \pi)^2} \mu^{2\epsilon} p^{-2-2 \epsilon}
\left[ {1 \over \epsilon} + {4\over 3} + {2\over3} \log 2 
\right] \;,
\label{int4HTL:3}
\\ 
\nonumber
{\rm Re} \int_Q {q^2 \over Q^2 r^2}
\left\langle {c^2 \over R_0^2 + r^2 c^2} \right\ranglec
\!\!&=&\!\!
{1 \over (4 \pi)^2} \mu^{2\epsilon} p^{-2 \epsilon}
\bigg[ {13-16\log2 \over 12 \epsilon}
\\&&
        +\; {29 \over 9} - {19\over18} \log2
                + {8\over3}\log^22  
- {4\over9} \pi^2 \bigg] \;,
\label{int4HTL:4}
\\ 
\left\langle\int_{Q}{q^2-p^2\over Q^2r^2(R_0^2+r^2c^2)}
\right\rangle_c
\!\!&=&\!\! {1\over(4\pi)^2}\mu^{2\epsilon} p^{-2 \epsilon}
\left[-{\pi^2\over3} \right] \;.
\label{last4d}
\eqa
The analytic continuation to $P_0 =-ip+\varepsilon$ is implied in these integrals and in all the four-dimensional integrals in the remainder of this subsection.

We proceed to describe the evaluation of the integrals (\ref{int4HTL:1}) and (\ref{int4HTL:2}). The integral over $Q_0$ can be evaluated by introducing a Feynman parameter to combine $Q^2$ and $R_0^2 + r^2 c^2$ into a single denominator:
\bqa
\int_Q {1 \over Q^2 (R_0^2 + r^2 c^2)} &&
\nonumber \\
&& \hspace{-4cm}
=\; {1\over4} \int_0^1 dx
\int_{\bf r}\left[ (1-x+xc^2) r^2 
+ 2(1-x) {\bf r} \!\cdot\! {\bf p}
        + (1-x)^2 p^2 - i \varepsilon \right]^{-3/2} \;,
\label{fp:1}
\eqa
where we have carried out the analytic continuation to $P_0 =-ip+\varepsilon$. Integrating over ${\bf r}$ and then over the Feynman parameter, we get a ${}_2F_1$ hypergeometric function with argument $1-c^2$:
\bqa
\int_Q {1 \over Q^2 (R_0^2 + r^2 c^2)}
\!\!&=&\!\! {1\over (4\pi)^2} (e^\gamma \mu^2)^\epsilon
        p^{-2\epsilon} {(1)_\epsilon \over \epsilon}
        e^{i \pi \epsilon} {(1)_{-2\epsilon} (1)_{-\epsilon} \over (2)_{-3\epsilon}}
\nonumber
\\ 
&& \times
        (1-c^2)^{-\epsilon}
      	F\left( { {3\over2}-2\epsilon , 1-\epsilon
                \atop 2-3\epsilon } \Bigg| 1-c^2 \right) \;.
\label{int4HTL:12Q}
\eqa
The subsequent weighted averages over $c$ give ${}_3F_2$ hypergeometric functions with argument $1$:
\bqa \nonumber
\int_Q {1 \over Q^2}
\left\langle {c^2 \over R_0^2 + r^2 c^2} \right\ranglec
\!\!&=&\!\!
{1 \over (4\pi)^2} (e^\gamma \mu^2)^\epsilon
        p^{-2\epsilon} {(1)_\epsilon \over \epsilon}
{1 \over 3} e^{i \pi \epsilon}
{ ({3\over2})_{-\epsilon} (1)_{-2\epsilon} (1)_{-2\epsilon}
        \over ({5\over2})_{- 2\epsilon} (2)_{-3\epsilon} }
\\
&& \times
F\left({ 1-2\epsilon , {3\over2}-2\epsilon , 1-\epsilon
        \atop {5\over2}-2\epsilon , 2-3\epsilon } \Bigg| 1 \right) \;,
\\ \nonumber
\int_Q {1 \over Q^2}
\left\langle {c^2 (1-c^2) \over R_0^2 + r^2 c^2} \right\ranglec
\!\!&=&\!\!
{1 \over (4\pi)^2} (e^\gamma \mu^2)^\epsilon
        p^{-2\epsilon}  {(1)_\epsilon \over \epsilon}
{2 \over 15} e^{i \pi \epsilon}
{ ({3\over2})_{-\epsilon} (1)_{-2\epsilon} (2)_{-2\epsilon}
        \over ({7\over2})_{- 2\epsilon} (2)_{-3\epsilon} }
\\
&& \times
F\left( { 2-2\epsilon \;\;{3\over2}-2\epsilon , 1-\epsilon
        \atop {7\over2}-2\epsilon , 2-3\epsilon } \Bigg| 1 \right) \;.
\eqa
After expanding in powers of $\epsilon$, the real part is (\ref{int4HTL:2}).

The integral (\ref{int4HTL:3}) has a factor of $1/r^2$ in the integrand. After using (\ref{fp:1}), it is convenient to use a second Feynman parameter to combine $(1-x+xc^2)r^2$ with the other denominator before integrating over ${\bf r}$:
\bqa
\int_Q {1 \over Q^2 r^2 (R_0^2 + r^2 c^2)}
\!\!&=&\!\! {3\over8} \int_0^1 dx \, (1-x+xc^2) \int_0^1 dy \, y^{1/2}
\nonumber
\\ \nonumber
&& \hspace{-1cm}
\times\int_{\bf r}\left[ (1-x+xc^2) r^2 
+ 2y(1-x) {\bf r} \!\cdot\! {\bf p}
        + y(1-x)^2 p^2 - i \varepsilon \right]^{-5/2} \;.
\\
\label{fp-2}
\eqa
After integrating over ${\bf r}$ and then $y$, we obtain ${}_2F_1$ hypergeometric functions with arguments $x(1-c^2)$. The integral over $x$ gives a ${}_2F_1$ hypergeometric function with argument $1-c^2$:
\bqa \nonumber
\int_Q {1 \over Q^2 r^2 (R_0^2 + r^2 c^2)}
\!\!&=&\!\!
{1\over (4\pi)^2} (e^\gamma \mu^2)^\epsilon
p^{-2-2\epsilon} {(1)_\epsilon \over \epsilon}
\left\{ {(-{1\over2})_{-\epsilon} (1)_{-\epsilon}
        \over ({1\over2})_{-2\epsilon}}
\right.\\ &&\left. \hspace{-1cm}
-\; {3 \over 2(1+2 \epsilon)} e^{i \pi \epsilon}
{(1)_{-2\epsilon} (1)_{-\epsilon} \over (1)_{-3\epsilon}} (1-c^2)^{-\epsilon}
        F\left( { {1\over2}-2\epsilon , -\epsilon
                \atop -3\epsilon } \Bigg| 1-c^2 \right)
        \right\} \;.
\nonumber \\
\label{int4HTL:3Q}
\eqa
After averaging over $c$, we get a hypergeometric functions with argument 1:
\bqa \nonumber
\int_Q {1 \over Q^2 r^2}
\left\langle {c^2 \over R_0^2 + r^2 c^2} \right\ranglec
\!\!&=&\!\!
{1 \over (4\pi)^2} (e^\gamma \mu^2)^\epsilon
p^{-2-2\epsilon}
{(1)_\epsilon \over \epsilon}
\left\{ {1 \over 3-2\epsilon} \,
        { (-{1\over2})_{-\epsilon} (1)_{-\epsilon}
                \over ({1\over2})_{- 2\epsilon} }
\right.\\ && \left.
-\; {1 \over 2} e^{i \pi \epsilon} \,
{ (-{1\over2})_{-\epsilon} (1)_{-2\epsilon} (2)_{-2\epsilon}
        \over ({5\over2})_{- 2\epsilon} (1)_{-3\epsilon} }
F \left( { 1-2\epsilon , {1\over2}-2\epsilon , -\epsilon
        \atop {5\over2}-2\epsilon , -3\epsilon } \Bigg| 1 \right)
\right\} \;.
\nonumber \\
\label{int4HTL:3Qc}
%\\ &&\nonumber \hspace{-5mm}
%\int_Q {1 \over Q^2 r^2}
%\left\langle {1\over R_0^2 + r^2 c^2} \right\ranglec
%=
%{1 \over (4\pi)^2} (e^\gamma \mu^2)^\epsilon
%p^{-2-2\epsilon}
%{(1)_\epsilon \over \epsilon}
%\\ &&\nonumber\hspace{-3mm}\times
%\left\{ %{1 \over 3-2\epsilon} \,
%        { (-{1\over2})_{-\epsilon} ({3\over2})_{-\epsilon}
%                \over ({1\over2})_{- 2\epsilon} }
%\;-\; {1 \over 2} e^{i \pi \epsilon} \,
%{(1)_{-2\epsilon}^2 (1)_{-\epsilon}
%        \over ({3\over2})_{- 2\epsilon} (1)_{-3\epsilon} }
%\right.\\ && \left. \hspace{19mm} \times
%F \left( { 1-2\epsilon , {1\over2}-2\epsilon , -\epsilon
%        \atop {3\over2}-2\epsilon , -3\epsilon } \Bigg| 1 \right)
%\right\} \;.
%\label{int4HTL:3Qcex}
\eqa
%
%{[\bf jmp] mod 2nd eq}
After expanding in powers of $\epsilon$, the real part is (\ref{int4HTL:3}).

To evaluate the integral (\ref{int4HTL:4}), it is convenient to first express it as the sum of three integrals by expanding the factor of $q^2$ in the numerator as $q^2 = p^2 + 2 {\bf p} \cdot {\bf r} + r^2$:
\bqa
\int_Q {q^2 \over Q^2 r^2 (R_0^2 + r^2 c^2)}
\;=\; \int_Q
\left( {p^2 \over r^2} + 2 {{\bf p} \cdot {\bf r} \over r^2} + 1 \right)
{1 \over Q^2 (R_0^2 + r^2 c^2)} \;.
\eqa
To evaluate the integral with ${\bf p} \cdot {\bf r}$ in the numerator, we first combine the denominators using Feynman parameters as in (\ref{fp-2}). After integrating over ${\bf r}$ and then $y$, we obtain ${}_2F_1$ hypergeometric functions with arguments $x(1-c^2)$. The integral over $x$ gives ${}_2F_1$ hypergeometric functions with arguments $1-c^2$:
\bqa
\int_Q {{\bf p} \cdot {\bf r}
        \over Q^2 r^2 (R_0^2 + r^2 c^2)}
\!\!&=&\!\!
{1\over (4\pi)^2} (e^\gamma \mu^2)^\epsilon
p^{-2\epsilon} {(1)_\epsilon \over 2\epsilon^2}
\left\{ - {({3\over2})_{-\epsilon} (1)_{-\epsilon}
        \over ({3\over2})_{-2\epsilon}}
\right. \nonumber \\ && \left.
+\; e^{i \pi \epsilon}
{(1)_{-2\epsilon} (1)_{-\epsilon} \over (1)_{-3\epsilon}} (1-c^2)^{-\epsilon}
        F\left( { {3\over2}-2\epsilon , -\epsilon
                \atop 1-3\epsilon } \Bigg| 1-c^2 \right)
        \right\} \;.
\nonumber \\
\label{int4HTL:5Q}
\eqa
After averaging over $c$, we get a hypergeometric function with argument 1:
\bqa
\int_Q {{\bf p} \cdot {\bf r} \over Q^2 r^2}
\left\langle {c^2 \over R_0^2 + r^2 c^2} \right\ranglec
\!\!&=&\!\!
{1 \over (4\pi)^2} (e^\gamma \mu^2)^\epsilon
p^{-2\epsilon}
{(1)_\epsilon \over 2\epsilon^2}
\left\{ - {1 \over 3-2\epsilon} \,
        { ({3\over2})_{-\epsilon} (1)_{-\epsilon}
                \over ({3\over2})_{- 2\epsilon} }
\right. \nonumber \\ && \left.
+\; {1 \over 3} e^{i \pi \epsilon}
{ ({3\over2})_{-\epsilon} (1)_{-2\epsilon} (1)_{-2\epsilon}
        \over ({5\over2})_{- 2\epsilon} (1)_{-3\epsilon} }
F \left( { 1-2\epsilon , {3\over2}-2\epsilon , -\epsilon
        \atop {5\over2}-2\epsilon , 1-3\epsilon } \Bigg| 1 \right)
\right\} \;.
\nonumber \\
\label{int4HTL:5Qc}
\eqa
After expanding in powers of $\epsilon$, the real part is
\bqa 
{\rm Re} \int_Q {{\bf p} \cdot {\bf r} \over Q^2 r^2}
\left\langle {c^2 \over R_0^2 + r^2 c^2} \right\ranglec
\!\!&=&\!\!
{1 \over (4 \pi)^2} \mu^{2\epsilon} p^{-2 \epsilon}
\left[ {-1 + \log 2 \over 3\epsilon}
\right.
\nonumber
\\
&& \left.
-\; {20\over9} + {14\over 9} \log2 -{2\over3} \log^22
        + {\pi^2\over 36} \right] \;.
\label{int4HTL:5}
\eqa
Combining this with (\ref{int4HTL:1}) and (\ref{int4HTL:2}), we obtain the integral (\ref{int4HTL:4}).

To evaluate the integral~(\ref{last4d}), we first express the numerator as a sum of two integrals whose averages have been calculated:
\bqa\nonumber
\left\langle\int_{Q}{q^2-p^2\over Q^2r^2(R_0^2+r^2c^2)}
\right\rangle_x
\!\!&=&\!\!
\left\langle\int_{Q}{2{\bf p}\cdot{\bf r}+r^2\over Q^2r^2(R_0^2+r^2c^2)} 
\right\rangle_x \\ \nonumber
&=&\!\! 
{1 \over (4\pi)^2} (e^\gamma \mu^2)^\epsilon
p^{-2\epsilon}
{(1)_\epsilon \over \epsilon}
\left\{ -{1\over\epsilon}
        { ({3\over2})_{-\epsilon} (1)_{-\epsilon}
                \over({3\over2})_{- 2\epsilon}}
\right.
\\ \nonumber
&& \left.
+\; e^{i \pi \epsilon} \,
{ (1)_{-\epsilon} (1)_{-2\epsilon}
        \over(1)_{-3\epsilon} }{1\over\epsilon}
(1-c^2)^{-\epsilon}F \left( { -\epsilon , {3\over2}-2\epsilon
        \atop 1-3\epsilon } \Bigg| 1-c^2 \right)
\right.\\ \nonumber
&&
\left.
+\; e^{i \pi \epsilon} \,
{ (1)_{-\epsilon} (1)_{-2\epsilon}
        \over(2)_{-3\epsilon} }
(1-c^2)^{-\epsilon}
F \left( { 1-\epsilon , {3\over2}-2\epsilon
        \atop 2-3\epsilon } \Bigg| 1-c^2 \right)
\right\} \;.
\\
\eqa
The two hypergeometric functions are now combined into a single hypergeometric functions, which yields
\bqa\nonumber
\left\langle\int_{Q}{2{\bf p}\cdot{\bf r}+r^2\over Q^2r^2(R_0^2+r^2c^2)} 
\right\rangle_x 
\!\!&=&\!\!
{1 \over (4\pi)^2} (e^\gamma \mu^2)^\epsilon
p^{-2\epsilon}
{(1)_\epsilon \over \epsilon^2}
\left\{ - { ({3\over2})_{-\epsilon} (1)_{-\epsilon}
                \over ({3\over2})_{- 2\epsilon} }
\right.\\ \nonumber&&\left.
+\; e^{i \pi \epsilon} \,
{ (1)_{-\epsilon} (2)_{-2\epsilon}
        \over(2)_{-3\epsilon}}
(1-c^2)^{-\epsilon}
F \left( { -\epsilon , {3\over2}-2\epsilon
        \atop 2-3\epsilon } \Bigg| 1-c^2 \right) \right\}
	\;.
\\
\eqa
Averaging over $c$, yields
\bqa\nonumber
&& \hspace{-0.8cm}
\left\langle\int_{Q}{2{\bf p}\cdot{\bf r}+r^2\over Q^2r^2(R_0^2+r^2c^2)} 
\right\rangle_{c,x} 
\;=\;
{1 \over (4\pi)^2} (e^\gamma \mu^2)^\epsilon
p^{-2\epsilon}
{1\over\epsilon^2}
{(1)_\epsilon(1)_{-\epsilon}({3\over2})_{-\epsilon} \over
({3\over2})_{-2\epsilon}}
\left[ -1
+e^{i \pi \epsilon}{(1)_{-2\epsilon}\over(1)_{-\epsilon}^2} 
\right] \;.
\\
\eqa
Expansion in powers of $\epsilon$, yields Eq.~(\ref{last4d}).

%%%%%%%%%%%%%%%%%%%%%%%%%%%%%%%%%%%%%%%%%%%%%%%%%%%%%%%%%%%%%
%
%	Include File:			DON'T COMPILE !!!
%
%%%%%%%%%%%%%%%%%%%%%%%%%%%%%%%%%%%%%%%%%%%%%%%%%%%%%%%%%%%%%

\chapter{Hypergeometric Functions}
\label{app:hyper}

The generalized hypergeometric function of type $_pF_q$ is an analytic function of one variable with $p+q$ parameters. In our case, the parameters are functions of $\epsilon$, so the list of parameters sometimes gets lengthy and the standard notation for these functions becomes cumbersome. We therefore introduce a more concise notation:
\begin{equation}
F\left( { \alpha_1,\alpha_2,\ldots,\alpha_p
        \atop \beta_1,\ldots,\beta_q } \Bigg| z \right)
\; \equiv \;
{}_pF_q(\alpha_1,\alpha_2,\ldots,\alpha_p;\beta_1,\ldots,\beta_q;z) \;.
\end{equation}
The generalized hypergeometric function has a power series representation:
\begin{equation}
F\left( { \alpha_1,\alpha_2,\ldots,\alpha_p
        \atop \beta_1,\ldots,\beta_q } \Bigg| z \right)
\;=\; \sum_{n=0}^\infty{ (\alpha_1)_n (\alpha_2)_n \cdots (\alpha_p)_n
        \over (\beta_1)_n \cdots (\beta_q)_n n! } z^n
        \;,
\label{ps-pFq}
\end{equation}
where $(a)_b$ is Pochhammer's symbol:
\begin{equation}
(a)_b \;=\; {\Gamma(a+b) \over \Gamma(a)} \;.
\label{Poch}
\end{equation}
The power series converges for $|z|<1$. For $z=1$, it converges if ${\rm Re} \,s > 0$, where
\begin{equation}
s \;=\; \sum_{i=1}^{p-1} \beta_i -  \sum_{i=1}^p \alpha_i \;.
\label{s-def}
\end{equation}
The hypergeometric function of type $_{p+1}F_{q+1}$ has an integral representation in terms of the hypergeometric function of type $_pF_q$:
\bqa
&& \int_0^1 dt \, t^{\nu-1} (1-t)^{\mu-1} \,
F\left( { \alpha_1,\alpha_2,\ldots,\alpha_p
        \atop \beta_1,\ldots,\beta_q  } \Bigg| tz \right)
\;=\; { \Gamma(\mu) \Gamma(\nu) \over \Gamma(\mu+\nu)} \,
F\left( { \alpha_1,\alpha_2,\ldots,\alpha_p,\nu
        \atop \beta_1,\ldots,\beta_q,\mu+\nu} \Bigg| z \right) \;.
\label{int-pFq}
\nonumber \\
\eqa
If a hypergeometric function has an upper and lower parameter that are equal, both parameters can be deleted:
\begin{equation}
F\left( { \alpha_1,\alpha_2,\ldots,\alpha_p,\nu
        \atop \beta_1,\ldots,\beta_q, \nu} \Bigg| z \right)
\;=\; F\left( { \alpha_1,\alpha_2,\ldots,\alpha_p
        \atop \beta_1,\ldots,\beta_q } \Bigg| z \right) \;.
\end{equation}

The simplest hypergeometric function is the one of type $_1F_0$.
It can be expressed in an analytic form:
\begin{equation}
{}_1F_0(\alpha; \, ;z) \;=\; (1-z)^{-\alpha} \;.
\end{equation}
The next simplest hypergeometric functions are those of type $_2F_1$. They satisfy transformation formulas that allow an $_2F_1$ with argument $z$ to be expressed in terms of an $_2F_1$ with argument $z/(z-1)$ or as a sum of two $_2F_1$'s with arguments $1-z$ or $1/z$ or $1/(1-z)$. The hypergeometric functions of type $_2F_1$ with argument $z=1$ can be evaluated analytically in terms of gamma functions:
\begin{equation}
F\left( { \alpha_1, \alpha_2 \atop \beta_1 } \Bigg| 1 \right)
\;=\; { \Gamma(\beta_1) \Gamma(\beta_1 - \alpha_1 - \alpha_2)
        \over \Gamma(\beta_1 - \alpha_1) \Gamma(\beta_1 - \alpha_2) } \;.
\label{2F1-1}
\end{equation}
The hypergeometric function of type $_3F_2$ with argument $z=1$ can be expressed as a $_3F_2$ with argument $z=1$ and different parameters \cite{3F2}:
\bqa\nonumber
&& F\left( { \alpha_1, \alpha_2, \alpha_3 \atop \beta_1, \beta_2 } \Bigg| 1 \right)
\;=\; { \Gamma(\beta_1) \Gamma(\beta_2) \Gamma(s)
        \over \Gamma(\alpha_1+s) \Gamma(\alpha_2+s) \Gamma(\alpha_3)} \,
F\left( { \beta_1-\alpha_3, \beta_2-\alpha_3, s
        \atop \alpha_1+s, \alpha_2+s } \Bigg| 1 \right) \;,
\\
\label{3F2-1}
\eqa
where $s = \beta_1 + \beta_2 - \alpha_1 - \alpha_2 - \alpha_3$. If all the parameters of a $_3F_2$ are integers and half-odd integers, this identity can be used to obtain equal numbers of half-odd integers among the upper and lower parameters. If the parameters of a $_3F_2$ reduce to integers and half-odd integers in the limit $\epsilon \to 0$, the use of this identity simplifies the expansion of the hypergeometric functions in powers of $\epsilon$ .

The  most important integration formulas involving $_2F_1$ hypergeometric functions is (\ref{int-pFq}) with $p=2$ and $q=1$. Another useful integration formula is
\begin{eqnarray}
\int_0^1 dt \, t^{\nu-1} (1-t)^{\mu-1} \,
F\left( { \alpha_1,\alpha_2
        \atop \beta_1 } \Bigg| {t \over 1-t} z \right)
\!\!&=&\!\!
 { \Gamma(\mu) \Gamma(\nu) \over \Gamma(\mu+\nu)} \,
F\left( { \alpha_1,\alpha_2,\nu
        \atop \beta_1,1-\mu } \Bigg| -z \right)
\nonumber \\ \nonumber
&& \hspace{-6cm}
+\;
{ \Gamma(\alpha_1+\mu) \Gamma(\alpha_2+\mu) \Gamma(\beta_1) \Gamma(-\mu)
        \over \Gamma(\alpha_1) \Gamma(\alpha_2) \Gamma(\beta_1+\mu) } \,
(-z)^\mu \, 
F\left( { \alpha_1+\mu,\alpha_2+\mu,\nu+\mu
        \atop \beta_1+\mu,1+\mu} \Bigg| -z \right) \;.
\\
\label{int-2F1}
\end{eqnarray}
This is derived by first inserting the integral representation for $_2F_1$ in (\ref{int-pFq}) with integration variable $t'$ and then evaluating the integral over $t$ to get a $_2F_1$ with argument $1+t'z$.  After using a transformation formula to change the argument to $-t'z$, the remaining integrals over $t'$ are evaluated using (\ref{int-pFq}) to get $_3F_2$'s with arguments $-z$.

For the calculation of two-loop thermal integrals involving HTL averages, we require the expansion in powers of $\epsilon$ for hypergeometric functions of type $_pF_{p-1}$ with argument 1 and parameters that are linear in $\epsilon$. If the power series representation (\ref{ps-pFq}) of the hypergeometric function is convergent at $z=1$ for $\epsilon=0$, this can be accomplished simply by expanding the summand in powers of $\epsilon$ and then evaluating the sums. If the power series is divergent, we must make subtractions on the sum before expanding in powers of $\epsilon$. The convergence properties of the power series at $z=1$ is determined by the variable $s$ defined in (\ref{s-def}). If $s>0$, the power series converges. If $s\to 0$ in the limit $\epsilon \to 0$, only one subtraction is necessary to make the sum convergent:
\begin{eqnarray}
F\left( { \alpha_1,\alpha_2,\ldots,\alpha_p
        \atop \beta_1,\ldots,\beta_{p-1} } \Bigg| 1 \right)
\!\!&=&\!\!
{ \Gamma(\beta_1) \cdots \Gamma(\beta_{p-1}) \over
        \Gamma(\alpha_1) \Gamma(\alpha_2) \cdots \Gamma(\alpha_p) }
\zeta(s+1)
\nonumber
\\\nonumber
&& \hspace{-2cm}
+\; \sum_{n=0}^\infty
\left( { (\alpha_1)_n (\alpha_2)_n \cdots (\alpha_p)_n
        \over (\beta_1)_n \cdots (\beta_q)_n n! }
- { \Gamma(\beta_1) \cdots \Gamma(\beta_{p-1}) \over
        \Gamma(\alpha_1) \Gamma(\alpha_2) \cdots \Gamma(\alpha_p)}
        (n+1)^{-s-1} \right)
        \;.
\\        
\end{eqnarray}
If $s\to -1$ in the limit $\epsilon \to 0$, two subtractions are necessary to make the sum convergent:
\begin{eqnarray}
F\left( { \alpha_1,\alpha_2,\ldots,\alpha_p
        \atop \beta_1,\ldots,\beta_{p-1} } \Bigg| 1 \right)
\!\!&=&\!\!
{ \Gamma(\beta_1) \cdots \Gamma(\beta_{p-1}) \over
        \Gamma(\alpha_1) \Gamma(\alpha_2) \cdots \Gamma(\alpha_p)}
\left[ \zeta(s+1) + t \, \zeta(s+2) \right]
\nonumber \\ &&
+\; \sum_{n=0}^\infty
\bigg( { (\alpha_1)_n (\alpha_2)_n \cdots (\alpha_p)_n
        \over (\beta_1)_n \cdots (\beta_q)_n n! }
        - { \Gamma(\beta_1) \cdots \Gamma(\beta_{p-1}) \over
        \Gamma(\alpha_1) \Gamma(\alpha_2) \cdots \Gamma(\alpha_p)}
\nonumber
\\
&& \hspace{1.5cm} \times
\left[  (n+1)^{-s-1} + t \, (n+1)^{-s-2} \right]
\bigg) \;,
\end{eqnarray}
where $t$ is given by
\begin{equation}
t \;=\; \sum_{i=1}^p {(\alpha_i-1)(\alpha_i-2)\over2}
- \sum_{i=1}^{p-1} {(\beta_i-1)(\beta_i-2) \over2}  \;.
\end{equation}

The expansion of a $_pF_{p-1}$ hypergeometric function in powers of $\epsilon$ is particularly simple if in the limit $\epsilon \to 0$ all its parameters are integers or half-odd-integers, with equal numbers of half-odd-integers among the upper and lower parameters. If the power series representation for such a hypergeometric function is expanded in powers of $\epsilon$, the terms in the summand will be rational functions of $n$, possibly multiplied by factors of the polylogarithm function $\psi(n+a)$ or its derivatives. The terms  in the sums can often be simplified by using the obvious identity
\begin{equation}
\sum_{n=0}^\infty \left[ f(n) - f(n+k) \right]
\;=\; \sum_{i=0}^{k-1} f(i) \;.
\end{equation}
The sums over $n$ of rational functions of $n$ can be evaluated by applying the partial fraction decomposition and then using identities such as
\begin{eqnarray}
\sum_{n=0}^\infty \left({1 \over  n+a} - {1\over n+b} \right)
\!\!&=&\!\! \psi(b) - \psi(a)  \;,
\\
\sum_{n=0}^\infty {1 \over (n+a)^2} \!\!&=&\!\! \psi'(a) \;.
\end{eqnarray}
The sums of polygamma functions of $n+1$ or $n+{1\over2}$ divided by $n+1$ or $n+{1\over2}$ can be evaluated using
\begin{eqnarray} 
&&\hspace{-1cm}
\sum_{n=0}^\infty
\left( {\psi(n+1) \over n+1}
        - {\log(n+1) \over n+1} \right)
\;=\; - {1\over2} \gamma^2 - {\pi^2 \over 12} 
- \gamma_1 \;,
\\ 
&&\hspace{-1cm}
\sum_{n=0}^\infty
\left( {\psi(n+1) \over n+{1\over 2}}
        - {\log(n+1) \over n+1} \right)
\;=\;  - {1\over2} (\gamma + 2 \log 2)^2 
+ {\pi^2 \over 12} - \gamma_1 \;,
\\ 
&&\hspace{-1cm}
\sum_{n=0}^\infty
\left( {\psi(n+{1\over2}) \over n+1}
        - {\log(n+1) \over n+1} \right)
\;=\; - {1\over2} \gamma^2 - 4 \log 2 + 2 \log^2 2
        - {\pi^2 \over 12} - \gamma_1 \;,
\\ 
&&\hspace{-1cm}
\sum_{n=0}^\infty
\left( {\psi(n+{1\over2}) \over n+{1\over 2}}
        - {\log(n+1) \over n+1} \right)
\;=\;
 - {1\over2} (\gamma + 2 \log 2)^2 
- {\pi^2 \over 4} - \gamma_1 \;,
\end{eqnarray}
where $\gamma_1$ is Stieltje's first gamma constant defined in (\ref{zeta}). The sums of polygamma functions of $n+1$ or $n+{1\over2}$ can be evaluated using
\begin{eqnarray}
&&\hspace{-1cm}
\sum_{n=0}^\infty
\left( \psi(n+1) - \log(n+1)  + {1 \over 2(n+1)} \right)
\;=\; {1\over2} + {1\over2} \gamma -{1\over 2} \log(2 \pi) \;,
\\
&&\hspace{-1cm}
\sum_{n=0}^\infty
\left( \psi(n+\mbox{$1\over2$}) - \log(n+1)  + {1 \over n+1} \right)
\;=\; {1\over2}\gamma  - \log2 -{1\over 2} \log(2 \pi) \;.
\end{eqnarray}

We also need the expansions in $\epsilon$ of some integrals of $_2F_1$ hypergeometric functions of $y$ that have a factor of $|1-2y|$. For example, the following two integrals are needed to obtain (\ref{avecx:4}):
\begin{eqnarray}
&& \hspace{-1cm}
\int_0^1 dy \, y^{-2\epsilon} (1-y)^{1+\epsilon} |1-2y| \,
F\left( { 1-\epsilon,\epsilon
        \atop -3\epsilon} \Bigg| y \right) 
\;=\; {1\over 6} + \left( {2\over9} + {4\over9} \log 2 \right) \epsilon \;,
\label{Fabs-1}
\\
&& \hspace{-2cm} \int_0^1 dy \, y^{1+\epsilon} (1-y)^{1+\epsilon} |1-2y| \,
F\left( { 2+2\epsilon,1+\epsilon
        \atop 2+3\epsilon} \Bigg| y \right) 
\;=\; {1\over 4} + \left( {7\over12} + {2\over3} \log 2 \right) \epsilon \;.
\label{Fabs-2}
\end{eqnarray}
These integrals can be evaluated by expressing them in the form
\begin{eqnarray}
&& \int_0^1 dy \, y^{\nu-1} (1-y)^{\mu-1} |1-2y| \,
F\left( { \alpha_1,\alpha_2
        \atop \beta_1} \Bigg| y \right)
\nonumber \\
&& \hspace{-5mm}
=\; \int_0^1 dy \, y^{\nu-1} (1-y)^{\mu-1}  (2y-1) \,
F\left( { \alpha_1,\alpha_2
        \atop \beta_1} \Bigg| y \right)
\nonumber \\
&& \hspace{0mm}
+\; 2 \int_0^{1\over2} dy \, y^{\nu-1} (1-y)^{\mu-1}  (1-2y) \;
F\left( { \alpha_1,\alpha_2
        \atop \beta_1} \Bigg| y \right)  \;.
\end{eqnarray}
The evaluation of the first integral on the right side gives $_3F_2$ hypergeometric functions with argument 1. The integrals from 0 to {$1\over2$ can be evaluated by expanding the power series representation (\ref{ps-pFq}) of the hypergeometric function in powers of $\epsilon$. The resulting series can be summed analytically and then the integral over $y$ can be evaluated.

%app-e.tex include

\bibliographystyle{utphys}
\bibliography{thesis,mcy5v2,phd_main_reading_db}

\end{document}